\documentclass{jfp1}
\expandafter\let\csname equation*\endcsname\relax
\expandafter\let\csname endequation*\endcsname\relax
\newtheorem{lemma}{Lemma}[section]

\newtheorem{theorem}{Theorem}[section]

\newtheorem{corollary}{Corollary}[section]

\usepackage{alltt}
\usepackage{amsfonts}
\usepackage{amsmath}
\usepackage{array}
\usepackage{color}
\usepackage{environ}
\usepackage{etoolbox}
\usepackage[T1]{fontenc}
\usepackage{hyperref}
\usepackage{stmaryrd}
\usepackage{tikz}
\usetikzlibrary{decorations.pathreplacing}
\usepackage{times}

\usepackage{mathpartir}
\usepackage{olingrammar}
\usepackage{plstx}
\usepackage{trfrac}
\usepackage{ct}	

\input{decls}
\newcommand{\FigAbstractSyntax}{
  \begin{figure}
    \begin{grammar}
      \nt{\expr}{\Expr} & & \grmk{Expressions} \\
      \noalt & \var & \grmk{Variable reference} \\
      & \ttsexp{\text{\tt array}}{\ttparens{\sequence{\nat}} \; \sequence{\atom}}
      & \grmk{Array, containing atoms} \\
      & \ttsexp{\text{\tt array}}{\ttparens{\sequence{\nat}} \; \type}
      & \grmk{Empty array, with its atom type} \\
      & \ttsexp{\text{\tt frame}}{\ttparens{\sequence{\nat}} \; \sequence{\expr}}
      & \grmk{Frame, containing array cells} \\
      & \ttsexp{\text{\tt frame}}{\ttparens{\sequence{\nat}} \; \sequence{\expr}}
      & \grmk{Empty frame, with its cell type} \\
      & \ttsexp{\expr_f}{\sequence{\expr_a}}
      & \grmk{Term application} \\
      & \ttsexp{\text{\tt t-app}}{\expr \; \sequence{\type}}
      & \grmk{Type application} \\
      & \ttsexp{\text{\tt i-app}}{\expr \; \sequence{\idx}}
      & \grmk{Index application} \\
      & \ttsexp{\text{\tt unbox}}{\ttparens{\sequence{\var_i} \; \var_e \; \expr_s} \; \expr_b}
      & \grmk{Let-binding box contents} \\
      \nt{\val}{\Val} & \var \gramalt \ttsexp{\text{\tt array}}{\ttparens{\sequence{\nat}} \; \sequence{\atval}} & \grmk{Values} \\
      \nt{\atom}{\Atom} & & \grmk{Atoms} \\
      \noalt & \baseval & \grmk{Base value} \\
      & \primop & \grmk{Primitive operator} \\
      & \lam{\sequence{\notevar{\var}{\type}}}{\expr} & \grmk{Term abstraction} \\
      & \tlam{\sequence{\notevar{\var}{\kind}}}{\val} & \grmk{Type abstraction} \\
      & \ilam{\sequence{\notevar{\var}{\sort}}}{\val} & \grmk{Index abstraction} \\
      & \dsum{\sequence{\idx}}{\expr}{\type} & \grmk{Boxed array} \\
      \nt{\atval}{\Atval} &
      \baseval \gramalt \primop \gramalt \lam{\sequence{\notevar{\var}{\type}}}{\expr}
      \gramalt \tlam{\sequence{\notevar{\var}{\kind}}}{\val}
      & \grmk{Atomic values} \\
      & \ilam{\sequence{\notevar{\var}{\sort}}}{\val} \gramalt \dsum{\sequence{\idx}}{\val}{\type}  & \\
      \nt{\type}{\Type} & & \grmk{Types} \\
      \noalt & \var & \grmk{Type variable} \\
      & \basetype & \grmk{Base type} \\
      & \typearray{\type}{\idx} & \grmk{Array} \\
      & \typefun{\sequence{\type}}{\type'} & \grmk{Function} \\
      & \typeuniv{\sequence{\notevar{\var}{\kind}}}{\type} & \grmk{Universal} \\
      & \typedprod{\sequence{\notevar{\var}{\sort}}}{\type} & \grmk{Dependent product} \\
      & \typedsum{\sequence{\notevar{\var}{\sort}}}{\type} & \grmk{Dependent sum} \\
      \nt{\kind}{\Kind} & \kindarray \gramalt \kindatom & \grmk{Kinds} \\
      \nt{\idx}{\Idx} & & \grmk{Type indices} \\
      \noalt & \var & \grmk{Type variable} \\
      & \nat & \grmk{Single dimension} \\
      & \idxshape{\sequence{\idx}} & \grmk{Sequence of dimensions)} \\
      & \idxadd{\sequence{\idx}} & \grmk{Adding dimensions} \\
      & \idxappend{\sequence{\idx}} & \grmk{Appending shapes} \\
      \nt{\sort}{\Sort} & \sortshp \gramalt \sortdim & \grmk{Index sorts} \\
      \nt{\primop}{\Primop} &
      \text{\tt +} \gramalt \text{\tt -} \gramalt \text{\tt *} \gramalt \text{\tt /} \gramalt
      \text{\tt append} \gramalt \text{\tt reduce} \gramalt \text{\tt iota} \gramalt \text{\tt ...}
      & \grmk{Primitive operators} \\
      \nt{\func}{\Func} & \primop \gramalt \expr & \grmk{Functions} \\
      \nt{\term}{\Term} & \atom \gramalt \lam{\sequence{\notevar{\var}{\type}}}{\expr} & \grmk{Terms} \\
    \end{grammar}
    \caption{Core Remora grammar}
    \label{fig:AbstractSyntax}
  \end{figure}
}

\newcommand{\FigSortingRules}{
  \begin{figure}
    \fbox{$\sortof{\sEnv}{\idx}{\sort}$}
    \begin{mathpar}
      \infr[S-Nat]
      {\nat \in \mathbb{N}}
      {\sortof
        {\sEnv}
        {\nat}
        {\sortdim}}
      \and
      \infr[S-Var]
      {\parens{\hassort{\var}{\sort}} \in \sEnv}
      {\sortof
        {\sEnv}
        {\var}
        {\sort}}
      \and
      \infr[S-Shape]
      {\seqpremise{j}
        {\sortof
          {\sEnv}
          {\idx_j}
          {\sortdim}}}
      {\sortof
        {\sEnv}
        {\idxshape{\sequence{\idx}}}
        {\sortshp}}
      \and
      \infr[S-Plus]
      {\seqpremise{j}
        {\sortof
          {\sEnv}
          {\idx_j}
          {\sortdim}}}
      {\sortof
        {\sEnv}
        {\idxadd{\sequence{\idx}}}
        {\sortdim}}
      \and
      \infr[S-Append]
      {\seqpremise{j}
        {\sortof
          {\sEnv}
          {\idx_j}
          {\sortshp}}}
      {\sortof
        {\sEnv}
        {\idxappend{\sequence{\idx}}}
        {\sortshp}}
    \end{mathpar}
    \caption{Sorting rules}
    \label{fig:SortingRules}
  \end{figure}
}

\newcommand{\FigKindingRules}{
  \begin{figure}
    \fbox{$\kindof{\sEnv}{\kEnv}{\type}{\kind}$}
    \begin{mathpar}
      \infr[K-Var]
      {\parens{\haskind{\var}{\kind}} \in \kEnv}
      {\kindof{\sEnv}{\kEnv}{\var}{\kind}}
      \and
      \infr[K-Base]
      {\quad}
      {\kindof{\sEnv}{\kEnv}{\basetype}{\kindatom}}
      \and
      \infr[K-Fun]
      {\seqpremise{j}{\kindof{\sEnv}{\kEnv}{\type_j}{\kindarray}}
        \\\\
        \kindof{\sEnv}{\kEnv}{\type'}{\kindarray}}
      {\kindof{\sEnv}{\kEnv}{\typefun{\sequence{\type}}{\type'}}{\kindatom}}
      \and
      \infr[K-Univ]
      {\kindof{\sEnv}{\kEnv, \sequence{\haskind{\var}{\kind}}}{\type}{\kindarray}}
      {\kindof{\sEnv}{\kEnv}{\typeuniv{\sequence{\notevar{\var}{\kind}}}{\type}}{\kindatom}}
      \and
      \infr[K-Pi]
      {\kindof{\sEnv, \sequence{\hassort{\var}{\sort}}}{\kEnv}{\type}{\kindarray}}
      {\kindof{\sEnv}{\kEnv}{\typedprod{\sequence{\notevar{\var}{\sort}}}{\type}}{\kindatom}}
      \and
      \infr[K-Sigma]
      {\kindof{\sEnv, \sequence{\hassort{\var}{\sort}}}{\kEnv}{\type}{\kindarray}}
      {\kindof{\sEnv}{\kEnv}{\typedsum{\sequence{\notevar{\var}{\sort}}}{\type}}{\kindatom}}
      \and
      \infr[K-Array]
      {\sortof{\sEnv}{\idx}{\sortshp}
       \\
       \kindof{\sEnv}{\kEnv}{\type}{\kindatom}}
      {\kindof{\sEnv}{\kEnv}{\typearray{\type}{\idx}}{\kindarray}}
      \rulename{K-Array}
    \end{mathpar}
    \caption{Kinding rules}
    \label{fig:KindingRules}
  \end{figure}
}

\newcommand{\FigTypingRules}{
  \begin{figure}
    \fbox{$\typeof{\sEnv}{\kEnv}{\tEnv}{\term}{\type}$}
    \begin{mathpar}
      \infr[T-Op]
      {\ }
      {\typeof{\sEnv}{\kEnv}{\tEnv}{\primop}{\Sigref{\primop}}}
      \and
      \infr[T-Var]
      {\parens{\hastype{\var}{\type}} \in \tEnv}
      {\typeof{\sEnv}{\kEnv}{\tEnv}{\var}{\type}}
      \and
      \infr[T-Eqv]
      {\typeof{\sEnv}{\kEnv}{\tEnv}{\term}{\type'}\\
       \teqv{\type}{\type'}}
      {\typeof{\sEnv}{\kEnv}{\tEnv}{\term}{\type}}
      \and
      \infr[T-Array]
      {\seqpremise{j}{\typeof
          {\sEnv}{\kEnv}{\tEnv}
          {\atom_j}{\type}}
        \\\\
        \kindof{\sEnv}{\kEnv}{\type}{\kindatom}
        \\\\
        \mathit{Length}\llb \sequence{\atom} \rrb = \prod{\sequence{\nat}}}
      {\parbox{0.35\textwidth}
        {\({\sEnv};{\kEnv};{\tEnv}\vdash\)
          \({\arrlit{\sequence{\atom}}{\sequence{\nat}}}\)\\
          \(:{\typearray{\type}{\idxshape{\sequence{\nat}}}}\)}}
      \and
      \infr[T-0A]
      {\kindof{\sEnv}{\kEnv}{\type}{\kindatom}
        \\\\
        0 \in \sequence{\nat}}
      {\parbox{0.30\textwidth}
        {\({\sEnv};{\kEnv};{\tEnv}\vdash\)
          \({\emptyarrlit{\type}{\sequence{\nat}}}\)\\
          \(:{\typearray{\type}{\idxshape{\sequence{\nat}}}}\)}}
      \and
      \infr[T-Frame]
      {\seqpremise{j}{\typeof
          {\sEnv}{\kEnv}{\tEnv}
          {\expr_j}{\typearray{\type}{\idx}}}
        \\\\
        \kindof{\sEnv}{\kEnv}{\typearray{\type}{\idx}}{\kindarray}
        \\\\
        \mathit{Length}\llb \sequence{\expr} \rrb = \prod{\sequence{\nat}}}
      {\parbox{0.35\textwidth}
        {\({\sEnv};{\kEnv};{\tEnv}\vdash\)
          \({\frm{\sequence{\expr}}{\sequence{\nat}}}\)\\
          \(:{\typearray{\type}{\idxappend{\idxshape{\sequence{\nat}} \; \idx}}}\)}}
      \and
      \infr[T-0F]
      {\kindof{\sEnv}{\kEnv}{\type}{\kindatom}
        \\\\
        \sortof{\sEnv}{\idx}{\sortshp}
        \\
        0 \in \sequence{\nat}}
      {\parbox{0.38\textwidth}
        {\({\sEnv};{\kEnv};{\tEnv}\vdash\)
          \({\emptyfrm{\typearray{\type}{\idx}}{\sequence{\nat}}}\)\\
          \(:{\typearray{\type}{\idxappend{\idxshape{\sequence{\nat}} \; \idx}}}\)}}
      \and
      \infr[T-Lam]
      {\typeof
        {\sEnv}{\kEnv}{\tEnv, \sequence{\hastype{\var}{\type}}}
        {\expr}
        {\type'}
        \\\\
        \seqpremise{j}{\kindof{\sEnv}{\kEnv}{\type}{\kindarray}}}
      {\parbox{0.35\textwidth}
        {\({\sEnv};{\kEnv};{\tEnv}\vdash\)
          \({\lam{\sequence{\notevar{\var}{\type}}}{\expr}}\)\\
          \(:{\typefun{\sequence{\type}}{\type'}}\)}}
      \and
      \infr[T-TLam]
      {\typeof
        {\sEnv}{\kEnv, \sequence{\haskind{\var}{\kind}}}{\tEnv}
        {\val}
        {\type}}
      {\parbox{0.30\textwidth}
        {\({\sEnv};{\kEnv};{\tEnv}\vdash\)
          \({\tlam{\sequence{\notevar{\var}{\kind}}}{\val}}\)\\
          \(:{\typeuniv{\sequence{\notevar{\var}{\kind}}}{\type}}\)}}
      \and
      \infr[T-ILam]
      {\typeof
        {\sEnv, \sequence{\haskind{\var}{\sort}}}{\kEnv}{\tEnv}
        {\val}
        {\type}}
      {\parbox{0.30\textwidth}
        {\({\sEnv};{\kEnv};{\tEnv}\vdash\)
          \({\ilam{\sequence{\notevar{\var}{\sort}}}{\val}}\)\\
          \(:{\typedprod{\sequence{\notevar{\var}{\sort}}}{\type}}\)}}
      \and
      \infr[T-Box]
      {\seqpremise{j}{\sortof{\sEnv}{\idx}{\sort}}
        \\\\
        \kindof{\sEnv}{\kEnv}
        {\typedsum{\sequence{\notevar{\var}{\sort}}}{\type}}
        {\kindatom}
        \\\\
        \typeof
        {\sEnv}{\kEnv}{\tEnv}
        {\expr}
        {\seqsubst{\type}{\var}{\idx}}}
      {\parbox{0.45\textwidth}
        {\({\sEnv};{\kEnv};{\tEnv}\vdash\)
          \({\dsum
            {\sequence{\idx}}
            {\expr}
            {\typedsum{\sequence{\notevar{\var}{\sort}}}{\type}}}\)\\
          \(:{\typedsum{\sequence{\notevar{\var}{\sort}}}{\type}}\)}}
      \and
      \infr[T-TApp]
      {\typeof
        {\sEnv}{\kEnv}{\tEnv}
        {\expr}
        {\typearray{\typeuniv{\sequence{\notevar{\var}{\kind}}}{\typearray{\type_u}{\idx_u}}}{\idx_f}}
        \\
        \seqpremise{j}{\kindof{\sEnv}{\kEnv}{\type_j}{\kind_j}}}
      {\typeof
        {\sEnv}{\kEnv}{\tEnv}
        {\tapp{\expr}{\sequence{\type}}}
        {\typearray{\seqsubst{\type_u}{\var}{\type}}{\idxappend{\idx_f\;\idx_u}}}}
      \and
      \infr[T-IApp]
      {\typeof
        {\sEnv}{\kEnv}{\tEnv}
        {\expr}
        {\typearray{\typedprod{\sequence{\notevar{\var}{\sort}}}{\typearray{\type_p}{\idx_p}}}{\idx_f}}
        \\
        \seqpremise{j}{\sortof{\sEnv}{\idx_j}{\sort_j}}}
      {\typeof
        {\sEnv}{\kEnv}{\tEnv}
        {\iapp{\expr}{\sequence{\idx}}}
        {\typearray
          {\seqsubst{\type_p}{\var}{\idx}}
          {\idxappend{\idx_f\;\seqsubst{\idx_p}{\var}{\idx}}}}}
      \and
      \infr[T-Unbox]
      {\typeof
        {\sEnv}{\kEnv}{\tEnv}
        {\expr_s}
        {\typearray{\typedsum{\sequence{\notevar{\var_i'}{\sort}}}{\type_s}}
          {\idx_s}}
        \\
        \typeof
        {\sEnv,\sequence{\hassort{\var_i}{\sort}}}
        {\kEnv}
        {\tEnv,\hastype{\var_e}{\seqsubst{\type_s}{\var_i'}{\var_i}}}
        {\expr_b}
        {\typearray{\type_b}{\idx_b}}
        \\
        \kindof{\sEnv}{\kEnv}{\typearray{\type_b}{\idx_b}}{\kindarray}}
      {\typeof
        {\sEnv}{\kEnv}{\tEnv}
        {\dproj{\sequence{\var_i}}{\var_e}{\expr_s}{\expr_b}}
        {\typearray{\type_b}{\idxappend{\idx_s \; \idx_b}}}}
      \rulename{T-Unbox}
      \and
      \infr[T-App]
      {\typeof
        {\sEnv}{\kEnv}{\tEnv}
        {\expr_f}
        {\typearray
          {\typefun
            {\sequence{\typearray{\type}{\idx}}}
            {\typearray{\type'}{\idx'}}}
          {\idx_f}}
        \\
        \sequence{\typeof
          {\sEnv}{\kEnv}{\tEnv}
          {{\expr_a}}
          {\typearray{\type}{\idxappend{{\idx_a} \; {\idx}}}}}
        \\
        \idx_p = \displaystyle\bigsqcup\left\{ \idx_f \; \sequence{\idx_a} \right\}}
      {\typeof
        {\sEnv}{\kEnv}{\tEnv}
        {\app{\expr_f}{\sequence{\expr_a}}}
        {\typearray{\type'}{\idxappend{\idx_p \; \idx'}}}}
    \end{mathpar}
    \caption{Typing rules}
    \label{fig:TypingRules}
  \end{figure}
}

\newcommand{\FigTypeEqv}{
 \begin{figure}
   \fbox{$\teqv{\type}{\type'}$}
   \begin{mathpar}
     \infr[TEqv-Refl]
     {\ }
     {\teqv{\type}{\type}}
     \and
     \infr[TEqv-Array]
     {\teqv{\type}{\type'}
       \\
       \ieqv{\idx}{\idx'}
     }
     {\teqv
       {\typearray{\type}{\idx}}
       {\typearray{\type'}{\idx'}}}
     \and
     \infr[TEqv-Fn]
     {\seqpremise{j}{\teqv
         {{\type_i}_j}
         {{\type_i'}_j}}
       \\
       \teqv
       {\type_o}
       {\type_o'}}
     {\teqv
       {\typefun{\sequence{\type_i}}{\type_o}}
       {\typefun{\sequence{\type_i'}}{\type_o'}}}
     \and
     \infr[TEqv-Univ]
     {\teqv
       {\seqsubst{\type}{\var}{\var_f}}
       {\seqsubst{\type'}{\var'}{\var_f}}
       \fresh{\sequence{\var_f}}}
     {\teqv
       {\typeuniv{\sequence{\notevar{\var}{\kind}}}{\type}}
       {\typeuniv{\sequence{\notevar{\var'}{\kind}}}{\type'}}}
     \and
     \infr[TEqv-Pi]
     {\teqv
       {\seqsubst{\type}{\var}{\var_f}}
       {\seqsubst{\type'}{\var'}{\var_f}}
       \fresh{\sequence{\var_f}}}
     {\teqv
       {\typedprod{\sequence{\notevar{\var}{\sort}}}{\type}}
       {\typedprod{\sequence{\notevar{\var'}{\sort}}}{\type'}}}
     \and
     \infr[TEqv-Sigma]
     {\teqv
       {\seqsubst{\type}{\var}{\var_f}}
       {\seqsubst{\type'}{\var'}{\var_f}}
       \fresh{\sequence{\var_f}}}
     {\teqv
       {\typedsum{\sequence{\notevar{\var}{\sort}}}{\type}}
       {\typedsum{\sequence{\notevar{\var'}{\sort}}}{\type'}}}
   \end{mathpar}
   \caption{Type equivalence}
   \label{fig:TypeEqv}
 \end{figure}
}

\newcommand{\FigOverlapAxiom}{
  \begin{figure}
    \begin{center}
      \begin{tikzpicture}
        \path (-5.25, 1.5) node (aEdit) {$a$};
        \draw (-5, 1.05) -- (-0.05, 1.05) -- (-0.05, 1.95) -- (-5, 1.95) -- cycle;
        \path (2.25, 1.5) node (bEdit) {$b$};
        \draw (0.05, 1.05) -- (2, 1.05) -- (2, 1.95) -- (0.05, 1.95) -- cycle;
        \path (-5.25, 0.5) node (cEdit) {$c$};
        \draw (-5, 0.05) -- (-2.05, 0.05) -- (-2.05, 0.95) -- (-5, 0.95) -- cycle;
        \path (2.25, 0.5) node (dEdit) {$d$};
        \draw (-1.95, 0.05) -- (2, 0.05) -- (2, 0.95) -- (-1.95, 0.95) -- cycle;
        \path (-1, -0.5) node (wEdit) {$w$};
        \draw [decorate,decoration={brace},xshift=0pt,yshift=-4pt] (0, 0) -- (-2, 0);
        \draw [dashed] (-2, 0) -- (0, 0) -- (0, 2) -- (-2, 2) -- cycle;
      \end{tikzpicture}
    \end{center}
    \caption{Overlap axiom, visualized: $w$ is the overlapping portion of $a$ and $d$.}
    \label{fig:OverlapAxiom}
  \end{figure}
}

\newcommand{\FigListMetafns}{
  \begin{figure}
    \[\mathit{Split}_n\llb \ttparens{a_1, \dots, a_m} \rrb =
    \ttparens{\ttparens{a_1, \dots, a_n}, \ttparens{a_{n+1}, \dots, a_{2n}}, \dots, \ttparens{a_{m-n+1}, \dots, a_m}}\]
    \[\mathit{Rep}_n \llb \ttparens{ a_1, \dots, a_m} \rrb =
    \ttparens{a_{1,1}, \dots, a_{1,n}, \dots, a_{m,1}, \dots, a_{m,n}}
    \quad \text{where } a_{i,j}=a_i\]
    \[\mathit{Concat} \llb \ttparens{\ttparens{a_{1,1}, \dots, a_{1,n}}, \dots, \ttparens{a_{m,1}, \dots, a_{m,n}}} \rrb =
    \ttparens{a_{1, 1}, \dots, a_{1, n}, \dots, a_{m, 1}, \dots, a_{m, n}}\]
    \[\mathit{Transpose}\ttparens{\ttparens{a_{1,1}, \dots, a_{1,n}}, \dots, \ttparens{a_{m,1}, \dots, a_{m,n}}} =
    \ttparens{\ttparens{a_{1,1}, \dots, a_{m,1}}, \dots, \ttparens{a_{n,1}, \dots, a_{m,n}}}\]
    \caption{List-processing metafunctions}
    \label{fig:Metafunctions}
  \end{figure}
}

\newcommand{\FigDynamicSemantics}{
  \begin{figure}
    {\begin{alltt}
        ((array ($\sequence{\nat_f}$)
        $\sequence{\atval_f}$)%
        $^{\typearray{\typefun
            {\sequence{\typearray
                {\type_i}
                {\idxshape{\sequence{\nat_i}}}}}
            {\type_o}}
          {\idxshape{\sequence{\nat_f}}}}$\\
        \0(array ($\sequence{\nat_a} \sequence{\nat_i}$)
        $\sequence{\atval_a}$)%
        $^{\typearray
          {\type_i}
          {\idxshape{\sequence{\nat_a} \sequence{\nat_i}}}}$
        $\dots$)
        \\$\mapsto_{\mathit{lift}}$\\
        ((array ($\sequence{\nat_p}$)\\
        \hspace*{10ex}%
        $\mathit{Concat}\llb
        \mathit{Rep}_{\nat_{\mathit{fe}}}\llb
        \mathit{Split}_{1}\llb \sequence{\atval_f}
        \rrb\rrb\rrb$)%
        $^{\typearray{\typefun
            {\sequence{\typearray
                {\type_i}
                {\idxshape{\sequence{\nat_i}}}}}
            {\type_o}}
          {\idxshape{\sequence{\nat_p}}}}$\\
        \0(array ($\sequence{\nat_p} \sequence{\nat_i}$) \\
        \hspace*{10ex}%
        $\mathit{Concat}\llb
        \mathit{Rep}_{\nat_{\mathit{ae}}}\llb
        \mathit{Split}_{\nat_{\mathit{ac}}}\llb \sequence{\atval_a}
        \rrb\rrb\rrb$)%
        $^{\typearray
          {\type_i}
          {\idxshape{\sequence{\nat_p} \sequence{\nat_i}}}}$
        $\dots$)
        \\\normalfont where\\
        Not all of $\parens{\sequence{\nat_f}}, \sequence{\parens{\sequence{\nat_a}}}$ are equal \\
        \begin{tabular}{rclrcl}
          $\sequence{\nat_p}$
          & = &
                $\displaystyle\bigsqcup\llb \parens{\sequence{\nat_f}} \;
                \sequence{\parens{\sequence{\nat_a}}} \rrb$ &
          $\nat_{\mathit{fe}}$
          & = &
                $\frac
                {\displaystyle\prod{\pseq{\nat_p}}}
                {\displaystyle\prod{\pseq{\nat_f}}}$ \\
          $\sequence{\nat_{\mathit{ae}}}$
          & = &
                $\sequence{\frac
                {\displaystyle\prod{\pseq{\nat_p}}}
                {\displaystyle\prod{\pseq{\nat_a}}}}$ &
          $\sequence{\nat_{\mathit{ac}}}$
          & = &
                $\sequence{\parens{
                \displaystyle\prod{\parens{\sequence{n_i}}}}}$ \\
        \end{tabular}
      \end{alltt}}\smallskip
    {\begin{alltt}
        ((array ($\sequence{\nat_f}$) $\sequence{\atval_f}$)%
        $^{\typearray{\typefun
            {\sequence{\typearray{\type_i}{\idxshape{\sequence{\nat_i}}}}}
            {\type_o}}
          {\idxshape{\sequence{\nat_f}}}}$\\
        \0(array ($\sequence{\nat_f}$ $\sequence{\nat_i}$) $\sequence{\atval_a}$)%
        $^{\typearray
          {\type_i}
          {\idxshape{\sequence{\nat_f} \; \sequence{\nat_i}}}}$
       $\dots$)
       \\$\mapsto_{\mathit{map}}$\\
       (frame ($\sequence{\nat_f}$)\\
       \0\0\0\0\0\0\0((array () $\atval_f$)%
       $^{\typearray
         {\typefun
           {\sequence{\typearray{\type_i}{\idxshape{\sequence{\nat_i}}}}}
           {\type_o}}
         {\idxshape{}}}$\\
       \0\0\0\0\0\0\0\0(array ($\sequence{\nat_i}$) $\sequence{\atval_c}$)%
       $^{\typearray
         {\type_i}
         {\idxshape{\sequence{\nat_i}}}}$%
       $\dots$)$^{\type_o} \dots$)
       \\\normalfont where\\
       \begin{tabular}{rcl}
         $\sequence{\nat_c}$
         & = &
               $\sequence{\parens{\prod{\sequence{\nat_i}}}}$ \\
         $\sequence{\parens{\sequence{\parens{\sequence{\atval_c}}}}}$
         & = &
               $\mathit{Transpose} \llb \sequence{\mathit{Split}_{\nat_c} \llb \sequence{\atval_a} \rrb} \rrb$ \\
         $\mathit{Length} \llb \sequence{\nat_f} \rrb$ & > & 0 \\
       \end{tabular}
      \end{alltt}\smallskip}
    {\begin{alltt}
        ((array ()
        (\ttlm{$\sequence{\notevar{\var}{\type}}$}
        $\expr$))
        $\sequence{\val^{\type}}$)\\
        $\mapsto_{\mathit{\beta}}$
        $\seqsubst{\expr}{\var}{\val^{\type}}$
    \end{alltt}}\smallskip
    {\begin{alltt}
        (t-app
        (array ($\sequence{\nat}$)
        (T\ttlm{$\sequence{\notevar{\var}{\kind}}$}
        $\expr$) $\dots$)
        $\sequence{\type}$)\\
        $\mapsto_{\mathit{t\beta}}$
        (frame ($\sequence{\nat}$)
        $\seqsubst{\expr}{\var}{\type}$ $\dots$)
    \end{alltt}}\smallskip
    {\begin{alltt}
        (i-app
        (array ($\sequence{\nat}$)
        (I\ttlm{$\sequence{\notevar{\var}{\sort}}$}
        $\expr$) $\dots$)
        $\sequence{\idx}$)\\
        $\mapsto_{\mathit{i\beta}}$
        (frame ($\sequence{\nat}$)
        $\seqsubst{\expr}{\var}{\idx}$ $\dots$)
    \end{alltt}}\smallskip
    {\begin{alltt}
        (frame ($\sequence{\nat}$)
        (array ($\sequence{\nat'}$) $\atval$ $\dots$)
        $\dots$)%
        $^{\typearray{\type}{\idxshape{\sequence{\nat}\sequence{\nat'}}}}$\\
        $\mapsto_{\mathit{collapse}}$
        (array ($\sequence{\nat}\,\sequence{\nat'}$)
        $\mathit{Concat}\llb
        \sequence{\parens{\sequence{\atval}}}\rrb$)%
        $^{\typearray{\type}{\idxshape{\sequence{\nat}\sequence{\nat'}}}}$
    \end{alltt}}\smallskip
    {\begin{alltt}
        (unbox
        ($\sequence{\var_i}$ $\var_e$
        (array ($\sequence{\nat_s}$) $\sequence{\text{\tt(box $\sequence{\idx}$ $\val$ $\type$)}}$))
        $\expr$)\\
        $\mapsto_{\mathit{unbox}}$
        (frame ($\sequence{\nat_s}$) $\multisubst
        {\expr}
        {\sequence{\singlesubst{\var_i}{\idx},},
          \singlesubst{\var_e}{\val}}$)
    \end{alltt}}
    \caption{Dynamic semantics for Remora}
    \label{fig:DynamicSemantics}
  \end{figure}
}

\newcommand{\FigErasedAbstractSyntax}{
  \begin{figure}
    \begin{grammar}
      \nt{\eexpr}{\erased{\Expr}} & & \grmk{Type-erased expressions} \\
      \noalt & \var & \grmk{Variable reference} \\
      & \ttsexp{\text{\tt array}}{\ttparens{\sequence{\nat}} \; \sequence{\eatom}}
      & \grmk{Array, containing atoms} \\
      & \ttsexp{\text{\tt frame}}{\idx \; \sequence{\eexpr}}
      & \grmk{Frame, containing sub-arrays} \\
      & \ttsexp{\eexpr_f}{\sequence{\ttsexp{\eexpr_a}{\idx_a}} \; \idx_r}
      & \grmk{Term application} \\
      & \ttsexp{\text{\tt i-app}}{\eexpr_f \; \sequence{\idx_a} \; \idx_r}
      & \grmk{Index application} \\
      & \ttsexp{\text{\tt unbox}}{\ttparens{\sequence{\var_i} \; \var_e \; \eexpr_s} \; \eexpr_b \; \idx_b}
      & \grmk{Let-binding box contents} \\
      \nt{\eatom}{\erased{\Atom}} & & \grmk{Type-erased atoms} \\
      \noalt & \baseval & \grmk{Base value} \\
      & \efunc & \grmk{Function} \\
      & \ttsexp{{\tt I}\ttlm{\sequence{\var}}}{\evalue} & \grmk{Index abstraction} \\
      & \ttsexp{\text{\tt box}}{\sequence{\idx} \; \eexpr} & \grmk{Boxed array} \\
      \nt{\efunc}{\erased{\Func}} & & \grmk{Type-erased functions} \\
      & \primop & \grmk{Primitive operator} \\
      & \ttsexp{\ttlm{\sequence{\var}}}{\eexpr} & \grmk{Term abstraction} \\
      \nt{\evalue}{\erased{\Val}} & & \grmk{Type-erased values} \\
      \noalt & \var \\
      & \ttsexp{\text{\tt array}}{\ttparens{\sequence{\nat}} \; \sequence{\eatvalue}} \\
      \nt{\eatvalue}{\erased{\Atval}} & & \grmk{Type-erased atomic values} \\
      \noalt & \baseval \\
      & \efunc \\
      & \ttsexp{{\tt I}\ttlm{\sequence{\var}}}{\evalue} \\
      & \ttsexp{\text{\tt box}}{\sequence{\idx} \; \evalue} \\
      \nt{\ectxt}{\erased{\Ctxt}} & & \grmk{Type-erased evaluation contexts} \\
      \noalt & \hole \\
      & \ttsexp{\text{\tt array}}
      {\ttparens{\sequence{\nat}} \;
        \sequence{\eatvalue} \; \ttsexp{\text{\tt box}}{\sequence{\idx} \; \ectxt} \; \sequence{\eatom}} \\
      & \ttsexp{\text{\tt frame}}{\idx \; \sequence{\evalue} \; \ectxt \; \sequence{\eexpr}} \\
      & \ttsexp{\ectxt}{\sequence{\ttsexp{\eexpr_a}{\idx_a}} \; \idx_r} \\
      & \ttsexp{\eexpr_f}
      {\sequence{\ttsexp{\evalue_a}{\idx_a}} \;
        \ttsexp{\ectxt}{\idx_a} \;
        \sequence{\ttsexp{\eexpr_a}{\idx_a}} \; \idx_r} \\
      & \ttsexp{\text{\tt i-app}}{\ectxt \; \sequence{\idx_a} \; \idx_r} \\
      & \ttsexp{\text{\tt unbox}}{\ttparens{\sequence{\var_i} \; \var_e \; \ectxt} \; \eexpr_b \; \idx_b} \\
      & \ttsexp{\text{\tt unbox}}{\ttparens{\sequence{\var_i} \; \var_e \; \evalue_s} \; \ectxt \; \idx_b} \\
      \nt{\eterm}{\erased{\Term}} & \eexpr \gramalt \eatom & \grmk{Type-erased terms} \\
    \end{grammar}
    \caption{Abstract syntax for type-erased Remora}
    \label{fig:ErasedAbstractSyntax}
  \end{figure}}

\newcommand{\FigErasedSemantics}{
  \begin{figure}
    {\begin{alltt}
        ((array ($\sequence{\nat_f}$) $\sequence{\eatvalue_f}$)
        \\\0((array ($\sequence{\nat_a}\sequence{\nat_i}$) $\sequence{\eatvalue_a}$)
        $\idxshape{\sequence{\nat_i}}$)
        $\dots$
        \\\0$\idx_r$)
        \\$\mapsto_{\mathit{lift}}$\\
        ((array ($\sequence{\nat_p}$)
        $\mathit{Concat}\llb
        \mathit{Rep}_{\nat_{\mathit{fe}}}\llb
        \mathit{Split}_{1}\llb
        \sequence{\eatvalue_f}
        \rrb\rrb\rrb$)
        \\\0((array ($\sequence{\nat_p}\;\sequence{\nat_i}$)
        $\mathit{Concat}\llb
        \mathit{Rep}_{\nat_{\mathit{ae}}}\llb
        \mathit{Split}_{\nat_{\mathit{ac}}}\llb
        \sequence{\eatvalue_a}
        \rrb\rrb\rrb$)
        $\idxshape{\sequence{\nat_i}}$) $\dots$
        \\\0$\idx_r$)
        \\\normalfont where\\
        Not all of $\parens{\sequence{\nat_f}}, \sequence{\parens{\sequence{\nat_a}}}$ are equal \\
        \begin{tabular}{rclrcl}
          $\sequence{\nat_p}$
          & = &
                $\displaystyle\bigsqcup\llb \parens{\sequence{\nat_f}} \;
                \sequence{\parens{\sequence{\nat_a}}} \rrb$ &
                                                              $\nat_{\mathit{fe}}$
          & = &
                $\frac
                {\displaystyle\prod{\pseq{\nat_p}}}
                {\displaystyle\prod{\pseq{\nat_f}}}$ \\
          $\sequence{\nat_{\mathit{ae}}}$
          & = &
                $\sequence{\frac
                {\displaystyle\prod{\pseq{\nat_p}}}
                {\displaystyle\prod{\pseq{\nat_a}}}}$ &
                                                        $\sequence{\nat_{\mathit{ac}}}$
          & = &
                $\sequence{\parens{
                \displaystyle\prod{\parens{\sequence{n_i}}}}}$ \\
        \end{tabular}
      \end{alltt}} \smallskip
    {\begin{alltt}
        ((array ($\sequence{\nat_f}$) $\sequence{\eatvalue_f}$)
        \\\0((array ($\sequence{\nat_f}\sequence{\nat_i}$) $\sequence{\eatvalue_a}$)
        $\idxshape{\sequence{\nat_i}}$) $\dots$
        \\\0$\idx_r$)
        \\$\mapsto_{\mathit{map}}$\\
        (frame $\idx_r$
        \\\0((array () $\eatvalue_f$)
        ((array ($\sequence{\nat_i}$) $\sequence{\eatvalue_c}$) $\idxshape{\sequence{\nat_i}}$) $\idx_c$) $\dots$)
        \\\normalfont where\\
        \begin{tabular}{rcl}
          $\sequence{\nat_c}$
          & = &
                $\sequence{\parens{\prod{\sequence{\nat_i}}}}$ \\
          $\sequence{\parens{\sequence{\parens{\sequence{\atval_c}}}}}$
          & = &
                $\mathit{Transpose} \llb \sequence{\mathit{Split}_{\nat_c} \llb \sequence{\atval_a} \rrb} \rrb$ \\
          $\mathit{Length} \llb \sequence{\nat_f} \rrb$ & > & 0 \\
          $\sequence{\idx_c}$
          & = &
                $\sequence{\parens{\idx_r \thmonus \idxshape{\sequence{\nat_f}}}}$ \\
        \end{tabular}
      \end{alltt}} \smallskip
    {\begin{alltt}
        ((array () (\ttlm{$\sequence{\var}$} $\eexpr$))
        ((array ($\sequence{\nat_i}$) $\evalue$) $\idxshape{\sequence{\nat_i}}$) $\dots$ $\idx_r$)
        \\$\mapsto_{\mathit{\beta}}$
        $\seqsubst{\eexpr}{\var}{\evalue}$
      \end{alltt}} \smallskip
    {\begin{alltt}
        (i-app (array ($\sequence{\nat_f}$) (i\ttlm{$\sequence{\var}$} $\eexpr$) $\dots$)
        $\sequence{\idx_a}$ $\idx_r$)
        \\$\mapsto_{\mathit{i\beta}}$
        (frame $\idx_r$ $\sequence{\seqsubst{\eexpr}{\var}{\idx_a}}$)
      \end{alltt}} \smallskip
    {\begin{alltt}
        (frame $\idxshape{\sequence{\nat}}$
        (array ($\sequence{\nat'}$) $\atval$ $\dots$)
        $\dots$)
        \\$\mapsto_{\mathit{collapse}}$
        (array ($\sequence{\nat}\sequence{\nat'}$)
        $\mathit{Concat}\llb
        \sequence{\parens{\sequence{\atval}}}\rrb$)
      \end{alltt}} \smallskip
    {\begin{alltt}
        (unbox ($\sequence{\var_i}$ $\var_e$ (array ($\sequence{\nat_s}$) (box $\sequence{\idx}$ $\evalue$))) $\eexpr$ $\idx_b$)
        \\$\mapsto_{\mathit{unbox}}$
        (frame (++ (Shp $\sequence{\nat_s}$) $\idx_b$) $\multisubst
        {\expr}
        {\sequence{\singlesubst{\var_i}{\idx},},
          \singlesubst{\var_e}{\evalue}}$)
      \end{alltt}}
     \caption{Dynamic semantics for erased Remora}
    \label{fig:ErasedSemantics}
 \end{figure}
}

\newcommand{\ErasureDefn}{
  \begin{figure}
    \begin{align*}
      \EraseE{\arrlit{\sequence{\atom}}{\sequence{\nat}}^{\type_r}}
      &= \arrlit{\sequence{\EraseA{\atom}}}{\sequence{\nat}} \\
      \EraseE{\frm{\sequence{\expr}}{\sequence{\nat}}^{\type_r}}
      &= \frm{\sequence{\EraseE{\expr}}}{\EraseT{\type_r}} \\
      \EraseE{\app{\expr_f^{\typearray{\typefun{\sequence{\type_i}}{\type_o}}{\idx_f}}}{\sequence{\expr_a}}^{\type_r}}
      &= \app{\EraseE{\expr_f}}{\sequence{\ttparens{\EraseE{\expr_a} \; \EraseT{\type_i}}}\;\EraseT{\type_r}} \\
      \EraseE{\tapp{\expr_f}{\sequence{\type_a}}^{\type_r}}
      &= \iapp{\EraseE{\expr_f}}{\sequence{\EraseT{\type_a}} \; \EraseT{\type_r}} \\
      \EraseE{\iapp{\expr_f}{\sequence{\idx_a}}^{\type_r}}
      &= \iapp{\EraseE{\expr_f}}{\sequence{\idx_a}\; \EraseT{\type_r}} \\
      \EraseE{\dproj{\sequence{\var_i}}{\var_e}{\expr_s}{\expr_b^{\type_b}}}
      &= \dproj{\sequence{\var_i}}{\var_e}{\EraseE{\expr_s}}{\EraseE{\expr_b} \; \EraseT{\type_b}} \\
    \end{align*}
    \begin{align*}
      \EraseA{\primop} &= \primop \\
      \EraseA{\baseval} &= \baseval \\
      \EraseA{\lam{\sequence{\notevar{\var}{\type}}}{\expr}}
                       &= \lam{\sequence{\var}}{\EraseE{\expr}} \\
      \EraseA{\tlam{\sequence{\notevar{\var}{\kind}}}{\val}}
                       &= \ilam{\sequence{\var}}{\EraseE{\val}} \\
      \EraseA{\ilam{\sequence{\notevar{\var}{\sort}}}{\val}}
                       &= \ilam{\sequence{\var}}{\EraseE{\val}} \\
      \EraseA{\dsum{\sequence{\idx}}{\expr}{\type}}
                       &= \dsum{\sequence{\idx}}{\EraseE{\expr}}{} \\
    \end{align*}
    \begin{align*}
      \EraseT{\var} &= \var \\
      \EraseT{\typearray{\type}{\idx}} &= \idx \\
      \EraseT{\type} &= \idxshape{} \quad\text{otherwise} \\
    \end{align*}
    \[
    \EraseTrm{\atom} = \EraseA{\atom} \qquad
    \EraseTrm{\expr} = \EraseE{\expr}
    \]
    \caption{Type erasure for Remora}
    \label{fig:ErasureDefn}
  \end{figure}
}

\newcommand{\ContextErasure}{
  \begin{figure}
    \begin{align*}
      &\EraseC{\hole} = \hole \\
      &\EraseC{\arrlit{\sequence{\atval}\;\dsum{\sequence{\idx}}{\ctxt}{\type}\;\sequence{\atom}}{\sequence{\nat}}}
      \\ &\qquad= \arrlit{\sequence{\EraseA{\atval}}\;
           \dsum{\sequence{\idx}}{\EraseC{\ctxt}}{}\;
           \sequence{\EraseA{\atom}}}{\sequence{\nat}} \\
      &\EraseC{\frm{\sequence{\val}\;\ctxt\;\sequence{\expr}}{\sequence{\nat}}^{\type_r}}
      \\ &\qquad= \frm{\sequence{\EraseE{\val}}\;\EraseC{\ctxt}\;\sequence{\EraseE{\expr}}}{\EraseT{\type_r}} \\
      &\EraseC{\app{\ctxt^{\typearray{\typefun{\sequence{\type_i}}{\type_o}}{\idx_f}}}{\sequence{\expr_a}}^{\type_r}}
      \\ &\qquad= \app{\EraseC{\ctxt}}{\sequence{\ttparens{\EraseE{\expr_a}\;\EraseT{\type_i}}}\;\EraseT{\type_r}} \\
      &\EraseC{\app
        {\expr_f^{\typearray{\typefun{\sequence{\type_1}\;\type_2\;\sequence{\type_3}}{\type_o}}{\idx_f}}}
        {\sequence{\val_1}\;\ctxt\;\sequence{\expr_3}}^{\type_r}}
      \\ &\qquad= \app{\EraseE{\expr_f}}
           {\sequence{\ttparens{\EraseE{\val_a}\;\EraseT{\type_1}}}\;
           \ttparens{\EraseC{\ctxt}\;\EraseT{\type_2}}\;
           \sequence{\ttparens{\EraseE{\expr_3}\;\EraseT{\type_3}}}} \\
      &\EraseC{\tapp{\ctxt}{\sequence{\type_a}}^{\type_r}}
      \\ &\qquad= \iapp{\EraseC{\ctxt}}{\sequence{\EraseT{\type_a}}\;\EraseT{\type_r}} \\
      &\EraseC{\iapp{\ctxt}{\sequence{\idx_a}}^{\type_r}}
      \\ &\qquad= \iapp{\EraseC{\ctxt}}{\sequence{\idx_a}\;\EraseT{\type_r}} \\
      &\EraseC{\dproj{\sequence{\var_i}}{\var_e}{\ctxt}{\expr_b}}
      \\ &\qquad= \dproj{\sequence{\var_i}}{\var_e}{\EraseC{\ctxt}}{\EraseE{\expr_b}} \\
    \end{align*}
    \caption{Type-erasing Remora evaluation contexts}
    \label{fig:ContextErasure}
  \end{figure}
}

\newcommand{\libfun}[2]{#1 & \parbox[t]{0.75\textwidth}{#2}}
\newcommand{\FigArrayOps}{
  \renewcommand{\arraystretch}{2}
  \begin{figure}
    \begin{tabular}{r l}
      {\bf Function} & {\bf Type}
      \\
      \libfun{{\tt head}, {\tt tail}}
      {\tt (-> ((Arr t (++ (Shp (+ 1 d)) s)))\\
      \phantom{D-> }(Arr t s))}
      \\
      \libfun{{\tt behead}, {\tt curtail}}
      {\tt (-> ((Arr t (++ (Shp (+ 1 d)) s)))\\
      \phantom{D-> }(Arr t (++ (Shp d) s)))}
      \\
      \libfun{{\tt length}}
      {\tt (-> ((Arr t (++ (Shp d) s)))\\
      \phantom{D-> }(Arr Num (Shp)))}
      \\
      \libfun{{\tt shape}, {\tt ravel}}
      {\tt (-> ((Arr t s))\\
      \phantom{D-> }(Arr (Sigma ((d Dim)) (Arr Num (Shp d)))\\
      \phantom{D-> DArr }(Shp)))}
      \\
      \libfun{{\tt append}}
      {\tt (-> ((Arr t (++ (Shp m) s))\\
      \phantom{D-> D}(Arr t (++ (Shp n) s)))\\
      \phantom{D-> }(Arr t (++ (Shp (+ m n)) s)))}
      \\
      \libfun{{\tt reverse}}
      {\tt (-> ((Arr t (++ (Shp d) s)))\\
      \phantom{D-> }(Arr t (++ (Shp d) s)))}
      \\
      \libfun{{\tt rotate}}
      {\tt (-> ((Arr t (++ (Shp d) s))\\
      \phantom{D-> D}(Arr Num (Shp)))\\
      \phantom{D-> }(Arr t (++ (Shp d) s)))}
      \\
      \libfun{{\tt fold}}
      {\tt (-> ((Arr (-> ((Arr t s) T) T) (Shp))\\
      \phantom{D-> D}T\\
      \phantom{D-> D}(Arr t (++ (Shp d) s)))\\
      \phantom{D-> }T)}
      \\
      \libfun{{\tt reduce}}
      {\tt (-> ((Arr (-> ((Arr t s) (Arr t s))\\
      \phantom{D-> DDArr D-> }(Arr t s))\\
      \phantom{D-> DDArr }(Shp))\\
      \phantom{D-> D}(Arr t (++ (Shp (+ 1 d)) s)))\\
      \phantom{D-> }(Arr t s))}
      \\
      \libfun{{\tt scan}}
      {\tt (-> ((Arr (-> ((Arr u r) (Arr t s)) (Arr u r)) (Shp))\\
      \phantom{D-> D}(Arr u r)\\
      \phantom{D-> D}(Arr t (++ (Shp d) s)))\\
      \phantom{D-> }(Arr u (++ (Shp d) r)))}
      \\
      \libfun{{\tt filter}}
      {\tt (-> ((Arr Bool d)\\
      \phantom{D-> D}(Arr t (++ (Shp d) s)))\\
      \phantom{D-> }(Arr (Sigma ((k Dim)) (Arr t (++ (Shp k) s))) (Shp)))}
      \\
      \libfun{{\tt read-nums}}
      {\tt (-> () (Arr (Sigma ((k Dim)) (Arr Num (Shp k))) (Shp)))}
      \\
      \libfun{{\tt iota}}
      {\tt (-> ((Arr Num (Shp d)))\\
      \phantom{D-> }(Arr (Sigma ((s Shape)) (Arr Num s)) (Shp)))}
      \\
      \libfun{{\tt reshape}}
      {\tt (-> ((Arr Num (Shp d))\\
      \phantom{D-> D}(Arr t r))\\
      \phantom{D-> }(Arr (Sigma ((s Shape)) (Arr Num s)) (Shp)))}
    \end{tabular}
    \vspace{1em}
    \caption{Common array-manipulation primitive operations and their Remora types}
    \label{fig:ArrayOps}
  \end{figure}
  \renewcommand{\arraystretch}{1}
}
\newcommand{\FigIotaVariants}{
  \renewcommand{\arraystretch}{2}
  \begin{figure}
    \begin{tabular}{r l}
      {\bf Function} & \hspace{2em}{\bf Type}
      \\
      \libfun{{\tt iota}}
      {\tt (-> ((Arr Num (Shp d)))\\
      \phantom{D-> }(Arr (Sigma ((s Shape)) (Arr Num s)) (Shp)))}
      \\
      \libfun{{\tt iota/v}}
      {\tt (-> ((Arr Num (Shp)))\\
      \phantom{D-> }(Arr (Sigma ((d Dim)) (Arr t (Shp d))) (Shp)))}
      \\
      \libfun{{\tt iota/s}}
      {\tt (Pi ((s Shape))\\
      \phantom{D}(-> () (Arr Num s)))}
      \\
      \libfun{{\tt iota/w}}
      {\tt (-> ((Arr t s))\\
      \phantom{D-> }(Arr Num s))}
    \end{tabular}
    \vspace{1em}
    \caption{Types for {\tt iota} and its variants}
    \label{fig:IotaVariants}
  \end{figure}
  \renewcommand{\arraystretch}{1}
}

\def\pftype{\sketchproof} 

\begin{document}

\title[The Semantic Foundations of Rank Polymorphism]
{The Semantics of Rank Polymorphism}
\author[Justin Slepak, Olin Shivers and Panagiotis Manolios]
{Justin Slepak, Olin Shivers and Panagiotis Manolios \\ Northeastern University}


\maketitle

\begin{abstract}
Iverson's APL and its descendants (such as J, K and FISh)
are examples of the family of ``rank-polymorphic'' programming languages.
The principal control mechanism of such languages is the
general lifting of functions that operate on arrays of rank (or dimension)
$r$ to operate on arrays of any higher rank $r' > r$.
We present a core, functional language, Remora,
that captures this mechanism,
and develop both a formal, dynamic semantics for the language,
and an accompanying static, rank-polymorphic type system for the language.
Critically, the static semantics captures
the shape-based lifting mechanism of the language.
We establish the usual progress and preservation properties for
the type system, showing that it is sound,
which means that ``array shape'' errors cannot occur at run time
in a well-typed program.
Our type system uses dependent types,
including an existential type abstraction
which permits programs to operate on arrays whose shape or rank is
computed dynamically;
however, it is restricted enough to permit static type checking.

The rank-polymorphic computational paradigm is unusual in that the
types of arguments affect the dynamic execution of the program---they
are what drive the rank-polymorphic distribution of a function across
arrays of higher rank.
To highlight this property,
we additionally present a dynamic semantics for a \emph{partially erased}
variant of the fully-typed language
and show that a computation performed with a fully-typed term stays
in lock step with the computation performed with its partially erased term.
The residual types thus precisely characterise the type information that
is needed by the dynamic semantics,
a property useful for the (eventual) construction of efficient compilers for
rank-polymorphic languages.

\end{abstract}

\section{Introduction}

The essence of the rank-polymorphic programming model
is implicitly treating all operations as \emph{aggregate} operations,
usable on arrays with arbitrarily many dimensions.
The model was first introduced by Iverson with the language APL \cite{APL}.
Over time, Iverson continued to develop this programming model,
making it gradually more flexible,
eventually leading to the creation of J \cite{J} as a successor to APL.
The boon APL offered programmers
was a notation without loops or recursion:
Programs would automatically follow a control-flow structure
appropriate for the data being consumed.
The nature of the implicit iteration structure
could be modified using second-order operators,
such as folding, scanning, or operating over a moving window.
These second-order operators would directly reveal
all loop-carried data dependences.

In this sense,
other languages demanded that
unnecessary work be put into both compilers and user programs.
The programmer would be expected to
write the program's iteration structure explicitly;
in many languages this entails describing a particular serial encoding
of what is fundamentally parallelizable computation.
The compiler must then perform intricate static analysis
to see past the programmer's overspecified iteration schedule.

The design of APL earned a Turing award for Iverson \cite{IversonTuringLecture}
as well as a mention in an earlier Turing lecture \cite{BackusTuringLecture},
praising it for showing the basis of a solution to the ``ven Neumann bottleneck.''
However APL's subsequent development proceded largely in isolation
from mainstream programming-language research.
The APL family of languages 
painted itself into a corner with design decisions
such as requiring functions to take only one or two arguments
and making parsing dependent on values assigned at run time.
As a result,
APL compilers were forced to support only a subset of the language
(such as Budd's compiler \cite{BuddCompiler})
or to operate on small sections of code,
alternating between executing each line of the program and compiling the next one
\cite{APL3000Compiler}.
What we gain from
the rank-polymorphic programming model's
natural friendliness to parallelism,
we can easily lose by continually interrupting the program
to return control to a line-at-a-time compiler.
Limiting the compiler to operating over a narrow window of code
can also eliminate opportunities for code transformations like fusion,
forcing unnecessary materialization of large arrays.

The tragedy of rank-polymorphic programming
does not end at forgone opportunities for performance.
Despite the convenience of rank polymorphism
for writing array-processing code%
---a common task in many application domains---%
APL and its close descendants do not see widespread use.
There is enough desire for implicitly aggregate computation
to support user communities for systems such as
NumPy \cite{Numpy} and MATLAB \cite{MATLAB},
which do not follow as principled or as flexible a rule
for matching functions with aggregate arguments%
\footnote{For example, operations which already expect aggregate data%
---for example if the programmer writes a function
to compute the norm of a vector or the determinant of a matrix---%
do not always lift easily to consume even higher-dimensional arguments}.
However, programmers are driven away from APL itself by features such as
obtuse syntax,
restrictions on function arity,
poor support for naming things,
and a limited universe of atomic data to populate the arrays \cite{WhatsWrong}.

Our goal is to study rank polymorphism itself
without getting bogged down by APL's other baggage.
A formal semantics of rank polymorphism
is the essential groundwork for understanding
how rank-polymorphic programs ought to behave,
how they should be compiled,
and how they can be safely transformed to reduce execution cost.
To that end, we develop Remora,
a language which integrates rank polymorphism with typed $\lambda$-calculus.

A key problem impeding static compilation of rank-polymorphic programs
is identifying the implicit iteration structure at each function application.
Even without obscuring the programming model
with the idiosyncratic special case behavior APL accreted,
rank polymorphism itself has seemed ``too dynamic'' for good static compilation
due to having its control structure derived from computed data.
The old style of line-at-a-time compilation
relied on inspecting functions and data at run time
to decide what loop structure to emit.

Remora's answer to the problem of finding the iteration space
is a type system which describes the shapes of arrays
and thereby identifies the implicit iteration space for each function application.
In order for types to provide enough detail about array shapes,
we use a restricted form of dependent typing,
in the style of Dependent ML \cite{DependentML}.
In Dependent ML, types are not parameterized over arbitrary program terms
but over a much more restricted language.
For Remora, our language of type indices
consists of natural numbers, describing individual dimensions,
and sequences of natural numbers, describing array shapes.

Past work on applying dependent types to computing with arrays
has focused on ensuring the safety of accessing individual array elements
\cite{DMLBounds, Qube}.
Bounds checking array indices is essential
in a programming model where extracting a single element
is the only elimination form for arrays,
but the rank-polymorphic programming model generally eschews this operation.
Instead, arrays are consumed whole,
and function application itself serves as the elimination form for arrays.

In Remora, a function's type describes the shapes of the expected argument arrays,
called the ``cells,''
and the type of the atomic data inside the array.
The typing rule for function application
is responsible for identifying the ``frame'' shape,
\ie, the iteration structure derived from the non-cell dimensions.
Type soundness means that
the type system produces more than a safety guarantee:
conclusions it draws about the iteration structure
can be used to correctly compile the program.
Our type system is flexible enough to express polymorphism over the cell shape,
such as a determinant function that can operate on square matrix cells of any size.
It can also handle functions whose output shape is not determined by input shape alone,
such as reading a vector of unknown size from user input
or generating an array of caller-specified shape.

We begin with an overview of the rank-polymorphic programming model,
written as a programming tutorial for an untyped variant of Remora.
After developing the intuition for rank polymorphism,
we present a formal description of Remora's core language.
This includes Remora's abstract syntax,
the language of type indices with its associated theory,
the static semantics which identifies array shapes and iteration spaces,
a type-driven dynamic semantics,
and a type-soundness theorem linking the static and dynamic semantics.
Since our formal presentation is intrinsically typed,
we also include an algorithm for partial type erasure,
to characterize which type-level information is truly necessary to keep at run time.
A bisimulation argument connects
the dynamic semantics of explicitly-typed Remora to that of erased Remora.

\section{Formalism}
\label{sec:Formalism}

We present a formal description of Core Remora,
which describes the control-flow mechanism used for computing on arrays.
While the nested-vector shorthand used earlier is convenient for human use,
this formalism explicitly distinguishes atoms from arrays.

The basic design goal for the language of types and indices
is to describe the program's control structure,
for a compiler's benefit.
This requires a detailed description of array shapes.
Knowing only the number of axes an array has is insufficient because
good mapping from source to hardware---%
\eg, whether to emit vector instructions, invoke a GPGPU kernel, or fork separate parallel threads---%
depends more on the actual sizes of individual axes than on how many there are.
We use indexed types, in the style of Dependent ML:
rather than allowing types to be parameterized over arbitrary terms,
they are parameterized over a limited language of type indices.
Remora's index language consists of natural numbers, representing individual dimensions,
and sequences of naturals, representing array shapes
(or fragments of shapes).
So the type of an array has the form
{\tt (Arr $\type$ $\idx$)},
where $\type$ identifies the type of the array's atoms,
and $\idx$ describes the array's shape.
This includes enough detail for the type system to describe
how the function and argument arrays align in function application.
It also grants the ability to statically detect arrays that cannot be properly aligned.

Fixed-size computation,
requiring every function to exactly specify its argument and result sizes,
is far too restrictive for practical use.
Programmers should not have to write a separate {\tt vector-mean} function
for every possible length of vector their programs might use.
So the index language must permit variables,
and the type language must allow universal quantification.
This is phrased as a dependent product:
$\typedprod{\sequence{\notevar{\var}{\sort}}}{\type}$.
Each $\var$ is marked with its sort $\sort$,
which specifies whether $\var$ ranges over
individual dimensions ($\sort = \sortdim$)
or sequences ($\sort = \sortshp$).
Now we can give a type to {\tt vector-mean}:

\begin{alltt}
(Pi ((n Dim))
  (-> ((Arr Float (Shp n)))
      (Arr Float (Shp))))
\end{alltt}
This function will lift to operate on higher-rank arrays of {\tt Float}s,
effectively behaving as a minor-axis mean function.
Having {\tt Shape} variables in addition to {\tt Dim} variables
allows us to type a major-axis mean function as well:

\begin{alltt}
(Pi ((c Shape) (n Dim))
  (-> ((Arr Float (++ (Shp n) c)))
      (Arr Float c)))
\end{alltt}

Combined with parametric polymorphism,
where type variables can be quantified separately over
the kinds $\kindatom$ and $\kindarray$,
we now have a lot of flexibility in describing a function's behavior.
For example, {\tt append} stitches two arrays together along their major axis.
This requires that
the $(n-1)$-dimensional pieces of each $n$-dimensional array
have the same type
(\ie, they must share the same $\kindatom$-kinded type variable),
but we must introduce separate index variables
(of sort $\sortdim$)
for the arguments' major axes.
The type we give to {\tt append} is

\begin{alltt}
(Pi ((c Shape) (m Dim) (n Dim))
  (Forall ((a Atom))
    (-> ((Arr a (++ (Shp m) c))
         (Arr a (++ (Shp n) c)))
        (Arr a (++ (Shp (+ m n)) c)))))
\end{alltt}
We have index variables {\tt m} and {\tt n} to stand for
each argument's first dimension
and {\tt c} to denote the rest of their shapes.
Quantifying over the type variable {\tt a} allows {\tt append} to work
independent of the type of atoms its arguments contain.
Our result type's first dimension is the sum of the arguments' first dimensions,
but it has the same ``remainder'' shape {\tt c}.

Quantifying over $\kindarray$ types is a convenience---%
it is not strictly necessary.
Concrete $\kindarray$-kinded types must be of the form
$\typearray{\type}{\idx}$,
so polymorphic types of the form
$\typeuniv{\notevar{\var}{\kindarray}}{\type}$
could be rewritten with fresh variables $\var_s$ and $\var_a$ as

\begin{alltt}
(Pi ((\(\var\sb{s}\) Shape))
  (Forall ((\(\var\sb{a}\) Atom))
    \(\subst{\type}{\var}{\typearray{\var\sb{a}}{\var\sb{s}}}\)))
\end{alltt}

Even having escaped the confines of fixed-size computation,
we so far only have functions whose result shape depends solely on its arguments' shapes.
Common utility functions such as {\tt iota} and {\tt filter}
have result shapes which depend on the actual run-time data they receive.
We can solve this limitation using existential quantification.
A dependent sum type,
\(\typedsum{\sequence{\notevar{\var}{\sort}}}{\type}\),
conceptually represents a tuple
containing indices whose respective sorts are $\sequence{\sort}$
and an array whose type $\type$ may depend on those indices.
This is the type-level description of a {\tt box},
the atomic wrapper around an arbitrary array.
Such types can encode arrays whose dimensions are not all known.
For example,
{\tt (Sigma ((n Dim)) (Arr Int (Shp n)))}
can describe any vector of integers without giving its specific length.
Note that the type still requires the underlying array to have rank 1.
No scalar or matrix or higher-ranked array can
fit the pattern specified by $\typearray{\text{\tt Int}}{\idxshape{\text{\tt n}}}$,
the dependent sum's body.
Use of dependent sums offers a lot of freedom in stating
what shape information is known precisely and what is hidden.
At one extreme,
{\tt (Sigma ((s Shape)) (Arr Int s))}
could contain an array of absolutely any shape.
We can also write more detailed descriptions,
\eg, floating-point matrices with exactly three rows and at least two columns:

\begin{alltt}
(Sigma ((c Dim))
  (Arr Float
       (Shp 3 (+ 2 c))))
\end{alltt}
Quantifying over shapes allows us to describe
arrays where only specific axes are known,
such as boolean arrays whose leading axis has length 10:

\begin{alltt}
(Sigma ((s Shape))
  (Arr Bool
       (++ (Shp 10) s)))
\end{alltt}

Typing boxes as dependent sums also permits controlled access to ragged arrays,
which are typed as arrays of boxes.
Consider a vector of 20 strings of varying lengths:

\begin{alltt}
(Arr (Sigma ((len Dim))
       (Arr Char (Shp len)))
     (Shp 20))
\end{alltt}
Any function written to operate on a box of the appropriate type,
in this case containing {\tt Char} vectors of completely unknown length,
can be safely lifted to operate on this vector of strings.
Separating the lifting over the outer dimensions
from the lifting over inner, existentially hidden dimensions
reflects an important consideration for code generation:
ragged dimensions in a type identify when implicit parallelism is \emph{irregular},
in contrast with the strictly regular parallelism offered in box-free code.
Raggedness is not restricted to the minor axis
because a box's type can still specify some exact dimensions,
as in
\begin{alltt}
(Arr (Sigma ((l Dim))
        (Arr Char (Shp l 80)))
     (Shp 20))
\end{alltt}
Here we have a vector of 20 documents,
each of which is a character array
containing an unknown number of 80-character lines.

\subsection{Syntax}
\FigAbstractSyntax
The grammar for Core Remora is given in Figure \ref{fig:AbstractSyntax}.
Term-level syntax is divided into atoms,
noted as $\atom$,
and expressions,
noted as $\expr$.
Expressions produce arrays,
which contain atoms.
For the most part, atom terms perform only trivial computation.
This rule applies to base values,
noted as $\baseval$;
primitive operators,
noted as $\primop$;
and $\lambda$-abstractions,
which may abstract over terms, types, and type indices.
As an exception,
a box gives an atomic view of an array of any shape
and may therefore perform any computation to compute its contents.
A box hides part of its contents' shape,
using a dependent sum.
It existentially quantifies type indices,
but an explicit type annotation is required.
A box built from the index {\tt 3} and a $3\times3$ matrix
could be meant, for example,
as an unspecified-length vector containing 3-vectors,
with type
{\tt (Sigma ((n Dim)) (Arr Int (Shp n 3)))}
or as a square matrix of unspecified size,
with type
{\tt (Sigma ((n Dim)) (Arr Int (Shp n n)))}
$\typedsum
{\notevar{\text{\tt n}}{\sortdim}}
{\typearray{\text{\tt Int}}{\idxshape{\text{\tt n n}}}}$;.

An array can be written as a literal,
with its shape and individual atoms listed directly.
It can also be written in nested form
as a frame containing cells (its subexpressions)
arranged in the specified shape.
For example, the matrix
$\left[\begin{smallmatrix}
    1 & 2 \\ 3 & 4
\end{smallmatrix}\right]$
can be written as the literal
$$\arrlit{\text{\tt 1 2 3 4}}{\text{\tt 2 2}}$$
or as a vector frame of vector literal cells:
$$\text{\tt (frame (2) (array (2) 1 2) (array (2) 3 4))}$$
The frame notation allows construction of arrays from unevaluated cells.
An empty array (\ie, one with a zero in its shape)
must be written with the type its elements are meant to have.
An empty vector of integers is a different value than
an empty vector of booleans,
and they inhabit different types.

Term, type, and index abstractions can be applied to
zero or more expressions, types, or indices.
The body of the abstraction must itself be an expression,
\ie, all functions produce arrays as their results.

Consuming a box let-binds its index- and term-level contents.
Suppose we have {\tt M},
a boxed square matrix of unspecified size.
Unboxing {\tt M} as in
$\dproj{\text{\tt l}}{\text{\tt a}}{\text{\tt M}}{\expr}$
lets us use the index variable {\tt l} and term variable {\tt a}
within $\expr$, the body.

Types include base types
(noted as $\basetype$),
functions, arrays,
universal types,
and dependent products and sums.
Universals specify the kind of each type argument,
and dependent products and sums specify the sort of each index argument.
Types are classified as either $\kindatom$ or $\kindarray$.
Type indices are naturals and sequences of naturals,
with addition and appending as the only operators.
They are classified into sorts, $\sortdim$ and $\sortshp$.

The grammar in Figure \ref{fig:AbstractSyntax} does not
require any specific set of primitive operators, base types, and base values.
An example collection of array-manipulation primitives and their types
is given in Figure \ref{fig:ArrayOps}.
For readability, we elide the enclosing {\tt Pi} and {\tt Forall} forms.
Most of these primitives perform some operation along the argument's major axis.
For example, {\tt head} extracts the first scalar of a vector, the first row of a matrix, \etc
This means that the argument shape must have
one dimension more than the result shape,
and that extra dimension must be nonzero.
This is expressed in the type of {\tt head}
by giving the argument shape {\tt (++ (+ 1 d) s)},
\ie, a single dimension which is 1 plus any arbitrary natural
followed by any arbitrary sequence of naturals.
In taking one scalar from a 3-vector,
we would instantiate {\tt d} as the dimension {\tt 2}
and {\tt s} as the empty shape {\tt (Shp)}.
If we want to extract the first plane of a $5 \times 6 \times 7$ array,
we use {\tt 4} for {\tt d} and {\tt (Shp 6 7)} for {\tt s}.

Since these operations work along the major axis,
we can use other axes instead
by instantiating them differently.
Suppose {\tt mtx} is the matrix {\tt (array (3 2) 0 1 2 3 4 5)},
which has type {\tt (Arr Num (Shp 3 2))}.
Then {\tt (t-app (i-app head 2 (Shp 2)) Num)}
is a function which extracts the first row
of a $(1+2) \times 2$ (\ie, $3 \times 2$) matrix.
So {\tt ((t-app (i-app head 2 (Shp 2)) Num) mtx)}
evaluates to {\tt (array (2) 0 1)}.
Instead, consider {\tt (t-app (i-app head 1 (Shp)) Num)}.
This is a function with input type {\tt (Arr Num (Shp 2))}
and output type {\tt (Arr Num (Shp))}.
It extracts the first scalar of a 2-vector.
When applied to {\tt mtx},
this function \emph{lifts}
to extract the first scalar from \emph{each} 2-vector,
gathering the results as
{\tt (frame (3) (array () 0) (array () 2) (array () 4))}.
Evaluation proceeds, reducing this to {\tt (array (3) 0 2 4)},
the first column of {\tt mtx}.

Several primitives must return boxed arrays
because the type system cannot keep track of enough information
to fully describe the result shape.
As an extreme example,
{\tt read-nums} reads a vector of numbers from user input,
and there is no way of knowing until run time
how long a vector the user will enter.
In other cases, the necessity of boxing
comes from a limit on the type system's expressive power.
The {\tt ravel} function produces a vector
whose atoms are all those of the argument array,
laid out in row-major order.
The length of the {\tt ravel} of some array
is fully determined by that array's shape:
it is the product of all of its dimensions.
However the undecidability of Peano arithmetic
would interfere with type checking
(not to mention future efforts on type inference).
Since ``product of all dimensions'' is not expressible in Presburger arithmetic,
we instead have {\tt ravel} return a boxed vector.

Boxing is not limited to vectors.
For example, {\tt filter} uses a vector of booleans
to decide which parts of an array to retain.
Since the number of true entries in that vector is unknown,
the size of the result's major axis is also unknown.
The resulting {\tt Sigma} type
existentially quantifies only that one dimension,
and leaves the remaining dimensions externally visible.

\FigArrayOps

The {\tt iota} functions and their variants ,
described in Figure \ref{fig:IotaVariants},
form a useful case study on
what invariants can be expressed in Remora's type system.
These functions produce arrays whose atoms are
successive natural numbers starting from 0,
such as {\tt (array (2 3) 0 1 2 3 4 5)},
representing the matrix
$\left[\begin{smallmatrix}
    0 & 1 & 2 \\ 3 & 4 & 5
\end{smallmatrix}\right]$.
The argument to {\tt iota}
is a vector of numbers specifying the result array's shape.
Since this vector can be dynamically computed,
we cannot give any specific shape for {\tt iota}'s return type.
Instead, {\tt iota} must return a box
with existentially quantified shape.
Recall that boxing arrays allows functions with data-dependent result shape to lift safely,
since applying {\tt iota} to {\tt (array (2 2) 3 3 4 4)}
must produce a $3 \times 3$ matrix and a $4 \times 4$ matrix
as its two result cells.

Variants on {\tt iota}
allow the programmer to
communicate more detailed knowledge to the type system.
When the result is meant to be a vector,
{\tt iota/v} takes that vector's length as the argument.
The resulting box is typed as a vector of unknown length
rather than an array of completely unknown shape.
Knowing that we have a vector of numbers rather than any arbitrary array
means, for example,
that summing the box's contents with {\tt reduce}
is certain to produce a scalar.
We can therefore type the following function as
consuming and producing \emph{non-boxed} scalar numbers:
\begin{alltt}
(\l{(n (Arr Num (Shp)))}
  (unbox (len nums ((array () iota/v) n))
    ((t-app (i-app (array () reduce) len (Shp)) Num)
     +
     ((t-app (i-app (array () append) 1 len (Shp)) Num)
      (array () 0)
     nums))))
\end{alltt}
In a more programmer-friendly surface language,
with automatic instantiation of polymorphic functions%
\footnote{This inference problem is beyond the scope of this paper.}
and conversion of bare atoms to scalar arrays,
this might be written as:
\begin{alltt}
(\l{(n (Arr Num (Shp)))}
  (unbox (len nums (iota/v n))
    (reduce + (append [0] nums))))
\end{alltt}

Alternatively, the programmer might prefer to use {\tt iota/s}
to pass the desired result shape as a type index
rather than as a term-level vector.
In that case, there is no need to box the result array.
In the automatic-instantiation shorthand,
{\tt iota/s} may be stylistically awkward,
calling for the variant {\tt iota/w}, which
takes an extra array argument as a ``shape witness''
rather than instantiating at a shape index.
Producing a number array whose shape matches some existing array {\tt xs}
could then be written as {\tt (iota/w xs)}
instead of {\tt ((i-app iota/s shape-of-xs))}.

\FigIotaVariants

The {\tt reshape} function behaves similarly to {\tt iota},
except that the atoms in the result array
are drawn from the second argument,
repeating them cyclically if necessary.
So using {\tt reshape}
with the shape specification {\tt (array (2) 3 2)}
and the vector {\tt (array (5) 1 2 3 4 5)}
produces the $3 \times 2$ matrix {\tt (array (3 2) 1 2 3 4 5 1)}.
Like {\tt iota},
{\tt reshape} benefits from alternative ways
for the programmer to specify the result shape.

\subsection{Theory of type indices} \label{IdxTheoryProse}
Type indices, given in program syntax as $\idx$,
represent individual dimensions, taken from $\mathbb{N}$,
and array shapes, taken from the free monoid on $\mathbb{N}$.
The theory of the free monoid on $\mathbb{N}$ includes as axioms
the associativity of adding naturals and appending sequences
as well as unique identity elements for addition (zero)
and appending (the scalar shape, $\thempty$):
\[0 + i = i + 0 = i\]
\[(i + j) + k = i + (j + k)\]
\[\thempty \thappend a = a \thappend \thempty = a\]
\[(a \thappend b) \thappend c = a \thappend (b \thappend c)\]

As the \emph{free} monoid,
it also follows an equidivisibility rule which states that
if two uses of the append operator give the same result,
there is some completing subsequence,
representing the overlap between each use's larger argument
(demonstrated in Figure \ref{fig:OverlapAxiom}):
\[ a \thappend b = c \thappend d \implies
\exists w . \left( a \thappend w = c\wedge w \thappend d = b \right)
\vee \left( c \thappend w = a \wedge w \thappend b = d \right) \]

\FigOverlapAxiom

A free monoid (on any set of generators) also has a homomorphism to
the monoid formed by $\mathbb{N}$ under addition,
with the property that only the free monoid's identity element
can be mapped to 0.
This can be axiomatized with one additional function symbol $L$:
\[L(a) = 0 \implies a = \thempty \]
\[L(a \thappend b) = L(a) + L(b) \]

Using equidivisibility and the homomorphism to the additive $\mathbb{N}$ monoid,
we can define a partial operator $\thmonus$ for prefix subtraction:
$a \thmonus b = c$ iff $b \thappend c = a$.
For example, $[3,4,5,6] \thmonus [3,4] = [5,6]$,
whereas $[3,4,5,6] \thmonus [4]$ is undefined.

Type checking only requires a very restricted fragment of this theory.
Pairs of indices are only checked for equality in isolation from each other,
and no information about an index
(other than its sort)
is given in the program.
So the check is for the validity of a single equality---%
no connectives or quantifiers needed.
This fragment can be decided efficiently by
comparing indices written in canonical form.
Two $\sortdim$s which are equal must simplify to sums with
the same constant component
and the same coefficient on corresponding variables.
For example,
\[\idxadd{\text{\tt x y 5 x}} = \idxadd{\idxadd{\text{\tt x x}} \text{\tt\ 5 y}}\]
is valid
because both simplify to $2x+y+5$,
whereas
\[\idxadd{\text{\tt q 5 y}} = \idxadd{\idxadd{\text{\tt x x}} \text{\tt\ 5 y}}\]
is false for any interpretation which
does not assign $q$ to twice the value assigned to $x$
(and thus is not valid).

To decide the validity of an equality on $\sortshp$s
(\ie, sequences of naturals),
we can again test by conversion to a canonical form:
a sequence is written out as the concatenation of
single $\sortdim$s and $\sortshp$ variables.
Sorting rules guarantee that
the individual elements of a sequence are natural numbers,
and associativity permits nested appends to be collapsed away.
Thus the index
\[\idxappend%
{\idxshape{\text{\tt 2\ }\idxadd{\text{\tt x 5 x}}}
  \text{\tt\ }
  \idxappend{\text{\tt\ d\ } \idxshape{\text{\tt3}}}}\]
canonicalizes to
\[\idxappend%
{\idxshape{\text{\tt 2}}
  \text{\tt\ }
  \idxshape{\idxadd{\text{\tt x x 5}}}
  \text{\tt\ d\ }
  \idxshape{\text{\tt 3}}}\]
To show that this process does produce a canonical form,
consider two shapes in this form which differ,
and focus on the leftmost differing position
in their respective lists of appended components.
If they are syntactically different singleton shapes---%
their respective contents are two different canonicalized naturals---%
then an assignment under which those naturals differ
will also make the full shapes differ at this position.
If one is a singleton $\idxshape{\idx}$ and the other a variable {\tt s}
(of sort $\sortshp$),
then an interpretation which assigns
the variables in $\idx$ such that its components sum to $n$
may also assign {\tt s} to be the shape $\idxshape{\idxadd{n \text{\tt\ 1}}}$.
Again, an interpretation forces the shapes to be unequal.
Finally, if this position has variables {\tt s} and {\tt t},
choose an interpretation mapping
{\tt s} to $\idxshape{\text{\tt 1}}$
and {\tt t} to $\idxshape{\text{\tt 2}}$
to produce unequal interpretations of the whole shapes.

Although type checking itself only requires this canonicalization process,
constraint-based type inference would call for a more sophisticated solver
due to the use of existential variables for choosing pieces of an index.

In order to describe when arguments' shapes are compatible,
it is useful to impose a lattice structure on the universe of shapes.
The lattice is built with the order $\sqsubseteq$ meaning that one shape is a prefix of another;
a $\top$ is added to represent the join of incompatible shapes
(we already have $\bot = \thempty$,
as the empty shape is a prefix of every shape).
For shapes $s_0$ and $s_1$,
we have $s_0 \sqcup s_1 \not= \top$ if and only if
$s_0 \sqsubseteq s_1$ or $s_1 \sqsubseteq s_0$.
Generalizing to arbitrary finite joins,
$\displaystyle\bigsqcup\left\{ \sequence{s} \right\} \not= \top$
implies that the shapes $\sequence{s}$ are totally ordered,
and the lattice structure means the shapes' join is one of the shapes themselves.

\subsection{Static Semantics}
\label{ssec:StaticSemantics}
Typing Core Remora uses a three-part environment structure:
$\sEnv$ is a partial function mapping index variables to sorts;
$\kEnv$ maps type variables to their kinds;
and $\tEnv$ maps term variables to their types.
The stratification of Dependent ML-style types
allows indices to be checked using only the sort environment
and types using only the sort and kind environments.
Following the definition of each judgment form,
we give a handful of lemmas
which will be needed for a type soundness argument in Subsection \ref{SoundnessSection}.
The well-formedness judgments
each come with a lemma stating that
the judgment gives a unique result to each well-formed term
and that unique result is preserved by
substituting well-formed assignments for free variables.
When we show type soundness for Remora,
these results will be needed to prove the preservation lemma.
Uniqueness of typing is particularly important for Remora,
where the implicit iteration in function application
(including index and type abstractions)
is driven by the types ascribed to the function and argument expressions.
Well-defined program behavior relies on having
a unique decomposition of each array into a frame of cells.

\subsubsection{Sorting}
\FigSortingRules
Figure \ref{fig:SortingRules} defines the sorting judgment,
$\sortof{\sEnv}{\idx}{\sort}$,
which states that in sort environment $\sEnv$,
the index $\idx$ has sort $\sort$.
Natural number literals have sort $\sortdim$.
A sequence of indices is a $\sortshp$,
provided that every element of the sequence is a $\sortdim$.
Addition is used on $\sortdim$ arguments to produce a $\sortdim$.
$\sortshp$ arguments may be appended,
to form another $\sortshp$.
Variables may be bound at either sort,
but they can only be introduced into the environment by
index abstraction and unboxing terms---%
the index language itself has no binding forms.

We give two results about the well-behaved nature of the sorting rules:
No index inhabits both sorts (in the same environment),
and replacing an index's variables with appropriately-sorted indices
does not change the sort.

\begin{lemma}[Uniqueness of sorting]
  \label{UniqSort}
  If $\sortof{\sEnv}{\idx}{\sort}$
  and $\sortof{\sEnv}{\idx}{\sort'}$,
  then $\sort = \sort'$.
\end{lemma}
\begin{sproof}[\BODY]
  No non-variable index form is compatible with multiple sorting rules,
  so they can only have whichever sort their one compatible rule concludes.
  It remains to show that uniqueness holds for variables.
  Since $\sEnv$ is a well-defined partial function,
  mapping variables to sorts,
  $\sEnv(\var)$ can only have one value.
  If $\sEnv(\var) = \sort$ and $\sEnv(\var) = \sort'$, $\sort = \sort'$.
\end{sproof}

\newcommand{\idxsubst}[1]{\subst{#1}{\var}{\idx_\var}}
\begin{lemma}[Preservation of sorts under index substitution]
  \label{IISub}
  If $\sortof{\sEnv,\hassort{\var}{\sort_\var}}{\idx}{\sort}$
  and $\sortof{\sEnv}{\idx_\var}{\sort_\var}$
  then $\sortof{\sEnv}{\idxsubst{\idx}}{\sort}$.
\end{lemma}
\begin{sproof}[This is straightforward induction on the original sort derivation.]
  We use induction on the sort derivation
  $\sortof{\sEnv,\hassort{\var}{\sort_\var}}{\idx}{\sort}$.
  \paragraph{Case {\sc Nat}}
  \[\infr[S-Nat]
  {\nat \in \mathbb{N}}
  {\sortof{\sEnv,\hassort{\var}{\sort_1}}{\nat}{\sortdim}}\]
  Since $\idx_0 = \nat$, $\idxsubst{\idx} = \idx_0$,
  so the original derivation is still valid.
  \paragraph{Case {\sc Var}}
  \[\infr[S-Var]
  {\parens{\hassort{\var'}{\sort}} \in \sEnv}
  {\sortof{\sEnv,\hassort{\var}{\sort_\var}}{\var'}{\sort}}\]
  Then $\idx$ has the form $\var'$.
  If $\var' = \var$, then $\idxsubst{\idx} = \idx_1$,
  and (by Lemma \ref{UniqSort}, uniqueness of sorting) $\sort = \sort_\var$.
  By assumption, $\sortof{\sEnv}{\idx_\var}{\sort_\var}$,
  so the equality of $\sort$ and $\sort_\var$ implies $\sortof{\sEnv}{\idx_\var}{\sort}$.
  If $\var' \not= \var$, then $\idxsubst{\idx} = \var$,
  and we use the same {\sc S-Var} derivation we started with.
  \paragraph{Case {\sc Shape}}
  \[\infr[S-Shape]
  {\sequence{\sortof{\sEnv,\hassort{\var}{\sort_\var}}{\idx_d}{\sortdim}}}
  {\sortof{\sEnv,\hassort{\var}{\sort_\var}}{\idxshape{\sequence{\idx_d}}}{\sortshp}}\]
  Then the induction hypothesis implies that
  $\sortof{\sEnv}{\idxsubst{\idx_d}}{\sortdim}$
  for each of the $\sequence{\idx_d}$.
  So applying {\sc S-Shape} derives
  \[\infr[S-Shape]
  {\sequence{\sortof{\sEnv}{\idxsubst{\idx_d}}{\sortdim}}}
  {\sortof{\sEnv}{\idxshape{\sequence{\idxsubst{\idx_d}}}}{\sortshp}}\]
  \paragraph{Case {\sc Plus}}
  \[\infr[S-Plus]
  {\sequence{\sortof{\sEnv,\hassort{\var}{\sort_\var}}{\idx_d}{\sortdim}}}
  {\sortof{\sEnv,\hassort{\var}{\sort_\var}}{\idxadd{\sequence{\idx_d}}}{\sortdim}}\]
  By the induction hypothesis,
  $\sortof{\sEnv}{\idxsubst{\idx_d}}{\sortdim}$
  for each of $\sequence{\idx_d}$.
  Then we use their derivations to construct
  \[\infr[S-Plus]
  {\sequence{\sortof{\sEnv}{\idxsubst{\idx_d}}{\sortdim}}}
  {\sortof{\sEnv}{\idxadd{\sequence{\idxsubst{\idx_d}}}}{\sortdim}}\]
  \paragraph{Case {\sc Append}}
  \[\infr[S-Append]
  {\sequence{\sortof{\sEnv,\hassort{\var}{\sort_\var}}{\idx_s}{\sortshp}}}
  {\sortof{\sEnv,\hassort{\var}{\sort_\var}}{\idxappend{\sequence{\idx_s}}}{\sortshp}}\]
  The induction hypothesis gives
  $\sortof{\sEnv}{\idxsubst{\idx_s}}{\sortshp}$
  for each of $\sequence{\idx_s}$.
  Then we derive
  \[\infr[S-Append]
  {\sequence{\sortof{\sEnv}{\idxsubst{\idx_s}}{\sortshp}}}
  {\sortof{\sEnv}{\idxappend{\sequence{\idxsubst{\idx_s}}}}{\sortshp}}\]
\end{sproof}

\subsubsection{Kinding}
\FigKindingRules
Kinding rules are given in figure \ref{fig:KindingRules}.
The $\kindarray$ kind is only ascribed to
types built by the array type constructor
and type variables bound at that kind.
The array type constructor requires as its arguments
an $\kindatom$ type and a $\sortshp$ index.
Base types are fundamental, non-aggregate types,
such as {\tt Float} or {\tt Bool},
so they are $\kindatom$s.
Function types have kind $\kindatom$,
but their input and output types must be $\kindarray$s.
This reflects the rule that application is performed on arrays,
and the function produces an array result.
Similarly, universal types and dependent products,
describing type and index abstractions,
must have an $\kindarray$ as their body,
while they themselves are $\kindatom$s.
This rules out types whose inhabitants would have to be syntactically illegal
due to containing expressions instead of atoms as their bodies.
Since boxes present arrays as atoms,
dependent sum types also have an $\kindarray$ body
and are kinded as $\kindatom$s.
A universal type adds bindings for its quantified type variables to $\kEnv$.
Dependent products and sums do the same for their index variables in $\sEnv$.

As with sorting of indices,
we expect a well-kinded type to inhabit only a single kind (fixing a particular environment).
The kinding system should also allow free index or type variables to be
replaced with appropriately sorted or kinded indices or types
without changing the original type's kind.

\begin{lemma}[Uniqueness of kinding]
  \label{UniqKind}
  If $\kindof{\sEnv}{\kEnv}{\type}{\kind}$
  and $\kindof{\sEnv}{\kEnv}{\type}{\kind'}$,
  then $\kind = \kind'$.
\end{lemma}
\begin{sproof}[\BODY]
  As with uniqueness of sorting,
  no non-variable type is compatible with multiple kinding rules.
  Since all kinding rules except for {\sc K-Var}
  ascribe a specific kind,
  the only remaining case is for type variables.
  The kind environment $\kEnv$ is a well-defined partial function,
  so $\kEnv(\var) = \kind$ and $\kEnv(\var) = \kind'$ imply $\kind = \kind'$.
\end{sproof}

\begin{lemma}[Preservation of kinds under index substitution]
  \label{ITSub}
  If $\kindof{\sEnv,\hassort{\var}{\sort}}{\kEnv}{\type}{\kind}$
  and $\sortof{\sEnv}{\idx}{\sort}$
  then $\kindof{\sEnv}{\kEnv}{\idxsubst{\type}}{\kind}$.
\end{lemma}
\begin{sproof}[This is straightforward induction on the original kind derivation.]
  We use induction on the kind derivation
  $\kindof{\sEnv,\hassort{\var}{\sort}}{\kEnv}{\type}{\kind}$.
  \paragraph{Case {\sc Var}}
  \[\infr[K-Var]
  {\haskind{\var'}{\kind} \in \kEnv}
  {\kindof{\sEnv,\hassort{\var}{\sort}}{\kEnv}{\var'}{\kind}}\]
  Since $\var$ cannot appear free in $\type$,
  $\idxsubst{\var'} = \var'$,
  so the same kind is derivable:
  \[\infr[K-Var]
  {\haskind{\var'}{\kind} \in \kEnv}
  {\kindof{\sEnv}{\kEnv}{\var'}{\kind}}\]
  \paragraph{Case {\sc Base}}
  \[\infr[K-Base]
  {\ }
  {\kindof{\sEnv,\hassort{\var}{\sort}}{\kEnv}{\basetype}{\kind}}\]
  Again, $\var$ cannot appear free in $\type$,
  so we derive
  \[\infr[K-Base]
  {\ }
  {\kindof{\sEnv}{\kEnv}{\basetype}{\kind}}\]
  \paragraph{Case {\sc Fun}}
  \[\infr[K-Fun]
  {\sequence{\kindof{\sEnv,\hassort{\var}{\sort}}{\kEnv}{\type_i}{\kindarray}}
    \qquad \kindof{\sEnv,\hassort{\var}{\sort}}{\kEnv}{\type_o}{\kindarray}}
  {\kindof{\sEnv,\hassort{\var}{\sort}}{\kEnv}
    {\typefun{\sequence{\type_i}}{\type_o}}
    {\kindarray}}\]
  By the induction hypothesis, we have
  $\kindof{\sEnv}{\kEnv}{\idxsubst{\type_i}}{\kindarray}$ for each of $\sequence{\type_i}$,
  and $\kindof{\sEnv}{\kEnv}{\idxsubst{\type_o}}{\kindarray}$.
  Then we can derive
  \[\infr[K-Fun]
  {\sequence{\kindof{\sEnv}{\kEnv}{\idxsubst{\type_i}}{\kindarray}}
    \qquad \kindof{\sEnv}{\kEnv}{\idxsubst{\type_o}}{\kindarray}}
  {\kindof{\sEnv}{\kEnv}
    {\typefun{\sequence{\idxsubst{\type_i}}}{\idxsubst{\type_o}}}
    {\kindarray}}\]
  \paragraph{Case {\sc Univ}}
  \[\infr[K-Univ]
  {\kindof{\sEnv,\hassort{\var}{\sort}}{\kEnv, \sequence{\haskind{\var_u}{\kind_u}}}{\type_u}{\kindarray}}
  {\kindof{\sEnv,\hassort{\var}{\sort}}{\kEnv}
    {\typeuniv{\sequence{\notevar{\var_u}{\kind_u}}}{\type_u}}
    {\kindatom}}\]
  The induction hypothesis implies
  $\kindof{\sEnv}{\kEnv, \sequence{\haskind{\var_u}{\kind_u}}}{\idxsubst{\type_u}}{\kindarray}$.
  So we derive
  \[\infr[K-Univ]
  {\kindof{\sEnv}{\kEnv, \sequence{\haskind{\var_u}{\kind_u}}}{\idxsubst{\type_u}}{\kindarray}}
  {\kindof{\sEnv,\hassort{\var}{\sort}}{\kEnv}
    {\typeuniv{\sequence{\notevar{\var_u}{\kind_u}}}{\idxsubst{\type_u}}}
    {\kindatom}}\]
  \paragraph{Case {\sc Pi}}
  \[\infr[K-Pi]
  {\kindof{\sEnv,\hassort{\var}{\sort}, \sequence{\hassort{\var_p}{\sort_p}}}{\kEnv}
    {\type_p}{\kindarray}}
  {\kindof{\sEnv,\hassort{\var}{\sort}}{\kEnv}
    {\typedprod{\sequence{\notevar{\var_p}{\sort_p}}}{\type_p}}
    {\kindatom}}\]
  By the induction hypothesis,
  $\kindof{\sEnv, \sequence{\hassort{\var_p}{\sort_p}}}{\kEnv}{\idxsubst{\type_p}}{\kindarray}$.
  Note that, following Barendregt's convention, $\sequence{\var_p}$ are all unique and distinct from $\var$.
  Then we construct the derivation
  \[\infr[K-Pi]
  {\kindof{\sEnv, \sequence{\hassort{\var_p}{\sort_p}}}{\kEnv}
    {\idxsubst{\type_p}}{\kindarray}}
  {\kindof{\sEnv}{\kEnv}
    {\typedprod{\sequence{\notevar{\var_p}{\sort_p}}}{\idxsubst{\type_p}}}
    {\kindatom}}\]
  \paragraph{Case {\sc Sigma}}
  \[\infr[K-Sigma]
  {\kindof{\sEnv,\hassort{\var}{\sort}, \sequence{\hassort{\var_s}{\sort_s}}}{\kEnv}
    {\type_s}{\kindarray}}
  {\kindof{\sEnv,\hassort{\var}{\sort}}{\kEnv}
    {\typedsum{\sequence{\notevar{\var_s}{\sort_s}}}{\type_s}}
    {\kindatom}}\]
  As in the {\sc K-Pi} case, the induction hypothesis gives
  $\kindof{\sEnv, \sequence{\hassort{\var_s}{\sort_s}}}{\kEnv}{\idxsubst{\type_s}}{\kindarray}$.
  Then we can derive
  \[\infr[K-Sigma]
  {\kindof{\sEnv, \sequence{\hassort{\var_s}{\sort_s}}}{\kEnv}
    {\idxsubst{\type_s}}{\kindarray}}
  {\kindof{\sEnv,\hassort{\var}{\sort}}{\kEnv}
    {\typedsum{\sequence{\notevar{\var_s}{\sort_s}}}{\idxsubst{\type_s}}}
    {\kindatom}}\]
  \paragraph{Case {\sc Array}}
  \[\infr[K-Array]
  {\kindof{\sEnv,\hassort{\var}{\sort}}{\kEnv}{\type_a}{\kindatom}
    \qquad \sortof{\sEnv,\hassort{\var}{\sort}}{\idx_a}{\sortshp}}
  {\kindof{\sEnv,\hassort{\var}{\sort}}{\kEnv}
    {\typearray{\type_a}{\idx_a}}
    {\kindarray}}\]
  The induction hypothesis implies
  $\kindof{\sEnv}{\kEnv}{\idxsubst{\type_a}}{\kindatom}$,
  and Lemma \ref{IISub} (substitution in an index) implies
  $\sortof{\sEnv}{\idxsubst{\idx_a}}{\sortshp}$.
  So we then have
  \[\infr[K-Array]
  {\kindof{\sEnv}{\kEnv}{\idxsubst{\type_a}}{\kindatom}
    \qquad \sortof{\sEnv}{\idxsubst{\idx_a}}{\sortshp}}
  {\kindof{\sEnv,\hassort{\var}{\sort}}{\kEnv}
    {\typearray{\idxsubst{\type_a}}{\idxsubst{\idx_a}}}
    {\kindarray}}\]
\end{sproof}

\newcommand{\typesubst}[1]{\subst{#1}{\var}{\type_\var}}
\begin{lemma}[Preservation of kinds under type substitution]
  \label{TTSub}
  If $\kindof{\sEnv}{\kEnv,\haskind{\var}{\kind_\var}}{\type}{\kind}$
  and $\kindof{\sEnv}{\kEnv}{\type_\var}{\kind_\var}$
  then $\kindof{\sEnv}{\kEnv}{\typesubst{\type}}{\kind}$.
\end{lemma}
\begin{sproof}[This is also provable by induction on the kind derivation for $\type$.]
  We use induction on the kind derivation
  $\kindof{\sEnv}{\kEnv,\haskind{\var}{\kind_\var}}{\type}{\kind}$.
  \paragraph{Case {\sc Var}, $\type = \var$}
  \[\infr[K-Var]
  {\parens{\haskind{\var}{\kind}} \in \kEnv, \haskind{\var}{\kind_\var}}
  {\kindof{\sEnv}{\kEnv, \haskind{\var}{\kind_\var}}{\var}{\kind}}\]
  Then $\typesubst{\type} = \type_\var$,
  which has kind $\kind_\var$ by assumption.

  \paragraph{Case {\sc Var}, $\type = \var' \not= \var$}
  \[\infr[K-Var]
  {\parens{\haskind{\var'}{\kind}} \in \kEnv, \haskind{\var}{\kind_\var}}
  {\kindof{\sEnv}{\kEnv, \haskind{\var}{\kind_\var}}{\var'}{\kind}}\]
  Then $\typesubst{\type} = \var'$,
  and $\var' \not= \var$ implies $\haskind{\var'}{\kind} \in \kEnv$.
  So we have
  \[\infr[K-Var]
  {\parens{\haskind{\var'}{\kind}} \in \kEnv}
  {\kindof{\sEnv}{\kEnv}{\var'}{\kind}}\]

  \paragraph{Case {\sc Base}}
  \[\infr[K-Base]
  {\ }{\kindof{\sEnv}{\kEnv, \haskind{\var}{\kind_\var}}{\basetype}{\kindatom}}\]
  Since $\var$ cannot appear free in $\type$,
  $\typesubst{\type} = \type = \basetype$.
  The kinding rule {\sc K-Base} applies in any environment,
  so we have
  \[\infr[K-Base]{\ }{\kindof{\sEnv}{\kEnv}{\basetype}{\kindatom}}\]

  \paragraph{Case {\sc Fun}}
  \[\infr[K-Fun]
  {\sequence{\kindof{\sEnv}{\kEnv, \haskind{\var}{\kind_\var}}{\type_i}{\kindarray}}
    \qquad
  \kindof{\sEnv}{\kEnv, \haskind{\var}{\kind_\var}}{\type_o}{\kindarray}}
  {\kindof{\sEnv}{\kEnv, \haskind{\var}{\kind_\var}}{\typefun{\sequence{\type_i}}{\type_o}}{\kindatom}}\]
  By the induction hypothesis,
  $\kindof{\sEnv}{\kEnv}{\typesubst{\type_i}}{\kindarray}$
  for each of $\sequence{\type_i}$,
  and $\kindof{\sEnv}{\kEnv}{\typesubst{\type_o}}{\kindarray}$.
  Then we derive
  \[\infr[K-Fun]
  {\sequence{\kindof{\sEnv}{\kEnv}
      {\typesubst{\type_i}}
      {\kindarray}}
    \qquad \kindof{\sEnv}{\kEnv}
    {\typesubst{\type_o}}
    {\kindarray}}
  {\kindof
    {\sEnv}{\kEnv}
    {\typefun{\sequence{\typesubst{\type_i}}}{\typesubst{\type_o}}}
    {\kindatom}}\]

  \paragraph{Case {\sc Univ}}
  \[\infr[K-Univ]
  {\kindof
    {\sEnv}{\kEnv, \haskind{\var}{\kind_\var}, \sequence{\haskind{\var_u}{\kind_u}}}
    {\type_u}{\kindarray}}
  {\kindof
    {\sEnv}{\kEnv, \haskind{\var}{\kind_\var}}
    {\typeuniv{\sequence{\notevar{\var_u}{\kind_u}}}{\type_u}}{\kindatom}}\]
  The induction hypothesis implies
  $\kindof{\sEnv}{\kEnv, \sequence{\haskind{\var_u}{\kind_u}}}{\typesubst{\type_u}}{\kindarray}$
  (again, Barendregt's convention promises that $\var \not\in \sequence{\var_u}$).
  This leads to the derivation
  \[\infr[K-Univ]
  {\kindof
    {\sEnv}{\kEnv, \sequence{\haskind{\var_u}{\kind_u}}}
    {\typesubst{\type_u}}{\kindarray}}
  {\kindof
    {\sEnv}{\kEnv}
    {\typeuniv{\sequence{\notevar{\var_u}{\kind_u}}}{\typesubst{\type_u}}}{\kindatom}}\]

  \paragraph{Case {\sc Pi}}
  \[\infr[K-Pi]
  {\kindof{\sEnv, \sequence{\hassort{\var_p}{\sort_p}}}{\kEnv, \haskind{\var}{\kind_\var}}
    {\type_p}{\kindarray}}
  {\kindof{\sEnv}{\kEnv, \haskind{\var}{\kind_\var}}
    {\typedprod{\sequence{\notevar{\var_p}{\sort_p}}}{\type_p}}
    {\kindatom}}\]
  By the induction hypothesis,
  $\kindof{\sEnv, \sequence{\hassort{\var_p}{\sort_p}}}{\kEnv}{\typesubst{\type_p}}{\kindarray}$.
  Then applying {\sc K-Pi} derives
  \[\infr[K-Pi]
  {\kindof{\sEnv, \sequence{\hassort{\var_p}{\sort_p}}}{\kEnv}
    {\typesubst{\type_p}}{\kindarray}}
  {\kindof{\sEnv}{\kEnv}
    {\typedprod{\sequence{\notevar{\var_p}{\sort_p}}}{\typesubst{\type_p}}}
    {\kindatom}}\]

  \paragraph{Case {\sc Sigma}}
  \[\infr[K-Sigma]
  {\kindof{\sEnv, \sequence{\hassort{\var_p}{\sort_p}}}{\kEnv, \haskind{\var}{\kind_\var}}
    {\type_s}{\kindarray}}
  {\kindof{\sEnv}{\kEnv, \haskind{\var}{\kind_\var}}
    {\typedsum{\sequence{\notevar{\var_s}{\sort_s}}}{\type_s}}
    {\kindatom}}\]
  The induction hypothesis gives
  $\kindof{\sEnv, \sequence{\hassort{\var_s}{\sort_s}}}{\kEnv}{\typesubst{\type_s}}{\kindarray}$.
  We then derive
  \[\infr[K-Sigma]
  {\kindof{\sEnv, \sequence{\hassort{\var_s}{\sort_s}}}{\kEnv}
    {\typesubst{\type_s}}{\kindarray}}
  {\kindof{\sEnv}{\kEnv}
    {\typedsum{\sequence{\notevar{\var_s}{\sort_s}}}{\typesubst{\type_s}}}
    {\kindatom}}\]
 
  \paragraph{Case {\sc Array}}
  \[\infr[K-Array]
  {\kindof{\sEnv}{\kEnv, \haskind{\var}{\kind_\var}}{\type_a}{\kindatom}
    \qquad \sortof{\sEnv}{\idx_a}{\sortshp}}
  {\kindof{\sEnv}{\kEnv, \haskind{\var}{\kind_\var}}{\typearray{\type_a}{\idx_a}}{\kindarray}}\]
  By the induction hypothesis,
  $\kindof{\sEnv}{\kEnv}{\typesubst{\type_a}}{\kindatom}$.
  Recycling the sort derivation for $\idx_a$ as is, we derive
  \[\infr[K-Array]
  {\kindof{\sEnv}{\kEnv}{\typesubst{\type_a}}{\kindatom}
    \qquad \sortof{\sEnv}{\idx_a}{\sortshp}}
  {\kindof{\sEnv}{\kEnv}{\typearray{\typesubst{\type_a}}{\idx_a}}{\kindarray}}\]
\end{sproof}

\subsubsection{Typing}
\FigTypingRules

The typing rules in Figure \ref{fig:TypingRules} relate
a full environment
($\sEnv$ mapping index variables to sorts,
$\kEnv$ mapping type variables to kinds,
and $\tEnv$ mapping term variables to types),
a term (whether an atom or an expression),
and its type under that environment.
Since an array type might have its shape described in multiple different ways,
\eg, a vector of length 6 or a vector of length $1+5$,
the {\sc T-Eqv} rule makes reference to a type equivalence judgment
(presented in full detail in \S\ref{TypeEqvProse})
which reconciles such differences according to
the algebraic theory of type indices
(presented earlier in \S\ref{IdxTheoryProse}).

The signature $\Signature$,
referenced in the {\sc T-Op} rule,
is a function mapping from primitive operators to their types.
For example, $\Sigref{\text{\tt +}}$ is
{\tt (-> ((Arr Num (Shp)) (Arr Num (Shp))) (Arr Num (Shp)))},
meaning {\tt +} is an operator
which consumes two scalar numbers
and produces one scalar number.

Array literals and nested frames both include a length check:
the number of atoms or cells must be equal to
the product of the given dimensions.
In the case of empty arrays, the length matching condition is fulfilled
if and only if the array has a {\tt 0} as one of its dimensions.
Term, type, and index abstractions all
bind their arguments' names in the appropriate environment.

Typing function application starts by identifying
the type of the expression in function position.
It must be an array of functions,
and the array's entire shape $\idx_f$ is treated as the function frame.
The function input types, also arrays,
specify the element type and cell shape for each argument.
Each cell shape $\idx$ must be a suffix of the shape
of the corresponding actual argument;
the remainder $\idx_a$ is the argument's frame.
The maximum of these frames under prefix ordering
(where $[2\, 3] \sqsubseteq [2\, 3\, 2]$
but $[2\, 3] \not\sqsubseteq [6\, 3\, 2]$)
is the principal frame $\idx_p$.
That is, the function and argument arrays will all be
lifted so as to have $\idx_p$ as their frames when the program runs.
Then $\idx_p$ is used as the frame around the function's output type
to give the result type for this function application.

Type and index application also require arrays in function position,
but they can skip prefix comparison
as type and index arguments do not come in arrays
that must be split into frames of cells.
Thus the function's frame shape $\idx_f$ passes through unaltered,
and arguments are substituted into the body type $\type_b$
to produce the resulting array's element type.

When constructing a box,
a dependent-sum type annotation is provided.
The box's index components must match their declared sorts,
and substituting them into the body of the dependent sum type
must produce a type that matches the box's array component.
Unboxing requires that $\expr_\mathit{box}$, the expression being destructed,
be a dependent sum.
The unbox form names the sum's index and array components
and adds them to the sort and type environments
when checking $\expr_\mathit{body}$.
Although the index components are in scope while checking the body,
information hidden by the existential is not permitted to leak out:
The end result type $\tau_\mathit{body}$ must be well-formed
without relying on the extended sort environment.
Unboxing a frame of boxes (scalars)
produces a frame of result cells,
similar to lifting function application.

Anticipating a progress lemma,
we prove a canonical-forms lemma for Remora's typing rules.
Following the atom/array distinction,
we have separate lemmas for atoms and arrays.
Although an atom can contain an array
if that atom is a box,
we avoid mutual dependence between the lemmas
by not making any claim about the syntactic structure
of the box's contents.

\begin{lemma}[Canonical forms for atomic values]
  \label{CFAtom}
  Let $\atval$ be a well-typed atomic value,
  that is, $\typeof{\emptyEnv}{\emptyEnv}{\emptyEnv}{\atval}{\type}$.
  \begin{enumerate}
    {\cfitem
      {\atval}
      {\typefun{\sequence{\type_i}}{\type_o}}
      {\primop$ or $\lam{\sequence{\notevar{\var}{\type_i}}}{\expr}}}
    {\cfitem
      {\atval}
      {\typeuniv{\sequence{\notevar{\var}{\kind}}}{\type_u}}
      {\tlam{\sequence{\notevar{\var_u}{\kind}}}{\val}}}
    {\cfitem
      {\atval}
      {\typedprod{\sequence{\notevar{\var}{\sort}}}{\type_p}}
      {\ilam{\sequence{\notevar{\var_p}{\sort}}}{\val}}}
    {\cfitem[,\\ with
        \(\teqv{\type}
               {\typedsum{\sequence{\notevar{\var_b}{\sort}}}{\type_b'}}\)]
      {\atval}
      {\typedsum{\sequence{\notevar{\var}{\sort}}}{\type_b}}
      {\dsum
        {\sequence{\idx}}
        {\val_b}
        {\typedsum{\sequence{\notevar{\var_b}{\sort}}}{\type_b'}}}}
    {\cfitem
      {\atval}
      {\basetype}
      {\baseval}}
  \end{enumerate}
\end{lemma}
\begin{sproof}[The type derivation may end with {\sc T-Eqv},
so we consider the subderivation prior to all final {\sc T-Eqv} instances.
We then examine which typing rules can ascribe a type of the right form
and the identify what form the term must take to match those rule.]
  A type derivation may end with a chain of uses of the {\sc T-Eqv} rule,
  but this chain must then be preceded by a use of some other rule.
  We consider the last non-{\sc T-Eqv} rule used in the derivation;
  the type ascribed by this rule must then be equivalent to
  the type ascribed by the full derivation.
  Types of two different forms (\eg, a function and a dependent sum)
  cannot be equivalent because no type equivalence rule can relate them.
  \begin{enumerate}
  \item{The only rules capable of ascribing a function type to an atom
    are {\sc T-Op} and {\sc T-Lam},
    which apply to atomic values of the form
    $\primop$ and
    $\lam{\sequence{\notevar{\var}{\type_i}}}{\expr}$ respectively.}
  \item{Ascribing this type to an atom requires {\sc T-TLam},
    which applies only to an atomic value of the form
    $\tlam{\sequence{\notevar{\var_u}{\kind}}}{\val}$.}
  \item{Ascribing this type to an atom requires {\sc T-ILam},
    which applies only to an atomic value of the form
    $\ilam{\sequence{\notevar{\var_p}{\sort}}}{\val}$.}
  \item{Ascribing this type to an atom requires {\sc T-Box},
    which applies only to an atomic value of the form
    \({\dsum
        {\sequence{\idx}}
        {\val_b}
        {\typedsum{\sequence{\notevar{\var_b}{\sort}}}{\type_b'}}}\).
      {\sc T-Box} ascribes the annotated type,
      but the full derivation may produce any equivalent type.}
  \item{Only {\sc T-Base} can ascribe base type to an atom,
      and it applies only to literals of the appropriate type of base value.}
  \end{enumerate}
\end{sproof}

\begin{lemma}[Canonical forms for arrays]
  \label{CFArray}
  Let $\val$ be a well-typed value,
  that is, $\typeof{\emptyEnv}{\emptyEnv}{\emptyEnv}{\val}{\type}$,
  \begin{enumerate}
    {\cfarritem
      {\val}
      {\typefun{\sequence{\type_i}}{\type_o}}
      {\func}}
    {\cfarritem
      {\val}
      {\typeuniv{\sequence{\notevar{\var}{\kind}}}{\type_u}}
      {\tlam{\sequence{\notevar{\var_u}{\kind}}}{\val_u}}}
    {\cfarritem
      {\val}
      {\typedprod{\sequence{\notevar{\var}{\sort}}}{\type_p}}
      {\ilam{\sequence{\notevar{\var_p}{\sort}}}{\val_u}}}
    {\cfarritem[,\\ with
        \(\teqv{\type}
               {\typedsum{\sequence{\notevar{\var_b}{\sort}}}{\type_b'}}\)]
      {\val}
      {\typedsum{\sequence{\notevar{\var}{\sort}}}{\type_b}}
      {\dsum
        {\sequence{\idx}}
        {\val_b}
        {\typedsum{\sequence{\notevar{\var_b}{\sort}}}{\type_b}}}}
    {\cfarritem[,\\ with
        $\typeof{\emptyEnv}{\emptyEnv}{\emptyEnv}{\baseval}{\basetype}$
        for each of $\sequence{\baseval}$]
      {\val}
      {\basetype}
      {\baseval}}
  \end{enumerate}
\end{lemma}
\begin{sproof}[This proceeds like the proof for Lemma \ref{CFAtom}.]
  As for Lemma \ref{CFAtom},
  the type derivation for $\val$ must at some point use a rule other than {\sc T-Eqv}
  to ascribe a type to $\val$,
  and any subsequent use of {\sc T-Eqv} must preserve the form of that type.
  All of the types considered here have the form $\typearray{\type_a}{\idx_a}$;
  array types must be ascribed by {\sc T-Array}.
  So $\val$ must be an array literal---$\arrlit{\sequence{\atval}}{\sequence{\nat}}$---%
  and we can shift to considering the forms of the atoms $\sequence{\atval}$.
  Each case in this lemma corresponds to a particular case in Lemma \ref{CFAtom}.
\end{sproof}

\subsubsection{Type equivalence} \label{TypeEqvProse}
\FigTypeEqv
Remora's typing rules rely on a type-equivalence relation
defined in Figure \ref{fig:TypeEqv}.
The equivalence relation is essentially $\alpha$-equivalence
augmented with a check as to whether array shapes are guaranteed to be equal.
We expect the relation $\teqv{}{}$ actually to be an equivalence relation,
\ie, reflexive, symmetric, and transitive.
Only reflexivity has its own inference rule,
so we now show symmetry and transitivity.

\begin{lemma}[Symmetry of $\teqv{}{}$]
  \label{SymTEqv}
  If $\teqv{\type}{\type'}$,
  then $\teqv{\type'}{\type}$.
\end{lemma}
\begin{sproof}[This follows via straightforward induction on the derivation of $\teqv{\type_0}{\type_1}$.]
  We use a straightforward inductive argument on the derivation of $\teqv{\type_0}{\type_1}$.
  \paragraph{Case {\sc Refl}:}
  \[\infr[TEqv-Refl]{\ }{\teqv{\type}{\type}}\]
  Since $\type$ and $\type'$ are equal,
  the obligation is to show $\teqv{\type}{\type}$,
  which we already have.
  \paragraph{Case {\sc Array}:}
  \[\infr[TEqv-Array]
  {\teqv{\type_a}{\type_a'} \qquad \ieqv{\idx_a}{\idx_a'}}
  {\teqv{\typearray{\type_a}{\idx_a}}{\typearray{\type_a'}{\idx_a'}}}\]
  Index equality is symmetric, so $\ieqv{\idx_a'}{\idx_a}$.
  The induction hypothesis gives $\teqv{\type_a'}{\type_a}$.
  We then apply {\sc TEqv-Array} to construct
  \[\infr[TEqv-Array]
  {\teqv{\type_a'}{\type_a} \qquad \ieqv{\idx_a'}{\idx_a}}
  {\teqv{\typearray{\type_a'}{\idx_a'}}{\typearray{\type_a}{\idx_a}}}\]
  \paragraph{Case {\sc Fn}:}
  \[\infr[TEqv-Fn]
  {\sequence{\teqv{\type_i}{\type_i'}}
    \qquad
  \teqv{\type_o}{\type_o'}}
  {\teqv{\typefun{\sequence{\type_i}}{\type_o}}{\typefun{\sequence{\type_i'}}{\type_o'}}}\]
  The induction hypothesis gives
  $\teqv{\type_i}{\type_i'}, \dots, \teqv{\type_o}{\type_o'}$,
  so {\sc TEqv-Fn} produces
  \[\infr[TEqv-Fn]
  {\sequence{\teqv{\type_i'}{\type_i}}
    \qquad
  \teqv{\type_o'}{\type_o}}
  {\teqv{\typefun{\sequence{\type_i'}}{\type_o'}}{\typefun{\sequence{\type_i}}{\type_o}}}\]
  \paragraph{Case {\sc Univ}:}
  \[\infr[TEqv-Univ]
  {\teqv
    {\seqsubst{\type_s}{\var}{\var_f}}
    {\seqsubst{\type_s'}{\var'}{\var_f}}
    \fresh{\sequence{\var_f}}}
  {\teqv
    {\typeuniv{\sequence{\notevar{\var}{\sort}}}{\type_u}}
    {\typeuniv{\sequence{\notevar{\var'}{\sort}}}{\type_u'}}}\]
  The induction hypothesis gives
  $\teqv
  {\seqsubst{\type_u'}{\var'}{\var_f}}
  {\seqsubst{\type_u}{\var}{\var_f}}$,
  so we construct
  \[\infr[TEqv-Univ]
  {\teqv
    {\seqsubst{\type_s'}{\var'}{\var_f}}
    {\seqsubst{\type_s}{\var}{\var_f}}
    \fresh{\sequence{\var_f}}}
  {\teqv
    {\typeuniv{\sequence{\notevar{\var'}{\sort}}}{\type_u'}}
    {\typeuniv{\sequence{\notevar{\var}{\sort}}}{\type_u}}}\]
  \paragraph{Case {\sc Pi}:}
  \[\infr[TEqv-Pi]
  {\teqv
    {\seqsubst{\type_s}{\var}{\var_f}}
    {\seqsubst{\type_s'}{\var'}{\var_f}}
    \fresh{\sequence{\var_f}}}
  {\teqv
    {\typedprod{\sequence{\notevar{\var}{\sort}}}{\type_p}}
    {\typedprod{\sequence{\notevar{\var'}{\sort}}}{\type_p'}}}\]
  The induction hypothesis gives
  $\teqv
  {\seqsubst{\type_s'}{\var'}{\var_f}}
  {\seqsubst{\type_s}{\var}{\var_f}}$.
  Then we derive
  \[\infr[TEqv-Pi]
  {\teqv
    {\seqsubst{\type_p'}{\var'}{\var_f}}
    {\seqsubst{\type_p}{\var}{\var_f}}
    \fresh{\sequence{\var_f}}}
  {\teqv
    {\typedprod{\sequence{\notevar{\var'}{\sort}}}{\type_p'}}
    {\typedprod{\sequence{\notevar{\var}{\sort}}}{\type_p}}}\]
  \paragraph{Case {\sc Sigma}:}
  \[\infr[TEqv-Sigma]
  {\teqv
    {\seqsubst{\type_s}{\var}{\var_f}}
    {\seqsubst{\type_s'}{\var'}{\var_f}}
    \fresh{\sequence{\var_f}}}
  {\teqv
    {\typedsum{\sequence{\notevar{\var}{\sort}}}{\type_s}}
    {\typedsum{\sequence{\notevar{\var'}{\sort}}}{\type_s'}}}\]
  The induction hypothesis gives
  $\teqv
  {\seqsubst{\type_s'}{\var'}{\var_f}}
  {\seqsubst{\type_s}{\var}{\var_f}}$,
  so we construct
  \[\infr[TEqv-Sigma]
  {\teqv
    {\seqsubst{\type_s'}{\var'}{\var_f}}
    {\seqsubst{\type_s}{\var}{\var_f}}
    \fresh{\sequence{\var_f}}}
  {\teqv
    {\typedsum{\sequence{\notevar{\var'}{\sort}}}{\type_s'}}
    {\typedsum{\sequence{\notevar{\var}{\sort}}}{\type_s}}}\]
\end{sproof}

\begin{lemma}[Transitivity of $\teqv{}{}$]
  \label{TransTEqv}
  If $\teqv{\type_0}{\type_1}$ and $\teqv{\type_1}{\type_2}$,
  then $\teqv{\type_0}{\type_2}$.
\end{lemma}
\begin{sproof}[This follows from induction on
  the derivations of $\teqv{\type_0}{\type_1}$ and $\teqv{\type_1}{\type_2}$.
  Since both derivations $\type_1$,
  the structure of the equivalence rules prohibits the derivations
  from ending with different rules
  (other than {\sc TEqv-Refl}, which passes that structural requirement on to its premises).]
  We use induction on the derivations of $\teqv{\type_0}{\type_1}$ and $\teqv{\type_1}{\type_2}$.
  Since both derivations involve $\type_1$,
  the structure of the equivalence rules prohibits the derivations from ending with different rules,
  unless one is {\sc TEqv-Refl}.
  All non-reflexivity rules place incompatible restrictions on the types they find equivalent.
  If either derivation is by {\sc TEqv-Refl},
  then $\type_1$ is either $\type_0$ or $\type_2$,
  so the conclusion $\teqv{\type_0}{\type_2}$ is the same as one of the assumptions.
  In the remaining cases, both derivations end with the same rule.
  \paragraph{Case {\sc Array}:}
  \[
  \infr[TEqv-Array]
  {\teqv{\type_0'}{\type_1'} \qquad {\ieqv{\idx_0}{\idx_1}}}
  {\teqv{\typearray{\type_0'}{\idx_0}}{\typearray{\type_1'}{\idx_1}}}\]
  \ctrand
  \[\infr[TEqv-Array]
  {\teqv{\type_1'}{\type_2'} \qquad {\ieqv{\idx_1}{\idx_2}}}
  {\teqv{\typearray{\type_1'}{\idx_1}}{\typearray{\type_2'}{\idx_2}}}
  \]
  Since $\teqv{\type_0'}{\type_1'}$ and $\teqv{\type_1'}{\type_2'}$,
  the induction hypothesis implies $\teqv{\type_0'}{\type_2'}$.
  Transitivity of equality (on indices) implies $\ieqv{\idx_0}{\idx_2}$.
  So we can derive
  \[\infr[TEqv-Array]
  {\teqv{\type_0'}{\type_2'} \qquad {\ieqv{\idx_0}{\idx_2}}}
  {\teqv{\typearray{\type_0'}{\idx_0}}{\typearray{\type_2'}{\idx_2}}}\]
  \paragraph{Case {\sc Fn}:}
  \[\infr[TEqv-Fn]
  {\sequence{\teqv{\type_{i_0}}{\type_{i_1}}}
    \qquad {\teqv{\type_{o_0}}{\type_{o_1}}}}
  {\teqv
    {\typefun{\sequence{\type_{i_0}}}{\type_{o_0}}}
    {\typefun{\sequence{\type_{i_1}}}{\type_{o_1}}}}\]
  \ctrand
  \[\infr[TEqv-Fn]
  {\sequence{\teqv{\type_{i_1}}{\type_{i_2}}}
    \qquad {\teqv{\type_{o_1}}{\type_{o_2}}}}
  {\teqv
    {\typefun{\sequence{\type_{i_1}}}{\type_{o_1}}}
    {\typefun{\sequence{\type_{i_2}}}{\type_{o_2}}}}\]
  We have equivalent argument types, via the induction hypothesis---%
  each $\sequence{\teqv{\type_{i_0}}{\type_{i_1}}}$ and $\sequence{\teqv{\type_{i_1}}{\type_{i_2}}}$
  implies $\sequence{\teqv{\type_{i_0}}{\type_{i_2}}}$.
  We also have equivalent result types, $\sequence{\teqv{\type_{o_0}}{\type_{o_2}}}$.
  Then {\sc TEqv-Fn} gives
  \[\infr[TEqv-Fn]
  {\sequence{\teqv{\type_{i_0}}{\type_{i_2}}} \qquad {\teqv{\type_{o_0}}{\type_{o_2}}}}
  {\teqv{\typefun{\sequence{\type_{i_0}}}{\type_{o_0}}}{\typefun{\sequence{\type_{i_2}}}{\type_{o_2}}}}\]
  \paragraph{Case {\sc Univ}:}
  \[\infr[TEqv-Univ]
  {\teqv
    {\seqsubst{\type_{u_0}}{\var_0}{\var_f}}
    {\seqsubst{\type_{u_1}}{\var_1}{\var_f}}}
  {\teqv
    {\typeuniv{\sequence{\notevar{\var_0}{\kind}}}{\type_{u_0}}}
    {\typeuniv{\sequence{\notevar{\var_1}{\kind}}}{\type_{u_1}}}}\]
  \ctrand
  \[\infr[TEqv-Univ]
  {\teqv
    {\seqsubst{\type_{u_1}}{\var_1}{\var_f}}
    {\seqsubst{\type_{u_2}}{\var_2}{\var_f}}}
  {\teqv
    {\typeuniv{\sequence{\notevar{\var_1}{\kind}}}{\type_{u_1}}}
    {\typeuniv{\sequence{\notevar{\var_2}{\kind}}}{\type_{u_2}}}}\]
  The induction hypothesis relates
  ${\seqsubst{\type_{u_0}}{\var_0}{\var_f}}$ and ${\seqsubst{\type_{u_2}}{\var_2}{\var_f}}$,
  so we can construct the derivation
  \[\infr[TEqv-Univ]
  {\teqv
    {\seqsubst{\type_{u_0}}{\var_0}{\var_f}}
    {\seqsubst{\type_{u_2}}{\var_2}{\var_f}}}
  {\teqv
    {\typeuniv{\sequence{\notevar{\var_0}{\kind}}}{\type_{u_0}}}
    {\typeuniv{\sequence{\notevar{\var_2}{\kind}}}{\type_{u_2}}}}\]
  \paragraph{Case {\sc Pi}:}
  \[\infr[TEqv-Pi]
  {\teqv
    {\seqsubst{\type_{p_0}}{\var_0}{\var_f}}
    {\seqsubst{\type_{p_1}}{\var_1}{\var_f}}}
  {\teqv
    {\typedprod{\sequence{\notevar{\var_0}{\sort}}}{\type_{p_0}}}
    {\typedprod{\sequence{\notevar{\var_1}{\sort}}}{\type_{p_1}}}}\]
  \ctrand
  \[\infr[TEqv-Pi]
  {\teqv
    {\seqsubst{\type_{p_1}}{\var_1}{\var_f}}
    {\seqsubst{\type_{p_2}}{\var_2}{\var_f}}}
  {\teqv
    {\typedprod{\sequence{\notevar{\var_1}{\kind}}}{\type_{u_1}}}
    {\typedprod{\sequence{\notevar{\var_2}{\kind}}}{\type_{u_2}}}}\]
  The induction hypothesis implies
  ${\seqsubst{\type_{u_0}}{\var_0}{\var_f}}$ and ${\seqsubst{\type_{u_2}}{\var_2}{\var_f}}$
  are equivalent,
  so we can apply {\sc TEqv-Pi}:
  \[\infr[TEqv-Pi]
  {\teqv
    {\seqsubst{\type_{u_0}}{\var_0}{\var_f}}
    {\seqsubst{\type_{u_2}}{\var_2}{\var_f}}}
  {\teqv
    {\typedprod{\sequence{\notevar{\var_0}{\kind}}}{\type_{u_0}}}
    {\typedprod{\sequence{\notevar{\var_2}{\kind}}}{\type_{u_2}}}}\]
  \paragraph{Case {\sc Sigma}:}
  \[\infr[TEqv-Sigma]
  {\teqv
    {\seqsubst{\type_{s_0}}{\var_0}{\var_f}}
    {\seqsubst{\type_{s_1}}{\var_1}{\var_f}}}
  {\teqv
    {\typedsum{\sequence{\notevar{\var_0}{\sort}}}{\type_{s_0}}}
    {\typedsum{\sequence{\notevar{\var_1}{\sort}}}{\type_{s_1}}}}\]
  \ctrand
  \[\infr[TEqv-Sigma]
  {\teqv
    {\seqsubst{\type_{s_1}}{\var_1}{\var_f}}
    {\seqsubst{\type_{s_2}}{\var_2}{\var_f}}}
  {\teqv
    {\typedsum{\sequence{\notevar{\var_1}{\sort}}}{\type_{s_1}}}
    {\typedsum{\sequence{\notevar{\var_2}{\sort}}}{\type_{s_2}}}}\]
  By the induction hypothesis,
  $\teqv{\seqsubst{\type_{s_0}}{\var_0}{\var_f}}{\seqsubst{\type_{s_2}}{\var_2}{\var_f}}$,
  allowing us to derive
  \[\infr[TEqv-Sigma]
  {\teqv
    {\seqsubst{\type_{s_0}}{\var_0}{\var_f}}
    {\seqsubst{\type_{s_2}}{\var_2}{\var_f}}}
  {\teqv
    {\typedsum{\sequence{\notevar{\var_0}{\sort}}}{\type_{s_0}}}
    {\typedsum{\sequence{\notevar{\var_2}{\sort}}}{\type_{s_2}}}}\]
\end{sproof}

\begin{theorem}
  \label{EqvTEqv}
  $\teqv{}{}$ is an equivalence relation.
\end{theorem}

A type-equivalence relation should not cross kind boundaries.
Violation of this principle would allow
use of {\sc T-Eqv} to ascribe an ill-kinded type to a well-typed term.
It follows directly from inspection of the equivalence rules that
they will not relate an $\kindatom$ with an $\kindarray$,
but correct use of type and index variables remains to be proven.
To that end, we show that two equivalent types
will be ascribed the same kind by the same environment.

\begin{lemma}
  \label{TEqvRespectsKinds}
  If $\kindof{\sEnv}{\kEnv}{\type}{\kind}$
  and $\teqv{\type}{\type'}$,
  then $\kindof{\sEnv}{\kEnv}{\type'}{\kind}$.
\end{lemma}
\begin{sproof}[This result is proven induction on
  the derivation of $\teqv{\type}{\type'}$.
  In each case,
  the induction hypothesis converts a kind derivation
  for some fragment of $\type$
  into a kind derivation for a corresponding fragment of $\type'$
  (and similar for index fragments).]
  We use induction on the derivation of
  $\teqv{\type}{\type'}$.
  \paragraph{Case {\sc Refl}:}
  \[{\infr[TEqv-Refl]
    {\ }
    {\teqv{\type}{\type}}}\]
  Then $\type = \type'$,
  so the kind derivation for $\type$ applies to $\type'$ as well.
  \paragraph{Case {\sc Array}:}
  \[{\infr[TEqv-Array]
    {\teqv{\type_a}{\type_a'}
    \qquad
    \ieqv{\idx_a}{\idx_a'}}
    {\teqv
      {\typearray{\type_a}{\idx_a}}
      {\typearray{\type_a'}{\idx_a'}}}}\]
  The kind derivation for $\type$ must have the form
  \[{\infr[K-Array]
    {\kindof{\sEnv}{\kEnv}{\type_a}{\kindatom}
      \qquad
      \sortof{\sEnv}{\idx_a}{\sortshp}}
    {\kindof{\sEnv}{\kEnv}{\typearray{\type_a}{\idx_a}}{\kindarray}}}\]
  Similarly, equivalent indices must be of the same sort
  (dimensions and shapes are distinct objects in our universe of type indices).
  Applying the induction hypothesis to our kinding result for $\type_a'$
  gives $\kindof{\sEnv}{\kEnv}{\type_a'}{\kindatom}$.
  Then we can derive for $\type'$:
  \[{\infr[K-Array]
    {\kindof{\sEnv}{\kEnv}{\type_a'}{\kindatom}
      \qquad
      \sortof{\sEnv}{\idx_a'}{\sortshp}}
    {\kindof{\sEnv}{\kEnv}{\typearray{\type_a'}{\idx_a'}}{\kindarray}}}\]
  \paragraph{Case {\sc Fn}:}
  \[{\infr[TEqv-Fn]
    {\sequence{\teqv{\type_i}{\type_i'}}
      \qquad
      \teqv{\type_o}{\type_o'}}
    {\teqv
      {\typefun{\sequence{\type_i}}{\type_o}}
      {\typefun{\sequence{\type_i'}}{\type_o'}}}}\]
  The kinding derivation for $\type$ has the form
  \[{\infr[K-Fun]
    {\sequence{\kindof{\sEnv}{\kEnv}{\type_i}{\kindarray}}
      \qquad
      \kindof{\sEnv}{\kEnv}{\type_o}{\kindarray}}
    {\kindof{\sEnv}{\kEnv}{\typefun{\sequence{\type_i}}{\type_o}}{\kindatom}}}\]
  By the induction hypothesis,
  we have kind derivations ascribing $\kindarray$
  to each of the $\sequence{\type_i}$ and $\type_o$.
  So we can then derive
  \[{\infr[K-Fun]
    {\sequence{\kindof{\sEnv}{\kEnv}{\type_i'}{\kindarray}}
      \qquad
      \kindof{\sEnv}{\kEnv}{\type_o'}{\kindarray}}
    {\kindof{\sEnv}{\kEnv}{\typefun{\sequence{\type_i'}}{\type_o'}}{\kindatom}}}\]
  \paragraph{Case {\sc Univ}:}
  \[{\infr[TEqv-Univ]
    {\teqv
      {\seqsubst{\type_u}{\var}{\var_f}}
      {\seqsubst{\type_u'}{\var'}{\var_f}}
      \fresh{\sequence{\var_f}}}
    {\teqv
      {\typeuniv{\sequence{\notevar{\var}{\kind}}}{\type_u}}
      {\typeuniv{\sequence{\notevar{\var'}{\kind}}}{\type_u'}}}}\]
  The kind derivation for $\type$ must have the form
  \[{\infr[K-Univ]
    {\kindof{\sEnv}{\kEnv, \sequence{\haskind{\var}{\kind}}}{\type_u}{\kindarray}}
    {\kindof{\sEnv}{\kEnv}
      {\typeuniv{\sequence{\notevar{\var}{\kind}}}{\type_u}}
      {\kindatom}}}\]
  The premise implies via $\alpha$-conversion in turn that
  $\kindof{\sEnv}{\kEnv, \sequence{\haskind{\var_f}{\kind}}}
  {\seqsubst{\type_u}{\var}{\var_f}}
  {\kindarray}$.
  By the induction hypothesis,
  $\kindof{\sEnv}{\kEnv, \sequence{\haskind{\var_f}{\kind}}}
  {\seqsubst{\type_u'}{\var'}{\var_f}}
  {\kindarray}$.
  Using $\alpha$-conversion again,
  converting $\seqsubst{\type_u'}{\var'}{\var_f}$ to
  $\seqsubst{\parens{\seqsubst{\type_u'}{\var'}{\var_f}}}{\var_f}{\var'}$,
  which is $\type_u'$,
  we get
  $\kindof{\sEnv}{\kEnv, \sequence{\haskind{\var'}{\kind}}}
  {\type_u'}
  {\kindarray}$.
  Then we can derive
  \[{\infr[K-Univ]
    {\kindof{\sEnv}{\kEnv, \sequence{\haskind{\var'}{\kind}}}
      {\type_u'}
      {\kindarray}}
    {\kindof{\sEnv}{\kEnv}
      {\typeuniv{\sequence{\notevar{\var'}{\kind}}}{\type_u'}}
      {\kindatom}}}\]
  \paragraph{Case {\sc Pi}:}
  \[{\infr[TEqv-Pi]
    {\teqv
      {\seqsubst{\type_p}{\var}{\var_f}}
      {\seqsubst{\type_p'}{\var'}{\var_f}}
      \fresh{\sequence{\var_f}}}
    {\teqv
      {\typedprod{\sequence{\notevar{\var}{\kind}}}{\type_p}}
      {\typedprod{\sequence{\notevar{\var'}{\kind}}}{\type_p'}}}}\]
  Deriving a kind for $\type$ must end with
  \[{\infr[K-Pi]
    {\kindof{\sEnv, \sequence{\hassort{\var}{\sort}}}{\kEnv}
      {\type_p}
      {\kindarray}}
    {\kindof{\sEnv}{\kEnv}
      {\typedprod{\sequence{\notevar{\var}{\kind}}}{\type_p}}
      {\kindatom}}}\]
  By $\alpha$-converting $\sequence{\var}$ to $\sequence{\var_f}$,
  we get the judgment
  $\kindof{\sEnv, \sequence{\hassort{\var_f}{\sort}}}{\kEnv}
  {\seqsubst{\type_p}{\var}{\var_f}}
  {\kindarray}$,
  implying via the induction hypothesis that
  $\kindof{\sEnv, \sequence{\hassort{\var_f}{\sort}}}{\kEnv}
  {\seqsubst{\type_p'}{\var'}{\var_f}}
  {\kindarray}$.
  Another round of $\alpha$-conversion gives
  $\kindof{\sEnv, \sequence{\hassort{\var'}{\sort}}}{\kEnv}
  {\type_p'}
  {\kindarray}$,
  so we derive
  \[{\infr[K-Pi]
    {\kindof{\sEnv, \sequence{\hassort{\var'}{\sort}}}{\kEnv}
      {\type_p'}
      {\kindarray}}
    {\kindof{\sEnv}{\kEnv}
      {\typedprod{\sequence{\notevar{\var'}{\kind}}}{\type_p'}}
      {\kindatom}}}\]
  \paragraph{Case {\sc Sigma}:}
  \[{\infr[TEqv-Sigma]
    {\teqv
      {\seqsubst{\type_s}{\var}{\var_f}}
      {\seqsubst{\type_s'}{\var'}{\var_f}}
      \fresh{\sequence{\var_f}}}
    {\teqv
      {\typedsum{\sequence{\notevar{\var}{\kind}}}{\type_s}}
      {\typedsum{\sequence{\notevar{\var'}{\kind}}}{\type_s'}}}}\]
  We also have a kind derivation for $\type$
  \[{\infr[K-Sigma]
    {\kindof{\sEnv, \sequence{\hassort{\var}{\sort}}}{\kEnv}
      {\type_s}
      {\kindarray}}
    {\kindof{\sEnv}{\kEnv}
      {\typedsum{\sequence{\notevar{\var}{\kind}}}{\type_s}}
      {\kindatom}}}\]
  As in the previous two cases,
  applying the induction hypothesis under $\alpha$-conversion gives
  $\kindof{\sEnv, \sequence{\hassort{\var'}{\sort}}}{\kEnv}
  {\type_s'}
  {\kindarray}$,
  leading to the derivation
  \[{\infr[K-Sigma]
    {\kindof{\sEnv, \sequence{\hassort{\var'}{\sort}}}{\kEnv}
      {\type_s'}
      {\kindarray}}
    {\kindof{\sEnv}{\kEnv}
      {\typedsum{\sequence{\notevar{\var'}{\kind}}}{\type_s'}}
      {\kindatom}}}\]
\end{sproof}

We also expect type equivalence to be well-behaved under substitution.
Ultimately, substituting equivalent types or indices into equivalent types
ought to produce equivalent types.
Proving that result by induction on derivation of equivalence is straightforward
except for the {\sc Refl} case.

\begin{lemma}
  \label{EqlIdxEqvType}
  If $\ieqv{\idx}{\idx'}$,
  then for any index variable $\var$,
  $\teqv{\subst{\type}{\var}{\idx}}{\subst{\type}{\var}{\idx'}}$.
\end{lemma}
\begin{sproof}[This is provable using induction on the structure of $\type$.
  Only the case for arrays makes direct use of $\idx$ and $\idx'$;
  the other cases simply use the induction hypothesis
  to prove the premises of the derivation of
  $\teqv{\subst{\type}{\var}{\idx}}{\subst{\type}{\var}{\idx'}}$.]
  This is provable using induction on the structure of $\type$.
  Only the case for arrays makes direct use of $\idx$ and $\idx'$;
  the other cases simply use the induction hypothesis
  to prove the premises of the derivation of
  $\teqv{\subst{\type}{\var}{\idx}}{\subst{\type}{\var}{\idx'}}$.
  \paragraph{Case {\sc Array}:}
  $\type = \typearray{\type_a}{\idx_a}$,
  so the induction hypothesis implies $\teqv{\subst{\type_a}{\var}{\idx}}{\subst{\type_a}{\var}{\idx'}}$.
  $\ieqv{\subst{\idx_a}{\var}{\idx}}{\subst{\idx_a}{\var}{\idx'}}$
  follows from the substitution lemma of first-order logic.
  Then we can construct the type equivalence derivation
  \[\infr[TEqv-Array]
  {\teqv{\subst{\type_a}{\var}{\idx}}{\subst{\type_a}{\var}{\idx'}}
    \qquad
    \ieqv{\subst{\idx_a}{\var}{\idx}}{\subst{\idx_a}{\var}{\idx'}}}
  {\teqv
    {\typearray{\subst{\type_a}{\var}{\idx}}{\subst{\idx_a}{\var}{\idx}}}
    {\typearray{\subst{\type_a}{\var}{\idx'}}{\subst{\idx_a}{\var}{\idx'}}}}\]
\end{sproof}

\begin{lemma}
  \label{EqlTypeEqvType}
  If $\teqv{\type_{\var}}{\type_{\var}'}$,
  then for any type variable $\var$,
  $\teqv
  {\subst{\type}{\var}{\type_{\var}}}
  {\subst{\type}{\var}{\type_{\var}'}}$.
\end{lemma}
\begin{sproof}[\BODY]
  We use induction on the structure of $\type$.
  The cases for universals, dependent products, and dependent sums
  require instantiating the induction hypothesis 
  with a substitution of fresh type variables $\sequence{\var_f}$.
  For example, when $\type = \typeuniv{\sequence{\notevar{\var_u}{\kind}}}{\type_u}$,
  the induction hypothesis promises the equivalence of $\type_u$
  after substituting in $\sequence{\var_f}$ for $\sequence{\var_u}$
  and \emph{also} $\type_x$ or $\type_x'$ for $\var$.
\end{sproof}

\begin{theorem}
  \label{EqvTypeEqvType}
  If $\teqv{\type}{\type'}$ and $\teqv{\type_{\var}}{\type_{\var}'}$,
  then for any type variable $\var$,
  $\teqv
  {\subst{\type}{\var}{\type_{\var}}}
  {\subst{\type'}{\var}{\type_{\var}'}}$.
\end{theorem}
\begin{sproof}[We use induction on the derivation of $\teqv{\type}{\type'}$.
  In each case, the induction hypothesis provides equivalence derivations
  for corresponding fragments of
  $\subst{\type}{\var}{\type_{\var}}$ and $\subst{\type'}{\var}{\type_{\var}'}$,
  which can then be used to prove the substituted types themselves equivalent.]
  We use induction on the derivation of $\teqv{\type}{\type'}$.
  \paragraph{Case {\sc Refl}:}
  $\type = \type'$.
  By Lemma \ref{EqlTypeEqvType},
  $\teqv{\subst{\type}{\var}{\type_{\var}}}{\subst{\type}{\var}{\type_{\var}'}}$.
  \paragraph{Case {\sc Array}:}
  \[\infr[TEqv-Array]
  {\teqv{\type_a}{\type_a'} \qquad \ieqv{\idx_a}{\idx_a'}}
  {\teqv{\typearray{\type_a}{\idx_a}}{\typearray{\type_a'}{\idx_a'}}}\]
  The induction hypothesis gives
  $\teqv{\subst{\type_a}{\var}{\type_{\var}}}{\subst{\type_a'}{\var}{\type_{\var}'}}$,
  so {\sc TEqv-Array} derives
  \[\infr[TEqv-Array]
  {\teqv{\subst{\type_a}{\var}{\type_{\var}}}{\subst{\type_a'}{\var}{\type_{\var}'}}
    \qquad \ieqv{\idx_a}{\idx_a'}}
  {\teqv
    {\typearray{\subst{\type_a}{\var}{\type_{\var}}}{\idx_a}}
    {\typearray{\subst{\type_a'}{\var}{\type_{\var}'}}{\idx_a'}}}\]
  \paragraph{Case {\sc Fn}:}
  \[\infr[TEqv-Fn]
  {\sequence{\teqv{\type_i}{\type_i'}} \qquad \teqv{\type_o}{\type_o'}}
  {\teqv
    {\typefun{\sequence{\type_i}}{\type_o}}
    {\typefun{\sequence{\type_i}}{\type_o}}}\]
  The induction hypothesis gives derivations for
  equivalence of corresponding input types after substitution,
  $\sequence{\teqv{\subst{\type_i}{\var}{\type_{\var}}}{\subst{\type_i'}{\var}{\type_{\var}'}}}$,
  as well as equivalence for output types,
  ${\subst{\type_o}{\var}{\type_{\var}}}$
  ${\teqv{}{}}$
  ${\subst{\type_o'}{\var}{\type_{\var}'}}$.
  Then applying {\sc TEqv-Fn} gives
  \[\infr[TEqv-Fn]
  {\sequence{\teqv{\subst{\type_i}{\var}{\type_{\var}}}{\subst{\type_i'}{\var}{\type_{\var}'}}}
    \qquad \teqv{\subst{\type_o}{\var}{\type_{\var}}}{\subst{\type_o'}{\var}{\type_{\var}'}}}
  {\teqv
    {\typefun{\sequence{\subst{\type_i}{\var}{\type_{\var}}}}{\subst{\type_o}{\var}{\type_{\var}}}}
    {\typefun{\sequence{\subst{\type_i'}{\var}{\type_{\var}'}}}{\subst{\type_o'}{\var}{\type_{\var}'}}}}\]
  \paragraph{Case {\sc Univ}:}
  \[\infr[TEqv-Univ]
  {\teqv
    {\seqsubst{\type_u}{\var_u}{\var_f}}
    {\seqsubst{\type_u'}{\var_u'}{\var_f}}}
  {\teqv
    {\typeuniv{\sequence{\notevar{\var_u}{\kind}}}{\type_u}}
    {\typeuniv{\sequence{\notevar{\var_u'}{\kind}}}{\type_u'}}}\]
  If $\var$ is shadowed by $\sequence{\var_u}$
  but not by $\sequence{\var_u'}$ (or vice versa),
  then the universal types $\type$ and $\type'$ will not be equivalent,
  as a type variable is only equivalent to itself (via {\sc TEqv-Refl}).
  If $\var$ does not appear at all in both $\type_u$ and $\type_u'$,
  then substitution leaves them unchanged,
  and our required conclusion is an initial assumption.
  So we now assume that $\var$ is free in both $\type_u$ and $\type_u'$.
  When we substitute in $\type_{\var}$ and $\type_{\var}'$,
  the induction hypothesis implies
  $\teqv
  {\subst{\seqsubst{\type_u}{\var_u}{\var_f}}{\var}{\type_{\var}}}
  {\subst{\seqsubst{\type_u'}{\var_u'}{\var_f}}{\var}{\type_{\var}'}}$.
  These types are equal to
  $\seqsubst{\subst{\type_u}{\var}{\type_{\var}}}{\var_u}{\var_f}$
  and $\seqsubst{\subst{\type_u'}{\var}{\type_{\var}'}}{\var_u'}{\var_f}$ respectively,
  so we then use {\sc TEqv-Univ}
  \[\infr[TEqv-Univ]
  {\teqv
    {\seqsubst{\subst{\type_u}{\var}{\type_{\var}}}{\var_u}{\var_f}}
    {\seqsubst{\subst{\type_u'}{\var}{\type_{\var}'}}{\var_u'}{\var_f}}}
  {\parbox{0.6\textwidth}
    {\centering
      \({\typeuniv{\sequence{\notevar{\var_u}{\kind}}}{\subst{\type_u}{\var}{\type_{\var}}}}\)\\
      \(\teqv{}{}
      {\typeuniv{\sequence{\notevar{\var_u'}{\kind}}}{\subst{\type_u'}{\var}{\type_{\var}'}}}\)}}\]
  \paragraph{Case {\sc Pi}:}
  \[{\infr[TEqv-Pi]
    {\teqv
      {\seqsubst{\type_p}{\var_p}{\var_f}}
      {\seqsubst{\type_p'}{\var_p'}{\var_f}}}
    {\teqv
      {\typedprod{\notevar{\var_p}{\sort}}{\type_p}}
      {\typedprod{\notevar{\var_p'}{\sort}}{\type_p'}}}}\]
  Shadowing $\var$ is no longer a concern because
  it is a type variable,
  whereas the dependent product only binds index variables.
  Substituting $\type_{\var}$ and $\type_{\var}'$ into the $\alpha$-converted body types
  produces $\subst{\seqsubst{\type_p}{\var_p}{\var_f}}{\var}{\type_{\var}}$
  and $\subst{\seqsubst{\type_p'}{\var_p'}{\var_f}}{\var}{\type_{\var}'}$,
  which the induction hypothesis implies are equivalent.
  They are in turn equal to
  $\seqsubst{\subst{\type_p}{\var}{\type_{\var}}}{\var_p}{\var_f}$
  and $\seqsubst{\subst{\type_p'}{\var}{\type_{\var}'}}{\var_p'}{\var_f}$
  respectively.
  We then construct the derivation
  \[{\infr[TEqv-Pi]
    {\teqv
      {\seqsubst{\subst{\type_p}{\var}{\type_{\var}}}{\var_p}{\var_f}}
      {\seqsubst{\subst{\type_p'}{\var}{\type_{\var}'}}{\var_p'}{\var_f}}}
    {\teqv
      {\typedprod{\notevar{\var_p}{\sort}}{\subst{\type_p}{\var}{\type_{\var}}}}
      {\typedprod{\notevar{\var_p}{\sort}}{\subst{\type_p'}{\var}{\type_{\var}'}}}}}\]
  \paragraph{Case {\sc Sigma}:}
  This case proceeds as the {\sc Pi} case.
\end{sproof}

Having defined the typing judgment
and the type-equivalence relation on which it builds,
we can now prove the usual results about typing in Remora.
The {\sc T-Eqv} rule can allow many types to be ascribed to a single term,
but we will prove that an environment and term
can only map to a single equivalence class.

\begin{theorem}[Uniqueness of typing, up to equivalence]
  \label{UniqType}
  If $\typeof{\sEnv}{\kEnv}{\tEnv}{\term}{\type}$
  and $\typeof{\sEnv}{\kEnv}{\tEnv}{\term}{\type'}$,
  then $\teqv{\type}{\type'}$.
\end{theorem}
\begin{sproof}[This can be proven by induction on $\term$,
  showing that all derivations of
  $\typeof{\sEnv}{\kEnv}{\tEnv}{\term}{\type'}$
  must end with the same non-{\sc T-Eqv} rule
  (chosen according to the structure of $\term$)
  followed by 0 or more {\sc T-Eqv} instances,
  which keeps the result in the same equivalence class as $\type$.]
  Any derivation of $\typeof{\sEnv}{\kEnv}{\tEnv}{\term}{\type}$
  can be followed with zero or more additional uses of {\sc T-Eqv}
  or have any terminal uses of {\sc T-Eqv} stripped off
  to derive another type.
  Since type equivalence is transitive, any alternative type $\type'$ derived in this way
  is equivalent to $\type$.
  It remains to show, by induction on $\term$,
  that all derivations for $\typeof{\sEnv}{\kEnv}{\tEnv}{\term}{\type'}$ 
  ending in a non-{\sc Eqv} rule
  ascribe a $\type'$ equivalent to $\type$.
  \paragraph{Variable:}
  $\term = \var$,
  so the only applicable rule is {\sc T-Var},
  which ascribes $\tEnv(\var)$.
  \paragraph{Array:}
  $\term = \arrlit{\sequence{\atom}}{\sequence{\nat}}$.
  Only {\sc T-Array} can derive a type,
  so the derivations for $\type$ and $\type'$ are
  \[\infr[T-Array]
  {\sequence{\typeof{\sEnv}{\kEnv}{\tEnv}{\atom}{\type_a}}
    \qquad \kindof{\sEnv}{\kEnv}{\type_a}{\kindatom}
    \qquad \mathit{Length}\llb \sequence{\atom} \rrb = \prod{\sequence{\nat}}}
  {\typeof{\sEnv}{\kEnv}{\tEnv}
    {\arrlit{\sequence{\atom}}{\sequence{\nat}}}
    {\typearray{\type_a}{\idxshape{\sequence{\nat}}}}}\]
  \ctrand
  \[\infr[T-Array]
  {\sequence{\typeof{\sEnv}{\kEnv}{\tEnv}{\atom}{\type_a'}}
    \qquad \kindof{\sEnv}{\kEnv}{\type_a'}{\kindatom}
    \qquad \mathit{Length}\llb \sequence{\atom} \rrb = \prod{\sequence{\nat}}}
  {\typeof{\sEnv}{\kEnv}{\tEnv}
    {\arrlit{\sequence{\atom}}{\sequence{\nat}}}
    {\typearray{\type_a'}{\idxshape{\sequence{\nat}}}}}\]
  By the induction hypothesis,
  the derivations for each $\atom$ must produce equivalent types,
  \ie, $\teqv{\type_a}{\type_a'}$.
  Thus we can derive
  \[\infr[TEqv-Array]
  {\teqv{\type_a}{\type_a'}
    \qquad \ieqv{\idxshape{\sequence{\nat}}}{\idxshape{\sequence{\nat}}}}
  {\teqv
    {\typearray{\type_a}{\idxshape{\sequence{\nat}}}}
    {\typearray{\type_a'}{\idxshape{\sequence{\nat}}}}}\]
  \paragraph{Frame:}
  $\term = \frm{\sequence{\expr}}{\sequence{\nat}}$.
  Only {\sc T-Frame} is applicable, so the derivations are
  \[\infr[T-Frame]
  {\sequence{\typeof
      {\sEnv}{\kEnv}{\tEnv}
      {\expr}{\typearray{\type_a}{\idx_c}}}
    \qquad \kindof{\sEnv}{\kEnv}{\typearray{\type_a}{\idx_c}}{\kindarray}
    \\\\ \mathit{Length}\llb \sequence{\atom} \rrb = \prod{\sequence{\nat}}}
  {\typeof{\sEnv}{\kEnv}{\tEnv}
    {\frm{\sequence{\expr}}{\sequence{\nat}}}
    {\typearray
      {\type_a}
      {\idxappend{\idxshape{\sequence{\nat}} \; \idx_c}}}}\]
  \ctrand
  \[\infr[T-Frame]
  {\sequence{\typeof
      {\sEnv}{\kEnv}{\tEnv}
      {\expr}{\typearray{\type_a'}{\idx_c'}}}
    \qquad \kindof{\sEnv}{\kEnv}{\typearray{\type_a'}{\idx_c'}}{\kindarray}
    \\\\ \mathit{Length}\llb \sequence{\atom} \rrb = \prod{\sequence{\nat}}}
  {\typeof{\sEnv}{\kEnv}{\tEnv}
    {\frm{\sequence{\expr}}{\sequence{\nat}}}
    {\typearray
      {\type_a'}
      {\idxappend{\idxshape{\sequence{\nat}} \; \idx_c'}}}}\]
  By the induction hypothesis,
  $\teqv {\typearray{\type_a}{\idx_c}}{\typearray{\type_a'}{\idx_c'}}$.
  The derivation for this must either end with {\sc TEqv-Array}
  or be simply {\sc TEqv-Refl}.
  In either case, we have $\teqv{\type_a}{\type_a'}$
  and $\ieqv{\idx_c}{\idx_c'}$.
  The latter implies
  $\ieqv
  {\idxappend{\idxshape{\sequence{\nat}} \; \idx_c}}
  {\idxappend{\idxshape{\sequence{\nat}} \; \idx_c'}}$,
  which leads to the derivation
  \[\infr[TEqv-Array]
  {\teqv{\type_a}{\type_a'}
    \qquad
    \ieqv{\idxshape{\sequence{\nat}} \; \idx_c}{\idxshape{\sequence{\nat}} \; \idx_c'}}
  {\parbox{0.6\textwidth}
    {\centering
      \({\typearray{\type_a}{\idxappend{\idxshape{\sequence{\nat}} \; \idx_c}}}\)\\
      \(\teqv{}{}{\typearray{\type_a'}{\idxappend{\idxshape{\sequence{\nat}} \; \idx_c'}}}\)}}\]
  \paragraph{Empty array:}
  $\term = \emptyarrlit{\type_a}{\sequence{\nat}}$.
  The only possible non-{\sc Eqv} rule for ending a type derivation is
  \[\infr[T-EmptyA]
  {\kindof{\sEnv}{\kEnv}{\type_a}{\kindatom}
    \qquad 0 \in \sequence{\nat}}
  {\typeof{\sEnv}{\kEnv}{\tEnv}
    {\emptyarrlit{\type_a}{\sequence{\nat}}}
    {\typearray{\type_a}{\idxshape{\sequence{\nat}}}}}\]
  Note that there are no atoms for derivations to conclude
  have distinct (but equivalent) types.
  Instead, the exact atom type is specified in the term itself.
  Therefore all types derived for $\term$ by {\sc TEqv-EmptyA} are equal,
  thus equivalent.
  \paragraph{Empty frame:}
  $\term = \emptyfrm{\typearray{\type_a}{\idx}}{\sequence{\nat}}$.
  Similar to the empty array case,
  only one non-{\sc Eqv} end to the derivation is possible:
  \[\infr[T-EmptyF]
  {\kindof{\sEnv}{\kEnv}{\type_a}{\kindatom}
    \qquad \sortof{\sEnv}{\idx}{\sortshp}
    \qquad 0 \in \sequence{\nat}}
  {\typeof
    {\sEnv}{\kEnv}{\tEnv}
    {\emptyfrm{\typearray{\type_a}{\idx}}{\sequence{\nat}}}
    {\typearray{\type_a}{\idxshape{\sequence{\nat}}}}}\]
  Again, only one unique type is derivable.
  \paragraph{Application:}
  $\term = \app{\expr_f}{\sequence{\expr_a}}$
  The of a typing derivations for $\term$ end with
  \[\infr
  {\typeof
    {\sEnv}{\kEnv}{\tEnv}
    {\expr_f}
    {\typearray
      {\typefun
        {\sequence{\typearray{\type_i}{\idx_i}}}
        {\typearray{\type_o}{\idx_o}}}
      {\idx_f}}
    \\\\
    \sequence{\typeof
      {\sEnv}{\kEnv}{\tEnv}
      {\expr_a}
      {\typearray{\type_i}{\idxappend{\idx_a \; \idx_i}}}}
    \\\\
    \idx_p = \mathit{Max}\llb \idx_f \; \sequence{\idx_a} \rrb}
  {\typeof
    {\sEnv}{\kEnv}{\tEnv}
    {\app{\expr_f}{\sequence{\expr_a}}}
    {\typearray{\type_o}{\idxappend{\idx_p \; \idx_o}}}}\]
  \ctrand
  \[\infr
  {\typeof
    {\sEnv}{\kEnv}{\tEnv}
    {\expr_f}
    {\typearray
      {\typefun
        {\sequence{\typearray{\type_i'}{\idx_i'}}}
        {\typearray{\type_o'}{\idx_o'}}}
      {\idx_f'}}
    \\\\
    \sequence{\typeof
      {\sEnv}{\kEnv}{\tEnv}
      {\expr_a}
      {\typearray{\type_i'}{\idxappend{\idx_a' \; \idx_i'}}}}
    \\\\
    \idx_p' = \mathit{Max}\llb \idx_f' \; \sequence{\idx_a'} \rrb}
  {\typeof
    {\sEnv}{\kEnv}{\tEnv}
    {\app{\expr_f}{\sequence{\expr_a}}}
    {\typearray{\type_o'}{\idxappend{\idx_p' \; \idx_o'}}}}\]
  By the induction hypothesis, the types ascribed to $\expr_f$ are equivalent.
  Since their equivalence can only be concluded from {\sc TEqv-Fn} or {\sc TEqv-Refl},
  we know that corresponding input types---%
  $\typearray{\type_i}{\idx_i}$ and $\typearray{\type_i'}{\idx_i'}$---%
  are equivalent.
  This in turn implies the equivalence of corresponding atom types $\type_i$ and $\type_i'$
  and of corresponding argument cell shapes $\idx_i$ and $\idx_i'$
  (the array types could only be equivalent because of {\sc TEqv-Array} or {\sc TEqv-Refl}).
  For the same reason, we have
  equality of the function frames and of corresponding argument shapes.
  What we then need is equality of corresponding \emph{argument frames}.

  In the free monoid,
  $a \thappend b \theq a \thappend c$ implies $b \theq c$,
  and $b \thappend a \theq c \thappend a$ implies $b \theq c$
  (\nb, this does not hold in all monoids).
  In effect, if we know that prefixes match on two equal shapes,
  we can conclude that their suffixes match as well.
  Applied to our situation,
  this means that the equality of
  $\idxappend{\idx_a \; \idx_i}$ and $\idxappend{\idx_a' \; \idx_i'}$
  with equal $\idx_i$ and $\idx_i'$
  entails the equality of $\idx_a$ and $\idx_a'$.
  So any two frames derived for the same argument \emph{are} equal.
  Since all corresponding frames are equal,
  their respective maxima $\idx_p$ and $\idx_p'$ must also be equal.

  Returning to the derived function types,
  their equivalence also implies the equivalence of their output types.
  Since $teqv{\typearray{\type_o}{\idx_o}}{\typearray{\type_o'}{\idx_o'}}$,
  it must be the case that $\teqv{\type_o}{\type_o'}$
  and that $\ieqv{\idx_o}{\idx_o'}$.

  The equivalences $\teqv{\type_o}{\type_o'}$, $\ieqv{\idx_o}{\idx_o'}$, and $\ieqv{\idx_p}{\idx_p'}$
  lead to
  \[\infr
  {\teqv{\type_o}{\type_o'}
    \qquad \ieqv{\idxappend{\idx_p \; \idx_o}}{\idxappend{\idx_p' \; \idx_o'}}}
  {\teqv
    {\typearray{\type_o}{\idxappend{\idx_p \; \idx_o}}}
    {\typearray{\type_o'}{\idxappend{\idx_p' \; \idx_o'}}}}\]
  \paragraph{Type application:}
  $\term = \tapp{\expr_f}{\sequence{\type_a}}$.
  The derivations must have the form
  \[\infr[T-TApp]
  {\typeof{\sEnv}{\kEnv}{\tEnv}
    {\expr_f}
    {\typearray{\typeuniv{\sequence{\notevar{\var}{\kind}}}{\typearray{\type_u}{\idx_u}}}{\idx_f}}
    \\\\ \sequence{\kindof{\sEnv}{\kEnv}{\type_a}{\kind}}}
  {\typeof{\sEnv}{\kEnv}{\tEnv}
    {\tapp{\expr_f}{\sequence{\type_a}}}
    {\typearray{\seqsubst{\type_u}{\var}{\type_a}}{\idxappend{\idx_f\;\idx_u}}}}\]
  \ctrand
  \[\infr[T-TApp]
  {\typeof{\sEnv}{\kEnv}{\tEnv}
    {\expr_f}
    {\typearray{\typeuniv{\sequence{\notevar{\var'}{\kind}}}{\typearray{\type_u'}{\idx_u'}}}{\idx_f'}}
    \\\\ \sequence{\kindof{\sEnv}{\kEnv}{\type_a}{\kind}}}
  {\typeof{\sEnv}{\kEnv}{\tEnv}
    {\tapp{\expr_f}{\sequence{\type_a}}}
    {\typearray{\seqsubst{\type_u'}{\var'}{\type_a}}{\idxappend{\idx_f'\;\idx_u'}}}}\]
  From the induction hypothesis, we have
  ${\typearray{\typeuniv{\sequence{\notevar{\var}{\kind}}}{\typearray{\type_u}{\idx_u}}}{\idx_f}}$
  $\teqv{}{}$
  ${\typearray{\typeuniv{\sequence{\notevar{\var'}{\kind}}}{\typearray{\type_u'}{\idx_u'}}}{\idx_f'}}$,
  which then implies that
  ${\seqsubst{\type_u}{\var}{\var_f}}$
  $\teqv{}{}$
  ${\seqsubst{\type_u'}{\var'}{\var_f}}$.
  Because substituting equal types into equivalent types produces equivalent types
  ${\seqsubst{\parens{\seqsubst{\type_u}{\var}{\var_f}}}{\var_f}{\type_a}}$
  $\teqv{}{}$
  ${\seqsubst{\parens{\seqsubst{\type_u'}{\var'}{\var_f}}}{\var_f}{\type_a}}$.
  Applying both substitutions in order,
  we have
  $\teqv{\seqsubst{\type_u}{\var}{\type_a}}{\seqsubst{\type_u'}{\var'}{\type_a}}$.
  Equivalence of types derived for $\expr_f$ also implies
  equivalence of the result cell shapes $\idx_u$ and $\idx_u'$
  and of the frame shapes $\idx_f$ and $\idx_f'$.
  This means $\ieqv{\idxappend{\idx_f \; \idx_u}}{\idxappend{\idx_f' \; \idx_u'}}$.
  So we can derive
  \[\infr[TEqv-Array]
  {\teqv
    {\seqsubst{\type_u}{\var}{\type_a}}
    {\seqsubst{\type_u'}{\var'}{\type_a}}
    \\\\ \ieqv{\idxappend{\idx_f \; \idx_u}}{\idxappend{\idx_f' \; \idx_u'}}}
  {\parbox{0.42\textwidth}
    {\centering
      \({\typearray{\seqsubst{\type_u}{\var}{\type_a}}{\idxappend{\idx_f\;\idx_u}}}\)\\
      \(\teqv{}{}{\typearray{\seqsubst{\type_u'}{\var'}{\type_a}}{\idxappend{\idx_f'\;\idx_u'}}}\)}}\]
  \paragraph{Index application:}
  $\term = \iapp{\expr_f}{\sequence{\idx_a}}$.
  Non-{\sc Eqv} endings of the type derivations must have the form
  \[\infr[T-IApp]
  {\typeof
    {\sEnv}{\kEnv}{\tEnv}
    {\expr}
    {\typearray{\typedprod{\sequence{\notevar{\var}{\sort}}}{\typearray{\type_p}{\idx_p}}}{\idx_f}}
    \qquad
    \sequence{\sortof{\sEnv}{\idx_a}{\sort}}}
  {\typeof
    {\sEnv}{\kEnv}{\tEnv}
    {\iapp{\expr_f}{\sequence{\idx_a}}}
    {\typearray{\seqsubst{\type_p}{\var}{\idx}}{\idxappend{\idx_f\;\seqsubst{\idx_p}{\var}{\idx}}}}}\]
  In two typing derivations,
  $\expr_f$ is ascribed types
  $\typearray{\typedprod{\sequence{\notevar{\var}{\sort}}}{\typearray{\type_p}{\idx_p}}}{\idx_f}$
  and $\typearray{\typedprod{\sequence{\notevar{\var'}{\sort}}}{\typearray{\type_p'}{\idx_p'}}}{\idx_f'}$.
  The induction hypothesis implies that these two types are equivalent.
  Deriving this equivalence must use {\sc TEqv-Array},
  relating $\idx_f$ with $\idx_f'$
  and $\typearray{\type_p}{\idx_p}$ with $\typearray{\type_p'}{\idx_p'}$.
  By similar substitution-tracing arguments as in the type application case,
  we have $\teqv{\seqsubst{\type_p}{\var}{\idx_a}}{\seqsubst{\type_p'}{\var'}{\idx_a}}$
  and
  $\text{\sc Valid} \llbracket {\seqsubst{\idx_p}{\var}{\idx}}
  \theq {\seqsubst{\idx_p'}{\var}{\idx}} \rrbracket$,
  leading to
  \[\infr[TEqv-Pi]
  {\teqv
    {\seqsubst{\type_p}{\var}{\idx_a}}
    {\seqsubst{\type_p'}{\var'}{\idx_a}}
    \\\\
    \ieqv
    {\idxappend{\idx_f\;\seqsubst{\idx_p}{\var}{\idx}}}
    {\idxappend{\idx_f'\;\seqsubst{\idx_p'}{\var}{\idx}}}}
  {\parbox{0.55\textwidth}
    {\centering
      \({\typearray
        {\seqsubst{\type_p}{\var}{\idx}}
        {\idxappend{\idx_f\;\seqsubst{\idx_p}{\var}{\idx}}}}\)\\
      \(\teqv{}{}
      {\typearray
        {\seqsubst{\type_p'}{\var'}{\idx}}
        {\idxappend{\idx_f'\;\seqsubst{\idx_p'}{\var}{\idx}}}}\)}}\]
  \paragraph{Unboxing:}
  $\term = \dproj{\sequence{\var_i}}{\var_e}{\expr_s}{\expr_b}$.
  The only non-{\sc T-Eqv} rule which can type $\term$ is {\sc T-Unbox},
  so type derivations must end with
  \[\infr[]
  {\typeof
    {\sEnv}{\kEnv}{\tEnv}
    {\expr_s}
    {\typearray{\typedsum{\sequence{\notevar{\var_i'}{\sort}}}{\type_s}}{\idx_s}}
    \\\\
    \typeof
    {\sEnv,\sequence{\hassort{\var_i}{\sort}}}{\kEnv}{\tEnv,\hastype{\var_e}{\seqsubst{\type_s}{\var_i'}{\var_i}}}
    {\expr_b}
    {\typearray{\type_b}{\idx_b}}
    \qquad
    \kindof{\sEnv}{\kEnv}{\typearray{\type_b}{\idx_b}}{\kindarray}}
  {\typeof
    {\sEnv}{\kEnv}{\tEnv}
    {\dproj{\sequence{\var_i}}{\var_e}{\expr_s}{\expr_b}}
    {\typearray{\type_b}{\idxappend{\idx_s \; \idx_b}}}}\]
  By the induction hypothesis, any two types $\typearray{\type_b}{\idxappend{\idx_s \; \idx_b}}$
  and $\typearray{\type_b'}{\idxappend{\idx_s' \; \idx_b'}}$
  ascribed to $\expr_b$ using the extended environment by the second premise
  must be equivalent---%
  \ie, $\teqv{\typearray{\type_b}{\idxappend{\idx_s \; \idx_b}}}{\typearray{\type_b'}{\idxappend{\idx_s' \; \idx_b'}}}$.
  This is exactly the proof obligation.
  \paragraph{Abstraction:}
  $\term = \lam{\sequence{\notevar{\var}{\type_i}}}{\expr}$.
  Type derivations must end with {\sc T-Lam}:
  \[\infr[T-Lam]
  {\typeof
    {\sEnv}{\kEnv}{\tEnv, \sequence{\hastype{\var}{\type_i}}}
    {\expr}
    {\type_o}
    \\
    \sequence{\kindof{\sEnv}{\kEnv}{\type_i}{\kindarray}}}
  {\typeof
    {\sEnv}{\kEnv}{\tEnv}
    {\lam{\sequence{\notevar{\var}{\type_i}}}{\expr}}
    {\typefun{\sequence{\type_i}}{\type_o}}}\]
  For any two such derivations,
  the induction hypothesis implies that types $\type_o$ and $\type_o'$
  are equivalent, since they are both ascribed to $\expr$.
  Since the input types $\sequence{\type_i}$ are given explicitly in the term,
  we can derive
  \[\infr[]
  {\sequence{\infr[TEqv-Refl]{\ }{\teqv{\type_i}{\type_i}}}
    \qquad \teqv{\type_o}{\type_o'}}
  {\teqv
    {\typefun{\sequence{\type_i}}{\type_o}}
    {\typefun{\sequence{\type_i}}{\type_o'}}}\]
  \paragraph{Type abstraction:}
  $\term = \tlam{\sequence{\notevar{\var}{\kind}}}{\val}$.
  Non-{\sc Eqv} derivations must end with
  \[\infr[T-TLam]
  {\typeof
    {\sEnv}{\kEnv, \sequence{\haskind{\var}{\kind}}}{\tEnv}
    {\val}{\type_u}}
  {\typeof
    {\sEnv}{\kEnv}{\tEnv}
    {\tlam{\sequence{\notevar{\var}{\kind}}}{\val}}
    {\typeuniv{\sequence{\notevar{\var}{\kind}}}{\type_u}}}\]
  Any $\type_u$ and $\type_u'$ ascribed to $\val$, the body of the abstraction,
  must be equivalent,
  and substituting in the same fresh variables
  will preserve that equivalence,
  by Lemma \ref{EqlTypeEqvType}.
  So we can construct the derivation
  \[\infr[TEqv-Univ]
  {\teqv
    {\seqsubst{\type_u}{\var}{\var_f}}
    {\seqsubst{\type_u'}{\var}{\var_f}}}
  {\teqv
    {\typeuniv{\sequence{\notevar{\var}{\kind}}}{\type_u}}
    {\typeuniv{\sequence{\notevar{\var}{\kind}}}{\type_u'}}}\]
  \paragraph{Index abstraction:}
  $\term = \ilam{\sequence{\notevar{\var}{\sort}}}{\val}$.
  \[\infr[T-ILam]
  {\typeof
    {\sEnv, \sequence{\hassort{\var}{\sort}}}{\kEnv}{\tEnv}
    {\val}{\type_p}}
  {\typeof
    {\sEnv}{\kEnv}{\tEnv}
    {\tlam{\sequence{\notevar{\var}{\kind}}}{\val}}
    {\typeuniv{\sequence{\notevar{\var}{\kind}}}{\type_p}}}\]
  Similar to the type abstraction case,
  any types ascribed to the abstraction body must be equivalent,
  and $\alpha$-conversion will preserve that equivalence,
  leading to
  \[\infr[TEqv-Pi]
  {\teqv
    {\seqsubst{\type_p}{\var}{\var_f}}
    {\seqsubst{\type_p'}{\var}{\var_f}}}
  {\teqv
    {\typeuniv{\sequence{\notevar{\var}{\kind}}}{\type_p}}
    {\typeuniv{\sequence{\notevar{\var}{\kind}}}{\type_p'}}}\]
  \paragraph{Box construction:}
  $\term = \dsum{\sequence{\idx}}{\expr}{\type_s}$.
  Only $\type_s$ can be ascribed to $\term$ without ending the derivation with {\sc T-Eqv}.
  \paragraph{Base value or primitive operator:}
  $\term = \baseval$ or $\term = \primop$.
  Each base value and primitive operator has a single defined type.
\end{sproof}

We also require guarantees about substitution in terms:
replacing an index variable with an appropriately-sorted index,
a type variable with an appropriately-kinded type,
or a term variable with an appropriately-typed expression
should not change the type of the original term.
If substitution turns a term $\term$ with type $\type$
into $\term'$ with type $\type'$,
where $\teqv{\type}{\type'}$,
we can add a {\sc T-Eqv} at the end of the new type derivation
to conclude $\term'$ has type $\type$.
As such, we do not need to include an ``up to equivalence'' caveat
when stating the preservation of typing lemmas.

\begin{lemma}[Preservation of types under index substitution]
  \label{IESub}
  If $\typeof{\sEnv,\hassort{\var}{\sort}}{\kEnv}{\tEnv}{\term}{\type}$
  and $\sortof{\sEnv}{\idx}{\sort}$
  then $\typeof{\sEnv}{\kEnv}{\idxsubst{\tEnv}}
  {\idxsubst{\term}}{\idxsubst{\type}}$.
\end{lemma}
\begin{sproof}[This is straightforward induction on the derivation of
  $\typeof{\sEnv,\hassort{\var}{\sort}}{\kEnv}{\tEnv}{\term}{\type}$.]
  We use induction on the derivation of
  $\typeof{\sEnv,\hassort{\var}{\sort}}{\kEnv}{\tEnv}{\term}{\type}$.
  \paragraph{Case {\sc Eqv}}
  \[\infr[T-Eqv]
  {\typeof{\sEnv,\hassort{\var}{\sort}}{\kEnv}{\tEnv}{\term}{\type'}
    \qquad \teqv{\type}{\type'}}
  {\typeof{\sEnv,\hassort{\var}{\sort}}{\kEnv}{\tEnv}{\term}{\type}}\]
  By the induction hypothesis,
  $\typeof{\sEnv}{\kEnv}{\idxsubst{\tEnv}}
  {\idxsubst{\term}}
  {\idxsubst{\type'}}$,
  and preservation of equivalence (Lemma \ref{EqlIdxEqvType}) implies
  $\teqv{\idxsubst{\type}}{\idxsubst{\type'}}$.
  Thus we can build the derivation
  \[{\infr[T-Eqv]
    {\typeof{\sEnv}{\kEnv}{\idxsubst{\tEnv}}
      {\idxsubst{\term}}{\idxsubst{\type'}}
      \qquad \teqv{\idxsubst{\type}}{\idxsubst{\type'}}}
    {\typeof{\sEnv}{\kEnv}{\idxsubst{\tEnv}}
      {\idxsubst{\term}}{\idxsubst{\type}}}}\]
  \paragraph{Case {\sc Op}, {\sc Base}}
  These cases are trivial,
  since primitive operators and base values are only ascribed closed types.
  \paragraph{Case {\sc Var}}
  \[{\infr[T-Var]
    {\parens{\hastype{\var'}{\type}} \in \tEnv}
    {\typeof{\sEnv, \hassort{\var}{\sort}}{\kEnv}{\tEnv}{\var'}{\type}}}\]
  In this case, $\var$ cannot appear in $\term$,
  so $\idxsubst{\var'} = \var'$.
  Since $\var'$ is bound in $\tEnv$,
  we can construct a similar derivation
  with $\tEnv(\var')$ updated by substitution:
  \[{\infr[T-Var]
    {\parens{\hastype{\var'}{\idxsubst{\type}}}
      \in \idxsubst{\tEnv}}
    {\typeof{\sEnv}{\kEnv}
      {\idxsubst{\tEnv}}
      {\var'}{\idxsubst{\type}}}}\]
  \paragraph{Case {\sc Array}}
  \[{\infr[T-Array]
    {\sequence{\typeof{\sEnv, \hassort{\var}{\sort}}{\kEnv}{\tEnv}
        {\atom}{\type_a}}
      \qquad {\kindof{\sEnv, \hassort{\var}{\sort}}{\kEnv}
        {\type_a}{\kindatom}}
      \\\\ {\mathit{Length}\llb \sequence{\atom} \rrb
        = \prod{\sequence{\nat}}}}
    {\typeof{\sEnv,\hassort{\var}{\sort}}{\kEnv}{\tEnv}
      {\arrlit{\sequence{\atom}}{\sequence{\nat}}}
      {\typearray{\type_a}{\idxshape{\sequence{\nat}}}}}}\]
  By the induction hypothesis,
  $\typeof{\sEnv}{\kEnv}{\idxsubst{\tEnv}}{\idxsubst{\atom}}{\idxsubst{\type_a}}$
  for each of $\sequence{\atom}$.
  Preservation of kinds under substitution (Lemma \ref{ITSub}) implies
  $\kindof{\sEnv}{\kEnv}{\idxsubst{\type_a}}{\kindatom}$.
  We use these results to construct the derivation
  \[{\infr[T-Array]
    {\sequence{\typeof{\sEnv}{\kEnv}{\idxsubst{\tEnv}}
        {\idxsubst{\atom}}{\idxsubst{\type_a}}}
      \\\\ {\kindof{\sEnv}{\kEnv}
        {\idxsubst{\type_a}}{\kindatom}}
      \\\\ {\mathit{Length}\llb \sequence{\atom} \rrb
        = \prod{\sequence{\nat}}}}
    {\parbox{0.45\textwidth}
      {\centering
        \({\sEnv};{\kEnv};{\tEnv}\vdash\)
        \({\arrlit{\sequence{\idxsubst{\atom}}}{\sequence{\nat}}}\)\\
        \(:{\typearray{\idxsubst{\type_a}}{\idxshape{\sequence{\nat}}}}\)}}}\]
  \paragraph{Case {\sc Frame}}
  \[{\infr[T-Frame]
    {\sequence{\typeof{\sEnv, \hassort{\var}{\sort}}{\kEnv}{\tEnv}
        {\expr_a}
        {\typearray{\type_a}{\idx_a}}}
      \\\\ {\kindof{\sEnv, \hassort{\var}{\sort}}{\kEnv}
        {\typearray{\type_a}{\idx_a}}
        {\kindarray}}
      \\\\ {\mathit{Length}\llb \sequence{\expr_a} \rrb
        = \prod{\sequence{\nat}}}}
    {\parbox{0.45\textwidth}
      {\centering
        \({\sEnv, \hassort{\var}{\sort}};{\kEnv};{\tEnv}\vdash\)
        \({\frm{\sequence{\expr}}{\sequence{\nat}}}\)\\
        \(:{\typearray{\type_a}
          {\idxappend{\idxshape{\sequence{\nat}}
              \; \idx_a}}}\)}}}\]
  The induction hypothesis implies
  $\typeof{\sEnv}{\kEnv}{\idxsubst{\tEnv}}
  {\idxsubst{\expr}}{\typearray{\idxsubst{\type_a}}{\idxsubst{\idx_a}}}$
  for each of the $\sequence{\expr}$.
  As in the {\sc T-Array} case,
  we use preservation of kinds to determine that
  $\kindof{\sEnv}{\kEnv}
  {\typearray{\idxsubst{\type_a}}{\idxsubst{\idx_a}}}{\kindarray}$.
  Then we derive
  \[{\infr[T-Frame]
    {\sequence{\typeof{\sEnv}{\kEnv}{\idxsubst{\tEnv}}
        {\idxsubst{\expr_a}}
        {\typearray{\idxsubst{\type_a}}{\idxsubst{\idx_a}}}}
      \\\\ {\kindof{\sEnv}{\kEnv}
        {\typearray{\idxsubst{\type_a}}{\idxsubst{\idx_a}}}
        {\kindarray}}
      \\\\ {\mathit{Length}\llb \sequence{\expr_a} \rrb
        = \prod{\sequence{\nat}}}}
    {\parbox{0.6\textwidth}
      {\centering
        \({\sEnv};{\kEnv};{\idxsubst{\tEnv}}\vdash\)
        \({\frm{\sequence{\idxsubst{\expr}}}{\sequence{\nat}}}\)\\
        \(:{\typearray{\idxsubst{\type_a}}
          {\idxappend{\idxshape{\sequence{\nat}}
              \; \idxsubst{\idx_a}}}}\)}}}\]
  \paragraph{Case {\sc EmptyA}}
  \[{\infr[T-EmptyA]
    {\kindof{\sEnv, \hassort{\var}{\sort}}{\kEnv}{\type_a}{\kindatom}
      \qquad 0 \in \sequence{\nat}}
    {\parbox{0.4\textwidth}
      {\centering
        \({\sEnv, \hassort{\var}{\sort}};{\kEnv};{\tEnv}\vdash\)
        \({\emptyarrlit{\type_a}{\sequence{\nat}}}\)
        \(:{\typearray{\type_a}{\idxshape{\sequence{\nat}}}}\)}}}\]
  By preservation of kinds,
  $\kindof{\sEnv}{\kEnv}{\idxsubst{\type_a}}{\kindatom}$.
  This leads to the derivation
  \[{\infr[T-EmptyA]
    {\kindof{\sEnv}{\kEnv}{\idxsubst{\type_a}}{\kindatom}
      \qquad 0 \in \sequence{\nat}}
    {\parbox{0.51\textwidth}
      {\centering
        \({\sEnv};{\kEnv};{\idxsubst{\tEnv}}\vdash\)
        \({\emptyarrlit{\idxsubst{\type_a}}{\sequence{\nat}}}\)
        \(:{\typearray{\idxsubst{\type_a}}{\idxshape{\sequence{\nat}}}}\)}}}\]
  \paragraph{Case {\sc EmptyF}}
  \[{\infr[T-EmptyF]
    {\kindof{\sEnv, \hassort{\var}{\sort}}{\kEnv}{\type_a}{\kindatom}
      \\\\
      \sortof{\sEnv, \hassort{\var}{\sort}}{\idx_a}{\sortshp}
      \\ 0 \in \sequence{\nat}}
    {\parbox{0.5\textwidth}
      {\centering
        \({\sEnv, \hassort{\var}{\sort}};{\kEnv};{\tEnv}\vdash\)
        \({\emptyfrm{\typearray{\type_a}{\idx_a}}{\sequence{\nat}}}\)
        \(:{\typearray
          {\type_a}
          {\idxappend{\idxshape{\sequence{\nat}} \; \idx_a}}}\)}}}\]
  As before, preservation of kinds implies
  $\kindof{\sEnv}{\kEnv}{\idxsubst{\type_a}}{\kindatom}$.
  We also have, by preservation of sorts under substitution (Lemma \ref{IISub})
  $\sortof{\sEnv}{\idxsubst{\idx_a}}{\sortshp}$.
  So we can derive
  \[{\infr[T-EmptyF]
    {\kindof{\sEnv}{\kEnv}{\idxsubst{\type_a}}{\kindatom}
      \\\\
      \sortof{\sEnv}{\idxsubst{\idx_a}}{\sortshp}
      \\ 0 \in \sequence{\nat}}
    {\parbox{0.63\textwidth}
      {\centering
        \({\sEnv};{\kEnv};{\tEnv}\vdash\)
        \({\emptyfrm
          {\typearray{\idxsubst{\type_a}}{\idxsubst{\idx_a}}}
          {\sequence{\nat}}}\)
        \(:{\typearray
          {\idxsubst{\type_a}}
          {\idxappend{\idxshape{\sequence{\nat}} \; \idxsubst{\idx_a}}}}\)}}}\]
  \paragraph{Case {\sc Lam}}
  \[{\infr[T-Lam]
    {\typeof{\sEnv, \hassort{\var}{\sort}}{\kEnv}
      {\tEnv, \sequence{\hastype{\var_i}{\type_i}}}
      {\expr}{\type_o}
      \qquad \kindof{\sEnv, \hassort{\var}{\sort}}{\kEnv}
             {\type_i}{\kindarray}}
    {\typeof{\sEnv, \hassort{\var}{\sort}}{\kEnv}{\tEnv}
      {\lam{\sequence{\notevar{\var_i}{\type_i}}}{\expr}}
      {\typefun{\sequence{\type_i}}{\type_o}}}}\]
  The induction hypothesis gives us
  $\typeof{\sEnv}{\kEnv}
  {\idxsubst{\tEnv}, \sequence{\hastype{\var_i}{\idxsubst{\type_i}}}}
  {\idxsubst{\expr}}
  {\idxsubst{\type_o}}$.
  By preservation of kinds,
  $\kindof{\sEnv}{\kEnv}{\idxsubst{\tEnv}}{\kindarray}$
  for each of $\sequence{\type_i}$,
  which leads to the derivation
  \[{\infr[T-Lam]
    {\typeof{\sEnv}{\kEnv}
      {\idxsubst{\tEnv}, \sequence{\hastype{\var_i}{\idxsubst{\type_i}}}}
      {\idxsubst{\expr}}{\idxsubst{\type_o}}
      \\\\ \sequence{\kindof{\sEnv}{\kEnv}
        {\idxsubst{\type_i}}{\kindarray}}}
    {\parbox{0.6\textwidth}
      {\centering
        \({\sEnv};{\kEnv};{\idxsubst{\tEnv}}\vdash\)
        \({\lam{\sequence{\var_i}{\idxsubst{\type_i}}}{\idxsubst{\expr}}}\)
        \(:{\typefun{\sequence{\idxsubst{\type_i}}}{\idxsubst{\type_o}}}\)}}}\]
  \paragraph{Case {\sc TLam}}
  \[{\infr[T-TLam]
    {\typeof{\sEnv, \hassort{\var}{\sort}}
      {\kEnv, \sequence{\haskind{\var_u}{\kind}}}{\tEnv}
      {\val}{\type_u}}
    {\typeof{\sEnv, \hassort{\var}{\sort}}{\kEnv}{\tEnv}
      {\tlam{\sequence{\notevar{\var_u}{\kind}}}{\val}}
      {\typeuniv{\sequence{\notevar{\var_u}{\kind}}}{\type_u}}}}\]
  By the induction hypothesis,
  $\typeof{\sEnv}{\kEnv, \sequence{\haskind{\var_u}{\kind}}}
  {\idxsubst{\tEnv}}
  {\idxsubst{\val}}{\idxsubst{\type_u}}$,
  so we derive
  \[{\infr[T-TLam]
    {\typeof{\sEnv}
      {\kEnv, \sequence{\haskind{\var_u}{\kind}}}
      {\idxsubst{\tEnv}}
      {\idxsubst{\val}}{\idxsubst{\type_u}}}
    {\parbox{0.6\textwidth}
      {\centering
        \({\sEnv};{\kEnv};{\idxsubst{\tEnv}}\vdash\)
        \({\tlam{\sequence{\notevar{\var_u}{\kind}}}{\idxsubst{\val}}}\)
        \(:{\typeuniv{\sequence{\notevar{\var_u}{\kind}}}{\idxsubst{\type_u}}}\)}}}\]
  \paragraph{Case {\sc ILam}}
  \[{\infr[T-ILam]
    {\typeof{\sEnv, \hassort{\var}{\sort},
        \sequence{\haskind{\var_u}{\sort_p}}}
      {\kEnv}{\tEnv}
      {\val}{\type_p}}
    {\typeof{\sEnv, \hassort{\var}{\sort}}{\kEnv}{\tEnv}
      {\ilam{\sequence{\notevar{\var_p}{\sort_p}}}{\val}}
      {\typedprod{\sequence{\notevar{\var_p}{\sort_p}}}{\type_p}}}}\]
  Following Barendregt's convention,
  we assume that $\var$ does not appear in $\idx_{\var}$.
  The induction hypothesis gives
  $\typeof{\sEnv, \sequence{\hassort{\var_p}{\sort_p}}}{\kEnv}
  {\idxsubst{\tEnv}}
  {\idxsubst{\val}}{\idxsubst{\type_p}}$,
  leading to
  \[{\infr[T-ILam]
    {\typeof{\sEnv, \sequence{\hassort{\var_u}{\sort_u}}}{\kEnv}
      {\idxsubst{\tEnv}}
      {\idxsubst{\val}}{\idxsubst{\type_u}}}
    {\parbox{0.6\textwidth}
      {\centering
        \({\sEnv};{\kEnv};{\idxsubst{\tEnv}}\vdash\)
        \({\tlam{\sequence{\notevar{\var_u}{\kind}}}{\idxsubst{\val}}}\)
        \(:{\typeuniv{\sequence{\notevar{\var_u}{\kind}}}{\idxsubst{\type_u}}}\)}}}\]
  \paragraph{Case {\sc Box}}
  \[{\infr[T-Box]
    {\sequence{\sortof{\sEnv, \hassort{\var}{\sort}}{\idx_s}{\sort_s}}
      \qquad
      \kindof{\sEnv, \hassort{\var}{\sort}}{\kEnv}
      {\typedsum{\sequence{\notevar{\var_s}{\sort_s}}}{\type_s}}
      {\kindatom}
      \\
      \typeof{\sEnv, \hassort{\var}{\sort}}{\kEnv}{\tEnv}
      {\expr}{\seqsubst{\type_s}{\var_s}{\idx_s}}}
    {\parbox{0.7\textwidth}
      {\centering
        \({\sEnv, \hassort{\var}{\sort}};{\kEnv};{\tEnv}\vdash\)
        \({\dsum{\sequence{\idx_s}}{\expr}
          {\typedsum{\sequence{\notevar{\var_s}{\sort_s}}}{\type_s}}}\)
        \(:{\typedsum{\sequence{\notevar{\var_s}{\sort_s}}}{\type_s}}\)}}}\]
  By the induction hypothesis,
  $\typeof{\sEnv}{\kEnv}{\idxsubst{\tEnv}}
  {\idxsubst{\expr}}{\idxsubst{\seqsubst{\type_s}{\var_s}{\idx_s}}}$,
  which is equal to $\seqsubst{\idxsubst{\type_s}}{\var_s}{\idx_s}$.
  Preservation of sorts implies
  $\sortof{\sEnv}{\idxsubst{\idx_s}}{\sort_s}$
  for each of $\sequence{\idx_s}$,
  while preservation of kinds implies
  $\kindof{\sEnv}{\kEnv}
  {\typedsum{\sequence{\notevar{\var_s}{\sort_s}}}{\idxsubst{\type_s}}}
  {\kindatom}$.
  Applying {\sc T-Box} derives
  \[{\infr[T-Box]
    {\sequence{\sortof{\sEnv}{\idxsubst{\idx_s}}{\sort_s}}
      \\\\
      \kindof{\sEnv}{\kEnv}
      {\typedsum{\sequence{\notevar{\var_s}{\sort_s}}}{\idxsubst{\type_s}}}
      {\kindatom}
      \\\\
      \typeof{\sEnv}{\kEnv}{\idxsubst{\tEnv}}
      {\idxsubst{\expr}}
      {\seqsubst{\idxsubst{\type_s}}{\var_s}{\idx_s}}}
    {\parbox{0.72\textwidth}
      {\centering
        \({\sEnv};{\kEnv};{\idxsubst{\tEnv}}\vdash\)
        \({\dsum{\sequence{\idxsubst{\idx_s}}}{\idxsubst{\expr}}
          {\typedsum{\sequence{\notevar{\var_s}{\sort_s}}}{\idxsubst{\type_s}}}}\)
        \(:{\typedsum{\sequence{\notevar{\var_s}{\sort_s}}}{\idxsubst{\type_s}}}\)}}}\]
  \paragraph{Case {\sc TApp}}
  \[{\infr[T-TApp]
    {\typeof{\sEnv, \hassort{\var}{\sort}}{\kEnv}{\tEnv}
      {\expr}
      {\typearray{\typeuniv
          {\sequence{\notevar{\var_u}{\kind}}}
          {\typearray{\type_u}{\idx_u}}}{\idx_f}}
      \\\\
      \sequence{\kindof
        {\sEnv, \hassort{\var}{\sort}}{\kEnv}
        {\type_a}{\kind}}}
    {\typeof{\sEnv, \hassort{\var}{\sort}}{\kEnv}{\tEnv}
      {\tapp{\expr}{\sequence{\type_a}}}
      {\typearray
        {\seqsubst{\type_u}{\var_u}{\type_a}}
        {\idxappend{\idx_f \; \idx_u}}}}}\]
  The induction hypothesis gives
  ${\sEnv};{\kEnv};{\idxsubst{\tEnv}}\vdash$
  ${\idxsubst{\expr}}$ :
  {\tt
    (Arr (Forall (($\var_u$ $\kind$) $\sequence{}$)
    (Arr $\idxsubst{\type_u}$ $\idxsubst{\idx_u}$))
    ($\idxsubst{\idx_f}$))}.
  Preservation of kinds under index substitution (Lemma \ref{ITSub}) implies
  $\kindof{\sEnv}{\kEnv}{\idxsubst{\type_a}}{\kind}$
  for each of $\sequence{\type_a}$.
  So we derive
  \[{\infr[T-TApp]
    {\sequence{\kindof
        {\sEnv}{\kEnv}
        {\idxsubst{\type_a}}{\kind}}
      \quad
      \parbox{0.5\textwidth}
      {
        \({\sEnv};{\kEnv};{\idxsubst{\tEnv}}\vdash\)
        \({\idxsubst{\expr}}\)\\
        {\tt
          : (Arr (Forall (($\var_u$ $\kind$) $\sequence{}$)\\
          \phantom{: DArr DF}(Arr $\idxsubst{\type_u}$ $\idxsubst{\idx_u}$))\\
          \phantom{: DArr }($\idxsubst{\idx_f}$))}
      }}
    {\parbox{0.7\textwidth}
      {\centering
        \({\sEnv};{\kEnv};{\idxsubst{\tEnv}}\vdash\)
        \({\tapp{\expr}{\sequence{\idxsubst{\type_a}}}}\)
        \(:{\typearray
          {\seqsubst{\idxsubst{\type_u}}{\var_u}{\type_a}}
          {\idxappend{\idxsubst{\idx_f} \; \idxsubst{\idx_u}}}}\)}}}\]
  \paragraph{Case {\sc IApp}}
  \[{\infr[T-IApp]
    {\typeof{\sEnv, \hassort{\var}{\sort}}{\kEnv}{\tEnv}
      {\expr}
      {\typearray{\typedprod
            {\sequence{\notevar{\var_p}{\sort_p}}}
            {\typearray{\type_p}{\idx_p}}}{\idx_f}}
      \\\\
      \sequence{\sortof{\sEnv, \hassort{\var}{\sort}}
        {\idx_a}{\sort_p}}}
    {\parbox{0.6\textwidth}
      {\centering
        \({\sEnv, \hassort{\var}{\sort}};{\kEnv};{\tEnv}\vdash\)
        \({\iapp{\expr}{\sequence{\idx_a}}}\)
        \(:{\typearray
          {\seqsubst{\type_p}{\var_p}{\idx_a}}
          {\idxappend{\idx_f \; \seqsubst{\idx_p}{\var_p}{\idx_a}}}}\)}}}\]
  The induction hypothesis implies
  ${\sEnv};{\kEnv};{\idxsubst{\tEnv}}\vdash{\idxsubst{\expr}}$
  {\tt : (Arr (Pi (($\var_p$ $\sort_p$) $\sequence{}$) (Arr $\idxsubst{\type_p}$ $\idxsubst{\idx_p}$)) $\idxsubst{\idx_f}$)}.
  Preservation of sorts under index substitution (Lemma \ref{IISub}) implies
  $\sortof{\sEnv}{\idxsubst{\idx_a}}{\sort_a}$
  for each corresponding pair $\sequence{(\idx_a, \sort_a)}$.
  So we construct the derivation
  \[{\infr[T-IApp]
    {\sequence{\sortof{\sEnv}
        {\idxsubst{\idx_a}}{\sort_p}}
      \qquad
      \parbox{0.5\textwidth}
      {
        \({\sEnv};{\kEnv};{\idxsubst{\tEnv}}\vdash\)
        \({\idxsubst{\expr}}\)\\
        {\tt
          : (Arr (Forall (($\var_u$ $\kind$) $\sequence{}$)\\
          \phantom{: DArr DP}(Arr $\idxsubst{\type_p}$ $\idxsubst{\idx_p}$))\\
          \phantom{: DArr }($\idxsubst{\idx_f}$))}
      }}
    {\parbox{0.6\textwidth}
      {\centering
        \({\sEnv};{\kEnv};{\idxsubst{\tEnv}}\vdash\)
        \({\iapp{\expr}{\sequence{\idxsubst{\idx_a}}}}\)
        \(:{\typearray
          {\seqsubst{\idxsubst{\type_p}}{\var_p}{\idx_a}}
          {\idxappend{\idxsubst{\idx_f} \; \seqsubst{\idxsubst{\idx_p}}{\var_p}{\idx_a}}}}\)}}}\]
  \paragraph{Case {\sc App}}
  \[{\infr[T-App]
    {\typeof{\sEnv, \hassort{\var}{\sort}}{\kEnv}{\tEnv}
      {\expr_f}
      {\typearray
        {\typefun
          {\sequence{\typearray{\type_i}{\idx_i}}}
          {\typearray{\type_o}{\idx_o}}}
        {\idx_f}}
      \\\\
      {\sequence
        {\typeof{\sEnv, \hassort{\var}{\sort}}{\kEnv}{\tEnv}
          {\expr_a}
          {\typearray{\type_i}{\idxappend{\idx_a \; \idx_i}}}}}
      \qquad
      {\idx_p = \mathit{Max}\llb \idx_f \; \sequence{\idx_a} \rrb}}
    {\typeof{\sEnv, \hassort{\var}{\sort}}{\kEnv}{\tEnv}
      {\app{\expr_f}{\sequence{\expr_a}}}
      {\typearray{\type_o}{\idxappend{\idx_p \; \idx_o}}}}}\]
  By the induction hypothesis,
  ${\sEnv};{\kEnv};{\idxsubst{\tEnv}}\vdash$
  ${\idxsubst{\expr_f}}$
  {\tt :
    (Arr (-> ((Arr $\idxsubst{\type_i}$ $\idxsubst{\idx_i}$) $\sequence{}$)
    (Arr $\idxsubst{\type_o}$ $\idxsubst{\idx_o}$))
    $\idxsubst{\idx_f}$)}
  and for each of $\sequence{\expr_a}$, we have
  $\typeof{\sEnv}{\kEnv}{\idxsubst{\tEnv}}{\idxsubst{\expr_a}}
  {\typearray
    {\idxsubst{\type_i}}
    {\idxappend{\idxsubst{\idx_a} \; \idxsubst{\idx_i}}}}$.
  Since substituting into a shape is monotonic in the subbed-in index,
  \ie, prefix ordering commutes with substitution,
  $\mathit{Max}\llb \idxsubst{\idx_f} \; \sequence{\idxsubst{\idx_a}} \rrb
  = \idxsubst{\mathit{Max}\llb \idx_f \; \sequence{\idx_a} \rrb}
  = \idxsubst{\idx_p}$.
  Then we build the derivation
  \[{\infr[T-App]
    {\parbox{0.5\textwidth}
      {\({\sEnv};{\kEnv};{\idxsubst{\tEnv}}\vdash\)
        \({\idxsubst{\expr_f}}\)\\
        {\tt
          : (Arr (-> ((Arr $\idxsubst{\type_i}$ $\idxsubst{\idx_i}$)\\
          \phantom{: DArr D-> D}$\sequence{}$)\\
          \phantom{: DArr D-> }(Arr $\idxsubst{\type_o}$ $\idxsubst{\idx_o}$))\\
          \phantom{: DArr }$\idx_f$)}
      }
      \\\\\\
      {\sequence
        {\parbox{0.5\textwidth}
          {\centering
            \({\sEnv};{\kEnv};{\idxsubst{\tEnv}}\vdash\)
            \({\idxsubst{\expr_a}}\)\\
            \(:{\typearray
              {\idxsubst{\type_i}}
              {\idxappend{\idxsubst{\idx_a} \; \idxsubst{\idx_i}}}}\)}}}
      \\\\\\
      {\idxsubst{\idx_p} =
        \mathit{Max}\llb \idxsubst{\idx_f} \; \sequence{\idxsubst{\idx_a}} \rrb}}
    {\parbox{0.6\textwidth}
      {\centering
        \({\sEnv};{\kEnv};{\idxsubst{\tEnv}}\vdash\)
        \({\app{\idxsubst{\expr_f}}{\sequence{\idxsubst{\expr_a}}}}\)
        \(:{\typearray
          {\idxsubst{\type_o}}
          {\idxappend{\idxsubst{\idx_p} \; \idxsubst{\idx_o}}}}\)}}}\]
  \paragraph{Case {\sc Unbox}}
  \[{\infr[T-Unbox]
    {\typeof{\sEnv, \hassort{\var}{\sort}}{\kEnv}{\tEnv}
      {\expr_s}
      {\typearray{\typedsum{\sequence{\var_i'}{\sort_i}}{\type_s}}
        {\idx_s}}
      \\\\
      {\typeof{\sEnv, \hassort{\var}{\sort},
          \sequence{\hassort{\var_i}{\sort_i}}}
        {\kEnv}
        {\tEnv, \hastype{\var_e}{\seqsubst{\type_s}{\var_i'}{\var_i}}}
        {\expr_b}
        {\typearray{\type_b}{\idx_b}}}
      \\\\
      {\kindof{\sEnv, \hassort{\var}{\sort}}{\kEnv}
        {\typearray{\type_b}{\idx_b}}{\kindarray}}}
    {\typeof{\sEnv, \hassort{\var}{\sort}}{\kEnv}{\tEnv}
      {\dproj{\sequence{\var_i}}{\var_e}{\expr_s}{\expr_b}}
      {\typearray{\type_b}{\idxappend{\idx_s \; \idx_b}}}}}\]
  Per Barendregt's convention, we stipulate that $\var \not\in \sequence{\var_i'}$.
  Then the induction hypothesis gives both
  $\typeof{\sEnv}{\kEnv}{\idxsubst{\tEnv}}
  {\idxsubst{\expr_s}}
  {\typearray{\typedsum{\sequence{\var_i'}{\sort_i}}{\idxsubst{\type_s}}}
    {\idxsubst{\idx_s}}}$
  and
  $\typeof{\sEnv, \sequence{\hassort{\var_i}{\sort_i}}}{\kEnv}
  {\idxsubst{\tEnv}, \hastype{\var_e}{\idxsubst{\parens{\seqsubst{\type_s}{\var_i'}{\var_i}}}}}
  {\idxsubst{\expr_b}}
  {\idxsubst{\type_b}}$.
  By preservation of kinds,
  $\kindof{\sEnv}{\kEnv}{\idxsubst{\typearray{\type_b}{\idx_b}}}{\kindarray}$.
  So we can build the derivation
  \[{\infr[T-Unbox]
    {{\typeof{\sEnv}{\kEnv}{\idxsubst{\tEnv}}
        {\idxsubst{\expr_s}}
        {\typearray{\typedsum{\sequence{\var_i'}{\sort_i}}{\idxsubst{\type_s}}}
          {\idx_s}}}
      \\\\\\
      {\parbox{0.6\textwidth}
        {\centering
          \({\sEnv, \sequence{\hassort{\var_i}{\sort_i}}};{\kEnv};
          {\idxsubst{\tEnv}, \hastype{\var_e}{\idxsubst{\parens{\seqsubst{\type_s}{\var_i'}{\var_i}}}}}\vdash\)\\
          \({\idxsubst{\expr_b}}\)
          \(:{\idxsubst{\typearray{\type_b}{\idx_b}}}\)}}
      \\\\\\
      {\kindof{\sEnv}{\kEnv}{\idxsubst{\typearray{\type_b}{\idx_b}}}{\kindarray}}}
    {\parbox{0.7\textwidth}
      {\centering
        \({\sEnv};{\kEnv};{\idxsubst{\tEnv}}\vdash\)
        \({\dproj{\sequence{\var_i}}{\var_e}{\idxsubst{\expr_s}}{\idxsubst{\expr_b}}}\)\\
        \(:{\idxsubst{\typearray{\type_b}{\idx_b}}}\)}}}\]
\end{sproof}

\begin{lemma}[Preservation of types under type substitution]
  \label{TESub}
  If $\typeof{\sEnv}{\kEnv,\haskind{\var}{\kind}}{\tEnv}{\term}{\type}$
  and $\kindof{\sEnv}{\kEnv}{\type_\var}{\kind}$
  then $\typeof{\sEnv}{\kEnv}{\typesubst{\tEnv}}
  {\typesubst{\term}}
  {\typesubst{\type}}$.
\end{lemma}
\begin{sproof}[This is straightforward induction on the derivation of
  $\typeof{\sEnv}{\kEnv,\haskind{\var}{\kind}}{\tEnv}{\term}{\type}$.]
  We use induction on the derivation of
  $\typeof{\sEnv}{\kEnv,\haskind{\var}{\kind}}{\tEnv}{\term}{\type}$.
  \paragraph{Case {\sc Eqv}}
  \[{\infr[T-Eqv]
    {\typeof{\sEnv}{\kEnv, \haskind{\var}{\kind}}{\tEnv}{\term}{\type'}
      \qquad \teqv{\type}{\type'}}
    {\typeof{\sEnv}{\kEnv, \haskind{\var}{\kind}}{\tEnv}
      {\term}
      {\type}}}\]
  The induction hypothesis implies
  $\typeof{\sEnv}{\kEnv}{\typesubst{\tEnv}}
  {\typesubst{\term}}{\typesubst{\type'}}$.
  Preservation of equivalence (Lemma \ref{EqlTypeEqvType}) implies
  $\teqv{\typesubst{\type}}{\typesubst{\type'}}$.
  So we derive
  \[{\infr[T-Eqv]
    {\typeof{\sEnv}{\kEnv}{\typesubst{\tEnv}}{\typesubst{\term}}{\typesubst{\type'}}
      \qquad \teqv{\typesubst{\type}}{\typesubst{\type'}}}
    {\typeof{\sEnv}{\kEnv}{\typesubst{\tEnv}}
      {\typesubst{\term}}
      {\typesubst{\type}}}}\]
  \paragraph{Case {\sc Op}, {\sc Base}}
  Since $\var$ cannot appear free in $\term$ or $\type$,
  the goal, after substitution, is to ascribe $\type$ to $\term$
  in the diminished environment,
  which can be done because
  {\sc T-Op} and {\sc T-Base} do not depend on the contents of the environment.
  \paragraph{Case {\sc Var}}
  \[{\infr[T-Var]
    {\parens{\hastype{\var'}{\type}} \in \tEnv}
    {\typeof{\sEnv}{\kEnv, \haskind{\var}{\kind}}{\tEnv}
      {\var'}
      {\type}}}\]
  The type variable $\var$ cannot appear free in $\var'$ (a term variable),
  so $\typesubst{\var'} = \var'$.
  Applying substitution to $\tEnv$ maps $\tEnv(\var')$ into $\typesubst{\tEnv(\var')}$,
  so $\parens{\hastype{\var'}{\typesubst{\type}}} \in \typesubst{\tEnv}$.
  We then derive
  \[\infr[T-Var]
  {\parens{\hastype{\var'}{\typesubst{\type}}} \in \typesubst{\tEnv}}
  {\typeof{\sEnv}{\kEnv}{\typesubst{\tEnv}}
    {\var'}
    {\typesubst{\type}}}\]
  \paragraph{Case {\sc Array}}
  \[{\infr[T-Array]
    {\sequence{\typeof{\sEnv}{\kEnv, \haskind{\var}{\kind}}{\tEnv}
        {\atom}{\type_a}}
      \\\\
      \kindof{\sEnv}{\kEnv, \haskind{\var}{\kind}}{\type_a}{\kindatom}
      \qquad
      \mathit{Length}\llb \sequence{\atom} \rrb = \prod{\sequence{\nat}}}
    {\typeof{\sEnv}{\kEnv, \haskind{\var}{\kind}}{\tEnv}
      {\arrlit{\sequence{\atom}}{\sequence{\nat}}}
      {\typearray{\type_a}{\idxshape{\sequence{\nat}}}}}}\]
  The induction hypothesis implies
  $\typeof{\sEnv}{\kEnv}{\typesubst{\tEnv}}{\atom}{\typesubst{\type_a}}$
  for each of the $\sequence{\atom}$.
  By preservation of kinds (Lemma \ref{TTSub}),
  $\kindof{\sEnv}{\kEnv}{\typesubst{\type_a}}{\kindatom}$.
  This leads to the derivation
  \[{\infr[T-Array]
    {\sequence{\typeof{\sEnv}{\kEnv}{\typesubst{\tEnv}}
        {\typesubst{\atom}}{\typesubst{\type_a}}}
      \\\\
      \kindof{\sEnv}{\kEnv}{\typesubst{\type_a}}{\kindatom}
      \qquad
      \mathit{Length}\llb \sequence{\atom} \rrb = \prod{\sequence{\nat}}}
    {\parbox{0.6\textwidth}
      {\centering
        \({\sEnv};{\kEnv};{\typesubst{\tEnv}}\vdash\)
        \({\arrlit{\sequence{\typesubst{\atom}}}{\sequence{\nat}}}\)
        \(:{\typearray{\typesubst{\type_a}}{\idxshape{\sequence{\nat}}}}\)}}}\]
  \paragraph{Case {\sc Frame}}
  \[{\infr[T-Frame]
    {\sequence{\typeof{\sEnv}{\kEnv, \haskind{\var}{\kind}}{\tEnv}
        {\expr}{\typearray{\type_a}{\idx_a}}}
      \\\\
      {\kindof{\sEnv}{\kEnv, \haskind{\var}{\kind}}
        {\typearray{\type_a}{\idx_a}}
        {\kindarray}}
      \qquad
      \mathit{Length}\llb \sequence{\atom} \rrb = \prod{\sequence{\nat}}}
  {\typeof{\sEnv}{\kEnv, \haskind{\var}{\kind}}{\tEnv}
    {\frm{\sequence{\expr}}{\sequence{\nat}}}
    {\typearray{\type_a}{\idxappend{\idxshape{\sequence{\nat}} \; \idx_a}}}}}\]
  By the induction hypothesis,
  $\typeof{\sEnv}{\kEnv}{\typesubst{\tEnv}}{\typesubst{\expr}}{\typearray{\typesubst{\type_a}}{\idx_a}}$
  for each of the $\sequence{\expr}$.
  Preservation of kinds gives
  $\kindof{\sEnv}{\kEnv}{\typearray{\typesubst{\type_a}}{\idx_a}}{\kindarray}$.
  So we derive
  \[{\infr[T-Frame]
    {\sequence{\typeof{\sEnv}{\kEnv}{\typesubst{\tEnv}}{\typesubst{\expr}}{\typearray{\typesubst{\type_a}}{\idx_a}}}
      \\\\
      \kindof{\sEnv}{\kEnv}{\typearray{\typesubst{\type_a}}{\idx_a}}{\kindarray}
      \qquad
      \mathit{Length}\llb \sequence{\atom} \rrb = \prod{\sequence{\nat}}}
    {\parbox{0.6\textwidth}
      {\centering
        \({\sEnv};{\kEnv};{\typesubst{\tEnv}}\vdash\)
        \({\frm{\sequence{\typesubst{\expr}}}{\sequence{\nat}}}\)
        \(:{\typearray{\typesubst{\type_a}}{\idxappend{\idxshape{\sequence{\nat}} \; \idx_a}}}\)}}}\]
  \paragraph{Case {\sc EmptyA}}
  \[{\infr[T-EmptyA]
    {\kindof{\sEnv}{\kEnv, \haskind{\var}{\kind}}{\type_a}{\kindatom}
      \qquad 0 \in \sequence{\nat}}
    {\typeof{\sEnv}{\kEnv, \haskind{\var}{\kind}}{\tEnv}
      {\emptyarrlit{\type_a}{\sequence{\nat}}}
      {\typearray{\type_a}{\idxshape{\sequence{\nat}}}}}}\]
  Preservation of kinds implies
  $\kindof{\sEnv}{\kEnv}{\typesubst{\type_a}}{\kindatom}$,
  so we can derive
  \[{\infr[T-EmptyA]
    {\kindof{\sEnv}{\kEnv}{\typesubst{\type_a}}{\kindatom}
      \qquad 0 \in \sequence{\nat}}
    {\parbox{0.6\textwidth}
      {\centering
        \({\sEnv};{\kEnv};{\typesubst{\tEnv}}\vdash\)
        \({\emptyarrlit{\typesubst{\type_a}}{\sequence{\nat}}}\)\\
        \(:{\typearray{\typesubst{\type_a}}{\idxshape{\sequence{\nat}}}}\)}}}\]
  \paragraph{Case {\sc EmptyF}}
  \[{\infr[T-EmptyF]
    {\kindof{\sEnv}{\kEnv, \haskind{\var}{\kind}}
      {\typearray{\type_a}{\idx_a}}
      {\kindarray}
      \qquad 0 \in \sequence{\nat}}
    {\parbox{0.6\textwidth}
      {\centering
        \({\sEnv};{\kEnv, \haskind{\var}{\kind}};{\tEnv}\vdash\)
        \({\emptyfrm{\typearray{\type_a}{\idx_a}}{\sequence{\nat}}}\)
        \(:{\typearray{\type_a}{\idxappend{\idx_a \; \idxshape{\sequence{\nat}}}}}\)}}}\]
  Preservation of kinds implies
  $\kindof{\sEnv}{\kEnv}{\typearray{\typesubst{\type_a}}{\idx_a}}{\kindarray}$.
  so we can derive
  \[{\infr[T-EmptyF]
    {\kindof{\sEnv}{\kEnv}
      {\typearray{\typesubst{\type_a}}{\idx_a}}
      {\kindarray}
      \qquad 0 \in \sequence{\nat}}
    {\parbox{0.6\textwidth}
      {\centering
        \({\sEnv};{\kEnv, \haskind{\var}{\kind}};{\tEnv}\vdash\)
        \({\emptyfrm{\typearray{\typesubst{\type_a}}{\idx_a}}{\sequence{\nat}}}\)
        \(:{\typearray{\typesubst{\type_a}}{\idxappend{\idx_a \; \idxshape{\sequence{\nat}}}}}\)}}}\]
  \paragraph{Case {\sc Lam}}
  \[{\infr[T-Lam]
    {\typeof{\sEnv}{\kEnv, \haskind{\var}{\kind}}
      {\tEnv, \sequence{\hastype{\var_i}{\type_i}}}
      {\expr}{\type_o}}
    {\typeof{\sEnv}{\kEnv, \haskind{\var}{\kind}}{\tEnv}
      {\lam{\sequence{\notevar{\var_i}{\type_i}}}{\expr}}
      {\typefun{\sequence{\type_i}}{\type_o}}}}\]
  The induction hypothesis implies
  $\typeof{\sEnv}{\kEnv}{\typesubst{\tEnv}, \sequence{\hastype{\var_i}{\typesubst{\type_i}}}}
  {\typesubst{\expr}}{\typesubst{\type_o}}$
  (\nb,
  $\typesubst{\parens{\tEnv, \sequence{\hastype{\var_i}{\type_i}}}}
  = \typesubst{\tEnv}, \sequence{\hastype{\var_i}{\type_i}}$).
  Then applying {\sc T-Lam} derives
  \[{\infr[T-Lam]
    {\typeof{\sEnv}{\kEnv}
      {\typesubst{\tEnv}, \sequence{\hastype{\var_i}{\typesubst{\type_i}}}}
      {\typesubst{\expr}}{\typesubst{\type_o}}}
    {\parbox{0.6\textwidth}
      {\centering
        \({\sEnv};{\kEnv};{\typesubst{\tEnv}}\vdash\)
        \({\lam{\sequence{\notevar{\var_i}{\typesubst{\type_i}}}}{\typesubst{\expr}}}\)
        \(:{\typefun{\sequence{\typesubst{\type_i}}}{\typesubst{\type_o}}}\)}}}\]
  \paragraph{Case {\sc TLam}}
  \[{\infr[T-TLam]
    {\typeof{\sEnv}{\kEnv, \haskind{\var}{\kind}, \sequence{\haskind{\var_u}{\kind_u}}}{\tEnv}
      {\expr}
      {\type_u}}
    {\typeof{\sEnv}{\kEnv, \haskind{\var}{\kind}}{\tEnv}
      {\tlam{\sequence{\notevar{\var_u}{\kind_u}}}{\expr}}
      {\typeuniv{\sequence{\notevar{\var_u}{\kind_u}}}{\type_u}}}}\]
  By the induction hypothesis,
  $\typeof{\sEnv}{\kEnv, \sequence{\haskind{\var_u}{\kind_u}}}{\typesubst{\tEnv}}
  {\typesubst{\expr}}{\typesubst{\type_u}}$
  (per Barendregt's convention, $\var \not\in \sequence{\var_u}$).
  We then derive
  \[{\infr[T-TLam]
    {\typeof{\sEnv}{\kEnv, \sequence{\haskind{\var_u}{\kind_u}}}{\typesubst{\tEnv}}
      {\typesubst{\expr}}{\typesubst{\type_u}}}
    {\parbox{0.6\textwidth}
      {\centering
        \({\sEnv};{\kEnv};{\typesubst{\tEnv}}\vdash\)
        \({\tlam{\sequence{\notevar{\var_u}{\kind_u}}}{\typesubst{\expr}}}\)
        \(:{\typeuniv{\sequence{\notevar{\var_u}{\kind_u}}}{\typesubst{\type_u}}}\)}}}\]
  \paragraph{Case {\sc ILam}}
  \[{\infr[T-ILam]
    {\typeof{\sEnv, \sequence{\hassort{\var_p}{\sort_p}}}{\kEnv, \haskind{\var}{\kind}}{\tEnv}
      {\expr}
      {\type_p}}
    {\typeof{\sEnv}{\kEnv, \haskind{\var}{\kind}}{\tEnv}
      {\ilam{\sequence{\notevar{\var_p}{\sort_p}}}{\expr}}
      {\typedprod{\sequence{\notevar{\var_p}{\sort_p}}}{\type_p}}}}\]
  The induction hypothesis implies
  $\typeof{\sEnv, \sequence{\hassort{\var_p}{\sort_p}}}{\kEnv}{\typesubst{\tEnv}}
  {\typesubst{\expr}}{\typesubst{\type_p}}$.
  Then we can derive
  \[{\infr[T-ILam]
    {\typeof{\sEnv, \sequence{\hassort{\var_p}{\sort_p}}}{\kEnv}{\typesubst{\tEnv}}
      {\typesubst{\expr}}
      {\typesubst{\type_p}}}
    {\parbox{0.6\textwidth}
      {\centering
        \({\sEnv};{\kEnv};{\typesubst{\tEnv}}\vdash\)
        \({\ilam{\sequence{\notevar{\var_p}{\sort_p}}}{\typesubst{\expr}}}\)
        \(:{\typedprod{\sequence{\notevar{\var_p}{\sort_p}}}{\typesubst{\type_p}}}\)}}}\]
  \paragraph{Case {\sc Box}}
  \[{\infr[T-Box]
    {\sequence{\sortof{\sEnv}{\idx_s}{\sort_s}}
      \qquad
      {\kindof{\sEnv}{\kEnv, \haskind{\var}{\kind}}
        {\typedsum{\sequence{\notevar{\var_s}{\sort_s}}}{\type_s}}
        {\kindatom}}
      \\
      {\typeof{\sEnv}{\kEnv, \haskind{\var}{\kind}}{\tEnv}
        {\expr}
        {\seqsubst{\type_s}{\var_s}{\idx_s}}}}
    {\parbox{0.7\textwidth}
      {\centering
        \({\sEnv};{\kEnv, \haskind{\var}{\kind}};{\tEnv}\vdash\)
        \({\dsum
          {\sequence{\idx_s}}{\expr}
          {\typedsum{\sequence{\notevar{\var_s}{\sort_s}}}{\type_s}}}\)\\
        \(:{\typedsum{\sequence{\notevar{\var_s}{\sort_s}}}{\type_s}}\)}}}\]
  By the induction hypothesis,
  $\typeof{\sEnv}{\kEnv}{\typesubst{\tEnv}}{\expr}{\typesubst{\seqsubst{\type_s}{\var_s}{\idx_s}}}$.
  Preservation of kinds under type substitution gives a derivation for
  $\kindof{\sEnv}{\kEnv}
  {\typedsum{\sequence{\notevar{\var_s}{\sort_s}}}{\typesubst{\type_s}}}
  {\kindatom}$.
  The ascribed type ${\typesubst{\seqsubst{\type_s}{\var_s}{\idx_s}}}$
  is equal to $\multisubst{\type_s}{\sequence{\singlesubst{\var_s}{\idx_s}}, \singlesubst{\var}{\type_1}}$.
  We then derive
  \[{\infr[T-Box]
    {\sequence{\sortof{\sEnv}{\idx_s}{\sort_s}}
      \qquad
      {\kindof{\sEnv}{\kEnv}
        {\typedsum{\sequence{\notevar{\var_s}{\sort_s}}}{\typesubst{\type_s}}}
        {\kindatom}}
      \\
      {\typeof{\sEnv}{\kEnv}{\typesubst{\tEnv}}
        {\expr}
        {\multisubst{\type_s}
          {\sequence{\singlesubst{\var_s}{\idx_s}},
            \singlesubst{\var}{\type_1}}}}}
    {\parbox{0.85\textwidth}
      {\centering
        \({\sEnv};{\kEnv};{\typesubst{\tEnv}}\vdash\)
        \({\dsum
          {\sequence{\idx_s}}{\typesubst{\expr}}
          {\typedsum{\sequence{\notevar{\var_s}{\sort_s}}}{\typesubst{\type_s}}}}\)\\
        \(:{\typedsum{\sequence{\notevar{\var_s}{\sort_s}}}{\typesubst{\type_s}}}\)}}}\]
  \paragraph{Case {\sc TApp}}
  \[{\infr[T-TApp]
    {\typeof
      {\sEnv}{\kEnv, \haskind{\var}{\kind}}{\tEnv}
      {\expr}
      {\typearray{\typeuniv{\sequence{\notevar{\var_u}{\kind_u}}}{\typearray{\type_u}{\idx_u}}}{\idx_f}}
      \\\\
      \sequence{\kindof{\sEnv}{\kEnv, \haskind{\var}{\kind}}{\type_a}{\kind_u}}}
    {\typeof{\sEnv}{\kEnv, \haskind{\var}{\kind}}{\tEnv}
      {\tapp{\expr}{\sequence{\type_a}}}
      {\typearray{\seqsubst{\type_u}{\var_u}{\type_a}}{\idxappend{\idx_f\;\idx_u}}}}}\]
  The induction hypothesis gives
  ${\sEnv};{\kEnv};{\typesubst{\tEnv}}\vdash$
  ${\typesubst{\expr}}$ :
  {\tt (Arr (Forall (($\var_u$ $\kind_u$) $\sequence{}$) (Arr $\typesubst{\type_u}$ $\idx_u$)) $\idx_f$)}.
  Lemma \ref{TTSub} (preservation of kinds) implies for each of $\sequence{\type_a}$ that
  $\kindof{\sEnv}{\kEnv}{\typesubst{\type_a}}{\kind_u}$.
  Then we can derive
  \[{\infr[T-TApp]
    {\parbox{0.63\textwidth}
      {\centering
        \({\sEnv};{\kEnv};{\typesubst{\tEnv}}\vdash\)
        \({\typesubst{\expr}}\)
        \(:{\typearray
          {\typeuniv
            {\sequence{\notevar{\var_u}{\kind_u}}}
            {\typearray{\typesubst{\type_u}}{\idx_u}}}
          {\idx_f}}\)}
      \\\\\\
      \sequence{\kindof{\sEnv}{\kEnv}{\typesubst{\type_a}}{\kind_u}}}
    {\parbox{0.58\textwidth}
      {\centering
        \({\sEnv};{\kEnv};{\typesubst{\tEnv}}\vdash\)
        \({\tapp{\typesubst{\expr}}{\sequence{\typesubst{\type_a}}}}\)
        \(:{\typearray
          {\seqsubst{\typesubst{\type_u}}{\var_u}{\type_a}}
          {\idxappend{\idx_f \; \idx_u}}}\)}}}\]
  \paragraph{Case {\sc IApp}}
  \[{\infr[T-IApp]
    {\typeof
      {\sEnv}{\kEnv, \haskind{\var}{\kind}}{\tEnv}
      {\expr}
      {\typearray{\typedprod{\sequence{\notevar{\var_p}{\sort_p}}}{\typearray{\type_p}{\idx_p}}}{\idx_f}}
      \\\\
      \sequence{\sortof{\sEnv}{\idx_a}{\sort_p}}}
    {\parbox{0.6\textwidth}
      {\centering
        \({\sEnv};{\kEnv, \haskind{\var}{\kind}};{\tEnv}\vdash\)
        \({\iapp{\expr}{\sequence{\idx_a}}}\)\\
        \(:{\typearray{\seqsubst{\type_p}{\var_p}{\idx_a}}{\idxappend{\idx_f\;\seqsubst{\idx_p}{\var_p}{\idx_a}}}}\)}}}\]
  By the induction hypothesis,
  ${\sEnv};{\kEnv};{\typesubst{\tEnv}}\vdash$
  ${\typesubst{\expr}}$ :
  {\tt (Arr (Pi (($\var_p$ $\sort_p$) $\sequence{}$) (Arr $\typesubst{\type_p}$ $\idx_p$)) $\idx_f$)}.
  Applying {\sc T-IApp} produces the derivation
  \[{\infr[T-IApp]
    {\parbox{0.6\textwidth}
      {\centering
        \({\sEnv};{\kEnv};{\typesubst{\tEnv}}\vdash\)
        \({\typesubst{\expr}}\)
        \(:{\typearray
          {\typedprod
            {\sequence{\notevar{\var_p}{\sort_p}}}
            {\typearray{\typesubst{\type_p}}{\idx_p}}}
          {\idx_f}}\)}
      \\\\\\
      \sequence{\sortof{\sEnv}{\idx_a}{\sort_p}}}
    {\parbox{0.6\textwidth}
      {\centering
        \({\sEnv};{\kEnv};{\typesubst{\tEnv}}\vdash\)
        \({\iapp{\typesubst{\expr}}{\sequence{\idx_a}}}\)\\
        \(:{\typearray{\seqsubst{\type_p}{\var_p}{\idx_a}}{\idxappend{\idx_f\;\seqsubst{\idx_p}{\var_p}{\idx_a}}}}\)}}}\]
  \paragraph{Case {\sc App}}
  \[{\infr[T-App]
    {\typeof{\sEnv}{\kEnv, \haskind{\var}{\kind}}{\tEnv}
      {\expr_f}
      {\typearray
        {\typefun
          {\sequence{\typearray{\type_i}{\idx_i}}}
          {\typearray{\type_o}{\idx_o}}}
        {\idx_f}}
      \\\\
      {\sequence
        {\typeof{\sEnv}{\kEnv, \haskind{\var}{\kind}}{\tEnv}
          {\expr_a}
          {\typearray{\type_i}{\idxappend{\idx_a \; \idx_i}}}}}
      \qquad
      {\idx_p = \mathit{Max}\llb \idx_f \; \sequence{\idx_a} \rrb}}
    {\typeof{\sEnv}{\kEnv, \haskind{\var}{\kind}}{\tEnv}
      {\app{\expr_f}{\sequence{\expr_a}}}
      {\typearray{\type_o}{\idxappend{\idx_p \; \idx_o}}}}}\]
  The induction hypothesis gives derivations for
  ${\sEnv};{\kEnv, \haskind{\var}{\kind}};{\typesubst{\tEnv}}\vdash$
  ${\typesubst{\expr_f}}$ :
  {\tt (Arr (-> ((Arr $\typesubst{\type_i}$ $\idx_i$) $\sequence{}$) (Arr $\typesubst{\type_o}$ $\idx_o$)) $\idx_f$)}
  and $\typeof{\sEnv}{\kEnv, \haskind{\var}{\kind}}{\typesubst{\tEnv}}
  {\typesubst{\expr_a}}
  {\typearray{\typesubst{\type_i}}{\idxappend{\idx_a \; \idx_i}}}$
  for each of the $\sequence{\expr_a}$.
  Note that the frames of the function and argument arrays have not changed,
  so the principal frame $\idx_p$ is also unchanged.
  We then derive
  \[{\infr[T-App]
    {\parbox{0.7\textwidth}
      {\centering
        \({\sEnv};{\kEnv};{\typesubst{\tEnv}}\vdash\)
        \({\typesubst{\expr_f}}\)\\
        \(:{\typearray
          {\typefun
            {\sequence{\typearray{\typesubst{\type_i}}{\idx_i}}}
            {\typearray{\typesubst{\type_o}}{\idx_o}}}
          {\idx_f}}\)}
      \\\\\\
      {\sequence
        {\typeof{\sEnv}{\kEnv}{\typesubst{\tEnv}}
          {\typesubst{\expr_a}}
          {\typearray{\typesubst{\type_i}}{\idxappend{\idx_a \; \idx_i}}}}}
      \\\\\\
      {\idx_p = \mathit{Max}\llb \idx_f \; \sequence{\idx_a} \rrb}}
    {\parbox{0.6\textwidth}
      {\centering
        \({\sEnv};{\kEnv};{\typesubst{\tEnv}}\vdash\)
        \({\app{\typesubst{\expr_f}}{\sequence{\typesubst{\expr_a}}}}\)\\
        \(:{\typearray{\typesubst{\type_o}}{\idxappend{\idx_p \; \idx_o}}}\)}}}\]
  \paragraph{Case {\sc Unbox}}
  \[{\infr[T-Unbox]
    {\typeof{\sEnv}{\kEnv, \haskind{\var}{\kind}}{\tEnv}
      {\expr_s}
      {\typearray{\typedsum{\sequence{\notevar{\var_i'}{\sort_i}}}{\type_s}}
        {\idx_s}}
      \\\\
      {\typeof
        {\sEnv, \sequence{\hassort{\var_i}{\sort_i}}}
        {\kEnv, \haskind{\var}{\kind}}
        {\tEnv, \hastype{\var_e}{\seqsubst{\type_s}{\var_i'}{\var_i}}}
        {\expr_b}
        {\typearray{\type_b}{\idx_b}}}
      \\\\
      {\kindof{\sEnv}{\kEnv, \haskind{\var}{\kind}}
        {\typearray{\type_b}{\idx_b}}{\kindarray}}}
    {\typeof{\sEnv}{\kEnv, \haskind{\var}{\kind}}{\tEnv}
      {\dproj{\sequence{\var_i}}{\var_e}{\expr_s}{\expr_b}}
      {\typearray{\type_b}{\idxappend{\idx_s \; \idx_b}}}}}\]
  By the induction hypothesis,
  ${\sEnv};{\kEnv};{\typesubst{\tEnv}}\vdash$
  ${\typesubst{\expr_s}}$ :
  {\tt (Arr (Sigma (($\var_i'$ $\sort_i$) $\sequence{}$) $\typesubst{\type_s}$) $\idx_s$)},
  and
  $\typeof
  {\sEnv, \sequence{\hassort{\var_i}{\sort_i}}}
  {\kEnv}
  {\typesubst{\tEnv}, \hastype{\var_e}{\typesubst{\parens{\seqsubst{\type_s}{\var_i'}{\var_i}}}}}
  {\typesubst{\expr_b}}
  {\typesubst{\typearray{\type_b}{\idx_b}}}$.
  Preservation of kinds implies
  $\kindof{\sEnv}{\kEnv}{\typesubst{\typearray{\type_b}{\idx_b}}}{\kindarray}$.
  Then we can derive
  \[{\infr[T-Unbox]
    {\parbox{0.6\textwidth}
      {\centering
        \({\sEnv};{\kEnv};{\typesubst{\tEnv}}\vdash\)
        \({\typesubst{\expr_s}}\)\\
        \(:{\typearray{\typedsum{\sequence{\notevar{\var_i'}{\sort_i}}}{\typesubst{\type_s}}}
          {\idx_s}}\)}
      \\\\\\
      {\parbox{0.6\textwidth}
        {\centering
          \({\sEnv, \sequence{\hassort{\var_i}{\sort_i}}};
          {\kEnv};
          {\typesubst{\tEnv}, \hastype{\var_e}{\typesubst{\parens{\seqsubst{\type_s}{\var_i'}{\var_i}}}}}\)
          \(\vdash{\typesubst{\expr_b}}\)
          \(:{\typesubst{\typearray{\type_b}{\idx_b}}}\)}}
      \\\\\\
      {\kindof{\sEnv}{\kEnv}
        {\typesubst{\typearray{\type_b}{\idx_b}}}{\kindarray}}}
    {\parbox{0.75\textwidth}
      {\centering
        \({\sEnv};{\kEnv};{\typesubst{\tEnv}}\vdash\)
        \({\dproj{\sequence{\var_i}}{\var_e}{\typesubst{\expr_s}}{\typesubst{\expr_b}}}\)\\
        \(:{\typesubst{\typearray{\type_b}{\idxappend{\idx_s \; \idx_b}}}}\)}}}\]
\end{sproof}

\newcommand{\exprsubst}[1]{\subst{#1}{\var}{\expr_{\var}}}
\begin{lemma}[Preservation of types under term substitution]
  \label{EESub}
  If $\typeof{\sEnv}{\kEnv}{\tEnv,\hastype{\var}{\type_{\var}}}{\term}{\type}$
  and $\typeof{\sEnv}{\kEnv}{\tEnv}{\expr_{\var}}{\type_{\var}}$
  then $\typeof{\sEnv}{\kEnv}{\tEnv}
  {\exprsubst{\term}}
  {\type}$.
\end{lemma}
\begin{sproof}[We use induction on the derivation of
  $\typeof{\sEnv}{\kEnv}{\tEnv,\hastype{\var}{\type_{\var}}}{\term}{\type}$.]
  We use induction on the derivation of
  $\typeof{\sEnv}{\kEnv}{\tEnv,\hastype{\var}{\type_{\var}}}{\term}{\type}$.
  \paragraph{Case {\sc Eqv}}
  \[{\infr[T-Eqv]
    {\typeof{\sEnv}{\kEnv}{\tEnv, \hastype{\var}{\type_{\var}}}{\term}{\type'}
      \qquad \teqv{\type}{\type'}}
    {\typeof{\sEnv}{\kEnv}{\tEnv, \hastype{\var}{\type_{\var}}}
      {\term}
      {\type}}}\]
  By the induction hypothesis,
  $\typeof{\sEnv}{\kEnv}{\tEnv}{\exprsubst{\term}}{\type'}$.
  So we can derive
  \[{\infr[T-Eqv]
    {\typeof{\sEnv}{\kEnv}{\tEnv}{\exprsubst{\term}}{\type'}
      \qquad \teqv{\type}{\type'}}
    {\typeof{\sEnv}{\kEnv}{\tEnv, \hastype{\var}{\type_{\var}}}
      {\exprsubst{\term}}
      {\type}}}\]
  \paragraph{Case {\sc Var}}
  \[{\infr[T-Var]
    {\parens{\hastype{\var'}{\type}} \in \tEnv}
    {\typeof{\sEnv}{\kEnv}{\tEnv, \hastype{\var}{\type_{\var}}}
      {\var'}
      {\type}}}\]
  Suppose $\var' = \var$.
  Then $\exprsubst{\var'} = \expr_{\var}$,
  and by assumption, $\typeof{\sEnv}{\kEnv}{\tEnv}{\expr_{\var}}{\type_{\var}}$.
  Since the type environment maps $\var$ to both $\type$ and $\type_{\var}$,
  we know that $\type = \type_{\var}$.
  Therefore, $\typeof{\sEnv}{\kEnv}{\tEnv}{\expr_{\var}}{\type}$.
  Otherwise, $\var' \not= \var$, and $\exprsubst{\var'} = \var'$.
  The type environment still contains $\parens{\hastype{\var'}{\type}}$,
  so we can still derive
  \[{\infr[T-Var]
    {\parens{\hastype{\var'}{\type}} \in \tEnv}
    {\typeof{\sEnv}{\kEnv}{\tEnv}{\var'}{\type}}}\]
  \paragraph{Case {\sc Op}, {\sc Base}}
  The variable $\var$ cannot appear free in $\term$, so $\exprsubst{\term} = \term$.
  Neither {\sc T-Op} nor {\sc T-Base} depends on the environment,
  so the same rule which produced the original derivation
  can also derive $\typeof{\sEnv}{\kEnv}{\tEnv}{\term}{\type}$.
  \paragraph{Case {\sc Array}}
  \[{\infr[T-Array]
    {\sequence{\typeof{\sEnv}{\kEnv}{\tEnv, \hastype{\var}{\type_{\var}}}
        {\atom}{\type_a}}
    \\\\
    \kindof{\sEnv}{\kEnv}{\type_a}{\kindatom}
    \qquad
    \mathit{Length}\llb \sequence{\atom} \rrb = \prod{\sequence{\nat}}}
    {\typeof{\sEnv}{\kEnv}{\tEnv, \hastype{\var}{\type_{\var}}}
      {\arrlit{\sequence{\atom}}{\sequence{\nat}}}
      {\typearray{\type_a}{\idxshape{\sequence{\nat}}}}}}\]
  By the induction hypothesis,
  $\typeof{\sEnv}{\kEnv}{\tEnv}{\exprsubst{\atom}}{\type_a}$
  for each of $\sequence{\atom}$.
  So we can derive
  \[{\infr[T-Array]
    {\sequence{\typeof{\sEnv}{\kEnv}{\tEnv}
        {\exprsubst{\atom}}{\type_a}}
    \\\\
    \kindof{\sEnv}{\kEnv}{\type_a}{\kindatom}
    \qquad
    \mathit{Length}\llb \sequence{\atom} \rrb = \prod{\sequence{\nat}}}
    {\typeof{\sEnv}{\kEnv}{\tEnv}
      {\arrlit{\sequence{\exprsubst{\atom}}}{\sequence{\nat}}}
      {\typearray{\type_a}{\idxshape{\sequence{\nat}}}}}}\]
  \paragraph{Case {\sc Frame}}
  \[{\infr[T-Frame]
    {\sequence{\typeof{\sEnv}{\kEnv}{\tEnv, \hastype{\var}{\type_{\var}}}
        {\expr_a}{\typearray{\type_a}{\idx_a}}}
      \\\\
      {\kindof{\sEnv}{\kEnv}
        {\typearray{\type_a}{\idx_a}}
        {\kindarray}}
      \qquad
      \mathit{Length}\llb \sequence{\atom} \rrb = \prod{\sequence{\nat}}}
  {\parbox{0.6\textwidth}
    {\centering
      \({\sEnv};{\kEnv};{\tEnv, \hastype{\var}{\type_{\var}}}\vdash\)
      \({\frm{\sequence{\expr_a}}{\sequence{\nat}}}\)\\
      \(:{\typearray{\type_a}{\idxappend{\idxshape{\sequence{\nat}} \; \idx_a}}}\)}}}\]
  The induction hypothesis implies for each of $\sequence{\expr_a}$
  that $\typeof{\sEnv}{\kEnv}{\tEnv}{\expr_a}{\typearray{\type_a}{\idx_a}}$.
  This leads to
  \[{\infr[T-Frame]
    {\sequence{\typeof{\sEnv}{\kEnv}{\tEnv}
        {\exprsubst{\expr_a}}{\typearray{\type_a}{\idx_a}}}
      \\\\
      {\kindof{\sEnv}{\kEnv}
        {\typearray{\type_a}{\idx_a}}
        {\kindarray}}
      \qquad
      \mathit{Length}\llb \sequence{\atom} \rrb = \prod{\sequence{\nat}}}
  {\parbox{0.6\textwidth}
    {\centering
      \({\sEnv};{\kEnv};{\tEnv}\vdash\)
      \({\frm{\sequence{\exprsubst{\expr_a}}}{\sequence{\nat}}}\)\\
      \(:{\typearray{\type_a}{\idxappend{\idxshape{\sequence{\nat}} \; \idx_a}}}\)}}}\]
  \paragraph{Case {\sc EmptyA}, {\sc EmptyF}}
  For each of these rules,
  the only premise is a kind check,
  which  does not mention the conclusion's type environment.
  So $\typeof{\sEnv}{\kEnv}{\tEnv}{\exprsubst{\term}}{\type}$
  can be derived from the same premise.
  \paragraph{Case {\sc Lam}}
  \[{\infr[T-Lam]
    {\typeof{\sEnv}{\kEnv}
      {\tEnv, \hastype{\var}{\type_{\var}}, \sequence{\hastype{\var_i}{\type_i}}}
      {\expr}{\type_o}}
    {\typeof{\sEnv}{\kEnv}{\tEnv, \hastype{\var}{\type_{\var}}}
      {\lam{\sequence{\notevar{\var_i}{\type_i}}}{\expr}}
      {\typefun{\sequence{\type_i}}{\type_o}}}}\]
  The induction hypothesis gives
  $\typeof{\sEnv}{\kEnv}{\tEnv, \sequence{\hastype{\var_i}{\type_i}}}
  {\exprsubst{\expr}}{\type_o}$,
  so we derive
  \[{\infr[T-Lam]
    {\typeof{\sEnv}{\kEnv}
      {\tEnv, \sequence{\hastype{\var_i}{\type_i}}}
      {\exprsubst{\expr}}{\type_o}}
    {\typeof{\sEnv}{\kEnv}{\tEnv, \hastype{\var}{\type_{\var}}}
      {\lam{\sequence{\notevar{\var_i}{\type_i}}}{\exprsubst{\expr}}}
      {\typefun{\sequence{\type_i}}{\type_o}}}}\]
  \paragraph{Case {\sc TLam}}
  \[{\infr[T-TLam]
    {\typeof{\sEnv}{\kEnv, \sequence{\haskind{\var_u}{\kind_u}}}{\tEnv, \hastype{\var}{\type_{\var}}}
      {\expr}
      {\type_u}}
    {\typeof{\sEnv}{\kEnv}{\tEnv, \hastype{\var}{\type_{\var}}}
      {\tlam{\sequence{\notevar{\var_u}{\kind_u}}}{\expr}}
      {\typeuniv{\sequence{\notevar{\var_u}{\kind_u}}}{\type_u}}}}\]
  By the induction hypothesis,
  $\typeof{\sEnv}{\kEnv, \sequence{\haskind{\var_u}{\kind_u}}}{\tEnv}
  {\exprsubst{\expr}}{\type_u}$.
  We then derive
  \[{\infr[T-TLam]
    {\typeof{\sEnv}{\kEnv, \sequence{\haskind{\var_u}{\kind_u}}}{\tEnv}
      {\exprsubst{\expr}}
      {\type_u}}
    {\typeof{\sEnv}{\kEnv}{\tEnv}
      {\tlam{\sequence{\notevar{\var_u}{\kind_u}}}{\exprsubst{\expr}}}
      {\typeuniv{\sequence{\notevar{\var_u}{\kind_u}}}{\type_u}}}}\]
  \paragraph{Case {\sc ILam}}
  \[{\infr[T-ILam]
    {\typeof{\sEnv, \sequence{\hassort{\var_p}{\sort_p}}}{\kEnv}{\tEnv, \hastype{\var}{\type_{\var}}}
      {\expr}
      {\type_p}}
    {\typeof{\sEnv}{\kEnv}{\tEnv, \hastype{\var}{\type_{\var}}}
      {\ilam{\sequence{\notevar{\var_p}{\sort_p}}}{\expr}}
      {\typedprod{\sequence{\notevar{\var_p}{\sort_p}}}{\type_p}}}}\]
  The induction hypothesis implies
  $\typeof{\sEnv, \sequence{\hassort{\var_p}{\sort_p}}}{\kEnv}{\tEnv}
      {\exprsubst{\expr}}{\type_p}$,
      which leads to the derivation
  \[{\infr[T-ILam]
    {\typeof{\sEnv, \sequence{\hassort{\var_p}{\sort_p}}}{\kEnv}{\tEnv}
      {\exprsubst{\expr}}
      {\type_p}}
    {\typeof{\sEnv}{\kEnv}{\tEnv}
      {\ilam{\sequence{\notevar{\var_p}{\sort_p}}}{\exprsubst{\expr}}}
      {\typedprod{\sequence{\notevar{\var_p}{\sort_p}}}{\type_p}}}}\]
  \paragraph{Case {\sc Box}}
  \[{\infr[T-Box]
    {\sequence{\sortof{\sEnv}{\idx_s}{\sort_s}}
      \qquad
      {\kindof{\sEnv}{\kEnv}
        {\typedsum{\sequence{\notevar{\var_s}{\sort_s}}}{\type_s}}
        {\kindatom}}
      \\
      {\typeof{\sEnv}{\kEnv}{\tEnv, \hastype{\var}{\type_{\var}}}
        {\expr}
        {\seqsubst{\type_s}{\var_s}{\idx_s}}}}
    {\parbox{0.67\textwidth}
      {\centering
        \({\sEnv};{\kEnv};{\tEnv, \hastype{\var}{\type_{\var}}}\vdash\)
        \({\dsum
          {\sequence{\idx_s}}{\expr}
          {\typedsum{\sequence{\notevar{\var_s}{\sort_s}}}{\type_s}}}\)\\
        \(:{\typedsum{\sequence{\notevar{\var_s}{\sort_s}}}{\type_s}}\)}}}\]
  By the induction hypothesis,
  $\typeof{\sEnv}{\kEnv}{\tEnv}{\exprsubst{\expr}}{\seqsubst{\type_s}{\var_s}{\idx_s}}$.
  We then derive
  \[{\infr[T-Box]
    {\sequence{\sortof{\sEnv}{\idx_s}{\sort_s}}
      \qquad
      {\kindof{\sEnv}{\kEnv}
        {\typedsum{\sequence{\notevar{\var_s}{\sort_s}}}{\type_s}}
        {\kindatom}}
      \\
      {\typeof{\sEnv}{\kEnv}{\tEnv}
        {\exprsubst{\expr}}
        {\seqsubst{\type_s}{\var_s}{\idx_s}}}}
    {\parbox{0.67\textwidth}
      {\centering
        \({\sEnv};{\kEnv};{\tEnv}\vdash\)
        \({\dsum
          {\sequence{\idx_s}}{\exprsubst{\expr}}
          {\typedsum{\sequence{\notevar{\var_s}{\sort_s}}}{\type_s}}}\)\\
        \(:{\typedsum{\sequence{\notevar{\var_s}{\sort_s}}}{\type_s}}\)}}}\]
  \paragraph{Case {\sc TApp}}
  \[{\infr[T-TApp]
    {\typeof
      {\sEnv}{\kEnv}{\tEnv, \hastype{\var}{\type_{\var}}}
      {\expr}
      {\typearray{\typeuniv{\sequence{\notevar{\var_u}{\kind_u}}}{\typearray{\type_u}{\idx_u}}}{\idx_f}}
      \\\\
      \sequence{\kindof{\sEnv}{\kEnv}{\type_a}{\kind_u}}}
    {\typeof {\sEnv}{\kEnv}{\tEnv, \hastype{\var}{\type_{\var}}}
      {\tapp{\expr}{\sequence{\type_a}}}
      {\typearray{\seqsubst{\type_u}{\var_u}{\type_a}}{\idxappend{\idx_f\;\idx_u}}}}}\]
  The induction hypothesis gives a derivation for
  ${\sEnv};{\kEnv};{\tEnv}\vdash$
  ${\exprsubst{\expr}}$ :
  {\tt (Arr (Forall (($\var_u$ $\kind_u$) $\sequence{}$) (Arr $\type_u$ $\idx_u$)) $\idx_f$)}.
  We can then construct the derivation
  \[{\infr[T-TApp]
    {\typeof
      {\sEnv}{\kEnv}{\tEnv}
      {\exprsubst{\expr}}
      {\typearray{\typeuniv{\sequence{\notevar{\var_u}{\kind_u}}}{\typearray{\type_u}{\idx_u}}}{\idx_f}}
      \\\\
      \sequence{\kindof{\sEnv}{\kEnv}{\type_a}{\kind_u}}}
    {\typeof {\sEnv}{\kEnv}{\tEnv}
      {\tapp{\exprsubst{\expr}}{\sequence{\type_a}}}
      {\typearray{\seqsubst{\type_u}{\var_u}{\type_a}}{\idxappend{\idx_f\;\idx_u}}}}}\]
  \paragraph{Case {\sc IApp}}
  \[{\infr[T-IApp]
    {\typeof
      {\sEnv}{\kEnv}{\tEnv, \hastype{\var}{\type_{\var}}}
      {\expr}
      {\typearray{\typedprod{\sequence{\notevar{\var_p}{\sort_p}}}{\typearray{\type_p}{\idx_p}}}{\idx_f}}
      \\\\
      \sequence{\sortof{\sEnv}{\idx_a}{\sort_p}}}
    {\parbox{0.6\textwidth}
      {\centering
        \({\sEnv};{\kEnv};{\tEnv, \hastype{\var}{\type_{\var}}}\vdash\)
        \({\iapp{\expr}{\sequence{\idx_a}}}\)\\
        \(:{\typearray
          {\seqsubst{\type_p}{\var_p}{\idx_a}}
          {\idxappend{\idx_f\;\seqsubst{\idx_p}{\var_p}{\idx_a}}}}\)}}}\]
  By the induction hypothesis, we have
  ${\sEnv};{\kEnv};{\tEnv, \hastype{\var}{\type_{\var}}}\vdash$
  ${\exprsubst{\expr}}$ :
  {\tt (Arr (Pi (($\var_p$ $\sort_p$) $\sequence{}$) (Arr $\type_p$ $\idx_p$)) $\idx_f$)}.
  This leads to
  \[{\infr[T-IApp]
    {\typeof
      {\sEnv}{\kEnv}{\tEnv}
      {\exprsubst{\expr}}
      {\typearray{\typedprod{\sequence{\notevar{\var_p}{\sort_p}}}{\typearray{\type_p}{\idx_p}}}{\idx_f}}
      \\\\
      \sequence{\sortof{\sEnv}{\idx_a}{\sort_p}}}
    {\parbox{0.6\textwidth}
      {\centering
        \({\sEnv};{\kEnv};{\tEnv}\vdash\)
        \({\iapp{\exprsubst{\expr}}{\sequence{\idx_a}}}\)\\
        \(:{\typearray
          {\seqsubst{\type_p}{\var_p}{\idx_a}}
          {\idxappend{\idx_f\;\seqsubst{\idx_p}{\var_p}{\idx_a}}}}\)}}}\]
  \paragraph{Case {\sc App}}
  \[{\infr[T-App]
    {\typeof{\sEnv}{\kEnv}{\tEnv, \hastype{\var}{\type_{\var}}}
      {\expr_f}
      {\typearray
        {\typefun
          {\sequence{\typearray{\type_i}{\idx_i}}}
          {\typearray{\type_o}{\idx_o}}}
        {\idx_f}}
      \\\\
      {\sequence
        {\typeof{\sEnv}{\kEnv}{\tEnv, \hastype{\var}{\type_{\var}}}
          {\expr_a}
          {\typearray{\type_i}{\idxappend{\idx_a \; \idx_i}}}}}
      \qquad
      {\idx_p = \mathit{Max}\llb \idx_f \; \sequence{\idx_a} \rrb}}
    {\typeof{\sEnv}{\kEnv}{\tEnv, \hastype{\var}{\type_{\var}}}
      {\app{\expr_f}{\sequence{\expr_a}}}
      {\typearray{\type_o}{\idxappend{\idx_p \; \idx_o}}}}}\]
  The induction hypothesis implies
  ${\sEnv};{\kEnv};{\tEnv}\vdash$
  ${\exprsubst{\expr_f}}$ :
  {\tt (Arr (-> ((Arr $\type_i$ $\idx_i$) $\sequence{}$) (Arr $\type_o$ $\idx_o$)) $\idx_f$)}
  and
  $\typeof{\sEnv}{\kEnv}{\tEnv}
  {\exprsubst{\expr_a}}
  {\typearray{\type_i}{\idxappend{\idx_a \; \idx_i}}}$
  for each of $\sequence{\expr_a}$.
  Since the individual frames $\idx_f, \sequence{\idx_a}$ are unchanged,
  so is the principal frame $\idx_p$.
  Thus we derive
  \[{\infr[T-App]
    {\typeof{\sEnv}{\kEnv}{\tEnv}
      {\exprsubst{\expr_f}}
      {\typearray
        {\typefun
          {\sequence{\typearray{\type_i}{\idx_i}}}
          {\typearray{\type_o}{\idx_o}}}
        {\idx_f}}
      \\\\
      {\sequence
        {\typeof{\sEnv}{\kEnv}{\tEnv}
          {\exprsubst{\expr_a}}
          {\typearray{\type_i}{\idxappend{\idx_a \; \idx_i}}}}}
      \qquad
      {\idx_p = \mathit{Max}\llb \idx_f \; \sequence{\idx_a} \rrb}}
    {\typeof{\sEnv}{\kEnv}{\tEnv}
      {\app{\exprsubst{\expr_f}}{\sequence{\exprsubst{\expr_a}}}}
      {\typearray{\type_o}{\idxappend{\idx_p \; \idx_o}}}}}\]
  \paragraph{Case {\sc Unbox}}
  \[{\infr[T-Unbox]
    {\typeof{\sEnv}{\kEnv}{\tEnv, \hastype{\var}{\type_{\var}}}
      {\expr_s}
      {\typearray{\typedsum{\sequence{\var_i'}{\sort_i}}{\type_s}}
        {\idx_s}}
      \\\\
      {\typeof
        {\sEnv, \sequence{\hassort{\var_i}{\sort_i}}}
        {\kEnv}
        {\tEnv, \hastype{\var_e}{\seqsubst{\type_s}{\var_i'}{\var_i}}, \hastype{\var}{\type_{\var}}}
        {\expr_b}
        {\typearray{\type_b}{\idx_b}}}
      \\\\
      {\kindof{\sEnv}{\kEnv, \haskind{\var}{\kind}}
        {\typearray{\type_b}{\idx_b}}{\kindarray}}}
    {\typeof{\sEnv}{\kEnv}{\tEnv, \hastype{\var}{\type_{\var}}}
      {\dproj{\sequence{\var_i}}{\var_e}{\expr_s}{\expr_b}}
      {\typearray{\type_b}{\idxappend{\idx_s \; \idx_b}}}}}\]
  The induction hypothesis gives
  ${\sEnv};{\kEnv};{\tEnv}\vdash$
  ${\exprsubst{\expr_s}}$ :
  {\tt (Arr (Sigma (($\var_i'$ $\sort_i$) $\sequence{}$) $\type_s$) $\idx_s$)}
  and
  $\typeof{\sEnv, \sequence{\hassort{\var_i}{\sort_i}}}
  {\kEnv}
  {\tEnv, \hastype{\var_e}{\seqsubst{\type_s}{\var_i'}{\var_i}}}
  {\exprsubst{\expr_b}}{\typearray{\type_b}{\idx_b}}$.
  We then derive
  \[{\infr[T-Unbox]
    {\typeof{\sEnv}{\kEnv}{\tEnv}
      {\exprsubst{\expr_s}}
      {\typearray{\typedsum{\sequence{\var_i'}{\sort_i}}{\type_s}}
        {\idx_s}}
      \\\\
      {\typeof
        {\sEnv, \sequence{\hassort{\var_i}{\sort_i}}}
        {\kEnv}
        {\tEnv, \hastype{\var_e}{\seqsubst{\type_s}{\var_i'}{\var_i}}}
        {\exprsubst{\expr_b}}
        {\typearray{\type_b}{\idx_b}}}
      \\\\
      {\kindof{\sEnv}{\kEnv, \haskind{\var}{\kind}}
        {\type_b}{\kindarray}}}
    {\parbox{0.6\textwidth}
      {\centering
        \({\sEnv};{\kEnv};{\tEnv}\vdash\)
        \({\dproj{\sequence{\var_i}}{\var_e}{\exprsubst{\expr_s}}{\exprsubst{\expr_b}}}\)\\
        \(:{\typearray{\type_b}{\idxappend{\idx_s \; \idx_b}}}\)}}}\]
\end{sproof}

We call an environment well-formed,
noted as $\wfenv{\sEnv}{\kEnv}{\tEnv}$,
if for every binding $\hastype{\var}{\type} \in \tEnv$,
we can derive $\kindof{\sEnv}{\kEnv}{\type}{\kindarray}$.
This is the expected case,
rather than permitting $\type$ to have kind $\kindatom$,
because a lone variable is an expression and ought to stand for an array value.

When we show that the typing judgment only ascribes types of the appropriate kind,
the case for the {\sc T-Eqv} rule
relies on the earlier lemma that the type equivalence relation respects kinding,
\ie, two equivalent types will have the same kind
when checked in the same environment.

\begin{theorem}[Ascription of well-kinded types]
  \label{wellkinded}
  Given $\typeof{\sEnv}{\kEnv}{\tEnv}{\term}{\type}$ where $\wfenv{\sEnv}{\kEnv}{\tEnv}$:
\begin{itemize}
  \item{If $\term$ is an expression, then $\kindof{\sEnv}{\kEnv}{\type}{\kindarray}$}
  \item{If $\term$ is an atom, then $\kindof{\sEnv}{\kEnv}{\type}{\kindatom}$}
\end{itemize}
\end{theorem}
\begin{sproof}[This follows from induction on
  the derivation of $\typeof{\sEnv}{\kEnv}{\tEnv}{\term}{\type}$.
  It is not sufficient to point out that each typing rule ascribes a type
  whose form matches the appropriate kind.
  Elimination-form cases call for a little extra work.
  For {\sc Unbox}, the kind check on the result type
  is necessary to ensure that existentially quantified variables do not leak out.
  The {\sc App} case must ensure that index variables in the ascribed type
  actually appear in the environment.
  This is guaranteed because the principal frame is always
  chosen to be one of the function- or argument-position frames.]
  We use induction on the derivation of $\typeof{\sEnv}{\kEnv}{\tEnv}{\term}{\type}$.
  First, we prove the cases where $\term$ is an expression.
  \paragraph{Case {\sc Var}:}
  \[{\infr[T-Array]
    {\parens{\hastype{\var}{\type}} \in \tEnv}
    {\typeof{\sEnv}{\kEnv}{\tEnv}{\var}{\type}}}\]
  Since $\hastype{\var}{\type} \in \tEnv$, this follows directly from $\wfenv{\sEnv}{\kEnv}{\tEnv}$.
  \paragraph{Case {\sc Array}:}
  \[{\infr[T-Array]
    {\sequence{\typeof {\sEnv}{\kEnv}{\tEnv}{\atom_j}{\type_a}}
      \\\\
      \kindof{\sEnv}{\kEnv}{\type_a}{\kindatom}
      \\\\
      \mathit{Length}\llb \sequence{\atom} \rrb = \prod{\sequence{\nat}}}
    {\typeof{\sEnv}{\kEnv}{\tEnv}
      {\arrlit{\sequence{\atom}}{\sequence{\nat}}}
      {\typearray{\type_a}{\idxshape{\sequence{\nat}}}}}}\]
  We can derive
  \[{\infr[K-Array]
    {\kindof{\sEnv}{\kEnv}{\type_a}{\kindatom}
      \qquad
      {\infr[S-Shape]
        {\sequence{\infr[S-Nat]{\ }{\sortof{\sEnv}{\nat}{\sortdim}}}}
        {\sortof{\sEnv}{\idxshape{\sequence{\nat}}}{\sortshp}}}}
    {\kindof{\sEnv}{\kEnv}{\typearray{\type_a}{\idxshape{\sequence{\nat}}}}{\kindarray}}}\]
  \paragraph{Case {\sc Frame}:}
  \[{\infr[T-Frame]
    {\sequence{\typeof{\sEnv}{\kEnv}{\tEnv}
        {\expr}
        {\typearray{\type_c}{\idx_c}}}
      \\
      \kindof{\sEnv}{\kEnv}{\typearray{\type_c}{\idx_c}}{\kindarray}
      \\
      \mathit{Length}\llb \sequence{\expr} \rrb = \prod{\parens{\sequence{\nat}}}}
    {\typeof{\sEnv}{\kEnv}{\tEnv}
      {\frm{\sequence{\expr}}{\sequence{\nat}}}
      {\typearray{\type_c}{\idxappend{\idxshape{\sequence{\nat}}\;\idx_c}}}}}\]
  The second premise must be derived as
  \[{\infr[K-Array]
    {\sortof{\sEnv}{\idx_c}{\sortshp}
      \qquad
      \kindof{\sEnv}{\kEnv}{\type_c}{\kindatom}}
    {\kindof{\sEnv}{\kEnv}{\typearray{\type_c}{\idx_c}}{\kindarray}}}\]
  Using this knowledge about $\idx_c$,
  we can show the array type's shape is indeed a $\sortshp$:
  \[{\infr[S-Append]
    {\sortof{\sEnv}{\idx_c}{\sortshp}
      \qquad
      {\infr[S-Shape]
        {\sequence{\infr[S-Nat]
            {\ }
            {\sortof{\sEnv}{\nat}{\sortdim}}}}
        {\sortof{\sEnv}{\idxshape{\sequence{\nat}}}{\sortshp}}}}
    {\sortof{\sEnv}{\idxappend{\idxshape{\sequence{\nat}}\;\idx_c}}{\sortshp}}}\]
  Then we can derive
  \[{\infr[K-Array]
    {\kindof{\sEnv}{\kEnv}{\type_c}{\kindatom}
      \qquad
      \sortof{\sEnv}{\idxappend{\idxshape{\sequence{\nat}}\;\idx_c}}{\sortshp}}
    {\kindof{\sEnv}{\kEnv}
      {\typearray{\type_c}{\idxappend{\idxshape{\sequence{\nat}}\;\idx_c}}}
      {\kindarray}}}\]
  \paragraph{Case {\sc EmptyA}:}
  \[{\infr[T-EmptyA]
    {\kindof{\sEnv}{\kEnv}{\type_a}{\kindatom}
      \qquad
    0 \in \sequence{\nat}}
    {\typeof{\sEnv}{\kEnv}{\tEnv}
      {\emptyarrlit{\type_a}{\sequence{\nat}}}
      {\typearray{\type_a}{\idxshape{\sequence{\nat}}}}}}\]
  We derive
  \[{\infr[K-Array]
    {\kindof{\sEnv}{\kEnv}{\type_a}{\kindatom}
      \qquad
      {\infr[S-Shape]
        {\sequence{\infr[S-Nat]
            {\ }
            {\sortof{\sEnv}{\nat}{\sortdim}}}}
        {\sortof{\sEnv}{\idxshape{\sequence{\nat}}}{\sortshp}}}}
    {\kindof{\sEnv}{\kEnv}{\typearray{\type_a}{\idxshape{\sequence{\nat}}}}{\kindarray}}}\]
  \paragraph{Case {\sc EmptyF}:}
  \[{\infr[T-EmptyF]
    {\kindof{\sEnv}{\kEnv}{\type_a}{\kindatom}
      \qquad
      \sortof{\sEnv}{\idx}{\sortshp}
      \qquad
      0 \in \sequence{\nat}}
    {\typeof{\sEnv}{\kEnv}{\tEnv}
      {\emptyfrm{\typearray{\type_a}{\idx_c}}{\sequence{\nat}}}
      {\typearray{\type_a}{\idxappend{\idxshape{\sequence{\nat}}\;\idx_c}}}}}\]
  First, we show that the array type's shape has sort $\sortshp$:
  \[{\infr[S-Append]
    {\sortof{\sEnv}{\idx_c}{\sortshp}
      \qquad
      {\infr[S-Shape]
        {\sequence{\infr[S-Nat]
            {\ }
            {\sortof{\sEnv}{\nat}{\sortdim}}}}
        {\sortof{\sEnv}{\idxshape{\sequence{\nat}}}{\sortshp}}}}
    {\sortof{\sEnv}{\idxappend{\idxshape{\sequence{\nat}}\;\idx_c}}{\sortshp}}}\]
  Then we derive
  \[{\infr[K-Array]
    {\kindof{\sEnv}{\kEnv}{\type_a}{\kindatom}
      \\
      \sortof{\sEnv}{\idxappend{\idxshape{\sequence{\nat}}\;\idx_c}}{\sortshp}}
    {\kindof{\sEnv}{\kEnv}
      {\typearray{\type_a}{\idxappend{\idxshape{\sequence{\nat}}\;\idx_c}}}
      {\kindarray}}}\]
  \paragraph{Case {\sc TApp}:}
  \[{\infr[T-TApp]
    {\typeof{\sEnv}{\kEnv}{\tEnv}
      {\expr_f}
      {\typearray
        {\typeuniv{\sequence{\notevar{\var}{\kind}}}{\typearray{\type_u}{\idx_u}}}
        {\idx_f}}
      \\\\
      \sequence{\kindof{\sEnv}{\kEnv}{\type_a}{\kind}}}
    {\typeof{\sEnv}{\kEnv}{\tEnv}
      {\tapp{\expr_f}{\sequence{\type_a}}}
      {\typearray{\seqsubst{\type_u}{\var}{\type_a}}{\idxappend{\idx_f \; \idx_u}}}}}\]
  By the induction hypothesis, we have a kind derivation for the type of $\expr_f$:
  \[{\infr[K-Array]
    {\sortof{\sEnv}{\idx_f}{\sortshp}
      \qquad
      \kindof{\sEnv}{\kEnv}
      {\typeuniv
        {\sequence{\notevar{\var}{\kind}}}
        {\typearray{\type_u}{\idx_u}}}
      {\kindatom}}
    {\kindof{\sEnv}{\kEnv}
      {\typearray
        {\typeuniv
          {\sequence{\notevar{\var}{\kind}}}
          {\typearray{\type_u}{\idx_u}}}
        {\idx_f}}
      {\kindarray}}}\]
  The derivation for the second premise must have the following structure:
  \[{\infr[K-Univ]
    {\infr[K-Array]
      {\sortof{\sEnv}{\idx_u}{\sortshp}
        \\
        \kindof{\sEnv}{\kEnv, \sequence{\haskind{\var}{\kind}}}
        {\type_u}{\kindatom}}
      {\kindof{\sEnv}{\kEnv, \sequence{\haskind{\var}{\kind}}}
        {\typearray{\type_u}{\idx_u}}
        {\kindarray}}}
    {\kindof{\sEnv}{\kEnv}
      {\typeuniv
        {\sequence{\notevar{\var}{\kind}}}
        {\typearray{\type_u}{\idx_u}}}
      {\kindatom}}}\]
  Using the kind derivation for $\type_u$ in the extended environment,
  we apply Lemma \ref{TTSub} (type substitution preserves kinds) to get
  $\kindof{\sEnv}{\kEnv}{\seqsubst{\type_u}{\var}{\type_a}}{\kindatom}$.
  Then we can derive
  \[{\infr[K-Array]
    {\kindof{\sEnv}{\kEnv}{\seqsubst{\type_u}{\var}{\type_a}}{\kindatom}
      \qquad
      {\infr[S-Append]
        {\sortof{\sEnv}{\idx_f}{\sortshp}
          \\\\
          \sortof{\sEnv}{\idx_u}{\sortshp}}
        {\sortof{\sEnv}{\idxappend{\idx_f \; \idx_u}}{\sortshp}}}}
    {\kindof{\sEnv}{\kEnv}
      {\typearray
        {\seqsubst{\type_u}{\var}{\type_a}}
        {\idxappend{\idx_f \; \idx_u}}}
      {\kindarray}}}\]
  \paragraph{Case {\sc IApp}:}
  \[{\infr[T-IApp]
    {\typeof{\sEnv}{\kEnv}{\tEnv}
      {\expr_f}
      {\typearray
        {\typedprod
          {\sequence{\notevar{\var}{\sort}}}
          {\typearray{\type_p}{\idx_p}}}
        {\idx_f}}
      \qquad
      \sequence{\sortof{\sEnv}{\idx_a}{\sort}}}
    {\typeof{\sEnv}{\kEnv}{\tEnv}
      {\iapp{\expr_f}{\sequence{\idx_a}}}
      {\typearray
        {\seqsubst{\type_p}{\var}{\idx_a}}
        {\idxappend{\idx_f  \; \seqsubst{\idx_p}{\var}{\idx_a}}}}}}\]
  The induction hypothesis implies that the type of $\expr_f$ has kind $\kindarray$,
  giving us a derivation which ends as follows:
  \[{\infr[K-Array]
    {\sortof{\sEnv}{\idx_f}{\sortshp}
      \qquad
      {\kindof{\sEnv}{\kEnv}
        {\typedprod
          {\sequence{\notevar{\var}{\sort}}}
          {\typearray{\type_p}{\idx_p}}}
        {\kindatom}}}
    {\kindof{\sEnv}{\kEnv}
      {\typearray
        {\typedprod
          {\sequence{\notevar{\var}{\sort}}}
          {\typearray{\type_p}{\idx_p}}}
        {\idx_f}}
      {\kindarray}}}\]
  The derivation of the second premise must itself have this form:
  \[{\infr[K-Pi]
    {\infr[K-Array]
      {\sortof{\sEnv, \sequence{\hassort{\var}{\sort}}}{\idx_p}{\sortshp}
        \qquad
        \kindof{\sEnv, \sequence{\hassort{\var}{\sort}}}{\kEnv}{\type_p}{\kindatom}}
      {\kindof{\sEnv, \sequence{\hassort{\var}{\sort}}}{\kEnv}
        {\typearray{\type_p}{\idx_p}}
        {\kindarray}}}
    {\kindof{\sEnv}{\kEnv}
      {\typedprod
        {\sequence{\notevar{\var}{\sort}}}
        {\typearray{\type_p}{\idx_p}}}
      {\kindatom}}}\]
  Using Lemma \ref{IISub} (index substitution preserves sorts) implies
  $\sortof{\sEnv}{\seqsubst{\idx_p}{\var}{\idx_a}}{\sortshp}$.
  So we derive
  \[{\infr[S-Append]
    {\sortof{\sEnv}{\idx_f}{\sortshp}
      \qquad
      \sortof{\sEnv}{\seqsubst{\idx_p}{\var}{\idx_a}}{\sortshp}}
    {\sortof{\sEnv}{\idxappend{\idx_f  \; \seqsubst{\idx_p}{\var}{\idx_a}}}{\sortshp}}}\]
  Next, Lemma \ref{ITSub} (index substitution preserves kinds) gives us
  $\kindof{\sEnv}{\kEnv}{\seqsubst{\type_p}{\var}{\idx_a}}{\kindatom}$.
  We now have the pieces for the derivation
  \[{\infr[K-Array]
    {\kindof{\sEnv}{\kEnv}{\seqsubst{\type_p}{\var}{\idx_a}}{\kindatom}
      \qquad
      \sortof{\sEnv}{\idxappend{\idx_f  \; \seqsubst{\idx_p}{\var}{\idx_a}}}{\sortshp}}
    {\kindof{\sEnv}{\kEnv}
      {\typearray
        {\seqsubst{\type_p}{\var}{\idx_a}}
        {\idxappend{\idx_f  \; \seqsubst{\idx_p}{\var}{\idx_a}}}}
      {\kindarray}}}\]
  \paragraph{Case {\sc Unbox}:}
  \[{\infr[T-Unbox]
    {\typeof{\sEnv}{\kEnv}{\tEnv}
      {\expr_s}
      {\typearray{\typedsum{\sequence{\notevar{\var_i'}{\sort}}}{\type_s}}
        {\idx_s}}
      \\
    \typeof
    {\sEnv, \sequence{\hassort{\var_i}{\sort}}}
    {\kEnv}
    {\tEnv, \hastype{\var_e}{\seqsubst{\type_s}{\var_i'}{\var_i}}}
    {\expr_b}
    {\typearray{\type_b}{\idx_b}}
    \\
    {\kindof{\sEnv}{\kEnv}{\typearray{\type_b}{\idx_b}}{\kindarray}}
  }
    {\typeof{\sEnv}{\kEnv}{\tEnv}
      {\dproj{\sequence{\var_i}}{\var_e}{\expr_s}{\expr_b}}
      {\typearray{\type_b}{\idxappend{\idx_s \; \idx_b}}}}}\]
  The third premise must be derived by
  \[{\infr[K-Array]
    {\kindof{\sEnv}{\kEnv}{\type_b}{\kindatom} \qquad \sortof{\sEnv}{\idx_b}{\sortshp}}
    {\kindof{\sEnv}{\kEnv}{\typearray{\type_b}{\idx_b}}{\kindarray}}}\]
  The induction hypothesis gives us a kinding derivation for the box array type
  \[{\infr[K-Array]
    {\kindof{\sEnv,\sequence{\notevar{\var_i'}{\sort}}}{\kEnv}
      {\type_s}{\kindarray}
    \qquad
    {\sortof{\sEnv}{\idx_s}{\sortshp}}}
    {\kindof{\sEnv}{\kEnv}
      {\typearray{\typedsum{\sequence{\notevar{\var_i'}{\sort}}}{\type_s}}{\idx_s}}
      {\kindarray}}}\]
  From these derivations, we have the pieces needed to construct the kind derivation for the result type:
  \[\infr[K-Array]
  {\kindof{\sEnv}{\kEnv}{\type_b}{\kindatom}
    \qquad
    \infr[S-Append]
    {\sortof{\sEnv}{\idx_s}{\sortshp}
      \qquad
      \sortof{\sEnv}{\idx_b}{\sortshp}}
    {\sortof{\sEnv}{\idxappend{\idx_s \; \idx_b}}{\sortshp}}}
  {\kindof{\sEnv}{\kEnv}{\typearray{\type_b}{\idxappend{\idx_s \; \idx_b}}}{\kindarray}}\]
  \paragraph{Case {\sc App}:}
  \[{\infr[T-App]
    {\typeof{\sEnv}{\kEnv}{\tEnv}
      {\expr_f}
      {\typearray
        {\typefun
          {\sequence{\typearray{\type_i}{\idx_i}}}
          {\typearray{\type_o}{\idx_o}}}
        {\idx_f}}
      \\
      \sequence{\typeof{\sEnv}{\kEnv}{\tEnv}
        {\expr_a}
        {\typearray{\type_i}{\idxappend{\idx_a \; \idx_i}}}}
      \\
      \idx_p = \bigsqcup\parens{\idx_f, \sequence{\idx_a}}}
    {\typeof{\sEnv}{\kEnv}{\tEnv}
      {\app{\expr_f}{\sequence{\expr_a}}}
      {\typearray{\type_o}{\idxappend{\idx_p \; \idx_o}}}}}\]
  Since $\expr_f$ is well-typed,
  the induction hypothesis implies that
  its type has kind $\kindarray$.
  This derivation must have the form
  \[{\infr[K-Array]
    {\sortof{\sEnv}{\idx_f}{\sortshp}
      \\
      \kindof{\sEnv}{\kEnv}
      {\typefun
        {\sequence{\typearray{\type_i}{\idx_i}}}
        {\typearray{\type_o}{\idx_o}}}
      {\kindatom}}
    {\kindof{\sEnv}{\kEnv}
      {\typearray
        {\typefun
          {\sequence{\typearray{\type_i}{\idx_i}}}
          {\typearray{\type_o}{\idx_o}}}
        {\idx_f}}
      {\kindarray}}}\]
  The second premise must be derived via
  \[{\infr[K-Fun]
    {\sequence{\kindof{\sEnv}{\kEnv}{\typearray{\type_i}{\idx_i}}{\kindarray}}
      \\
      \kindof{\sEnv}{\kEnv}{\typearray{\type_o}{\idx_o}}{\kindarray}}
    {\kindof{\sEnv}{\kEnv}
      {\typefun
        {\sequence{\typearray{\type_i}{\idx_i}}}
        {\typearray{\type_o}{\idx_o}}}
      {\kindatom}}}\]
  Then the kinding derivation for each argument type $\typearray{\type_i}{\idx_i}$
  must include ascription of $\kindatom$ to $\type_i$ and $\sortshp$ to $\idx_i$
  (and similar for the result type $\typearray{\type_o}{\idx_o}$).
  Any index variables which appear in $\idx_p$ must also appear in at least one of
  $\idx_f, \sequence{\idx_a}$,
  so $\idx_p$ must be well-formed
  (\ie, $\sortof{\sEnv}{\idx_p}{\sortshp}$).
  We can then derive
  \[{\infr[K-Array]
    {\typeof{\sEnv}{\kEnv}{\type_o}{\kindatom}
      \qquad
      \sequence{\infr[S-Append]
        {\sortof{\sEnv}{\idx_p}{\sortshp}
          \qquad
          \sortof{\sEnv}{\idx_o}{\sortshp}}
        {\sortof{\sEnv}{\idxappend{\idx_p \; \idx_o}}{\sortshp}}}}
    {\kindof{\sEnv}{\kEnv}
      {\typearray{\type_o}{\idxappend{\idx_p \; \idx_o}}}
      {\kindarray}}}\]

  Now, we describe the atom cases.
  \paragraph{Case {\sc TLam}:}
  \[{\infr[T-TLam]
    {\typeof{\sEnv}{\kEnv, \sequence{\haskind{\var}{\kind}}}{\tEnv}
      {\expr_u}
      {\type_u}}
    {\typeof{\sEnv}{\kEnv}{\tEnv}
      {\tlam{\sequence{\notevar{\var}{\kind}}}{\expr_u}}
      {\typeuniv{\sequence{\notevar{\var}{\kind}}}{\type_u}}}}\]
  By the induction hypothesis,
  $\kindof{\sEnv}{\kEnv, \sequence{\haskind{\var}{\kind}}}{\type_u}{\kindarray}$,
  so we can derive
  \[{\infr[K-Univ]
    {\kindof{\sEnv}{\kEnv, \sequence{\haskind{\var}{\kind}}}{\type_u}{\kindarray}}
    {\kindof{\sEnv}{\kEnv}{\typeuniv{\sequence{\notevar{\var}{\kind}}}{\type_u}}{\kindatom}}}\]
  \paragraph{Case {\sc ILam}:}
  \[{\infr[T-ILam]
    {\typeof{\sEnv, \sequence{\hassort{\var}{\sort}}}{\kEnv}{\tEnv}
      {\expr_p}
      {\type_p}}
    {\typeof{\sEnv}{\kEnv}{\tEnv}
      {\ilam{\sequence{\notevar{\var}{\sort}}}{\expr_p}}
      {\typedprod{\sequence{\notevar{\var}{\sort}}}{\type_p}}}}\]
  The induction hypothesis implies
  $\kindof{\sEnv, \sequence{\hassort{\var}{\sort}}}{\kEnv}{\type_p}{\kindarray}$.
  We then derive
  \[{\infr[K-Pi]
    {\kindof{\sEnv, \sequence{\hassort{\var}{\sort}}}{\kEnv}{\type_p}{\kindarray}}
    {\kindof{\sEnv}{\kEnv}{\typedprod{\sequence{\notevar{\var}{\sort}}}{\expr_p}}{\kindatom}}}\]
  \paragraph{Case {\sc Box}:}
  \[{\infr[T-Box]
    {\sequence{\sortof{\sEnv}{\idx}{\sort}}
      \qquad
      \kindof{\sEnv}{\kEnv}
      {\typedsum{\sequence{\notevar{\var}{\sort}}}{\type_s}}
      {\kindatom}
      \\
      \typeof{\sEnv}{\kEnv}{\tEnv}
      {\expr_s}
      {\seqsubst{\type_s}{\var}{\idx}}}
    {\typeof{\sEnv}{\kEnv}{\tEnv}
      {\dsum
        {\sequence{\idx}}
        {\expr_s}
        {\typedsum{\sequence{\notevar{\var}{\sort}}}{\type_s}}}
      {\typedsum{\sequence{\notevar{\var}{\sort}}}{\type_s}}}}\]
  The required result is a premise in the original derivation.
  \paragraph{Case {\sc Lam}:}
  \[{\infr[T-Lam]
    {\typeof{\sEnv}{\kEnv}{\tEnv}{\expr_f}{\type_o}
      \qquad
      \sequence{\kindof{\sEnv}{\kEnv}{\type_i}{\kindarray}}}
    {\typeof{\sEnv}{\kEnv}{\tEnv}
      {\lam{\sequence{\notevar{\var}{\type_i}}}{\expr_f}}
      {\typefun{\sequence{\type_i}}{\type_o}}}}\]
  The induction hypothesis implies that
  $\kindof{\sEnv}{\kEnv}{\type_o}{\kindarray}$,
  so we can derive
  \[{\infr[K-Fun]
    {\sequence{\kindof{\sEnv}{\kEnv}{\type_i}{\kindarray}}
      \\
      \kindof{\sEnv}{\kEnv}{\type_o}{\kindarray}}
    {\kindof{\sEnv}{\kEnv}{\typefun{\sequence{\type_i}}{\type_o}}{\kindatom}}}\]

  In this final case, $\term$ may be an expression or an atom.
  \paragraph{Case {\sc Eqv}:}
  \[{\infr[T-Eqv]
    {\typeof{\sEnv}{\kEnv}{\tEnv}{\term}{\type'}
      \qquad
      \teqv{\type'}{\type}}
    {\typeof{\sEnv}{\kEnv}{\tEnv}{\term}{\type}}}\]
  The induction hypothesis gives
  $\kindof{\sEnv}{\kEnv}{\type'}{\kind}$,
  where $\kind = \kindatom$ if $\term$ is an atom
  and  $\kind = \kindarray$ if $\term$ is an expression.
  The required result ascribing the same kind to $\type$,
  that is
  $\kindof{\sEnv}{\kEnv}{\type}{\kind}$,
  follows directly from Lemma \ref{TEqvRespectsKinds}.
\end{sproof}

\subsection{Dynamic Semantics}
In the dynamic semantics for Remora,
the way function application is lifted to work on aggregate data
depends on the types of the function and argument terms.
Consulting type information avoids a ``hole'' in the semantics of
untyped array-oriented code,
where a frame whose shape includes a 0 dimension
evaluates to an array with indeterminate shape---%
there are no concrete cells whose shape can be used to
determine the overall shape of the resulting array.
Instead, the function's type tells us
the shape of the resulting cells,
even when there are zero such cells.

The small-step operational semantics,
given in Figure \ref{fig:DynamicSemantics},
assumes every atom or expression has been tagged with its type.
For example, $\beta$-reduction requires
that each atom in the function position array have input types $\sequence{\type}$
and that the argument arrays' types also match $\sequence{\type}$.
This matching is still subject to the type equivalence rules described in \S\ref{ssec:StaticSemantics},
\eg, a function tagged as having input type {\tt (Array Int (++ (Shp 3) (Shp 4)))}
can be applied to an argument tagged with type {\tt (Array Int (Shp 3 4))}.
Because every term now has type annotations attached,
we drop the ``empty'' array and frame syntactic forms.
Their replacements use the standard array and frame syntax
with an empty list of atoms or cells,
and the atom or cell type is implied by the expression's type annotation.

Several list-processing metafunctions are used in defining the reduction rules.
These metafunctions are defined in Figure \ref{fig:Metafunctions}.
$\mathit{Split}_{\nat}$ turns a list into a list of lists,
made up of the consecutive length-$\nat$ pieces of the original list.
For example, $\mathit{Split}_{3}\llb\text{\tt (1 2 3 4 5 6)}\rrb$
is {\tt ((1 2 3) (4 5 6))}.
$\mathit{Concat}$ flattens a list of lists into a single list,
effectively reversing a $\mathit{Split}$.
$\mathit{Rep}_{\nat}$ constructs a new list by
repeating each element of the original list $\nat$ times.
$\mathit{Rep}_2\llb\text {\tt (0 1)}\rrb$
is {\tt (0 0 1 1)}.
Used on nested lists,
the inner lists are treated atomically:
$\mathit{Rep}_2\llb\text {\tt ((1 2 3) (4 5 6))}\rrb$
is {\tt ((1 2 3) (1 2 3) (4 5 6) (4 5 6))}.
$\mathit{Transpose}$ takes a list of lists,
where the inner lists all have the same length,
and produces a new list of lists whose $i^{\mathit{th}}$ element
contains the $i^{\mathit{th}}$ elements of each original inner list.

\FigListMetafns

The reduction rules themselves are given in Figure \ref{fig:DynamicSemantics}.
Remora's function application is split into stages for
replicating cells to make frame shapes match ($\mathit{lift}$),
mapping the functions to corresponding argument cells ($\mathit{map}$),
and gathering the result cells back into an array ($\mathit{collapse}$).

Performing a $\mathit{lift}$ step identifies
the function array's frame,
the sequence $[\sequence{\nat_f}]$,
and each argument's frame,
$[\sequence{\nat_a}]$.
Then the sequence $[\sequence{\nat_p}]$ is chosen to be the largest frame
according to prefix ordering.
We require that at least one function or argument frame
be different from the principal frame---%
otherwise, a $\mathit{map}$ step would be appropriate instead.
Each argument's cell size $\nat_{\mathit{ac}}$ is
the product of the dimensions $[\sequence{\nat_{\mathit{in}}}]$
of the function's input type at that position;
the function array's cell size is always 1.
The number of replicas needed for each cell
($\nat_{\mathit{fe}}$ for the function and
$\nat_{\mathit{fa}}$ for each argument)
is determined by multiplying the dimensions
that must be added to each corresponding frame to produce the principal frame,
\ie, the principal frame minus whatever prefix was
already present in the original array's shape.
Given these numbers,
we split each array's atom list into its cells,
replicate those cells to match the new array shapes,
and then concatenate each array's replicated cells
to produce the new function and argument arrays.
Type annotations on the individual arrays update
to reflect their new shapes,
but the application form's type remains unchanged.

A $\mathit{map}$ step is possible when
every piece of a function application has the same frame shape.
Then the application becomes a {\tt frame} of application forms,
which themselves all have scalar principal frame.
This requires breaking each argument array's atom list
into its individual cells' atom lists,
then transposing to match the first cell of each argument with the first function,
the second cell of each argument with the second function,
and so on.

When function application has a scalar in function position,
and every argument array matches the function's corresponding input type,
then we can $\beta$-reduce or $\delta$-reduce.
$\beta$-reduction performs conventional $\lambda$-calculus substitution.
The $\delta$ rule uses a family of metafunctions,
each associated with a primitive operator.
No {\tt frame} construct is necessary in either result,
as this is the degenerate case of function lifting---%
the principal frame is scalar.

Applying type and index abstractions is handled by
the $\mathit{t\beta}$ and $\mathit{i\beta}$ rules.
The application frame is the shape of the array of type or index abstractions,
since there are no argument \emph{arrays}.
Every {\tt T$\lambda$} or {\tt I$\lambda$} is applied to
the full list of type or index arguments.
Substitution into the body of each abstraction
should be read as affecting type annotations as well as subterms:
if we are replacing the type variable {\tt T} with {\tt Int},
then {\tt x$^{\text{\tt (Array T (Shp 3))}}$} becomes {\tt x$^{\text{\tt (Array Int (Shp 3))}}$}.

Once a {\tt frame} has every one of its cells reduced to an {\tt array} literal,
the nested representation can be merged into a single literal.
In the case where one of $\sequence{\nat}$ is 0,
there will be no cells to examine to determine the cell dimensions $\sequence{\nat'}$,
so this information is taken from the type annotation on the {\tt frame} form.
The type annotation itself passes through unchanged.
The atom lists from the cells are concatenated
to produce the collapased array's atom list.

Destructing a {\tt box} with an {\tt unbox} form
behaves like a conventional {\tt let}.
The result is the body $\expr$,
where the index variables $\sequence{\var_i}$
are replaced with the {\tt box}'s indices $\sequence{\idx}$,
and the term variable $\var_e$ is replaced with the contained array $\val$.

\FigDynamicSemantics

\subsection{Type Soundness}
\label{SoundnessSection}
The value of a type soundness theorem for Remora is not only assurance
that well-typed programs do not suffer from shape-mismatching errors.
It also ensures that the types ascribed to program terms
accurately describe the shapes of the data those terms compute.
That is the guarantee that justifies a compiler's use of the type system
as a static analysis for array shape.

With supporting lemmas,
such as canonical forms and substitution,
already taken care of,
we now establish progress and preservation lemmas.
Since we have not committed to
a collection of primitive operators that are all total functions,
the progress lemma acknowledges the possibility of non-shape errors,
such as division by zero.
However, we do assume that any value returned by a primitive operator
inhabits that operator's output type.

\newcommand{\misapp}{\app{\arrlit{\primop}{}}{\sequence{\val}}}
\begin{lemma}[Progress]
  \label{Progress}
  Given an expression $\expr$ such that
  $\typeof{\emptyEnv}{\emptyEnv}{\emptyEnv}{\expr}{\type}$,
  one of the following holds:
  \begin{itemize}
  \item{$\expr$ is a value $\val$}
  \item{There exists $\expr'$ such that $\expr \mapsto \expr'$}
  \item{$\expr$ is
    $\ctxt\left[\misapp\right]$
    where $\primop$ is a partial function applied to
    appropriately-typed values outside its domain.}
  \end{itemize}
\end{lemma}
\begin{sproof}[We use induction on the derivation of
  $\typeof{\emptyEnv}{\emptyEnv}{\emptyEnv}{\expr}{\type}$.
  We consider only cases for typing rules which apply to expressions (as opposed to atoms).
  Since we do not reduce under a binder,
  our assumed type derivation ensures
  that the reducible subexpression of $\expr$
  is also typable using an empty environment.

  An {\tt array} form which is not already a value
  must have some non-value atom.
  That atom must itself contain a non-value expression,
  with its own type derivation.
  So the induction hypothesis implies that
  it can take a reduction step or is a mis-applied primitive operator.
  Similar reasoning applies to {\tt frame} forms:
  either we have a {\it collapse} redex,
  or some cell subexpression in the frame can make progress.

  An {\tt unbox} form can either
  make progress in the box position (via the induction hypothesis)
  or take an {\it unbox} step.
  Similarly, a type or index applications can make progress in function position
  or take $\mathit{t\beta}$ or $\mathit{i\beta}$ step.

  The function application case splits into subcases depending on
  whether the function and argument arrays are fully reduced
  and if so what their frame shapes are.
  If they are all value forms,
  we have all scalar frames (a $\beta$ or $\delta$ redex)
  or all identical non-scalar frames (a {\it map} redex),
  or non-identical prefix-compatible frames (a {\it lift} redex).
  Prefix-incompatible frames are ruled out by the type derivation.]
  We use induction on the derivation of
  $\typeof{\emptyEnv}{\emptyEnv}{\emptyEnv}{\expr}{\type}$.
  We consider only cases for typing rules which apply to expressions (as opposed to atoms).
  Note that a {\sc T-Var} derivation is impossible with an empty type environment.
  \paragraph{Case {\sc Array}}
  \[{\infr[T-Array]
    {\sequence{\typeof{\emptyEnv}{\emptyEnv}{\emptyEnv}{\atom}{\type_a}}
      \qquad
      \kindof{\emptyEnv}{\emptyEnv}{\type_a}{\kindatom}
      \qquad
      \mathit{Length}\llb \sequence{\atom} \rrb = \prod{\sequence{\nat}}}
    {\typeof{\emptyEnv}{\emptyEnv}{\emptyEnv}
      {\arrlit{\sequence{\atom}}{\sequence{\nat}}}
      {\typearray{\type_a}{\idxshape{\sequence{\nat}}}}}}\]
  We have two possibilities.
  One is that all of $\sequence{\atom}$ are atomic values, $\sequence{\atval}$,
  in which case $e = \typearray{\sequence{\atval}}{\sequence{\nat}}$ is a value.
  Otherwise, at least one $\atom$ is not an atomic value.
  The non-value must be of the form
  $\atom_s = \dsum{\sequence{\notevar{\var}{\sort}}}{\expr_s}{\type_a}$,
  with non-value $\expr_s$.
  The derivation of $\typeof{\emptyEnv}{\emptyEnv}{\emptyEnv}{\atom}{\type_a}$
  implies via the induction hypothesis that
  either there exists some $\expr_s'$ such that $\expr_s \mapsto \expr_s'$
  or $\expr_s$ has the form $\ctxt_s\left[\misapp\right]$
  with misapplied partial function $\primop$.
  We build the evaluation context
  $\ctxt = \arrlit{\sequence{\atom_0} \;
    \dsum{\sequence{\notevar{\var}{\sort}}}{\ctxt_s}{\type_a}
    \; \sequence{\atom_1}}
  {\sequence{\nat}}$
  where $\sequence{\atom_0}$ and $\sequence{\atom_1}$
  are the atoms appearing respectively before and after $\atom_s$ in $\expr$.
  If $\expr_s \mapsto \expr_s'$, then $\ctxt\left[\expr_s\right] \mapsto \ctxt\left[\expr_s'\right]$.
  Otherwise, $\expr = \ctxt\left[\misapp\right]$.

  \paragraph{Case {\sc Frame}}
  \[{\infr[T-Frame]
    {\sequence{\typeof{\emptyEnv}{\emptyEnv}{\emptyEnv}
        {\expr_a}
        {\typearray{\type_a}{\idx_a}}}
      \qquad
    \kindof{\emptyEnv}{\emptyEnv}{\typearray{\type_a}{\idx_a}}{\kindarray}
    \\\\
    \mathit{Length}\llb \sequence{\expr} \rrb = \prod{\sequence{\nat}}}
    {\typeof{\emptyEnv}{\emptyEnv}{\emptyEnv}
      {\frm{\sequence{\expr_a}}{\sequence{\nat}}}
      {\typearray{\type_a}{\idxappend{\idxshape{\sequence{\nat}} \; \idx_a}}}}}\]
  By the induction hypothesis,
  each of $\sequence{\expr_a}$ is a value, is reducible,
  or is a primitive operator misapplication
  with the form $\ctxt_a\left[\misapp\right]$.
  If they are all values, then
  each has the form $\arrlit{\sequence{\atval}}{\sequence{\nat'}}$, so
  $\expr \mapsto$
  {\tt (array ($\sequence{\nat}$ $\sequence{\nat'}$)
    $\mathit{Concat}\llbracket(\atval \sequence{})$ $\sequence{}$
    $\rrbracket$)}.
  Note that all of the array literals serving as cells in the frame must have the same shape,
  or their types would differ, making $\expr$ ill-typed.

  If some $\expr_e \in \sequence{\expr_a}$ is not a value,
  then the induction hypothesis implies that
  it is reducible to $\expr_e'$
  or it has the form $\ctxt_e\left[\misapp\right]$.
  We construct the evaluation context
  $\ctxt =$
  {\tt (frame ($\nat$ $\sequence{}$)
    $\expr_0$ $\sequence{}$ $\expr_e$ $\expr_1$ $\sequence{}$)},
  where $\parens{\sequence{\expr_a}} = \parens{\sequence{\expr_0} \; \expr_e \; \sequence{\expr_1}}$.
  If $\expr_e \mapsto \expr_e'$, then $\ctxt\left[\expr_e\right] \mapsto \ctxt\left[\expr_e'\right]$.
  Otherwise, $\expr = \ctxt\left[\misapp\right]$.

  \paragraph{Case {\sc TApp}}
  \[{\infr[T-TApp]
    {\typeof{\emptyEnv}{\emptyEnv}{\emptyEnv}
      {\expr_f}
      {\typearray{\typeuniv{\sequence{\notevar{\var}{\kind}}}{\typearray{\type_u}{\idx_u}}}{\idx_f}}
      \\\\
    \sequence{\kindof{\emptyEnv}{\emptyEnv}{\type_a}{\kind}}}
    {\typeof{\emptyEnv}{\emptyEnv}{\emptyEnv}
      {\tapp{\expr_f}{\sequence{\type_a}}}
      {\typearray{\seqsubst{\type_u}{\var}{\type_a}}{\idxappend{\idx_f \; \idx_u}}}}}\]
  By the induction hypothesis,
  $\expr_f$ is a value, is reducible, or misapplies a partial function.
  If $\expr_f$ is a value of type
  ${\typearray{\typeuniv{\sequence{\notevar{\var}{\kind}}}{\typearray{\type_u}{\idx_u}}}{\idx_f}}$,
  then Lemma \ref{CFArray} (canonical forms) implies
  that it is an array literal containing type abstractions---%
  \ie, $\expr_f = \arrlit{\sequence{{\tlam{\sequence{\notevar{\var_u}{\kind}}}{\val_u}}}}{\sequence{\nat}}$.
  This means we have a $\mathit{t\beta}$ redex.
  If $\expr_f$ is itself reducible to $\expr_f'$,
  then for some context $\ctxt_f$ and redex $\expr_r$,
  $\expr_f = \ctxt_f\left[\expr_r\right]$, and $\expr_r \mapsto \expr_r'$.
  Alternatively, $\expr_f = \ctxt_f\left[\misapp\right]$.
  Either way, construct $\ctxt = \tapp{\ctxt_f}{\sequence{\type_a}}$.
  In the first case, $\expr = \ctxt\left[\expr_r\right] \mapsto \ctxt\left[\expr_r'\right]$.
  In the second, $\expr = \ctxt\left[\misapp\right]$.
  \paragraph{Case {\sc IApp}}
  \[{\infr[T-IApp]
    {\typeof{\emptyEnv}{\emptyEnv}{\emptyEnv}
      {\expr_f}
      {\typearray{\typedprod{\sequence{\notevar{\var}{\sort}}}{\typearray{\type_p}{\idx_p}}}{\idx_f}}
      \qquad
      \sequence{\sortof{\emptyEnv}{\idx_a}{\sort}}}
    {\typeof{\emptyEnv}{\emptyEnv}{\emptyEnv}
      {\iapp{\expr_f}{\sequence{\idx_a}}}
      {\typearray{\seqsubst{\type_p}{\var}{\idx_a}}{\idxappend{\idx_f \; \seqsubst{\idx_p}{\var}{\idx_a}}}}}}\]
  As in the previous case,
  $\expr_f$ is a value, is reducible, or misapplies a partial function.
  If $\expr_f$ is a value, the canonical forms lemma implies that it has the form
  {\tt
    (array ($\nat$ $\sequence{}$) (I\cl (($\var_p$ $\sort$) $\sequence{}$) $\val_p$) $\sequence{}$)},
  which makes $\expr$ an $\mathit{i\beta}$ redex.
  Otherwise $\expr_f$ itself is reducible,
  \ie, of the form $\ctxt_f\left[\expr_r\right]$ for some rede $\expr_r$ reducing to $\expr_r'$,
  \emph{or} it is of the form $\ctxt_f\left[\misapp\right]$.
  Let the context $\ctxt = \iapp{\ctxt_f}{\sequence{\idx_a}}$.
  Then we have either $\expr = \ctxt\left[\expr_r\right] \mapsto \ctxt\left[\expr_r'\right]$
  or $\expr = \ctxt\left[\misapp\right]$.

  \paragraph{Case {\sc Unbox}}
  \[{\infr[T-Unbox]
    {\typeof{\emptyEnv}{\emptyEnv}{\emptyEnv}
      {\expr_s}
      {\typearray{\typedsum{\sequence{\notevar{\var_i'}{\sort}}}{\type_s}}{\idx_s}}
      \\\\
      \typeof
      {\sequence{\hassort{\var_i}{\sort}}}{\emptyEnv}{\hastype{\var_e}{\seqsubst{\type_s}{\var_i'}{\var_i}}}
      {\expr_b}
      {\typearray{\type_b}{\idx_b}}
      \\\\
      \kindof{\emptyEnv}{\emptyEnv}{\typearray{\type_b}{\idx_b}}{\kindarray}}
    {\typeof{\emptyEnv}{\emptyEnv}{\emptyEnv}
      {\dproj{\sequence{\var_i}}{\var_e}{\expr_s}{\expr_b}}
      {\typearray{\type_b}{\idxappend{\idx_s \; \idx_b}}}}}\]
  By the induction hypothesis,
  $\expr_s$ is a value, reducible, or a misapplication of a partial function.
  If $\expr_s$ is a value, the canonical forms lemma implies
  $\expr_s = \arrlit{\dsum{\sequence{\idx_s}}{\val}{\type}}{}$,
  so $\expr$ is an $\mathit{unbox}$ redex.
  If $\expr_s$ is reducible,
  \ie, it is $\ctxt_s\left[\expr_r\right]$ where the redex $\expr_r \mapsto \expr_r'$,
  then let $\ctxt = \dproj{\sequence{\var_i}}{\var_e}{\ctxt_s}{\expr_b}$.
  Thus $\expr = \ctxt\left[\expr_r\right] \mapsto \ctxt\left[\expr_r'\right]$.
  Otherwise, $\expr_s =$
  $\ctxt_s[$
  {\tt ((array () $\primop$) $\val$ $\sequence{}$)}
  $]$,
  so the same construction of $\ctxt$ gives
  $\expr = \ctxt\left[\misapp\right]$.
  \paragraph{Case {\sc App}}
  \[{\infr[T-App]
    {\typeof{\emptyEnv}{\emptyEnv}{\emptyEnv}{\expr_f}
      {\typearray
        {\typefun
          {\sequence{\typearray{\type_i}{\idx_i}}}
          {\typearray{\type_o}{\idx_o}}}
        {\idx_f}}
      \\\\
    \sequence{\typeof{\emptyEnv}{\emptyEnv}{\emptyEnv}
      {\expr_a}
      {\typearray{\type_i}{\idxappend{\idx_a \; \idx_i}}}}
    \qquad
    \idx_p = \mathit{Max}\llb \idx_f \; \sequence{\idx_a} \rrb}
    {\typeof{\emptyEnv}{\emptyEnv}{\emptyEnv}
      {\app{\expr_f}{\sequence{\expr_a}}}
      {\typearray{\type_o}{\idxappend{\idx_p \; \idx_o}}}}}\]
  Suppose $\expr_f$ is not a value.
  Then the induction hypothesis implies that
  \emph{either} $\expr_f \mapsto \expr_f'$,
  meaning $\expr_f = \ctxt_f\left[\expr_r\right]$
  for some redex $\expr_r \mapsto \expr_r'$
  \emph{or} $\expr_f = \ctxt_f\left[\misapp\right]$
  with incompatible but properly-typed arguments.
  Let $\ctxt = \app{\ctxt_f}{\sequence{\expr_a}}$.
  If $\expr_f \mapsto \expr_f'$, with $\expr_f = \ctxt_f\left[\expr_r\right]$,
  then $\expr = \ctxt\left[\expr_r\right] \mapsto \ctxt\left[\expr_r'\right]$.
  Otherwise, $\expr = \ctxt\left[\misapp\right]$.

  We assume from now that $\expr_f$ is a value.
  If any of $\sequence{\expr_a}$ is not a value,
  then a similar argument applies.
  Choose $\expr_c$ as the leftmost non-value argument,
  so that $\parens{\sequence{\expr_a}} = \parens{\sequence{\val_a} \; \expr_c \; \sequence{\expr_c'}}$.
  Then induction hypothesis implies that $\expr_c$ is a context $\ctxt_c$
  filled by either the redex $\expr_r$ or the misapplication $\misapp$.
  Then the context $\ctxt =$
  {\tt ($\expr_f$ $\val_a$ $\sequence{}$ $\ctxt_c$ $\expr'_c$ $\sequence{}$)}.
  If $\expr_c = \ctxt_x\left[\expr_r\right] \mapsto \ctxt_x\left[\expr_r'\right]$,
  then $\ctxt\left[\expr_r\right] \mapsto \ctxt\left[\expr_r'\right]$.
  If $\expr_c = \ctxt_x\left[\misapp\right]$, then $\expr = \ctxt\left[\misapp\right]$.

  Having addressed the cases where not all of $\expr_f, \sequence{\expr_a}$ are values,
  we now consider $\expr = \app{\val_f}{\sequence{\val_a}}$.
  By canonical forms, $\val_f$ has the form $\arrlit{\sequence{\func}}{\sequence{\nat_f}}$,
  and uniqueness of typing implies $\ieqv{\idx_f}{\idxshape{\sequence{\nat_f}}}$.
  We proceed by case analysis on $\idx_f, \sequence{\idx_a}$
  (\nb, per the typing derivation, they are all prefix-orderable).
  With one exception---where $\expr$ itself is misapplication of a partial function---%
  each line of argument leads to a particular applicable reduction rule.
  \paragraph{Subcase 1:} $\idx_f = \idxshape{}$, and each $\idx_a = \idxshape{}$.
  Then $\sequence{\nat_f}$ must be the empty sequence,
  and $\val_f$ must contain only a single atom, $\func$.
  So $\val_f = \arrlit{\func}{}$,
  and the respective types of the arguments $\sequence{\val_a}$ are
  $\sequence{\typearray{\type_i}{\idxappend{\idx_a \; \idx_i}}}$,
  or equivalently $\sequence{\typearray{\type_i}{\idx_i}}$.
  If $\func = \primop$,
  then we have the (partially annotated) expression
  $\expr =$
  {\tt ((array () $\primop^{\typefun
      {\sequence{\typearray{\type_i}{\idx_i}}}
      {\typearray{\type_o}{\idx_o}}}$)
    $\val_a^{\typearray{\type_i}{\idx_i}}$
    $\sequence{}$)}.
  If $\primop$ is a partial function
  with $\sequence{\val_a}$ as out-of-domain inputs,
  such as division by zero,
  then the third condition of the progress lemma holds.
  Otherwise, we have a $\delta$ redex.
  On the other hand, if
  $\func = \lam{\sequence{\notevar{\var}{\typearray{\type_i}{\idx_i}}}}{\expr_b}$,
  the annotated expression is
  $\app
  {\arrlit
    {\annotate
      [\typefun
      {\sequence{\typearray{\type_i}{\idx_i}}}
      {\typearray{\type_o}{\idx_o}}]
      {\lam{\sequence{\notevar{\var}{\typearray{\type_i}{\idx_i}}}}{\expr_b}}}{}}
  {\sequence{\annotate[\typearray{\type_i}{\idx_i}]{\val_a}}}$.
  This is a $\beta$ redex.
  \paragraph{Subcase 2:}
  $\idx_f = \idxshape{\sequence{\nat_f}}$,
  and each $\idx_a = \idxshape{\sequence{\nat_f}}$
  for nonempty sequence $\sequence{\nat_f}$.
  Then $\val_f$ is
  $${\annotate
    [\typearray
    {\typefun
      {\sequence{\typearray{\type_i}{\idxshape{\sequence{\nat_i}}}}}
      {\typearray{\type_o}{\idx_o}}}
    {\idxshape{\sequence{\nat_f}}}]
    {\arrlit{\sequence{\func}}{\sequence{\nat_f}}}}$$
  and it is applied to arguments
  $$\sequence{\val_a} =
  \sequence{\annotate
    [\typearray{\type_i}{\idxshape{\sequence{\nat_f} \; \sequence{\nat_i}}}]
    {\arrlit{\sequence{\atval}}{\sequence{\nat_f} \; \sequence{\nat_i}}}}$$
  This is a $\mathit{map}$ redex.
  \paragraph{Subcase 3:}
  $\idx_f, \sequence{\idx_a}$ are not all equal but still prefix orderable.
  The form of $\expr$ is similar to the previous subcase,
  except that each argument array $\val_a$ replaces $\sequence{\nat_f}$
  with its own particular frame shape.
  We then have a $\mathit{lift}$ redex.
\end{sproof}

\begin{lemma}[Preservation]
  \label{Preservation}
  Let $\sEnv,\kEnv,\tEnv$ be a well-formed environment, \ie, $\wfenv{\sEnv}{\kEnv}{\tEnv}$.
  If $\typeof{\sEnv}{\kEnv}{\tEnv}{\expr}{\type}$
  and $\expr \mapsto \expr'$
  then $\typeof{\sEnv}{\kEnv}{\tEnv}{\expr'}{\type}$.
\end{lemma}
\begin{sproof}[We use induction on the derivation of
  $\typeof{\emptyEnv}{\emptyEnv}{\emptyEnv}{\expr}{\type}$.
  An {\tt array} form which can take a reduction step
  must contain a reducible subexpression.
  Many typing rules give rise to subcases where the $\expr$ itself is not a redex
  but contains some subexpression $\expr_r$ which steps to $\expr_r'$.
  In these situations,
  the typing derivation for $\expr_r$ is included in that for $\expr$,
  so replacing that subderivation with one for $\expr_r'$
  (deriving the same type, according to the induction hypothesis)
  produces a derivation of
  $\typeof{\emptyEnv}{\emptyEnv}{\emptyEnv}{\expr'}{\type}$.
  
  The remaining nontrivial subcases each correspond to particular reduction rules.
  As in proving Progress,
  the {\sc T-App} case is split into subcases based on the function and argument frames.
  When frames are non-identical but prefix-compatible,
  the resulting {\it lift} reduction produces an application form with the same principal frame
  and thus the same result type.
  When we have identical non-scalar frames,
  the {\it map} reduction produces a {\tt frame} form
  whose frame shape is equal to the application form's principal frame
  and whose cell shape and atom type is the same as
  the function's return shape and atom type.
  This gives it a type equivalent to that of the {\it map} redex.
  With a scalar principal frame,
  we have a $\delta$ redex (trivial)
  or $\beta$ redex (follows from Lemma \ref{EESub},
  preservation of types under term substitution).
  Reasoning for type and index application forms is similar
  (via Lemma \ref{IESub} and Lemma \ref{TESub} respectively).
  A reducible {\tt unbox} form also
  substitutes a value in for a variable which is intended to have the same type,
  so Lemma \ref{EESub} again ensures the result type is $\type$.]
  We use induction on
  the derivation of
  $\typeof{\emptyEnv}{\emptyEnv}{\emptyEnv}{\expr}{\type}$.
  As with Lemma \ref{Progress} (progress),
  we only consider typing rules that can apply to an expression.
  We elide {\sc T-Var} because a variable does not reduce any further.
  \paragraph{Case {\sc Eqv}}
  \[{\infr[T-Eqv]
    {\typeof{\sEnv}{\kEnv}{\tEnv}{\expr}{\type'}
      \qquad \teqv{\type}{\type'}}
    {\typeof{\sEnv}{\kEnv}{\tEnv}{\expr}{\type}}}\]
  By the induction hypothesis,
  $\typeof{\sEnv}{\kEnv}{\tEnv}{\expr'}{\type'}$.
  Then applying {\sc T-Eqv} to that derivation produces
  \[{\infr[T-Eqv]
    {\typeof{\sEnv}{\kEnv}{\tEnv}{\expr'}{\type'}
      \qquad {\teqv{\type}{\type'}}}
    {\typeof{\sEnv}{\kEnv}{\tEnv}{\expr'}{\type}}}\]

  \paragraph{Case {\sc Array}}
  \[{\infr[T-Array]
    {\sequence{\typeof{\sEnv}{\kEnv}{\tEnv}{\atom}{\type_a}}
      \\ \kindof{\sEnv}{\kEnv}{\type_a}{\kindatom}
      \\\\ \mathit{Length}\llb \sequence{\atom} \rrb = \prod{\sequence{\nat}}}
    {\typeof{\sEnv}{\kEnv}{\tEnv}
      {\arrlit{\sequence{\atom}}{\sequence{\nat}}}
      {\typearray{\type_a}{\idxshape{\sequence{\nat}}}}}}\]
  An array literal is not itself a redex,
  so the only way for $\expr$ to reduce to $\expr'$ is
  if some atom $\atom_r$ in the array is reducible.
  Let $\parens{\sequence{\atom_0} \; \atom_r \; \sequence{\atom_1}}
    = \parens{\sequence{\atom}}$.
  The only atom form which may contain an expression
  that takes a reduction step is a box.
  So $\atom_r$ must have the form
  $\dsum
  {\sequence{\idx}}{\expr_s}
  {\typedsum{\sequence{\notevar{\var}{\sort}}}{\type_s}}$,
  where $\expr_s \mapsto \expr_s'$
  and
  $\teqv{\typedsum{\sequence{\notevar{\var}{\sort}}}{\type_s}}{\type_a}$.
  The typing derivation for $\atom_r$ must,
  except perhaps for use of {\sc T-Eqv},
  end with
  \[{\infr[T-Box]
    {\sequence{\sortof{\sEnv}{\idx}{\sort}}
      \qquad
      {\kindof{\sEnv}{\kEnv}{\typedsum{\sequence{\notevar{\var}{\sort}}}{\type_s}}{\kindatom}}
      \\
      \typeof{\sEnv}{\kEnv}{\tEnv}{\expr_s}
      {\seqsubst{\type_s}{\var}{\idx}}}
    {\typeof{\sEnv}{\kEnv}{\tEnv}
      {\dsum
        {\sequence{\idx}}
        {\expr_s}
        {\typedsum{\sequence{\notevar{\var}{\sort}}}{\type_s}}}
      {\typedsum{\sequence{\notevar{\var}{\sort}}}{\type_s}}}}\]
  The induction hypothesis implies that
  $\typeof{\sEnv}{\kEnv}{\tEnv}{\expr_s'}
  {\seqsubst{\type_s}{\var}{\idx}}$,
  so $\atom_r$ can be ascribed the same type
  ${\typedsum{\sequence{\notevar{\var}{\sort}}}{\type_s}}$,
  which is equivalent to $\type_a$.
  Since $\atom_r \mapsto \atom_r'$,
  which still has type $\type_a$,
  we derive
  \[{\infr[T-Array]
    {\hspace{5em}\sequence{\typeof{\sEnv}{\kEnv}{\tEnv}{\atom_0}{\type_a}}
      \\
      \typeof{\sEnv}{\kEnv}{\tEnv}{\atom_r'}{\type_a}
      \\\\
      \sequence{\typeof{\sEnv}{\kEnv}{\tEnv}{\atom_1}{\type_a}}
      \\
      \kindof{\sEnv}{\kEnv}{\type_a}{\kindatom}
      \hspace{5em}
      \\\\
      \mathit{Length}\llb \sequence{\atom} \rrb = \prod{\sequence{\nat}}}
    {\typeof{\sEnv}{\kEnv}{\tEnv}
      {\arrlit
        {\sequence{\atom}}
        {\sequence{\nat}}}
      {\typearray{\type_a}{\idxshape{\sequence{\nat}}}}}}\]
  \paragraph{Case {\sc Frame}}
  \[{\infr[T-Frame]
    {\sequence{\typeof{\sEnv}{\kEnv}{\tEnv}
        {\expr_a}{\typearray{\type_a}{\idx_a}}}
      \\
      \kindof{\sEnv}{\kEnv}{\typearray{\type_a}{\idx_a}}{\kindarray}
      \\
      \mathit{Length}\llb \sequence{\expr} \rrb = \prod{\sequence{\nat_f}}}
    {\typeof{\sEnv}{\kEnv}{\tEnv}
      {\frm{\expr_a}{\sequence{\nat_f}}}
      {\typearray{\type_a}{\idxappend{\idxshape{\sequence{\nat_f}} \; \idx_a}}}}}\]
  As in the {\sc T-Array} case,
  if $\expr$ itself is not a redex,
  then in order for it to reduce to $\expr'$,
  it must contain some reducible cell,
  \ie, there is some $\expr_r \in \sequence{\expr_a}$
  such that $\expr_r \mapsto \expr_r'$.
  Since the type derivation for $\frm{\expr_a}{\sequence{\nat_f}}$
  must ascribe the type $\typearray{\type_a}{\idx_a}$ to $\expr_r$,
  the induction hypothesis gives
  $\typeof{\sEnv}{\kEnv}{\tEnv}{\expr_r'}{\typearray{\type_a}{\idx_a}}$.
  We then patch that result into the type derivation we had.
  With $\sequence{\expr_a} =
  \parens{\sequence{\expr_0} \; \expr_r \; \sequence{\expr_1}}$,
  we derive
  \[{\infr[T-Frame]
    {\sequence{\typeof{\sEnv}{\kEnv}{\tEnv}
        {\expr_0}{\typearray{\type_a}{\idx_a}}}
      \\
      \typeof{\sEnv}{\kEnv}{\tEnv}
             {\expr_r}{\typearray{\type_a}{\idx_a}}
      \\
      \sequence{\typeof{\sEnv}{\kEnv}{\tEnv}
        {\expr_1}{\typearray{\type_a}{\idx_a}}}
      \\
      \kindof{\sEnv}{\kEnv}{\typearray{\type_a}{\idx_a}}{\kindarray}
      \\
      \mathit{Length}\llb \sequence{\expr} \rrb = \prod{\sequence{\nat_f}}}
    {\typeof{\sEnv}{\kEnv}{\tEnv}
      {\frm{\expr_a}{\sequence{\nat_f}}}
      {\typearray{\type_a}{\idxappend{\idxshape{\sequence{\nat_f}} \; \idx_a}}}}}\]
  If $\expr$ \emph{is} a redex,
  it must be a $\mathit{collapse}$ redex.
  Then $\expr$ is a frame of array literals, \ie,
  {\tt (frame
    ($\nat_f \sequence{}$)
    (array ($\nat_c \sequence{}$) $\atval \sequence{}$) $\sequence{}$)},
  and it must step to $\expr' =$
{\tt (array ($\nat_f \sequence{} \nat_c \sequence{}$)
$\mathit{Concat}\llbracket\sequence{(\sequence{\atval})}\rrbracket$)}.
  Each of $\expr$'s cells is a well-typed array literal,
  with type derivation:
  \[{\infr[T-Array]
    {\sequence{\typeof{\sEnv}{\kEnv}{\tEnv}
        {\atval}{\type_a}}
      \qquad
      \kindof{\sEnv}{\kEnv}{\type_a}{\kindatom}
      \\\\
      \mathit{Length}\llb\sequence{\atval}\rrb = \prod{\sequence{\nat_c}}}
    {\typeof{\sEnv}{\kEnv}{\tEnv}
      {\arrlit{\sequence{\atval}}{\sequence{\nat_c}}}
      {\typearray{\type_a}{\idxshape{\sequence{\nat_c}}}}}}\]
  The typing derivation for each cell in the {\tt frame}
  requires that $\mathit{Length}\llb\sequence{\atval}\rrb = \prod{\sequence{\nat_c}}$.
  Concatenating $\prod{\sequence{\nat_f}}$ sequences of atomic values
  each of which contains $\prod{\sequence{\nat_c}}$ elements
  gives a sequence whose length is $\prod{\parens{\sequence{\nat_f}\sequence{\nat_c}}}$.
  So we can derive a similar type for the collapsed array:
  \[{\infr[T-Array]
    {\sequence{\typeof{\sEnv}{\kEnv}{\tEnv}
        {\atval}{\type_a}}
      \\
      \kindof{\sEnv}{\kEnv}{\type_a}{\kindatom}
      \\\\
      \mathit{Length}\llb\mathit{Concat}\llb\sequence{\parens{\sequence{\atval}}}\rrb\rrb
      = \prod{\parens{\sequence{\nat_f}\sequence{\nat_c}}}}
    {\parbox{0.65\textwidth}
      {\centering ${\sEnv};{\kEnv};{\tEnv}\vdash
        {\arrlit
          {\mathit{Concat}\llb\sequence{\parens{\sequence{\atval}}}\rrb}
          {\sequence{\nat_f}\sequence{\nat_c}}}$
        $:{\typearray{\type_a}{\idxshape{\sequence{\nat_f}\sequence{\nat_c}}}}$}}}\]

  Strictly speaking, our goal is to ascribe the type
  {\tt (Arr $\type_a$ (++ (Shp $\nat_f \sequence{}$) $\idx_a$))}  $=$
  {\tt (Arr $\type_a$ (++ (Shp $\nat_f \sequence{}$) (Shp $\nat_c \sequence{}$)))}
  rather than the type we have just derived,
  but they are equivalent via {\sc TEqv-Array} because
  {\tt (++ (Shp $\nat_f \sequence{}$) (Shp $\nat_c \sequence{}$))
    $\theq$
    (Shp $\nat_f \sequence{} \nat_c \sequence{}$)}.
  So {\sc T-Eqv} completes the derivation for
${\sEnv};{\kEnv};{\tEnv}$
$\vdash$
$\expr'$
$:$
{\tt (Arr $\type_a$ (++ (Shp $\nat_f \sequence{}$) $\idx_a$))}.

  \paragraph{Case {\sc TApp}}
  \[{\infr[T-TApp]
    {\typeof{\sEnv}{\kEnv}{\tEnv}
      {\expr_f}
      {\typearray
        {\typeuniv
          {\sequence{\notevar{\var}{\kind}}}
          {\typearray{\type_u}{\idx_u}}}
        {\idx_f}}
      \qquad
      \sequence{\kindof{\sEnv}{\kEnv}{\type_a}{\kind}}}
    {\typeof{\sEnv}{\kEnv}{\tEnv}
      {\tapp{\expr_f}{\sequence{\type_a}}}
      {\typearray
        {\seqsubst{\type_u}{\var}{\type_a}}
        {\idxappend{\idx_f \; \idx_u}}}}}\]
  As in previous cases, if $\expr_f \mapsto \expr_f'$,
  the induction hypothesis gives us a derivation
  $$\typeof{\sEnv}{\kEnv}{\tEnv}
  {\expr_f'}
  {\typearray
    {\typeuniv
      {\sequence{\notevar{\var}{\kind}}}
      {\typearray{\type_u}{\idx_u}}}
    {\idx_f}}$$
  which we can then use for
  {\tt (t-app $\expr_f'$ $\type_a \sequence{}$)}.

  Otherwise, $\expr \mapsto \expr'$ is only possible
  if we have a $\mathit{t\beta}$ redex,
  with
  $\expr_f =$ {\tt (array ($\nat_f \sequence{}$) (T$\ttlm{\sequence{\ttparens{\var_u\; \kind}}}$ $\expr_u$) $\sequence{}$)}
  and $\idx_f = \idxshape{\sequence{\nat_f}}$.
  Following $\mathit{t\beta}$ reduction, $\expr' =$
  {\tt (frame ($\nat_f \sequence{}$) $\seqsubst{\expr_u}{\var_u}{\type_a} \sequence{}$)}.
  Since $\expr_f$ is a well-typed array of type abstractions,
  it must be the case that
  $\typeof{\sEnv}{\kEnv, \sequence{\haskind{\var_u}{\kind}}}{\tEnv}
  {\expr_u}
  {\typearray{\type_u}{\idx_u}}$
  for each of the abstraction bodies, $\sequence{\expr_u}$.
  Lemma \ref{TESub} (preservation of types under type substitution)
  implies that we can give $\seqsubst{\expr_u}{\var_u}{\type_a}$
  the ``same'' type as $\expr_u$,
  that is,
  $$\typeof{\sEnv}{\kEnv}{\seqsubst{\tEnv}{\var_u}{\type_a}}
  {\seqsubst{\expr_u}{\var_u}{\type_a}}
  {\typearray{\seqsubst{\type_u}{\var_u}{\type_a}}{\idx_u}}$$

  The type derivation for $\expr_f$ requires that
  the individual type abstractions' types all be
  equivalent to $\typeuniv{\sequence{\notevar{\var}{\kind}}}{\type_u}$
  despite possibly being written to bind different type names,
  so each resulting $\seqsubst{\type_u}{\var_u}{\type_a}$
  is equivalent to $\seqsubst{\type_u}{\var}{\type_a}$.
  Per Barendregt's convention,
  an abstraction's type variables $\sequence{\var_u}$
  are not used in the range of $\tEnv$,
  the type environment used in checking that abstraction
  (otherwise, the environment structure $\sEnv; \kEnv; \tEnv$
  would be ill-formed),
  so $\seqsubst{\tEnv}{\var_u}{\type_a} = \tEnv$.
  Theorem \ref{wellkinded} (ascription of well-kinded types)
  implies that the pieces used to form the result type,
  $\seqsubst{\type_u}{\var}{\type_a}$ and $\idx_u$,
  are well-formed at kind $\kindatom$ and sort $\sortshp$ respectively,
  so the cell type $\typearray{\seqsubst{\type_u}{\var}{\type_a}}{\idx_u}$
  has kind $\kindarray$.
  Finally, the number of result cells matches
  the number of atoms in $\expr_f$,
  which is the product of $\expr_f$'s dimensions $\sequence{\nat_f}$.
  Thus we can derive
  \[{\infr[T-Frame]
    {\sequence{
        \typeof{\sEnv}{\kEnv}{\tEnv}
        {\seqsubst{\expr_u}{\var_u}{\type_a}}
        {\typearray
          {\seqsubst{\type_u}{\var}{\type_a}}
          {\idx_u}}}
      \\\\
      \kindof
          {\sEnv}{\kEnv}
          {\typearray
            {\seqsubst{\type_u}{\var}{\type_a}}
            {\idx_u}}
          {\kindarray}
      \\\\
      \mathit{Length}\llb \sequence{\expr_u} \rrb = \prod{\sequence{\nat}}}
    {\parbox{0.6\textwidth}
      {\centering
        ${\sEnv};{\kEnv};{\tEnv}\vdash
        {\frm
          {\sequence{\seqsubst{\expr_u}{\var_u}{\type_a}}}
          {\sequence{\nat_f}}}$
        $: {\typearray
          {\seqsubst{\type_u}{\var}{\type_a}}
          {\idxappend{\idxshape{\sequence{\nat}} \; \idx_u}}}
        $}}}\]

  \paragraph{Case {\sc IApp}}
  \newcommand{\ibetasub}[1]{\seqsubst{#1}{\var}{\idx_a}}
  \newcommand{\ibetasubp}[1]{\seqsubst{#1}{\var'}{\idx_a}}
  \[\infr[T-IApp]
  {\typeof{\sEnv}{\kEnv}{\tEnv}
    {\expr_f}
    {\typearray
      {\typedprod
        {\sequence{\notevar{\var}{\sort}}}
        {\typearray{\type_p}{\idx_p}}}
      {\idx_f}}}
  {\typeof{\sEnv}{\kEnv}{\tEnv}
    {\iapp{\expr_f}{\idx_a}}
    {\typearray{\ibetasub{\type_p}}{\idxappend{\idx_f \; \ibetasub{\idx_p}}}}}\]
  If $\expr_f \mapsto \expr_f'$,
  then the induction hypothesis implies that
  ${\sEnv};{\kEnv};{\tEnv}\vdash$
  ${\expr_f'}$
  {\tt (Arr (Pi (($\var$ $\sort$) $\sequence{}$) (Arr $\type_p$ $\idx_p$)) $\idx_f$)}.
  Then the type derivation for $\expr_f'$ can be substituted into the original derivation for $\expr$,
  ascribing the same type
  to the reduced term
  ${\iapp{\expr_f'}{\sequence{\idx_a}}}$.

  Otherwise, $\expr$ must be an $\mathit{i\beta}$ redex,
  with $\expr_f = \arrlit{\sequence{\atval}}{\sequence{\nat_f}}$,
  each $\atval$ of the form $\ilam{\sequence{\notevar{\var'}{\sort}}}{\val}$,
  and $\ieqv{\idx_f}{\idxshape{\sequence{\nat_f}}}$.
  We must also be able to derive, for each $\atval$,
  $\typeof{\sEnv}{\kEnv}{\tEnv}
  {\atval}{\typedprod{\sequence{\notevar{\var}{\sort}}}{\typearray{\type_p}{\idx_p}}}$.
  Note that $\sequence{\var'}$ might not be the same variables as $\sequence{\var}$.
  If not, then the type derivation for $\atval$ must end with {\sc T-Eqv}
  relating the required array type $\typearray{\type_p}{\idx_p}$
  and an actually derived array type $\typearray{\type_p'}{\idx_p'}$.
  That is, use of {\sc T-Eqv} requires deriving both
  \[{\infr[T-ILam]
    {\typeof{\sEnv, \sequence{\hassort{\var'}{\sort}}}{\kEnv}{\tEnv}
      {\val}
      {\typearray{\type_p'}{\idx_p'}}}
    {\typeof{\sEnv}{\kEnv}{\tEnv}
      {\ilam{\sequence{\notevar{\var'}{\sort}}}{\val}}
      {\typedprod
        {\sequence{\notevar{\var'}{\sort}}}
        {\typearray{\type_p'}{\idx_p'}}}}}\]
  and, with free variables $\sequence{\var_f}$,
  \[{\infr[TEqv-Pi]
    {\infr[TEqv-Array]
      {\teqv
        {\seqsubst{\type_p}{\var}{\var_f}}
        {\seqsubst{\type_p'}{\var'}{\var_f}}
        \\\\
        \ieqv{\seqsubst{\idx_p}{\var}{\var_f}}{\seqsubst{\idx_p'}{\var'}{\var_f}}}
      {\teqv
        {\seqsubst{\typearray{\type_p}{\idx_p}}{\var}{\var_f}}
        {\seqsubst{\typearray{\type_p'}{\idx_p'}}{\var'}{\var_f}}}}
    {\teqv
      {\typedprod{\sequence{\notevar{\var}{\sort}}}{\typearray{\type_p}{\idx_p}}}
      {\typedprod{\sequence{\notevar{\var'}{\sort}}}{\typearray{\type_p'}{\idx_p'}}}}}\]

  We now ascribe a type to $\ibetasubp{\val}$.
  Previously, $\val$ was given the type $\typearray{\type_p'}{\idx_p'}$,
  so Lemma \ref{IESub} (preservation of types under index substitution) implies
  $\typeof{\sEnv}{\kEnv}{\ibetasubp{\tEnv}}
  {\ibetasubp{\val}}{\ibetasubp{\typearray{\type_p'}{\idx_p'}}}$.
  Well-formedness of the environment
  means that no free index variables are used in the range of $\tEnv$,
  so $\ibetasubp{\tEnv} = \tEnv$.
  This simplifies our typing result to
  $\typeof{\sEnv}{\kEnv}{\tEnv}
  {\ibetasubp{\val}}{\ibetasubp{\typearray{\type_p'}{\idx_p'}}}$.
  The ascribed type is equal to $\typearray{\ibetasubp{\type_p'}}{\ibetasubp{\idx_p'}}$.
  We must show that it is equivalent to
  $\typearray{\ibetasub{\type_p}}{\ibetasub{\idx_p}}$.

  We consider
  $\seqsubst{\seqsubst{\type_p}{\var}{\var_f}}{\var_f}{\idx_a}$
  and
  $\seqsubst{\seqsubst{\type_p'}{\var'}{\var_f}}{\var_f}{\idx_a}$,
  which can be simplified to
  $\seqsubst{\type_p}{\var}{\idx_a}$
  and
  $\seqsubst{\type_p'}{\var'}{\idx_a}$
  respectively.
  Since index substitution preserves type equivalence (per Lemma \ref{EqlIdxEqvType}),
  our earlier $\teqv{\seqsubst{\type_p}{\var}{\var_f}}{\seqsubst{\type_p'}{\var'}{\var_f}}$
  implies that $\teqv{\seqsubst{\type_p}{\var}{\idx_a}}{\seqsubst{\type_p'}{\var'}{\idx_a}}$.
  Similarly, preservation of index equality under substitution implies
  $\ieqv{\seqsubst{\idx_p}{\var}{\idx_a}}{\seqsubst{\idx_p'}{\var'}{\idx_a}}$.
  So we can derive the type equivalence
  \[{\infr[TEqv-Array]
    {\teqv
      {\seqsubst{\type_p}{\var}{\idx_a}}
      {\seqsubst{\type_p'}{\var'}{\idx_a}}
      \\\\
      \ieqv{\seqsubst{\idx_p}{\var}{\idx_a}}{\seqsubst{\idx_p'}{\var'}{\idx_a}}}
    {\teqv
      {\ibetasub{\typearray{\type_p}{\idx_p}}}
      {\ibetasubp{\typearray{\type_p'}{\idx_p'}}}}}\]
  which then allows us to derive for each of $\sequence{\val}$
  \[{\infr[T-Eqv]
    {\typeof{\sEnv}{\kEnv}{\tEnv}
      {\ibetasubp{\val}}
      {\ibetasub{\typearray{\type_p'}{\idx_p'}}}
      \\\\
      \teqv
      {\ibetasubp{\typearray{\type_p'}{\idx_p'}}}
      {\ibetasub{\typearray{\type_p}{\idx_p}}}}
    {\typeof{\sEnv}{\kEnv}{\tEnv}
      {\ibetasubp{\val}}
      {\ibetasub{\typearray{\type_p}{\idx_p}}}}}\]
  This enables a type derivation for
  $\expr' = \frm{\sequence{\ibetasubp{\val}}}{\sequence{\nat_f}}$:
  \[{\infr[T-Frame]
    {\sequence{\typeof{\sEnv}{\kEnv}{\tEnv}
        {\ibetasubp{\val}}
        {\ibetasub{\typearray{\type_p}{\idx_p}}}}
      \\\\
      \mathit{Length}\llb \sequence{\val} \rrb = \prod{\parens{\sequence{\nat_f}}}}
    {\typeof{\sEnv}{\kEnv}{\tEnv}
      {\expr'}
      {\typearray
        {\ibetasub{\type_p}}
        {\idxappend{\idxshape{\sequence{\nat_f}} \; \ibetasub{\idx_p}}}}}}\]

  \paragraph{Case {\sc Unbox}}
  \[{\infr[T-Unbox]
    {\typeof{\sEnv}{\kEnv}{\tEnv}{\expr_s}
      {\typearray{\typedsum{\sequence{\notevar{\var_i'}{\sort}}}{\type_s}}{\idx_s}}
      \\\\
      \typeof{\sEnv, \sequence{\hassort{\var_i}{\sort}}}
      {\kEnv}
      {\tEnv, \hastype{\var_e}{\seqsubst{\type_s}{\var_i'}{\var_i}}}
      {\expr_b}{\typearray{\type_b}{\idx_b}}
      \\\\
      \kindof{\sEnv}{\kEnv}{\typearray{\type_b}{\idx_b}}{\kindarray}}
    {\typeof{\sEnv}{\kEnv}{\tEnv}
      {\dproj{\sequence{\var_i}}{\var_e}{\expr_s}{\expr_b}}
      {\typearray{\type_b}{\idxappend{\idx_s \; \idx_b}}}}}\]
  If $\expr$ is reducible because $\expr_s \mapsto \expr_s'$,
  then the induction hypothesis implies
  $\typeof{\sEnv}{\kEnv}{\tEnv}{\expr_s'}
  {\typearray{\typedsum{\sequence{\notevar{\var_i'}{\sort}}}{\type_s}}{\idx_s}}$.
  This can be used to adapt the original type derivation for $\expr$
  to fit $\expr' = {\dproj{\sequence{\var_i}}{\var_e}{\expr_s'}{\expr_b}}$:
  \[{\infr[T-Unbox]
    {\typeof{\sEnv}{\kEnv}{\tEnv}{\expr_s'}
      {\typearray{\typedsum{\sequence{\notevar{\var_i'}{\sort}}}{\type_s}}{\idx_s}}
      \\\\
      \typeof{\sEnv, \sequence{\hassort{\var_i}{\sort}}}
      {\kEnv}
      {\tEnv, \hastype{\var_e}{\seqsubst{\type_s}{\var_i'}{\var_i}}}
      {\expr_b}{\typearray{\type_b}{\idx_b}}
      \\\\
      \kindof{\sEnv}{\kEnv}{\typearray{\type_b}{\idx_b}}{\kindarray}}
    {\typeof{\sEnv}{\kEnv}{\tEnv}
      {\dproj{\sequence{\var_i}}{\var_e}{\expr_s'}{\expr_b}}
      {\typearray{\type_b}{\idxappend{\idx_s \; \idx_b}}}}}\]
  Otherwise, $\expr$ must be an $\mathit{unbox}$ redex,
  where all of the following hold:
  \begin{enumerate}
  \item{$\expr_s$ is of the form
      $\arrlit{\sequence{\dsum{\sequence{\idx_s}}{\val_s}{\type_\sigma}}}{\sequence{\nat_s}}$}
  \item{$\type_\sigma =
      \teqv
      {\typedsum{\sequence{\notevar{\var_i''}{\sort}}}{\type_s'}}
      {\typedsum{\sequence{\notevar{\var_i'}{\sort}}}{\type_s}}$}
  \item{$\idx_s = \idxshape{\sequence{\nat_s}}$}
  \item{$\mathit{Length}\llb \sequence{\dsum{\sequence{\idx_s}}{\val_s}{\type_\sigma}} \rrb
      = \prod{\sequence{\nat_s}}$}
  \end{enumerate}

  The type derivation for $\expr_s$ must include
  ascription of $\type_\sigma$ to each box within the array:
  \[{\infr[T-Box]
    {\sequence{\sortof{\sEnv}{\idx_\sigma}{\sort}}
      \qquad
      {\kindof{\sEnv}{\kEnv}{\type_\sigma}{\kindatom}}
      \\
      \typeof{\sEnv}{\kEnv}{\tEnv}{\val_\sigma}{\seqsubst{\type_s'}{\var_i''}{\idx_\sigma}}}
    {\typeof{\sEnv}{\kEnv}{\tEnv}
      {\dsum{\sequence{\idx_\sigma}}{\val_\sigma}{\type_\sigma}}
      {\type_\sigma}}}\]

  Equivalence of $\type_\sigma$ and $\typedsum{\sequence{\notevar{\var_i'}{\sort}}}{\type_s}$
  means that each box's $\val_\sigma$ can also be typed as $\seqsubst{\type_s}{\var_i'}{\idx_\sigma}$.

  The resulting term $\expr'$ is
  $\multisubst{\expr_b}{\sequence{\singlesubst{\var_i}{\idx_\sigma}},\singlesubst{\var_e}{\val_\sigma}}
  = \subst{\seqsubst{\expr_b}{\var_i}{\idx_\sigma}}{\var_e}{\val_\sigma}$.
  Since we have a type derivation for $\expr_b$ in an extended environment,
  we will apply Lemmas \ref{IESub} and \ref{EESub}
  (preservation of types under index and expression substitution)
  to produce a type derivation in the original environment.

  Using the fact that each of $\sequence{\idx_\sigma}$ has its required sort
  (necessary for the previous derivation),
  Lemma \ref{IESub} gives us
  $\typeof{\sEnv}{\kEnv}{\seqsubst{\parens{\tEnv, \hastype{\var_e}{\seqsubst{\type_s}{\var_i'}{\var_i}}}}{\var_i}{\idx_\sigma}}
  {\seqsubst{\expr_b}{\var_i}{\idx_\sigma}}
  {\seqsubst{\parens{\seqsubst{\type_b}{\var_i}{\var_i'}}}{\var_i}{\idx_\sigma}}$.
  Well-formedness of the original environment means $\tEnv$ does not refer to the
  new index variables $\sequence{\var_i}$,
  which are instead bound by the {\tt unbox} expression.
  So the substituted environment
  $\seqsubst{\parens{\tEnv, \hastype{\var_e}{\seqsubst{\type_s}{\var_i'}{\var_i}}}}{\var_i}{\idx_\sigma}$
  is equal to $\tEnv, \hastype{\var_e}{\seqsubst{\type_s}{\var_i'}{\idx_\sigma}}$,
  turning the type derivation into
  $\typeof{\sEnv}{\kEnv}{\tEnv, \hastype{\var_e}{\seqsubst{\type_s}{\var_i'}{\idx_\sigma}}}
  {\seqsubst{\expr_b}{\var_i}{\idx_\sigma}}{\seqsubst{\type_b}{\var_i}{\var_i'}}$.
  The index variables $\sequence{\var_i}$ are not bound in $\sEnv$,
  but we still have $\kindof{\sEnv}{\kEnv}{\typearray{\type_b}{\idx_b}}{\kindarray}$.
  This means $\type_b$ and $\idx_b$ must be well-formed without those bindings,
  \ie, none of $\sequence{\var_i}$ appear free in $\type_b$ or $\idx_b$.
  Thus $\seqsubst{\typearray{\type_b}{\idx_b}}{\var_i}{\idx_s}$ is simply $\typearray{\type_b}{\idx_b}$.
  Our derivation now concludes
  $\typeof{\sEnv}{\kEnv}{\tEnv, \hastype{\var_e}{\seqsubst{\type_s}{\var_i'}{\idx_\sigma}}}
  {\seqsubst{\expr_b}{\var_i}{\idx_\sigma}}{\typearray{\type_b}{\idx_b}}$.

  Since $\typeof{\sEnv}{\kEnv}{\tEnv}{\val_s}{\seqsubst{\type_s}{\var_i'}{\idx_\sigma}}$,
  Lemma \ref{EESub} produces a derivation of
  $\typeof{\sEnv}{\kEnv}{\tEnv}
  {\subst{\seqsubst{\expr_b}{\var_i}{\idx_\sigma}}{\var_e}{\val_\sigma}}
  {\typearray{\type_b}{\idx_b}}$
  for each box's index contents $\sequence{\idx_\sigma}$ and term contents $\val_\sigma$.
  We can therefore type the result of the $\mathit{unbox}$ reduction step,
  {\tt (frame ($\nat_s$ $\sequence{}$)
    $\subst{\seqsubst{\expr_b}{\var_i}{\idx_\sigma}}{\var_e}{\val_\sigma}$ $\sequence{}$)}
  as follows:
  \[{\infr[T-Frame]
    {\sequence{\typeof{\sEnv}{\kEnv}{\tEnv}
        {\subst{\seqsubst{\expr_b}{\var_i}{\idx_\sigma}}{\var_e}{\val_\sigma}}
        {\typearray{\type_b}{\idx_b}}}
      \\\\
      \kindof{\sEnv}{\kEnv}{\typearray{\type_b}{\idx_b}}{\kindarray}
    \\\\
    \mathit{Length}\llb{\subst{\seqsubst{\expr_b}{\var_i}{\idx_\sigma}}{\var_e}{\val_\sigma}}\rrb
    = \prod{\sequence{\nat_s}}}
  {\parbox{0.6\textwidth}
    {\centering
      \({\sEnv};{\kEnv};{\tEnv}\vdash\)
      \({\frm{\subst{\seqsubst{\expr_b}{\var_i}{\idx_\sigma}}{\var_e}{\val_\sigma}}{\sequence{\nat_s}}}\)
      \({: \typearray{\type_b}{\idxappend{\idx_s}{\idx_b}}}\)}}
  }\]

  \paragraph{Case {\sc App}}
  \[{\infr[T-App]
    {\typeof{\sEnv}{\kEnv}{\tEnv}
      {\expr_f}
      {\typearray
        {\typefun
          {\sequence{\typearray{\type_i}{\idx_i}}}
          {\typearray{\type_o}{\idx_o}}}
        {\idx_f}}
      \\\\
      \sequence
      {\typeof{\sEnv}{\kEnv}{\tEnv}
        {\expr_a}
        {\typearray{\type_i}{\idxappend{\idx_a \; \idx_i}}}}
      \qquad
      \idx_p = \mathit{Max}\llb \idx_f \; \sequence{\idx_a} \rrb}
    {\typeof{\sEnv}{\kEnv}{\tEnv}
      {\app{\expr_f}{\sequence{\expr_a}}}
      {\typearray{\type_o}{\idxappend{\idx_p \; \idx_o}}}}}\]
  If either $\expr_f \mapsto \expr_f'$
  or there is some argument $\expr_A \in \sequence{\expr_a}$
  such that $\expr_A \mapsto \expr_A'$,
  then the induction hypothesis implies that
  the result $\expr_f'$ or $\expr_A'$ retains the same type as $\expr_f$ or $\expr_A$.
  The type derivation for $\expr$ can be updated
  with the new sub-derivation for $\expr_f$ or $\expr_A$
  to derive the same type for $\expr'$.

  Otherwise, $\expr$ must itself be a redex.
  The possible reductions for an application form
  are $\mathit{lift}$, $\mathit{map}$, $\beta$, and $\delta$.
  \paragraph{Subcase 1:}
  If $\expr$ is a $\mathit{lift}$ redex,
  then it has the form
  {\tt
    ((array ($\nat_f$ $\sequence{}$) $\atval_f$ $\sequence{}$)
    (array ($\nat_a$ $\sequence{}$ $\nat_i$ $\sequence{}$)
    $\atval_a$ $\sequence{}$) $\sequence{}$)}.
  In order to match the left hand side of the $\mathit{lift}$ rule,
  it must be the case that that
  $\ieqv{\idx_f}{\idxshape{\sequence{\nat_f}}}$
  and that for each argument's own type derivation,
  $\ieqv{\idx_a}{\idxshape{\sequence{\nat_a}}}$.
  The frame portion of each array's shape
  (\ie, $\sequence{\nat_f}$ for function position
  and $\sequence{\nat_a}$ for argument position)
  is replaced with the principal frame, $\sequence{\nat_p}$.
  The individual atoms used in the new function and arugment arrays
  all come from their corresponding original arrays
  and therefore retain the same types.
  That is, the atoms $\sequence{\atval_f'}$ in the new function array
  $\expr_f' = \arrlit
  {\mathit{Concat}\llb
    \mathit{Rep}_{\nat_{\mathit{fe}}}\llb
    \mathit{Split}_{1}\llb \sequence{\atval_f}
    \rrb\rrb\rrb}
  {\sequence{\nat_p}}$
  all have type
  $\typefun
  {\sequence{\typearray{\type_i}{\idx_i}}}
  {\typearray{\type_o}{\idx_o}}$,
  and the atoms $\sequence{\atval_a'}$ in any new argument array
  $\expr_a' = \arrlit
  {\mathit{Concat}\llb
    \mathit{Rep}_{\nat_{\mathit{ae}}}\llb
    \mathit{Split}_{\nat_{\mathit{ac}}}\llb \sequence{\atval_a}
    \rrb\rrb\rrb}
  {\sequence{\nat_p} \; \sequence{\nat_i}}$
  are all typed as the corresponding $\type_i$.
  While $\mathit{Split}$ breaks a sequence into subsequences
  (preserving the total number of elements)
  and $\mathit{Concat}$ merges a sequence-of-sequences into a single sequence
  (also preserving the total number of elements),
  $\mathit{Rep}_{\nat}$ produces $\nat$ copies of each element.
  The factor used by $\mathit{lift}$ reduction for replication of each array's cells is the quotient of
  the number of cells in the principal frame and the number of cells in the original array.
  Since the original frame shape is a prefix of the principal frame,
  divisibility is guaranteed.
  This also ensures that the number of atoms in each new argument array is
  $\nat_{ae}*\parens{\prod\sequence{\nat_a}}*\parens{\prod\sequence{\nat_i}}
  = \parens{\prod\sequence{\nat_p}}*\parens{\prod\sequence{\nat_i}}
  = \prod{\parens{\sequence{\nat_p}\;\sequence{\nat_i}}}$.
  Similarly, the number of atoms in the new function array is
  $\nat_{fe}*\parens{\prod\sequence{\nat_f}} = \prod{\sequence{\nat_p}}$.
  Based on the types and quantity of the function and argument arrays' atoms,
  we can derive
  \[{\infr[T-App]
    {\typeof{\sEnv}{\kEnv}{\tEnv}
      {\expr_f'}
      {\typearray
        {\typefun
          {\sequence{\typearray{\type_i}{\idx_i}}}
          {\typearray{\type_o}{\idx_o}}}
        {\idxshape{\sequence{\nat_p}}}}
      \\
      \sequence{\typeof{\sEnv}{\kEnv}{\tEnv}
        {\expr_a'}
        {\typearray{\type_i}{\idxappend{\idxshape{\sequence{\nat_p}} \; \idx_i}}}}
      \\\\
    \idx_p = \mathit{Max} \llb \idx_p \; \sequence{\idx_p} \rrb}
    {\typeof{\sEnv}{\kEnv}{\tEnv}
      {\expr'}
      {\typearray{\type_o}{\idxappend{\idx_p \; \idx_o}}}}}\]
  \paragraph{Subcase 2:}
  If $\expr$ is a $\mathit{map}$ redex,
  then it has the form
  {\tt
    ((array ($\sequence{\nat_f}$) $\sequence{\atom_f}$)
    (array ($\sequence{\nat_f}$ $\sequence{\nat_a}$) $\sequence{\atom_a}$) $\sequence{}$)}.
  Matching the left hand side of the $\mathit{map}$ rule requires that
  $\ieqv{\idx_i}{\idxshape{\sequence{\nat_i}}}$.
  Each argument $\expr_a = \arrlit{\sequence{\atom_a}}{\sequence{\nat_f} \; \sequence{\nat_a}}$
  has its sequence of atoms split into segments whose length is
  the product of $\sequence{\nat_i}$, the corresponding input type's dimensions.
  Transposition groups the first segment from each argument,
  then the second segment from each argument, and so on.
  So the nested sequence
  $\sequence{\parens{\sequence{\parens{\sequence{\atval_c}}}}}$
  has for each $j^{\text{th}}$ $\parens{\sequence{\parens{\sequence{\atval_c}}}}$
  a sequence of atoms corresponding to each of the function input types:
  the $(j,k)^{\text{th}}$ atom sequence contains the atoms which make up
  the $j^{\text{th}}$ cell of the $k^{\text{th}}$ argument.
  Since the length of this single $\sequence{\atval_c}$ is
  the product of the corresponding input type's dimensions
  and they all have the same type as the atoms in the corresponding input type,
  the array literal built from them,
  $\arrlit{\sequence{\atval_c}}{\sequence{\nat_i}}$
  has type $\typearray{\type_i}{\idxshape{\sequence{\nat_i}}}$.
  Each function array $\arrlit{\atval_f}{}$
  contains a single atom taken from the original $\expr_f$,
  implying that it must have type
  $\typefun
  {\sequence{\typearray{\type_i}{\idx_i}}}
  {\typearray{\type_o}{\idx_o}}$,
  which is equivalent to
  $\typefun
  {\sequence{\typearray{\type_i}{\idxshape{\sequence{\nat_i}}}}}
  {\typearray{\type_o}{\idx_o}}$.
  Since the single-cell argument arrays' types all match
  the singleton function array's input types
  with principal frame of $\idxshape{}$,
  each application form itself has type $\typearray{\type_o}{\idx_o}$.
  That is, each result-cell application form
  $\expr_c =$
  {\tt
    ((array () $\atval_f$)
    (array ($\sequence{\nat_i}$) $\sequence{\atval_c}$) $\sequence{}$)}
  in the resulting {\tt frame} form
  can be typed as:
  \[\infr[T-App]
  {\parbox{0.6\textwidth}
    {\centering
      \({\sEnv};{\kEnv};{\tEnv}\vdash\)
      ${\arrlit{\atval_f}{}}$\\
      $:{\typearray
        {\typefun
          {\sequence{\typearray{\type_i}{\idx_i}}}
          {\typearray{\type_o}{\idx_o}}}
        {\idxshape{}}}$}
    \\\\\\
    \sequence{\typeof{\sEnv}{\kEnv}{\tEnv}
      {\arrlit{\sequence{\atval_c}}{\sequence{\nat_i}}}
      {\typearray{\type_i}{\idx_i}}}
    \\\\\\
    \idx_p = \idxshape{}}
  {\typeof{\sEnv}{\kEnv}{\tEnv}
    {\expr_c}
    {\typearray{\type_o}{\idx_o}}}\]
  Recall that our goal type in this case is
  $\type =$
  $\typearray{\type_o}{\idxappend{\idx_p\;\idx_o}}$,
  where $\idx_p = \idxshape{\sequence{\nat_f}}$.
  That is, $\type$ is
  $\typearray{\type_o}{\idxappend{\idxshape{\sequence{\nat_f}}\;\idx_o}}$.
  Using the above derivations for the result cells $\sequence{\expr_c}$, we then derive
  \[{\infr[T-Frame]
    {{\sequence{\typeof{\sEnv}{\kEnv}{\tEnv}
          {\expr_c}
          {\typearray{\type_o}{\idx_o}}}}
      \\
      \kindof{\sEnv}{\kEnv}{\type}{\kindarray}}
    {\typeof{\sEnv}{\kEnv}{\tEnv}
      {\frm
        {\sequence{\expr_c}}
        {\sequence{\nat_f}}}
      {\typearray{\type_o}{\idxappend{\idxshape{\sequence{\nat_f}}\;\idx_o}}}}}\]
  \paragraph{Subcase 3:}
  If $\expr$ is a $\beta$ redex,
  then $\expr_f$ must have the form
  {\tt
    (array () (\cl (($\var$ (Arr $\type_i$ $\idx_i$)) $\sequence{}$) $\expr_0$))}.
  Then $\expr \mapsto_\beta \expr' = \seqsubst{\expr_o}{\var}{\expr_a}$.
  We also know for each of $\sequence{\expr_a}$ that
  $\typeof{\sEnv}{\kEnv}{\tEnv}{\expr_a}{\typearray{\type_i}{\idx_i}}$.
  In the type derivation for $\expr$ itself,
  we must have $\idx_f$ and every $\idx_a$ equal to $\idxshape{}$,
  so $\idx_p = \idxshape{}$.
  Thus
  $\teqv
  {\typearray{\type_o}{\idxappend{\idx_p \; \idx_o}}}
  {\typearray{\type_o}{\idx_o}}$
  The type of the function atom in $\expr_f$ is derived either by {\sc T-Lam},
  in which case, the derivation must ascribe $\typearray{\type_o}{\idx_o}$ to $\expr_o$,
  or by {\sc T-Eqv},
  in which case there must be some earlier use of {\sc T-Lam}
  which ascribes some type equivalent type to $\expr_o$.
  Either way, we have
  $\typeof{\sEnv}{\kEnv}{\tEnv, \sequence{\hastype{\var}{\typearray{\type_i}{\idx_i}}}}
  {\expr_o}
  {\typearray{\type_o}{\idx_o}}$.
  The substitution lemma (Lemma \ref{EESub}) then gives
  $\typeof{\sEnv}{\kEnv}{\tEnv}{\expr'}{\typearray{\type_o}{\idx_o}}$.
  \paragraph{Subcase 4:}
  If $\expr$ is a $\delta$ redex,
  then the required result follows from
  the requirements for primitive operators and their types.
\end{sproof}

\begin{theorem}[Type soundness]
  \label{TypeSoundness}
  If $\typeof{\emptyEnv}{\emptyEnv}{\emptyEnv}{\expr}{\type}$,
  then either
  $\expr$ diverges,
  there exists $\val$ such that $\expr \mapsto^* \val$
  and $\typeof{\emptyEnv}{\emptyEnv}{\emptyEnv}{\val}{\type}$,
  or there exist partial function $\primop$
  and appropriately-typed arguments $\sequence{\val}$
  such that
  $\expr \mapsto^* \ctxt\left[\misapp\right]$.
\end{theorem}
\begin{sproof}[\BODY]
  We argue coinductively using the sequence of reduction steps from $\expr$.
  For any well-typed $\expr$,
  Progress (Lemma \ref{Progress}) implies that either
  $\expr$ has the form $\val$,
  $\expr$ has the form $\ctxt[${\tt ((array () $\primop$) $\val$ $\sequence{}$)}$]$,
  or $\expr \mapsto \expr'$.
  In the first case,
  the reduction sequence terminates in a value,
  so we have $\expr \mapsto^* \val$.
  Furthermore, Preservation (Lemma \ref{Preservation}) implies that
  $\typeof{\emptyEnv}{\emptyEnv}{\emptyEnv}{\val}{\type}$
  In the second case,
  the reduction sequence terminates in a mis-applied primitive operator.
  In the third case,
  Preservation implies that $\typeof{\emptyEnv}{\emptyEnv}{\emptyEnv}{\expr'}{\type}$.
\end{sproof}

\section{Type Erasure}
The dynamic semantics given in Section \ref{sec:Formalism}
relies on ubiquitous type annotations
in order to determine how function application will proceed
or how a frame of sub-arrays should collapse to a single array.
While the possible case of constructing a frame
with no actual result cells whose shape can be inspected
can only be resolved by consulting a type annotation,
the types themselves contain more information than is strictly needed.
For example, it does not matter whether we are collapsing
an empty frame of functions,
an empty frame of integers,
or an empty frame of boxes.
The result shape is the same,
regardless of the type of the atoms contained within the cells.
All we truly need is the resulting shape (alternatively, the result cells' shape).
Similarly, evaluating a function application requires knowing
the expected cell shapes for the arguments,
but it could, in principle, be done
without knowing anything about their atoms.
Function application is still tagged with a result shape,
again to head off issues arising from mapping over an empty frame.

In a type-erased version of Remora,
we only need the term and index levels---%
the syntactic class of types is discarded.
The syntax for erased Remora is given in Figure \ref{fig:ErasedAbstractSyntax}.
Note that the grammar of type indices from Figure \ref{fig:AbstractSyntax} is still in use here,
although expressions, atoms, and their corresponding function and value-form subsets
are now replaced with type-erased versions.

\FigErasedAbstractSyntax

Evaluation in erased Remora proceeds similarly to explicit Remora.
A function-application form has a principal frame
chosen to be the largest of the function and argument frames,
and a $\mathit{lift}$ reduction replicates the function and argument arrays' atoms
to bring all of the frames into agreement.
The argument frames themselves are identified based on
the individual argument positions' cell-shape annotations,
rather than by inspecting a type annotation on the array in function position.
A $\mathit{map}$ reduction turns an application form where all pieces have the same frame
into a {\tt frame} form,
where the end-result shape matches the result shape tag on the original application.
Index application also maps over an array of index functions,
producing a {\tt frame} of substituted function bodies.
Since the type level has been eliminated,
there are no {\tt t$\lambda$} and {\tt t-app} forms
and no need for a $t\beta$ reduction rule.

\FigErasedSemantics

The translation from explicit Remora to erased Remora consists of three erasure functions:
$\EraseE{\cdot} : \mathit{Expr} \rightarrow \erased{\mathit{Expr}}$,
$\EraseA{\cdot} : \mathit{Atom} \rightarrow \erased{\mathit{Atom}}$,
and $\EraseT{\cdot} : \mathit{Type} \rightarrow \mathit{Index}$.
These functions are defined in Figure \ref{fig:ErasureDefn}.

\ErasureDefn

We also define $\EraseC{\cdot} : \Ctxt \rightarrow \erased{\Ctxt}$,
given in Figure \ref{fig:ContextErasure},
which is not needed for defining the erased form of an explicit Remora program
but is useful for demonstrating their equivalence.

\ContextErasure

Types in explicit Remora are turned into indices in erased Remora.
These indices are the dynamic residue of types,
in the same sense that term-level values are dynamic,
though the are still subject to a static discipline which governs their values
and their relation to the array values they describe.
Array types become just the shapes used to construct them,
whereas functions, universals, dependent sums and products, and base types
become the ``scalar'' shape.
Extracting the index components of all types means that type variables
can be turned into index variables,
which will stand for the index component of whatever type the variable originally stood for.
This translation captures exactly the information that a {\tt frame} form needs
in the event that there are no cells.
By extension, the term and index application forms
also get the bookkeeping information needed by
the {\tt frame}s they will eventually become.

For example, consider a function term whose type is
{\tt (-> (s (Arr t (Shp))) (Arr t (Shp k)))},
where {\tt s}, {\tt t}, and {\tt k} are
bound as {\tt Array}, {\tt Atom}, and {\tt Dim} respectively.
This function produces a vector of some statically uncertain length
containing atoms of uncertain type.
When we apply this function,
the explicitly typed application form
describes the resulting array's type.
If our arguments are a single {\tt s}
and a ${\tt n} \times 4$ matrix of numbers,
with {\tt n} also bound as a {\tt Dim},
the principal frame shape is {\tt (Shp n 4)}.
So we will have result type {\tt (Arr Num (Shp n 4 k))}.
Type-erasing the application form must still preserve
enough information to produce an array of the correct shape,
even if {\tt n} turns out to be 0,
leaving us with no result cells whose shape we can inspect.
However, the dynamic semantics does not rely on knowing that
the result array contains {\tt Num}s.
The binders for index variables {\tt n} and {\tt k},
which must be either {\tt I$\lambda$} or {\tt unbox},
are still present in the type-erased program,
since the indices they eventually bind to those variables
will affect the program's semantics.
The {\tt T$\lambda$}s which bind {\tt s} and {\tt t}
turn into {\tt I$\lambda$}s,
though the variable {\tt t} is never used in the type-erased program.
If any type argument was bound to {\tt s} in the original program,
we replace it with its shape.
All occurrences of {\tt s} from the original program
now stand for an array shape rather than a full array type.

We develop a bisimulation argument to show that
the behavior of an explicitly typed term matches the behavior of its erased form.
We define the space $S$ of machine states to be the sum of
the set of well-typed explicit Remora terms
and the set of their type-erased forms.
That is, $S = \wt{Expr} \uplus \erased{\wt{\Expr}}$,
where
$\wt{Expr} =
\{\expr \in \Expr | \typeof{\emptyEnv}{\emptyEnv}{\emptyEnv}{\expr}{\type}\}$
and
$\erased{\wt{Expr}} =
\{\EraseE{\expr} | \expr \in \wt{\Expr}\}$.
Transitions in the machine match the explicit and erased languages' respective $\mapsto$ relations.
We also define the ``erasure equivalence'' relation $\EraseEqv$ on machine states as
the equivalence closure of the relation imposed by $\EraseE{\cdot}$.
Before we show that $\EraseEqv$ is a bisimulation, several intermediate results are needed.

First, the bisimulation proof will in one case need to
reach deep into an expression to find the next redex.
A compositionality property of the erasure rule
will make it possible to reason about the redex and its reduced form
separately from the evaluation context in which it is embedded.

\begin{lemma}[Erasure in context]
  \label{ErasureInContext}
  Given an evaluation context $\ctxt$ and expression $\expr$,
  where ${\fillc{\ctxt}{\expr}}$ is well-typed,
  $\EraseE{\fillc{\ctxt}{\expr}} = \fillc{\EraseC{\ctxt}}{\EraseE{\expr}}$.
\end{lemma}
\begin{sproof}[This follows from straightforward induction on $\ctxt$.]
  We use induction on $\ctxt$.

\paragraph{Array literal containing unevaluated box:}
  \[\ctxt = \arrlit{\sequence{\atval}\,\dsum{\sequence{\idx}}{\ctxt'}{\type}\,\sequence{\atom}}{\sequence{\nat}}\]
  $\EraseE{\fillc
    {\arrlit{\sequence{\atval}\,\dsum{\sequence{\idx}}{\ctxt'}{\type}\,\sequence{\atom}}{\sequence{\nat}}}
    {\expr}}$ \\
  $= \EraseE{\arrlit
    {\sequence{\atval}\,\dsum{\sequence{\idx}}{\fillc{\ctxt'}{\expr}}{\type}\,\sequence{\atom}}
    {\sequence{\nat}}}$ \\
  $= \arrlit
  {\sequence{\EraseA{\atval}}\,\EraseA{\dsum{\sequence{\idx}}{\fillc{\ctxt'}{\expr}}{\type}}\,\sequence{\EraseA{\atom}}}
  {\sequence{\nat}}$ \\
  $= \arrlit
  {\sequence{\EraseA{\atval}}\,\dsum{\sequence{\idx}}{\EraseE{\fillc{\ctxt'}{\expr}}}{}\,\sequence{\EraseA{\atom}}}
  {\sequence{\nat}}$ \\
  $= \arrlit
  {\sequence{\EraseA{\atval}}\,\dsum{\sequence{\idx}}{\fillc{\EraseC{\ctxt'}}{\EraseE{\expr}}}{}\,\sequence{\EraseA{\atom}}}
  {\sequence{\nat}}$ \\
  $= \fillc{\arrlit
    {\sequence{\EraseA{\atval}}\,\dsum{\sequence{\idx}}{\EraseC{\ctxt'}}{}\,\sequence{\EraseA{\atom}}}
    {\sequence{\nat}}}
  {\EraseE{\expr}}$ \\
  $= \fillc{\EraseC{\arrlit
    {\sequence{\atval}\,\dsum{\sequence{\idx}}{\ctxt'}{\type}\,\sequence{\atom}}
    {\sequence{\nat}}}}
  {\EraseE{\expr}}$ \\

\paragraph{Frame containing unevaluated cells:}
  \[\ctxt = \notedfrm[\type_r]{\sequence{\val}\,\ctxt'\,\sequence{\expr'}}{\sequence{\nat}}\]
  $\EraseE{\fillc{\notedfrm[\type_r]{\sequence{\val}\,\ctxt'\,\sequence{\expr}}{\sequence{\nat}}}{\expr}}$ \\
  $= \EraseE{\notedfrm[\type_r]{\sequence{\val}\,\fillc{\ctxt'}{\expr}\,\sequence{\expr'}}{\sequence{\nat}}}$ \\
  $= \frm{\sequence{\EraseE{\val}}\,\EraseE{\fillc{\ctxt'}{\expr}}\,\sequence{\EraseE{\expr'}}}{\EraseT{\type_r}}$ \\
  $= \frm{\sequence{\EraseE{\val}}\,\fillc{\EraseC{\ctxt'}}{\EraseE{\expr}}\,\sequence{\EraseE{\expr'}}}{\EraseT{\type_r}}$ \\
  $= \fillc{\frm{\sequence{\EraseE{\val}}\,\EraseC{\ctxt'}\,\sequence{\EraseE{\expr'}}}{\EraseT{\type_r}}}{\EraseE{\expr}}$ \\
  $= \fillc{\EraseC{\notedfrm[\type_r]
      {\sequence{\val}\,\EraseC{\ctxt'}\,\sequence{\expr'}}
      {\sequence{\nat}}}}
  {\EraseE{\expr}}$ \\

\paragraph{Application with unevaluated function:}
  \[\ctxt =
  \notedapp[\type_r]
  {\annotate[\typearray{\typefun{\sequence{\type_i}}{\type_o}}{\idx_f}]{\ctxt'}}
  {\sequence{\expr_a}}\]
  $\EraseE{\fillc
    {\notedapp[\type_r]
      {\annotate[\typearray{\typefun{\sequence{\type_i}}{\type_o}}{\idx_f}]{\ctxt'}}
      {\sequence{\expr_a}}}
    {\expr}}$\\
  $= \EraseE{\notedapp[\type_r]
    {\fillc{\annotate[\typearray{\typefun{\sequence{\type_i}}{\type_o}}{\idx_f}]{\ctxt'}}{\expr}}
    {\sequence{\expr_a}}}$ \\
  $= \app
  {\EraseE{\fillc{\ctxt'}{\expr}}}
  {\sequence{\ttsexp{\EraseE{\expr_a}}{\EraseT{\type_i}}}\;\EraseT{\type_r}}$ \\
  $= \app
  {\fillc{\EraseC{\ctxt'}}{\expr}}
  {\sequence{\ttsexp{\EraseE{\expr_a}}{\EraseT{\type_i}}}\;\EraseT{\type_r}}$ \\
  $= \fillc{\app
    {\EraseC{\ctxt'}}
    {\sequence{\ttsexp{\EraseE{\expr_a}}{\EraseT{\type_i}}}\;\EraseT{\type_r}}}
  {\expr}$ \\
  $= \fillc{\EraseC{\app
      {\annotate[\typearray{\typefun{\sequence{\type_i}}{\type_o}}{\idx_f}]{\ctxt'}}
      {\sequence{\expr_a}}}}
  {\expr}$ \\

\paragraph{Application with unevaluated argument:}
  \[\ctxt =
  \notedapp[\type_r]
  {\annotate
    [\typearray{\typefun{\sequence{\type_1}\,\type_2\,\sequence{\type_3}}{\type_o}}{\idx_f}]
    {\expr_f}}
  {\sequence{\val_1}\,\ctxt'\,\sequence{\expr_3}}\]
  $\EraseE{\fillc
    {\notedapp[\type_r]
      {\annotate
        [\typearray{\typefun{\sequence{\type_1}\,\type_2\,\sequence{\type_3}}{\type_o}}{\idx_f}]
        {\expr_f}}
      {\sequence{\val_1}\,\ctxt'\,\sequence{\expr_3}}}
    {\expr}}$ \\
  $= \EraseE{\notedapp[\type_r]
    {\annotate
      [\typearray{\typefun{\sequence{\type_1}\,\type_2\,\sequence{\type_3}}{\type_o}}{\idx_f}]
      {\expr_f}}
    {\sequence{\val_1}\,\fillc{\ctxt'}{\expr}\,\sequence{\expr_3}}}$ \\
  $= \app
  {\EraseE{\expr_f}}
  {\sequence{\ttsexp{\EraseE{\val_1}}{\EraseT{\type_1}}}
    \,\ttsexp{\EraseE{\fillc{\ctxt'}{\expr}}}{\EraseT{\type_2}}
    \,\sequence{\ttsexp{\EraseE{\expr_3}}{\EraseT{\type_3}}}
    \;\EraseT{\type_r}}$ \\
  $= \app
  {\EraseE{\expr_f}}
  {\sequence{\ttsexp{\EraseE{\val_1}}{\EraseT{\type_1}}}
    \,\ttsexp{\fillc{\EraseC{\ctxt'}}{\EraseE{\expr}}}{\EraseT{\type_2}}
    \,\sequence{\ttsexp{\EraseE{\expr_3}}{\EraseT{\type_3}}}
    \;\EraseT{\type_r}}$ \\
  $= \fillc
  {\app
    {\EraseE{\expr_f}}
    {\sequence{\ttsexp{\EraseE{\val_1}}{\EraseT{\type_1}}}
      \,\ttsexp{\EraseC{\ctxt'}}{\EraseT{\type_2}}
      \,\sequence{\ttsexp{\EraseE{\expr_3}}{\EraseT{\type_3}}}
      \;\EraseT{\type_r}}}
  {\EraseE{\expr}}$ \\
  $= \fillc
  {\EraseC{\notedapp[\type_r]
      {\annotate
        [\typearray{\typefun{\sequence{\type_1}\,\type_2\,\sequence{\type_3}}{\type_o}}{\idx_f}]
        {\expr_f}}
      {\sequence{\val_1}\,\ctxt'\,\sequence{\expr_3}}}}
  {\EraseE{\expr}}$ \\

\paragraph{Type application:}
  \[\ctxt =
  \notedtapp[\type_r]
  {\ctxt'}
  {\sequence{\type_a}}\]
  $\EraseE{\fillc{\notedtapp[\type_r]
      {\ctxt'}
      {\sequence{\type_a}}}
    {\expr}}$ \\
  $\EraseE{\notedtapp[\type_r]
    {\fillc{\ctxt'}{\expr}}
    {\sequence{\type_a}}}$ \\
  $= \iapp
  {\EraseE{\fillc{\ctxt'}{\expr}}}
  {\sequence{\EraseT{\type_a}}\;\EraseT{\type_r}}$ \\
  $= \iapp
  {\fillc{\EraseC{\ctxt'}}{\EraseE{\expr}}}
  {\sequence{\EraseT{\type_a}}\;\EraseT{\type_r}}$ \\
  $= \fillc{\iapp
    {\EraseC{\ctxt'}}
    {\sequence{\EraseT{\type_a}}\;\EraseT{\type_r}}}
  {\EraseE{\expr}}$ \\
  $= \fillc{\EraseC{\notedtapp[\type_r]
      {\ctxt'}
      {\sequence{\type_a}}}}
  {\EraseE{\expr}}$ \\

\paragraph{Index application:}
  \[\ctxt =
  \notediapp[\type_r]
  {\ctxt'}
  {\sequence{\idx_a}}\]
  $\EraseE{\fillc
    {\notediapp[\type_r]
      {\ctxt'}
      {\sequence{\idx_a}}}
    {\expr}}$ \\
  $= \EraseE{\notediapp[\type_r]
    {\fillc{\ctxt'}{\expr}}
    {\sequence{\idx_a}}}$ \\
  $= \iapp
  {\EraseE{\fillc{\ctxt'}{\expr}}}
  {\sequence{\idx_a}\;\EraseT{\type_r}}$ \\
  $= \iapp
  {\fillc{\EraseC{\ctxt'}}{\EraseE{\expr}}}
  {\sequence{\idx_a}\;\EraseT{\type_r}}$ \\
  $= \fillc{\iapp
    {\EraseC{\ctxt'}}
    {\sequence{\idx_a}\;\EraseT{\type_r}}}
  {\EraseE{\expr}}$ \\
  $= \fillc{\EraseC{\notediapp[\type_r]
      {\ctxt'}
      {\sequence{\idx_a}}}}
  {\EraseE{\expr}}$ \\

\paragraph{Unboxing:}
  \[\ctxt =
  \dproj
  {\sequence{\var_i}}
  {\var_e}
  {\ctxt'}
  {\expr_b^{\type_b}}\]
  $\EraseE{\fillc
    {\dproj
      {\sequence{\var_i}}
      {\var_e}
      {\ctxt'}
      {\expr_b^{\type_b}}}
    {\expr}}$ \\
  $= \EraseE{\dproj
    {\sequence{\var_i}}
    {\var_e}
    {\fillc{\ctxt'}{\expr}}
    {\expr_b^{\type_b}}}$ \\
  $= \dproj
  {\sequence{\var_i}}
  {\var_e}
  {\EraseE{\fillc{\ctxt'}{\expr}}}
  {\EraseE{\expr_b^{\type_b}} \; \EraseT{\type_b}}$ \\
  $= \dproj
  {\sequence{\var_i}}
  {\var_e}
  {\fillc{\EraseC{\ctxt'}}{\EraseE{\expr}}}
  {\EraseE{\expr_b^{\type_b}} \; \EraseT{\type_b}}$ \\
  $= \fillc{\dproj
    {\sequence{\var_i}}
    {\var_e}
    {\EraseC{\ctxt'}}
    {\EraseE{\expr_b^{\type_b}} \; \EraseT{\type_b}}}
  {\EraseE{\expr}}$ \\
  $= \fillc{\EraseC{\dproj
      {\sequence{\var_i}}
      {\var_e}
      {\ctxt'}
      {\expr_b}}}
  {\EraseE{\expr}}$ \\
\end{sproof}

We will also rely on a series of lemmas showing that substitution commutes with erasure.
\newcommand{\eexprsubst}[1]{\subst{#1}{\var}{\EraseE{\expr_\var}}}
\newcommand{\etrmsubstE}[1]{\eexprsubst{\EraseTrm{#1}}}
\newcommand{\exprsubstE}[1]{\eexprsubst{\EraseE{#1}}}
\newcommand{\eatomsubstE}[1]{\eexprsubst{\EraseA{#1}}}
\begin{lemma}
  \label{EESubstErase}
  $\EraseTrm{\eexprsubst{\term}} = \etrmsubstE{\term}$
\end{lemma}
\begin{sproof}[This is straightforward induction on $\term$.]
We use induction on $\term$.
We skip the cases where $\var$ does not appear free in $\term$,
as substitution would not change $\term$.
\paragraph{Term abstraction:}
\[\term = \lam{\sequence{\notevar{\var_i}{\type}}}{\expr}\text{, where } \var \not\in \sequence{\var_i}\]
$\EraseTrm{\eexprsubst{\lam{\sequence{\notevar{\var_i}{\type}}}{\expr}}}$ \\
$= \EraseA{\eexprsubst{\lam{\sequence{\notevar{\var_i}{\type}}}{\expr}}}$ \\
$= \EraseA{\lam{\sequence{\notevar{\var_i}{\type}}}{\eexprsubst{\expr}}}$ \\
$= \lam{\sequence{\var_i}}{\EraseE{\eexprsubst{\expr}}}$ \\
$= \lam{\sequence{\var_i}}{\exprsubstE{\expr}}$ \\
$= \eexprsubst{\lam{\sequence{\var_i}}{\EraseE{\expr}}}$ \\
$= \eatomsubstE{\lam{\sequence{\notevar{\var_i}{\type}}}{\expr}}$ \\
$= \etrmsubstE{\lam{\sequence{\notevar{\var_i}{\type}}}{\expr}}$ \\

\paragraph{Type abstraction:}
\[\term = \tlam{\sequence{\notevar{\var_i}{\kind}}}{\val}\]
$\EraseTrm{\eexprsubst{\tlam{\sequence{\notevar{\var_i}{\kind}}}{\val}}}$ \\
$= \EraseA{\eexprsubst{\tlam{\sequence{\notevar{\var_i}{\kind}}}{\val}}}$ \\
$= \EraseA{\tlam{\sequence{\notevar{\var_i}{\kind}}}{\eexprsubst{\val}}}$ \\
$= \ilam{\sequence{\var_i}}{\EraseE{\eexprsubst{\val}}}$ \\
$= \ilam{\sequence{\var_i}}{\exprsubstE{\val}}$ \\
$= \eexprsubst{\ilam{\sequence{\var_i}}{\EraseE{\val}}}$ \\
$= \eatomsubstE{\tlam{\sequence{\notevar{\var_i}{\kind}}}{\val}}$ \\
$= \etrmsubstE{\tlam{\sequence{\notevar{\var_i}{\kind}}}{\val}}$ \\

\paragraph{Index abstraction:}
\[\term = \ilam{\sequence{\notevar{\var_i}{\sort}}}{\val}\]
$\EraseTrm{\eexprsubst{\ilam{\sequence{\notevar{\var_i}{\sort}}}{\val}}}$ \\
$= \EraseA{\eexprsubst{\ilam{\sequence{\notevar{\var_i}{\sort}}}{\val}}}$ \\
$= \EraseA{\ilam{\sequence{\notevar{\var_i}{\sort}}}{\eexprsubst{\val}}}$ \\
$= \ilam{\sequence{\var_i}}{\EraseE{\eexprsubst{\val}}}$ \\
$= \ilam{\sequence{\var_i}}{\exprsubstE{\val}}$ \\
$= \eexprsubst{\ilam{\sequence{\var_i}}{\EraseE{\val}}}$ \\
$= \eatomsubstE{\ilam{\sequence{\notevar{\var_i}{\sort}}}{\val}}$ \\
$= \etrmsubstE{\ilam{\sequence{\notevar{\var_i}{\sort}}}{\val}}$ \\

\paragraph{Box:}
\[\term = \dsum{\sequence{\idx}}{\expr_s}{\type}\]
$\EraseTrm{\eexprsubst{\dsum{\sequence{\idx}}{\expr_s}{\type}}}$ \\
$= \EraseA{\eexprsubst{\dsum{\sequence{\idx}}{\expr_s}{\type}}}$ \\
$= \EraseA{\dsum{\sequence{\idx}}{\eexprsubst{\expr_s}}{\type}}$ \\
$= \dsum{\sequence{\idx}}{\EraseE{\eexprsubst{\expr_s}}}{}$ \\
$= \dsum{\sequence{\idx}}{\exprsubstE{\expr_s}}{}$ \\
$= \eexprsubst{\dsum{\sequence{\idx}}{\EraseE{\expr_s}}{}}$ \\
$= \eatomsubstE{\dsum{\sequence{\idx}}{\expr_s}{\type}}$ \\
$= \etrmsubstE{\dsum{\sequence{\idx}}{\expr_s}{\type}}$ \\

\paragraph{Variable:}
\[\term = \var\]
$\EraseTrm{\eexprsubst{\var}}$ \\
$= \EraseE{\eexprsubst{\var}}$ \\
$= \EraseE{\expr_\var}$ \\
$= \eexprsubst{\var}$ \\
$= \exprsubstE{\var}$ \\
$= \etrmsubstE{\var}$ \\

\paragraph{Array:}
\[\term = \arrlit{\sequence{\atom}}{\sequence{\nat}}\]
$\EraseTrm{\eexprsubst{\arrlit{\sequence{\atom}}{\sequence{\nat}}}}$ \\
$= \EraseE{\eexprsubst{\arrlit{\sequence{\atom}}{\sequence{\nat}}}}$ \\
$= \EraseE{\arrlit{\sequence{\eexprsubst{\atom}}}{\sequence{\nat}}}$ \\
$= \arrlit{\sequence{\EraseA{\eexprsubst{\atom}}}}{\sequence{\nat}}$ \\
$= \arrlit{\sequence{\eatomsubstE{\atom}}}{\sequence{\nat}}$ \\
$= \eexprsubst{\arrlit{\sequence{\EraseA{\atom}}}{\sequence{\nat}}}$ \\
$= \exprsubstE{\arrlit{\sequence{\atom}}{\sequence{\nat}}}$ \\
$= \etrmsubstE{\arrlit{\sequence{\atom}}{\sequence{\nat}}}$ \\

\paragraph{Frame:}
\[\term = \notedfrm[\type_r]{\sequence{\expr_c}}{\sequence{\nat}}\]
$\EraseTrm{\eexprsubst{\notedfrm[\type_r]{\sequence{\expr_c}}{\sequence{\nat}}}}$ \\
$= \EraseE{\eexprsubst{\notedfrm[\type_r]{\sequence{\expr_c}}{\sequence{\nat}}}}$ \\
$= \EraseE{\notedfrm[\type_r]{\sequence{\eexprsubst{\expr_c}}}{\sequence{\nat}}}$ \\
$= \frm{\sequence{\EraseE{\eexprsubst{\expr_c}}}}{\EraseT{\type_r}}$ \\
$= \frm{\sequence{\exprsubstE{\expr_c}}}{\EraseT{\type_r}}$ \\
$= \eexprsubst{\frm{\sequence{\EraseE{\expr_c}}}{\EraseT{\type_r}}}$ \\
$= \exprsubstE{\notedfrm[\type_r]{\sequence{\expr_c}}{\sequence{\nat}}}$ \\
$= \etrmsubstE{\notedfrm[\type_r]{\sequence{\expr_c}}{\sequence{\nat}}}$ \\

\paragraph{Application:}
\[\term =\notedapp[\type_r]
{\annotate
  [\typearray{\typefun{\sequence{\type_i}}{\type_o}}{\idx_f}]
  {\expr_f}}
{\sequence{\expr_a}}\]
$\EraseTrm{\eexprsubst{\notedapp[\type_r]{\annotate
      [\typearray{\typefun{\sequence{\type_i}}{\type_o}}{\idx_f}]
      {\expr_f}}
    {\sequence{\expr_a}}}}$ \\
$= \EraseE{\eexprsubst{\notedapp[\type_r]{\annotate
      [\typearray{\typefun{\sequence{\type_i}}{\type_o}}{\idx_f}]
      {\expr_f}}
    {\sequence{\expr_a}}}}$ \\
$= \EraseE{\notedapp[\type_r]{\eexprsubst{\annotate
      [\typearray{\typefun{\sequence{\type_i}}{\type_o}}{\idx_f}]
      {\expr_f}}}
  {\sequence{\eexprsubst{\expr_a}}}}$ \\
$= \app
{\EraseE{\eexprsubst{\expr_f}}}
{\sequence{\ttsexp{\EraseE{\eexprsubst{\expr_a}}}{\EraseT{\type_i}}}\;\EraseT{\type_r}}$ \\
$= \app
{\exprsubstE{\expr_f}}
{\sequence{\ttsexp{\exprsubstE{\expr_a}}{\EraseT{\type_i}}}\;\EraseT{\type_r}}$ \\
$= \eexprsubst{\app
  {\EraseE{\expr_f}}
  {\sequence{\ttsexp{\EraseE{\expr_a}}{\EraseT{\type_i}}}\;\EraseT{\type_r}}}$ \\
$= \exprsubstE{\notedapp[\type_r]
  {\annotate
    [\typearray{\typefun{\sequence{\type_i}}{\type_o}}{\idx_f}]
    {\expr_f}}
  {\sequence{\expr_a}}}$ \\
$= \etrmsubstE{\notedapp[\type_r]
  {\annotate
    [\typearray{\typefun{\sequence{\type_i}}{\type_o}}{\idx_f}]
    {\expr_f}}
  {\sequence{\expr_a}}}$ \\

\paragraph{Type application:}
\[\term = \notedtapp[\type_r]{\expr_f}{\sequence{\type_a}}\]
$\EraseTrm{\eexprsubst{\notedtapp[\type_r]{\expr_f}{\sequence{\type_a}}}}$ \\
$= \EraseE{\eexprsubst{\notedtapp[\type_r]{\expr_f}{\sequence{\type_a}}}}$ \\
$= \EraseE{\notedtapp[\type_r]{\eexprsubst{\expr_f}}{\sequence{\type_a}}}$ \\
$= \iapp{\EraseE{\eexprsubst{\expr_f}}}{\sequence{\EraseT{\type_a}}\;\EraseT{\type_r}}$ \\
$= \iapp{\exprsubstE{\expr_f}}{\sequence{\EraseT{\type_a}}\;\EraseT{\type_r}}$ \\
$= \eexprsubst{\iapp{\EraseE{\expr_f}}{\sequence{\EraseT{\type_a}}\;\EraseT{\type_r}}}$ \\
$= \exprsubstE{\notedtapp[\type_r]{\expr_f}{\sequence{\type_a}}}$ \\
$= \etrmsubstE{\notedtapp[\type_r]{\expr_f}{\sequence{\type_a}}}$ \\

\paragraph{Index application:}
\[\term = \notediapp[\type_r]{\expr_f}{\sequence{\idx_a}}\]
$\EraseTrm{\eexprsubst{\notediapp[\type_r]{\expr_f}{\sequence{\idx_a}}}}$ \\
$= \EraseE{\eexprsubst{\notediapp[\type_r]{\expr_f}{\sequence{\idx_a}}}}$ \\
$= \EraseE{\notediapp[\type_r]{\eexprsubst{\expr_f}}{\sequence{\idx_a}}}$ \\
$= \iapp{\EraseE{\eexprsubst{\expr_f}}}{\sequence{\idx_a}\;\EraseT{\type_r}}$ \\
$= \iapp{\exprsubstE{\expr_f}}{\sequence{\idx_a}\;\EraseT{\type_r}}$ \\
$= \eexprsubst{\iapp{\EraseE{\expr_f}}{\sequence{\idx_a}\;\EraseT{\type_r}}}$ \\
$= \exprsubstE{\notediapp[\type_r]{\expr_f}{\sequence{\idx_a}}}$ \\
$= \etrmsubstE{\notediapp[\type_r]{\expr_f}{\sequence{\idx_a}}}$ \\

\paragraph{Unboxing:}
\[\term = \dproj{\sequence{\var_i}}{\var_e}{\expr_s}{\expr_b^{\type_b}}\]
$\EraseTrm{\eexprsubst{\dproj{\sequence{\var_i}}{\var_e}{\expr_s}{\expr_b^{\type_b}}}}$ \\
$= \EraseE{\eexprsubst{\dproj{\sequence{\var_i}}{\var_e}{\expr_s}{\expr_b^{\type_b}}}}$ \\
$= \EraseE{\dproj{\sequence{\var_i}}{\var_e}{\eexprsubst{\expr_s}}{\eexprsubst{\expr_b^{\type_b}}}}$ \\
$= \dproj{\sequence{\var_i}}{\var_e}{\EraseE{\eexprsubst{\expr_s}}}
{\EraseE{\eexprsubst{\expr_b^{\type_b}}}\;\EraseT{\type_b}}$ \\
$= \dproj{\sequence{\var_i}}{\var_e}{\exprsubstE{\expr_s}}{\exprsubstE{\expr_b^{\type_b}}\;\EraseT{\type_b}}$ \\
$= \eexprsubst{\dproj{\sequence{\var_i}}{\var_e}{\expr_s}{\expr_b\;\EraseT{\type_b}}}$ \\
$= \exprsubstE{\dproj{\sequence{\var_i}}{\var_e}{\expr_s}{\expr_b}}$ \\
$= \etrmsubstE{\dproj{\sequence{\var_i}}{\var_e}{\expr_s}{\expr_b}}$ \\

\end{sproof}

\newcommand{\etypesubst}[1]{\subst{#1}{\var}{\EraseT{\type_\var}}}
\begin{lemma}
  \label{TTSubstErase}
  $\EraseT{\typesubst{\type}} = \etypesubst{\EraseT{\type}}$
\end{lemma}
\begin{sproof}[This is straightforward induction on $\type$.]
  We use induction on $\type$.
  We elide the cases where $\var$ does not appear free in $\type$.
  \paragraph{Variable:}
  \[\type = \var\]
  $\EraseT{\typesubst{\var}}$ \\
  $= \EraseT{\type_\var}$ \\
  $= \etypesubst{\var}$ \\
  $= \etypesubst{\EraseT{\var}}$ \\

  \paragraph{Function:}
  \[\type = \typefun{\sequence{\type_i}}{\type_o}\]
  $\EraseT{\typesubst{\typefun{\sequence{\type_i}}{\type_o}}}$ \\
  $= \EraseT{\typefun{\sequence{\typesubst{\type_i}}}{\typesubst{\type_o}}}$ \\
  $= \idxshape{}$ \\
  $= \etypesubst{\EraseT{\typefun{\sequence{\type_i}}{\type_o}}}$ \\

  \paragraph{Universal:}
  \[\type = \typeuniv{\sequence{\notevar{\var_u}{\kind}}}{\type_u}\text{, where }\var \not\in \sequence{\var_u}\]
  $\EraseT{\typesubst{\typeuniv{\sequence{\notevar{\var_u}{\kind}}}{\type_u}}}$ \\
  $= \EraseT{\typeuniv{\sequence{\notevar{\var_u}{\kind}}}{\typesubst{\type_u}}}$ \\
  $= \idxshape{}$ \\
  $= \etypesubst{\EraseT{\typeuniv{\sequence{\notevar{\var_u}{\kind}}}{\type_u}}}$

  \paragraph{Dependent product:}
  \[\type = \typedprod{\sequence{\notevar{\var_p}{\sort}}}{\type_p}\]
  $\EraseT{\typesubst{\typedprod{\sequence{\notevar{\var_p}{\sort}}}{\type_p}}}$ \\
  $= \EraseT{\typedprod{\sequence{\notevar{\var_p}{\sort}}}{\typesubst{\type_p}}}$ \\
  $= \idxshape{}$ \\
  $= \etypesubst{\EraseT{\typedprod{\sequence{\notevar{\var_p}{\sort}}}{\type_p}}}$ \\

  \paragraph{Dependent sum:}
  \[\type = \typedsum{\sequence{\notevar{\var_p}{\sort}}}{\type_p}\]
  $\EraseT{\typesubst{\typedsum{\sequence{\notevar{\var_p}{\sort}}}{\type_p}}}$ \\
  $= \EraseT{\typedsum{\sequence{\notevar{\var_p}{\sort}}}{\typesubst{\type_p}}}$ \\
  $= \idxshape{}$ \\
  $= \etypesubst{\EraseT{\typedsum{\sequence{\notevar{\var_p}{\sort}}}{\type_p}}}$ \\

  \paragraph{Array:}
  \[\type = \typearray{\type_a}{\idx}\]
  $\EraseT{\typesubst{\typearray{\type_a}{\idx}}}$ \\
  $= \EraseT{\typearray{\typesubst{\type_a}}{\idx_a}}$ \\
  $= \idx_a$\qquad\text{(note: $\var$ is a type variable and cannot appear in $\idx_a$)} \\
  $= \etypesubst{\EraseT{\typearray{\type_a}{\idx}}}$

\end{sproof}

\begin{lemma}
  \label{TESubstErase}
  $\EraseTrm{\typesubst{\term}} = \etypesubst{\EraseTrm{\term}}$
\end{lemma}
\begin{sproof}[This is straightforward induction on $\term$.]
  We use induction on $\term$.
  We elide the cases where $\var$ does not appear free in $\term$.

  \paragraph{Term abstraction:}
  \[\term = \lam{\sequence{\notevar{\var_i}{\type}}}{\expr}\]
  $\EraseTrm{\typesubst{\lam{\sequence{\notevar{\var_i}{\type}}}{\expr}}}$ \\
  $= \EraseA{\typesubst{\lam{\sequence{\notevar{\var_i}{\type}}}{\expr}}}$ \\
  $= \EraseA{\lam{\sequence{\notevar{\var_i}{\type}}}{\typesubst{\expr}}}$ \\
  $= \lam{\sequence{\var_i}}{\EraseE{\etypesubst{\expr}}}$ \\
  $= \lam{\sequence{\var_i}}{\etypesubst{\EraseE{\expr}}}$ \\
  $= \etypesubst{\lam{\sequence{\var_i}}{\EraseE{\expr}}}$ \\
  $= \etypesubst{\EraseA{\lam{\sequence{\notevar{\var_i}{\type}}}{\expr}}}$ \\
  $= \etypesubst{\EraseTrm{\lam{\sequence{\notevar{\var_i}{\type}}}{\expr}}}$ \\

  \paragraph{Type abstraction:}
  \[\term = \tlam{\sequence{\notevar{\var_i}{\kind}}}{\val}\text{, where } \var \not\in \sequence{\var_i}\]
  $\EraseTrm{\typesubst{\tlam{\sequence{\notevar{\var_i}{\kind}}}{\val}}}$ \\
  $= \EraseA{\typesubst{\tlam{\sequence{\notevar{\var_i}{\kind}}}{\val}}}$ \\
  $= \EraseA{\tlam{\sequence{\notevar{\var_i}{\kind}}}{\typesubst{\val}}}$ \\
  $= \ilam{\sequence{\var_i}}{\EraseE{\typesubst{\val}}}$ \\
  $= \ilam{\sequence{\var_i}}{\etypesubst{\EraseE{\val}}}$ \\
  $= \etypesubst{\ilam{\sequence{\var_i}}{\EraseE{\val}}}$ \\
  $= \etypesubst{\EraseA{\tlam{\sequence{\notevar{\var_i}{\kind}}}{\val}}}$ \\
  $= \etypesubst{\EraseTrm{\tlam{\sequence{\notevar{\var_i}{\kind}}}{\val}}}$ \\

  \paragraph{Index abstraction:}
  \[\term = \ilam{\sequence{\notevar{\var_i}{\sort}}}{\val}\]
  $\EraseTrm{\typesubst{\ilam{\sequence{\notevar{\var_i}{\sort}}}{\val}}}$ \\
  $= \EraseA{\typesubst{\ilam{\sequence{\notevar{\var_i}{\sort}}}{\val}}}$ \\
  $= \EraseA{\ilam{\sequence{\notevar{\var_i}{\sort}}}{\typesubst{\val}}}$ \\
  $= \ilam{\sequence{\var_i}}{\EraseE{\typesubst{\val}}}$ \\
  $= \ilam{\sequence{\var_i}}{\etypesubst{\EraseE{\val}}}$ \\
  $= \etypesubst{\ilam{\sequence{\var_i}}{\EraseE{\val}}}$ \\
  $= \etypesubst{\EraseA{\ilam{\sequence{\notevar{\var_i}{\kind}}}{\val}}}$ \\
  $= \etypesubst{\EraseTrm{\ilam{\sequence{\notevar{\var_i}{\kind}}}{\val}}}$ \\

  \paragraph{Box:}
  \[\term = \dsum{\sequence{\idx}}{\expr_s}{\type}\]
  $\EraseTrm{\typesubst{\dsum{\sequence{\idx}}{\expr_s}{\type}}}$ \\
  $= \EraseA{\typesubst{\dsum{\sequence{\idx}}{\expr_s}{\type}}}$ \\
  $= \EraseA{{\dsum{\sequence{\idx}}{\typesubst{\expr_s}}{\typesubst{\type}}}}$ \\
  $= \dsum{\sequence{\idx}}{\EraseE{\typesubst{\expr_s}}}{}$ \\
  $= \dsum{\sequence{\idx}}{\etypesubst{\EraseE{\expr_s}}}{}$ \\
  $= \etypesubst{\dsum{\sequence{\idx}}{\EraseE{\expr_s}}{}}$ \\
  $= \etypesubst{\EraseA{\dsum{\sequence{\idx}}{\expr_s}{\type}}}$ \\
  $= \etypesubst{\EraseTrm{\dsum{\sequence{\idx}}{\expr_s}{\type}}}$ \\

  \paragraph{Array:}
  \[\term = \arrlit{\sequence{\atom}}{\sequence{\nat}}\]
  $\EraseTrm{\typesubst{\arrlit{\sequence{\atom}}{\sequence{\nat}}}}$ \\
  $= \EraseE{\typesubst{\arrlit{\sequence{\atom}}{\sequence{\nat}}}}$ \\
  $= \EraseE{\arrlit{\sequence{\typesubst{\atom}}}{\sequence{\nat}}}$ \\
  $= \arrlit{\sequence{\EraseA{\typesubst{\atom}}}}{\sequence{\nat}}$ \\
  $= \arrlit{\sequence{\etypesubst{\EraseA{\atom}}}}{\sequence{\nat}}$ \\
  $= \etypesubst{\arrlit{\sequence{\EraseA{\atom}}}{\sequence{\nat}}}$ \\
  $= \etypesubst{\EraseE{\arrlit{\sequence{\atom}}{\sequence{\nat}}}}$ \\
  $= \etypesubst{\EraseTrm{\arrlit{\sequence{\atom}}{\sequence{\nat}}}}$ \\

  \paragraph{Frame:}
  \[\term = \notedfrm[\type_r]{\sequence{\expr_c}}{\sequence{\nat}}\]
  $\EraseTrm{\typesubst{\notedfrm[\type_r]{\sequence{\expr_c}}{\sequence{\nat}}}}$ \\
  $= \EraseE{\typesubst{\notedfrm[\type_r]{\sequence{\expr_c}}{\sequence{\nat}}}}$ \\
  $= \EraseE{\notedfrm[\typesubst{\type_r}]{\sequence{\typesubst{\expr_c}}}{\sequence{\nat}}}$ \\
  $= \frm{\sequence{\EraseE{\typesubst{\expr_c}}}}{\EraseT{\typesubst{\type_r}}}$ \\
  $= \frm{\sequence{\etypesubst{\EraseE{\expr_c}}}}{\EraseT{\typesubst{\type_r}}}$%
  , by the induction hypothesis \\
  $= \frm{\sequence{\etypesubst{\EraseE{\expr_c}}}}{\etypesubst{\EraseT{\type_r}}}$%
  , by Lemma \ref{TTSubstErase} \\
  $= \etypesubst{\frm{\sequence{\EraseE{\expr_c}}}{\EraseT{\type_r}}}$ \\
  $= \etypesubst{\EraseE{\notedfrm[\type_r]{\sequence{\expr_c}}{\sequence{\nat}}}}$ \\
  $= \etypesubst{\EraseTrm{\notedfrm[\type_r]{\sequence{\expr_c}}{\sequence{\nat}}}}$ \\

  \paragraph{Application:}
  \[\term =
  \notedapp[\type_r]{\annotate
    [\typearray{\typefun{\sequence{\type_i}}{\type_o}}{\idx_f}]
    {\expr_f}}{\sequence{\expr_a}}\]
  $\EraseTrm{\typesubst{\notedapp[\type_r]{\annotate
        [\typearray{\typefun{\sequence{\type_i}}{\type_o}}{\idx_f}]
        {\expr_f}}{\sequence{\expr_a}}}}$ \\
  $= \EraseE{\typesubst{\notedapp[\type_r]{\annotate
        [\typearray{\typefun{\sequence{\type_i}}{\type_o}}{\idx_f}]
        {\expr_f}}{\sequence{\expr_a}}}}$ \\
  $= \EraseE{\notedapp[\typesubst{\type_r}]{\annotate
      [\typesubst{\typearray{\typefun{\sequence{\type_i}}{\type_o}}{\idx_f}}]
      {\typesubst{\expr_f}}}{\sequence{\typesubst{\expr_a}}}}$ \\
  $= \EraseE{\notedapp[\typesubst{\type_r}]{\annotate
      [\typearray{\typefun{\sequence{\typesubst{\type_i}}}{\typesubst{\type_o}}}{\idx_f}]
      {\typesubst{\expr_f}}}{\sequence{\typesubst{\expr_a}}}}$ \\
  $= \app{\EraseE{\typesubst{\expr_f}}}{\sequence
    {\ttsexp{\EraseE{\typesubst{\expr_a}}}{\EraseT{\typesubst{\type_i}}}}
    \;\EraseT{\typesubst{\type_r}}}$ \\
  $= \app{\etypesubst{\EraseE{\expr_f}}}{\sequence
    {\ttsexp{\etypesubst{\EraseE{\expr_a}}}{\EraseT{\typesubst{\type_i}}}}
    \;\EraseT{\typesubst{\type_r}}}$, by the induction hypothesis \\
  $= \app{\etypesubst{\EraseE{\expr_f}}}{\sequence
    {\ttsexp{\etypesubst{\EraseE{\expr_a}}}{\etypesubst{\EraseT{\type_i}}}}
    \;\etypesubst{\EraseE{\type_r}}}$, by Lemma \ref{TTSubstErase} \\
  $= \etypesubst{\app{\EraseE{\expr_f}}{\sequence
    {\ttsexp{\EraseE{\expr_a}}{\EraseT{\type_i}}}
    \;\EraseE{\type_r}}}$ \\
  $= \etypesubst{\EraseE{\notedapp[\type_r]{\annotate
        [\typearray{\typefun{\sequence{\type_i}}{\type_o}}{\idx_f}]
        {\expr_f}}{\sequence{\expr_a}}}}$ \\
  $= \etypesubst{\EraseTrm{\notedapp[\type_r]{\annotate
        [\typearray{\typefun{\sequence{\type_i}}{\type_o}}{\idx_f}]
        {\expr_f}}{\sequence{\expr_a}}}}$ \\

  \paragraph{Type application:}
  \[\term = \notedtapp[\type_r]{\expr_f}{\sequence{\type_a}}\]
  $\EraseTrm{\typesubst{\notedtapp[\type_r]{\expr_f}{\sequence{\type_a}}}}$ \\
  $= \EraseE{\typesubst{\notedtapp[\type_r]{\expr_f}{\sequence{\type_a}}}}$ \\
  $= \EraseE{\notedtapp[\typesubst{\type_r}]{\typesubst{\expr_f}}{\sequence{\typesubst{\type_a}}}}$ \\
  $= \iapp{\EraseE{\typesubst{\expr_f}}}{\sequence
    {\EraseT{\typesubst{\type_a}}}\;\EraseT{\typesubst{\type_r}}}$ \\
  $= \iapp{\etypesubst{\EraseE{\expr_f}}}{\sequence
    {\EraseT{\typesubst{\type_a}}}\;\EraseT{\typesubst{\type_r}}}$%
  , by the induction hypothesis \\
  $= \iapp{\etypesubst{\EraseE{\expr_f}}}{\sequence
    {\etypesubst{\EraseT{\type_a}}}\;\etypesubst{\EraseT{\type_r}}}$%
  , by Lemma \ref{TTSubstErase} \\
  $= \etypesubst{\iapp{\EraseE{\expr_f}}{\sequence
      {\EraseT{\type_a}}\;\EraseT{\type_r}}}$ \\
  $= \etypesubst{\EraseE{\notedtapp[\type_r]{\expr_f}{\sequence{\type_a}}}}$ \\
  $= \etypesubst{\EraseTrm{\notedtapp[\type_r]{\expr_f}{\sequence{\type_a}}}}$ \\

  \paragraph{Index application:}
  \[\term = \notediapp[\type_r]{\expr_f}{\sequence{\idx_a}}\]
  $\EraseTrm{\typesubst{\notediapp[\type_r]{\expr_f}{\sequence{\idx_a}}}}$ \\
  $= \EraseE{\typesubst{\notediapp[\type_r]{\expr_f}{\sequence{\idx_a}}}}$ \\
  $= \EraseE{\notediapp[\typesubst{\type_r}]{\typesubst{\expr_f}}{\sequence{\idx_a}}}$ \\
  $= \iapp{\EraseE{\typesubst{\expr_f}}}{\sequence{\idx_a}\;\EraseT{\typesubst{\type_r}}}$ \\
  $= \iapp{\etypesubst{\EraseE{\expr_f}}}{\sequence{\idx_a}\;\EraseT{\typesubst{\type_r}}}$%
  , by the induction hypothesis\\
  $= \iapp{\etypesubst{\EraseE{\expr_f}}}{\sequence{\idx_a}\;\etypesubst{\EraseT{\type_r}}}$%
  , by Lemma \ref{TTSubstErase}\\
  $= \etypesubst{\iapp{\EraseE{\expr_f}}{\sequence{\idx_a}\;\EraseT{\type_r}}}$ \\
  $= \etypesubst{\EraseE{\notediapp[\type_r]{\expr_f}{\sequence{\idx_a}}}}$ \\
  $= \etypesubst{\EraseTrm{\notediapp[\type_r]{\expr_f}{\sequence{\idx_a}}}}$ \\

  \paragraph{Unboxing:}
  \[\term = \dproj{\sequence{\var_i}}{\var_e}{\expr_s}{\expr_b^{\type_b}}\]
  $\EraseTrm{\typesubst{\dproj{\sequence{\var_i}}{\var_e}{\expr_s}{\expr_b^{\type_b}}}}$ \\
  $= \EraseE{\typesubst{\dproj{\sequence{\var_i}}{\var_e}{\expr_s}{\expr_b^{\type_b}}}}$ \\
  $= \EraseE{\dproj{\sequence{\var_i}}{\var_e}{\typesubst{\expr_s}}
    {\typesubst{\expr_b}^{\typesubst{\type_b}}}}$ \\
  $= \dproj{\sequence{\var_i}}{\var_e}{\EraseE{\typesubst{\expr_s}}}
  {\EraseE{\typesubst{\expr_b}} \; \EraseT{\typesubst{\type_b}}}$ \\
  $= \dproj{\sequence{\var_i}}{\var_e}{\EraseE{\typesubst{\expr_s}}}
  {\EraseE{\typesubst{\expr_b}} \; \etypesubst{\EraseT{\type_b}}}$ \\
  $= \dproj{\sequence{\var_i}}{\var_e}{\etypesubst{\EraseE{\expr_s}}}
  {\etypesubst{\EraseE{\expr_b}} \; \etypesubst{\EraseT{\type_b}}}$ \\
  $= \etypesubst{\dproj{\sequence{\var_i}}{\var_e}{\EraseE{\expr_s}}{\EraseE{\expr_b} \; \EraseT{\type_b}}}$ \\
  $= \etypesubst{\EraseE{\dproj{\sequence{\var_i}}{\var_e}{\expr_s}{\expr_b}}}$ \\
  $= \etypesubst{\EraseTrm{\dproj{\sequence{\var_i}}{\var_e}{\expr_s}{\expr_b}}}$ \\

\end{sproof}

\begin{lemma}
  \label{ITSubstErase}
  $\EraseT{\idxsubst{\type}} = \idxsubst{\EraseT{\type}}$
\end{lemma}
\begin{sproof}[This is straightforward induction on $\type$.]
  We use induction on $\type$.
  We elide the cases where $\var$ does not appear free in $\type$.
  \paragraph{Function:}
  \[\type = \typefun{\sequence{\type_i}}{\type_o}\]
  $\EraseT{\idxsubst{\typefun{\sequence{\type_i}}{\type_o}}}$ \\
  $= \EraseT{\typefun{\sequence{\idxsubst{\type_i}}}{\idxsubst{\type_o}}}$ \\
  $= \idxshape{}$ \\
  $= \idxsubst{\idxshape{}}$ \\
  $= \idxsubst{\EraseT{\typefun{\sequence{\type_i}}{\type_o}}}$ \\

  \paragraph{Universal:}
  \[\type = \typeuniv{\sequence{\notevar{\var_u}{\kind}}}{\type_u}\]
  $\EraseT{\idxsubst{\typeuniv{\sequence{\notevar{\var_u}{\kind}}}{\type_u}}}$ \\
  $= \EraseT{\typeuniv{\sequence{\notevar{\var_u}{\kind}}}{\idxsubst{\type_u}}}$ \\
  $= \idxshape{}$ \\
  $= \idxsubst{\idxshape{}}$ \\
  $= \idxsubst{\EraseT{\typeuniv{\sequence{\notevar{\var_u}{\kind}}}{\type_u}}}$ \\

  \paragraph{Dependent product:}
  \[\type = \typedprod{\sequence{\notevar{\var_p}{\sort}}}{\type_p}\text{, where }\var \not\in \sequence{\var_p}\]
  $\EraseT{\idxsubst{\typedprod{\sequence{\notevar{\var_p}{\sort}}}{\type_p}}}$ \\
  $= \EraseT{\typedprod{\sequence{\notevar{\var_p}{\sort}}}{\idxsubst{\type_p}}}$ \\
  $= \idxshape{}$ \\
  $= \idxsubst{\idxshape{}}$ \\
  $= \idxsubst{\EraseT{\typedprod{\sequence{\notevar{\var_p}{\sort}}}{\type_p}}}$ \\

  \paragraph{Dependent sum:}
  \[\type = \typedsum{\sequence{\notevar{\var_p}{\sort}}}{\type_p}\text{, where }\var \not\in \sequence{\var_p}\]
  $\EraseT{\idxsubst{\typedsum{\sequence{\notevar{\var_p}{\sort}}}{\type_p}}}$ \\
  $= \EraseT{\typedsum{\sequence{\notevar{\var_p}{\sort}}}{\idxsubst{\type_p}}}$ \\
  $= \idxshape{}$ \\
  $= \idxsubst{\idxshape{}}$ \\
  $= \idxsubst{\EraseT{\typedsum{\sequence{\notevar{\var_p}{\sort}}}{\type_p}}}$ \\

  \paragraph{Array:}
  \[\type = \typearray{\type_a}{\idx}\]
  $\EraseT{\idxsubst{\typearray{\type_a}{\idx}}}$ \\
  $= \EraseT{\typearray{\idxsubst{\type_a}}{\idxsubst{\idx}}}$ \\
  $= \idxsubst{\idx}$ \\
  $= \idxsubst{\EraseT{\typearray{\type_a}{\idx}}}$ \\

\end{sproof}

\begin{lemma}
  \label{IESubstErase}
  $\EraseTrm{\idxsubst{\term}} = \idxsubst{\EraseTrm{\term}}$
\end{lemma}
\begin{sproof}[This is straightforward induction on $\term$.]
  We use induction on $\term$.
  We elide the cases where $\var$ does not appear free in $\term$.

  \paragraph{Term abstraction:}
  \[\term = \lam{\sequence{\notevar{\var_i}{\type}}}{\expr}\]
  $\EraseTrm{\idxsubst{\lam{\sequence{\notevar{\var_i}{\type}}}{\expr}}}$ \\
  $= \EraseA{\idxsubst{\lam{\sequence{\notevar{\var_i}{\type}}}{\expr}}}$ \\
  $= \EraseA{\lam{\sequence{\notevar{\var_i}{\type}}}{\idxsubst{\expr}}}$ \\
  $= \lam{\sequence{\var_i}}{\EraseE{\idxsubst{\expr}}}$ \\
  $= \lam{\sequence{\var_i}}{\idxsubst{\EraseE{\expr}}}$ \\
  $= \idxsubst{\lam{\sequence{\var_i}}{\EraseE{\expr}}}$ \\
  $= \idxsubst{\EraseA{\lam{\sequence{\var_i}}{\expr}}}$ \\
  $= \idxsubst{\EraseTrm{\lam{\sequence{\var_i}}{\expr}}}$ \\

  \paragraph{Type abstraction:}
  \[\term = \tlam{\sequence{\notevar{\var_i}{\kind}}}{\val}\]
  Note: $\var$ is an index variable, while the $\sequence{\var_i}$ are type variables, so they do not shadow.
  $\EraseTrm{\idxsubst{\tlam{\sequence{\notevar{\var_i}{\kind}}}{\val}}}$ \\
  $= \EraseA{\idxsubst{\tlam{\sequence{\notevar{\var_i}{\kind}}}{\val}}}$ \\
  $= \EraseA{\tlam{\sequence{\notevar{\var_i}{\kind}}}{\idxsubst{\val}}}$ \\
  $= \ilam{\sequence{\var_i}}{\EraseE{\idxsubst{\val}}}$ \\
  $= \ilam{\sequence{\var_i}}{\idxsubst{\EraseE{\val}}}$ \\
  $= \idxsubst{\ilam{\sequence{\var_i}}{\EraseE{\val}}}$ \\
  $= \idxsubst{\EraseA{\tlam{\sequence{\notevar{\var_i}{\kind}}}{\val}}}$ \\
  $= \idxsubst{\EraseTrm{\tlam{\sequence{\notevar{\var_i}{\kind}}}{\val}}}$ \\

  \paragraph{Index abstraction:}
  \[\term = \ilam{\sequence{\notevar{\var_i}{\sort}}}{\val}\text{, where } \var \not\in \sequence{\var_i}\]
  $\EraseTrm{\idxsubst{\ilam{\sequence{\notevar{\var_i}{\sort}}}{\val}}}$ \\
  $= \EraseA{\idxsubst{\ilam{\sequence{\notevar{\var_i}{\sort}}}{\val}}}$ \\
  $= \EraseA{\ilam{\sequence{\notevar{\var_i}{\sort}}}{\idxsubst{\val}}}$ \\
  $= \ilam{\sequence{\var_i}}{\EraseE{\idxsubst{\val}}}$ \\
  $= \ilam{\sequence{\var_i}}{\idxsubst{\EraseE{\val}}}$ \\
  $= \idxsubst{\ilam{\sequence{\var_i}}{\EraseE{\val}}}$ \\
  $= \idxsubst{\EraseA{\ilam{\sequence{\notevar{\var_i}{\sort}}}{\val}}}$ \\
  $= \idxsubst{\EraseTrm{\ilam{\sequence{\notevar{\var_i}{\sort}}}{\val}}}$ \\

  \paragraph{Box:}
  \[\term = \dsum{\sequence{\idx}}{\expr_s}{\type}\]
  $\EraseTrm{\idxsubst{\dsum{\sequence{\idx}}{\expr_s}{\type}}}$ \\
  $= \EraseA{\idxsubst{\dsum{\sequence{\idx}}{\expr_s}{\type}}}$ \\
  $= \EraseA{\dsum{\sequence{\idxsubst{\idx}}}{\idxsubst{\expr_s}}{\idxsubst{\type}}}$ \\
  $= \dsum{\sequence{\idxsubst{\idx}}}{\EraseE{\idxsubst{\expr_s}}}{}$ \\
  $= \dsum{\sequence{\idxsubst{\idx}}}{\idxsubst{\EraseE{\expr_s}}}{}$ \\
  $= \idxsubst{\dsum{\sequence{\idx}}{\EraseE{\expr_s}}{}}$ \\
  $= \idxsubst{\EraseA{\dsum{\sequence{\idx}}{\expr_s}{\type}}}$ \\
  $= \idxsubst{\EraseTrm{\dsum{\sequence{\idx}}{\expr_s}{\type}}}$ \\

  \paragraph{Array:}
  \[\term = \arrlit{\sequence{\atom}}{\sequence{\nat}}\]
  $\EraseTrm{\idxsubst{\arrlit{\sequence{\atom}}{\sequence{\nat}}}}$ \\
  $= \EraseE{\idxsubst{\arrlit{\sequence{\atom}}{\sequence{\nat}}}}$ \\
  $= \EraseE{\arrlit{\sequence{\idxsubst\atom}}{\sequence{\nat}}}$ \\
  $= \arrlit{\sequence{\EraseA{\idxsubst{\atom}}}}{\sequence{\nat}}$ \\
  $= \arrlit{\sequence{\idxsubst{\EraseA{\atom}}}}{\sequence{\nat}}$ \\
  $= \idxsubst{\arrlit{\sequence{\EraseA{\atom}}}{\sequence{\nat}}}$ \\
  $= \idxsubst{\EraseE{\arrlit{\sequence{\atom}}{\sequence{\nat}}}}$ \\
  $= \idxsubst{\EraseTrm{\arrlit{\sequence{\atom}}{\sequence{\nat}}}}$ \\

  \paragraph{Frame:}
  \[\term = \notedfrm[\type_r]{\sequence{\expr_c}}{\sequence{\nat}}\]
  $\EraseTrm{\idxsubst{\notedfrm[\type_r]{\sequence{\expr_c}}{\sequence{\nat}}}}$ \\
  $= \EraseE{\idxsubst{\notedfrm[\type_r]{\sequence{\expr_c}}{\sequence{\nat}}}}$ \\
  $= \EraseE{\notedfrm[\idxsubst{\type_r}]{\sequence{\idxsubst{\expr_c}}}{\sequence{\nat}}}$ \\
  $= \frm{\sequence{\EraseE{\idxsubst{\expr_c}}}}{\EraseT{\idxsubst{\type_r}}}$ \\
  $= \frm{\sequence{\idxsubst{\EraseE{\expr_c}}}}{\EraseT{\idxsubst{\type_r}}}$%
  , by the induction hypothesis \\
  $= \frm{\sequence{\idxsubst{\EraseE{\expr_c}}}}{\idxsubst{\EraseT{\type_r}}}$%
  , by Lemma \ref{ITSubstErase} \\
  $= \idxsubst{\frm{\sequence{\EraseE{\expr_c}}}{\EraseT{\type_r}}}$ \\
  $= \idxsubst{\EraseE{\notedfrm[\type_r]{\sequence{\expr_c}}{\sequence{\nat}}}}$ \\
  $= \idxsubst{\EraseTrm{\notedfrm[\type_r]{\sequence{\expr_c}}{\sequence{\nat}}}}$ \\

  \paragraph{Application:}
  \[\term =
  \notedapp[\type_r]{\annotate
    [\typearray{\typefun{\sequence{\type_i}}{\type_o}}{\idx_f}]
    {\expr_f}}{\sequence{\expr_a}}\]
  $\EraseTrm{\idxsubst{\notedapp[\type_r]{\annotate
        [\typearray{\typefun{\sequence{\type_i}}{\type_o}}{\idx_f}]
        {\expr_f}}{\sequence{\expr_a}}}}$ \\
  $= \EraseE{\idxsubst{\notedapp[\type_r]{\annotate
        [\typearray{\typefun{\sequence{\type_i}}{\type_o}}{\idx_f}]
        {\expr_f}}{\sequence{\expr_a}}}}$ \\
  $= \EraseE{\notedapp[\idxsubst{\type_r}]{\annotate
      [\typearray{\typefun{\sequence{\idxsubst{\type_i}}}{\idxsubst{\type_o}}}{\idxsubst{\idx_f}}]
      {\idxsubst{\expr_f}}}{\sequence{\idxsubst{\expr_a}}}}$ \\
  $= \app{\EraseE{\idxsubst{\expr_f}}}{\sequence
    {\ttsexp{\EraseE{\idxsubst{\expr_a}}}{\EraseT{\idxsubst{\type_i}}}}\;\EraseT{\idxsubst{\type_r}}}$ \\
  $= \app{\idxsubst{\EraseE{\expr_f}}}{\sequence
    {\ttsexp{\idxsubst{\EraseE{\expr_a}}}{\EraseT{\idxsubst{\type_i}}}}\;\EraseT{\idxsubst{\type_r}}}$%
  , by the induction hypothesis \\
  $= \app{\idxsubst{\EraseE{\expr_f}}}{\sequence
    {\ttsexp{\idxsubst{\EraseE{\expr_a}}}{\idxsubst{\EraseT{\type_i}}}}\;\idxsubst{\EraseT{\type_r}}}$%
  , by Lemma \ref{TTSubstErase} \\
  $= \idxsubst{\app{\EraseE{\expr_f}}{\sequence
      {\ttsexp{\EraseE{\expr_a}}{\EraseT{\type_i}}}\;\EraseT{\type_r}}}$ \\
  $= \idxsubst{\EraseE{\notedapp[\type_r]{\annotate
        [\typearray{\typefun{\sequence{\type_i}}{\type_o}}{\idx_f}]
        {\expr_f}}{\sequence{\expr_a}}}}$ \\
  $= \idxsubst{\EraseTrm{\notedapp[\type_r]{\annotate
        [\typearray{\typefun{\sequence{\type_i}}{\type_o}}{\idx_f}]
        {\expr_f}}{\sequence{\expr_a}}}}$ \\

  \paragraph{Type application:}
  \[\term = \notedtapp[\type_r]{\expr_f}{\sequence{\type_a}}\]
  $\EraseTrm{\idxsubst{\notedtapp[\type_r]{\expr_f}{\sequence{\type_a}}}}$ \\
  $= \EraseE{\idxsubst{\notedtapp[\type_r]{\expr_f}{\sequence{\type_a}}}}$ \\
  $= \EraseE{\notedtapp[\idxsubst{\type_r}]{\idxsubst{\expr_f}}{\sequence{\idxsubst{\type_a}}}}$ \\
  $= \iapp{\EraseE{\idxsubst{\expr_f}}}{\sequence{\EraseT{\idxsubst{\type_a}}}\;\EraseT{\idxsubst{\type_r}}}$ \\
  $= \iapp{\idxsubst{\EraseE{\expr_f}}}{\sequence{\EraseT{\idxsubst{\type_a}}}\;\EraseT{\idxsubst{\type_r}}}$%
  , by the induction hypothesis \\
  $= \iapp{\idxsubst{\EraseE{\expr_f}}}{\sequence{\idxsubst{\EraseT{\type_a}}}\;\idxsubst{\EraseT{\type_r}}}$%
  , by Lemma \ref{TTSubstErase} \\
  $= \idxsubst{\iapp{\EraseE{\expr_f}}{\sequence{\EraseT{\type_a}}\;\EraseT{\type_r}}}$ \\
  $= \idxsubst{\EraseE{\notedtapp[\type_r]{\expr_f}{\sequence{\type_a}}}}$ \\
  $= \idxsubst{\EraseTrm{\notedtapp[\type_r]{\expr_f}{\sequence{\type_a}}}}$ \\

  \paragraph{Index application:}
  \[\term = \notediapp[\type_r]{\expr_f}{\sequence{\idx_a}}\]
  $\EraseTrm{\idxsubst{\notediapp[\type_r]{\expr_f}{\sequence{\idx_a}}}}$ \\
  $= \EraseE{\idxsubst{\notediapp[\type_r]{\expr_f}{\sequence{\idx_a}}}}$ \\
  $= \EraseE{\notediapp[\idxsubst{\type_r}]{\idxsubst{\expr_f}}{\sequence{\idxsubst{\idx_a}}}}$ \\
  $= \iapp{\EraseE{\idxsubst{\expr_f}}}{\sequence{\idxsubst{\idx_a}}\;\EraseT{\idxsubst{\type_r}}}$ \\
  $= \iapp{\idxsubst{\EraseE{\expr_f}}}{\sequence{\idxsubst{\idx_a}}\;\EraseT{\idxsubst{\type_r}}}$%
  , by the induction hypothesis \\
  $= \iapp{\idxsubst{\EraseE{\expr_f}}}{\sequence{\idxsubst{\idx_a}}\;\idxsubst{\EraseT{\type_r}}}$%
  , by Lemma \ref{TTSubstErase} \\
  $= \idxsubst{\iapp{\EraseE{\expr_f}}{\sequence{\idx_a}\;\EraseT{\type_r}}}$ \\
  $= \idxsubst{\EraseE{\notediapp[\type_r]{\expr_f}{\sequence{\idx_a}}}}$ \\
  $= \idxsubst{\EraseTrm{\notediapp[\type_r]{\expr_f}{\sequence{\idx_a}}}}$ \\

  \paragraph{Unboxing:}
  \[\term = \dproj{\sequence{\var_i}}{\var_e}{\expr_s}{\expr_b^{\type_b}}\text{, where }\var \not\in \sequence{\var_i}\]
  $\EraseTrm{\idxsubst{\dproj{\sequence{\var_i}}{\var_e}{\expr_s}{\expr_b^{\type_b}}}}$ \\
  $= \EraseE{\idxsubst{\dproj{\sequence{\var_i}}{\var_e}{\expr_s}{\expr_b^{\type_b}}}}$ \\
  $= \EraseE{\dproj{\sequence{\var_i}}{\var_e}{\idxsubst{\expr_s}}{\idxsubst{\expr_b}^{\idxsubst{\type_b}}}}$ \\
  $= \dproj{\sequence{\var_i}}{\var_e}{\EraseE{\idxsubst{\expr_s}}}
  {\EraseE{\idxsubst{\expr_b}} \; \EraseT{\idxsubst{\type_b}}}$ \\
  $= \dproj{\sequence{\var_i}}{\var_e}{\EraseE{\idxsubst{\expr_s}}}
  {\EraseE{\idxsubst{\expr_b}} \; \idxsubst{\EraseT{\type_b}}}$ \\
  $= \dproj{\sequence{\var_i}}{\var_e}{\idxsubst{\EraseE{\expr_s}}}
  {\idxsubst{\EraseE{\expr_b}} \; \idxsubst{\EraseT{\type_b}}}$ \\
  $= \idxsubst{\dproj{\sequence{\var_i}}{\var_e}{\EraseE{\expr_s}}{\EraseE{\expr_b} \; \EraseT{\type_b}}}$ \\
  $= \idxsubst{\EraseE{\dproj{\sequence{\var_i}}{\var_e}{\expr_s}{\expr_b^{\type_b}}}}$ \\
  $= \idxsubst{\EraseTrm{\dproj{\sequence{\var_i}}{\var_e}{\expr_s}{\expr_b^{\type_b}}}}$ \\

  \paragraph{Unboxing, with shadowed variable:}
  \[\term = \dproj{\sequence{\var_i}}{\var_e}{\expr_s}{\expr_b^{\type_b}}\text{, where }\var \in \sequence{\var_i}\]
  $\EraseTrm{\idxsubst{\dproj{\sequence{\var_i}}{\var_e}{\expr_s}{\expr_b^{\type_b}}}}$ \\
  $= \EraseE{\idxsubst{\dproj{\sequence{\var_i}}{\var_e}{\expr_s}{\expr_b^{\type_b}}}}$ \\
  $= \EraseE{\dproj{\sequence{\var_i}}{\var_e}{\idxsubst{\expr_s}}{\expr_b^{\type_b}}}$ \\
  $= \dproj{\sequence{\var_i}}{\var_e}{\EraseE{\idxsubst{\expr_s}}}{\EraseE{\expr_b} \; \EraseT{\type_b}}$ \\
  $= \dproj{\sequence{\var_i}}{\var_e}{\idxsubst{\EraseE{\expr_s}}}{\EraseE{\expr_b} \; \EraseT{\type_b}}$ \\
  $= \idxsubst{\dproj{\sequence{\var_i}}{\var_e}{\EraseE{\expr_s}}{\EraseE{\expr_b} \; \EraseT{\type_b}}}$ \\
  $= \idxsubst{\EraseE{\dproj{\sequence{\var_i}}{\var_e}{\expr_s}{\expr_b^{\type_b}}}}$ \\
  $= \idxsubst{\EraseTrm{\dproj{\sequence{\var_i}}{\var_e}{\expr_s}{\expr_b^{\type_b}}}}$ \\

\end{sproof}

\begin{lemma}[Values erase to values]
  \label{ValEraseToVal}
  For any well-typed term $\term$,
  \begin{itemize}
  \item If $\term$ has the form $\atval$, then $\EraseTrm{\term}$ has the form $\eatvalue$
  \item If $\term$ has the form $\val$, then $\EraseTrm{\term}$ has the form $\evalue$
  \end{itemize}
\end{lemma}
\begin{sproof}[We use induction on $\term$.
  The only nontrivial cases are {\tt box} and {\tt array} forms,
  which may be values or may contain incomplete computation.
  The contents of a {\tt box} value
  must itself be a value,
  which the induction hypothesis implies will erase to a value.
  Similarly, an {\tt array} value contains only atomic values,
  which erase to atomic values.]
  We use induction on $\term$.
  The result is trivial for all atom cases except for boxes,
  so we have only two cases left to consider.

  \paragraph{Box:}
  \[\term = \dsum{\sequence{\idx}}{\val}{\type}\]
  According to the grammar of Remora,
  $\val$ must have the form
  $\arrlit{\sequence{\atval}}{\sequence{\nat}}$ \\
  Then $\EraseTrm{\term}$
  $= \EraseA{\dsum{\sequence{\idx}}{\val}{\type}}$
  $= \dsum{\sequence{\idx}}{\EraseE{\val}}{}$.
  By the induction hypothesis,
  $\EraseE{\val}$ has the form $\evalue$,
  so $\EraseTrm{\term}$
  has the form $\dsum{\sequence{\idx}}{\evalue}{}$,
  which is a valid $\eatvalue$ form.

  \paragraph{Array literal:}
  \[\term = \arrlit{\sequence{\atval}}{\sequence{\nat}}\]
  The induction hypothesis implies that
  for each $\atval_i \in \sequence{\atval}$,
  $\EraseA{\atval_i}$ produces an erased atomic value $\eatvalue_i$.
  Therefore the erased term $\EraseTrm{\term}$
  $=$
  $\EraseE{\arrlit{\sequence{\atval}}{\sequence{\nat}}}$
  $=$
  {\tt (array ($\nat$ $\sequence{}$) $\EraseA{\atval}$ $\sequence{}$)}
  must have the form
  $\arrlit{\sequence{\eatvalue}}{\sequence{\nat}}$.
\end{sproof}

\begin{lemma}[Lockstep]
  \label{Lockstep}
  For any well-typed $\expr$,
  one of the following holds:
  \begin{itemize}
  \item{$\expr$ has the form $\val$, and $\EraseE{\expr}$ has the form $\evalue$}
  \item{$\expr \mapsto \expr'$, and $\EraseE{\expr} \mapsto \EraseE{\expr'}$}
  \item{$\expr \not \mapsto$, and $\EraseE{\expr} \not \mapsto$}
  \end{itemize}
\end{lemma}
\begin{sproof}[We prove this by induction on $\expr$.
  
  We rely on Lemma \ref{ErasureInContext} (erasure in context)
  when $\expr$ is a redex $\expr_r$ within an evaluation context $\ctxt$ other than $\hole$.
  If $\expr_r \not\mapsto$ because we have a mis-applied primitive operator,
  then the same is true for $\EraseE{\expr_r}$,
  so $\EraseE{\expr}$ is also an evaluation context around a mis-applied primitive operator.
  Otherwise, $\expr_r \mapsto \expr_r'$,
  and the induction hypothesis implies that $\EraseE{\expr_r} \mapsto \EraseE{\expr_r'}$.
  So $\EraseE{\expr} \mapsto \EraseE{\expr'}$,
  the erased context filled with $\expr_r'$.
  
  Values are handled by Lemma \ref{ValEraseToVal}.
  For the remaining cases---redexes---%
  straightforward symbol pushing shows that
  erased Remora's reduction rules follow those of Remora.]
  We prove this by induction on $\expr$.
  Note that if $\expr$ is not itself a redex or a value form,
  then the progress lemma implies that it must be
  an evaluation context filled with a redex.
  \paragraph{Value:}
  This case is exactly the expression case of Lemma \ref{ValEraseToVal}.

  \paragraph{Redex within nontrivial evaluation context:}
  \[\expr = \fillc{\ctxt}{\expr_r}\text{, where }\expr_r \mapsto \expr_r'\]
  Then $\expr \mapsto \expr' = \fillc{\ctxt}{\expr_r'}$.
  By Lemma \ref{ErasureInContext} (erasure in context),
  $\EraseE{\fillc{\ctxt}{\expr_r}} = \fillc{\EraseC{\ctxt}}{\EraseE{\expr_r}}$.
  The induction hypothesis implies that $\EraseE{\expr_r} \mapsto \EraseE{\expr_r'}$,
  so the full erased expression $\fillc{\EraseC{\ctxt}}{\EraseE{\expr_r}}$
  $ \mapsto \fillc{\EraseC{\ctxt}}{\EraseE{\expr_r'}}$.
  Erasure in context gives us
  $\fillc{\EraseC{\ctxt}}{\EraseE{\expr_r'}} = \EraseE{\fillc{\ctxt}{\expr_r'}}$.
  Therefore $\EraseE{\fillc{\ctxt}{\expr_r}} \mapsto \EraseE{\fillc{\ctxt}{\expr_r'}}$.

  \paragraph{Lift redex:}
  \newcommand{\liftfnannot}
  {\typearray
    {\typefun
      {\sequence{\typearray{\type_i}{\idxshape{\sequence{\nat_i}}}}}
      {\typearray{\type_o}{\idx_o}}}
    {\idxshape{\sequence{\nat_f}}}}
  \newcommand{\liftfn}
  {\notedarrlit[\liftfnannot]{\sequence{\atval_f}}{\sequence{\nat_f}}}
  \newcommand{\liftargannot}
  {\typearray
    {\type_i}
    {\idxshape{\sequence{\nat_a}\;\sequence{\nat_i}}}}
  \newcommand{\liftarg}
  {\notedarrlit[\liftargannot]{\sequence{\atval_a}}{\sequence{\nat_a}\,\sequence{\nat_i}}}
  \newcommand{\liftresultannot}
  {\typearray
    {\type_o}
    {\idxshape{\sequence{\nat_p}\;\sequence{\nat_o}}}}
  \newcommand{\liftedfnannot}
  {\typearray
    {\typefun
      {\sequence{\typearray{\type_i}{\idxshape{\sequence{\nat_i}}}}}
      {\typearray{\type_o}{\idx_o}}}
    {\idxshape{\sequence{\nat_p}}}}
  \newcommand{\liftedfncells}
  {\mathit{Concat}\llb
    \mathit{Rep}_{\nat_{\mathit{fe}}}\llb
    \mathit{Split}_1\llb
    \sequence{\atval_f}\rrb\rrb\rrb}
  \newcommand{\liftedfn}
  {\notedarrlit[\liftedfnannot]
    {\liftedfncells}
    {\sequence{\nat_p}}}
  \newcommand{\liftedargannot}
  {\typearray
    {\type_i}
    {\idxshape{\sequence{\nat_p}\;\sequence{\nat_i}}}}
  \newcommand{\liftedarg}
  {\notedarrlit[\liftedargannot]
    {\mathit{Concat}\llb
      \mathit{Rep}_{\nat_{\mathit{ae}}}\llb
      \mathit{Split}_{\nat_{\mathit{ac}}}\llb
      \sequence{\atval_a}\rrb\rrb\rrb}
    {\sequence{\nat_p}\,\sequence{\nat_i}}}
  \newcommand{\erasedliftfn}
  {\arrlit{\sequence{\EraseA{\atval_f}}}{\sequence{\nat_f}}}
  \newcommand{\erasedliftarg}
  {\arrlit{\sequence{\EraseA{\atval_a}}}{\sequence{\nat_a}\,\sequence{\nat_i}}}
  \newcommand{\erasedliftcell}{\idxshape{\sequence{\nat_i}}}
  \newcommand{\erasedliftresult}{\idxshape{\sequence{\nat_p}\;\sequence{\nat_o}}}
  \newcommand{\erasedliftedfn}
  {\arrlit
    {\mathit{Concat}\llb
      \mathit{Rep}_{\nat_{\mathit{fe}}}\llb
      \mathit{Split}_1\llb
      \sequence{\EraseA{\atval_f}}\rrb\rrb\rrb}
    {\sequence{\nat_p}}}
  \newcommand{\erasedliftedarg}
  {\arrlit
    {\mathit{Concat}\llb
      \mathit{Rep}_{\nat_{\mathit{ae}}}\llb
      \mathit{Split}_{\nat_{\mathit{ac}}}\llb
      \sequence{\EraseA{\atval_a}}\rrb\rrb\rrb}
    {\sequence{\nat_p}\,\sequence{\nat_i}}}
\begin{alltt}
\(\expr = \)(\(\liftfn\)\\
\phantom{\(\expr = \) }\(\liftarg\)\\
\phantom{\(\expr = \) }\(\cdots\))\(\sp{\liftresultannot}\)\\
\(\mapsto\sb{\mathit{lift}}\)\\
\(\expr' = \)((array ()\\
\phantom{\(\expr' = \) \ }\(\liftedfncells\))\(\sp{\liftedfnannot}\)\\
\phantom{\(\expr' = \) }\(\liftedarg\)\\
\phantom{\(\expr' = \) }\(\cdots\))\(\sp{\liftresultannot}\)
\end{alltt}
Then
\begin{alltt}
\(\EraseE{\expr} = \)(\(\erasedliftfn\)\\
\phantom{\(\EraseE{\expr} = \) }(\(\erasedliftarg\) \(\erasedliftcell\))\\
\phantom{\(\EraseE{\expr} = \) }\(\cdots\) \(\erasedliftresult\))
\end{alltt}
This is a $\mathit{lift}$ redex in Erased Remora, and it steps to
\begin{alltt}
\(\EraseE{\expr'} = \)(\(\erasedliftedfn\)\\
\phantom{\(\EraseE{\expr'} = \) }(\(\erasedliftedarg\) \(\erasedliftcell\))\\
\phantom{\(\EraseE{\expr'} = \) }\(\cdots\) \(\erasedliftresult\))
\end{alltt}
That is, $\EraseE{\expr} \mapsto_{\mathit{lift}} \EraseE{\expr'}$.

  \paragraph{Map redex:}
  \newcommand{\mapfn}{\liftfn}
  \newcommand{\mapargannot}
  {\typearray
    {\type_i}
    {\idxshape{\sequence{\nat_f}\;\sequence{\nat_i}}}}
  \newcommand{\maparg}
  {\notedarrlit[\mapargannot]{\sequence{\atval_a}}{\sequence{\nat_f}\,\sequence{\nat_i}}}
  \newcommand{\mapresultannot}
  {\typearray
    {\type_o}
    {\idxshape{\sequence{\nat_f}\;\sequence{\nat_o}}}}
  \newcommand{\mapfncell}
  {\notedarrlit[{\typearray
      {\typefun
        {\sequence{\typearray{\type_i}{\idxshape{\sequence{\nat_i}}}}}
        {\typearray{\type_o}{\idx_o}}}
      {\idxshape{}}}]
    {\atval_f}{}}
  \newcommand{\mapargcell}
  {\notedarrlit[{\typearray{\type_i}{\idxshape{\sequence{\nat_i}}}}]
    {\sequence{\atval_c}}{}}
  \newcommand{\mappedargannot}
  {\typearray
    {\type_i}
    {\idxshape{\sequence{\nat_i}}}}
  \newcommand{\resultcellannot}
  {\typearray
    {\type_o}
    {\idxshape{\sequence{\nat_o}}}}
  \newcommand{\erasedmapfn}{\erasedliftfn}
  \newcommand{\erasedmaparg}
  {\arrlit{\sequence{\EraseA{\atval_a}}}{\sequence{\nat_f}\,\sequence{\nat_i}}}
  \newcommand{\erasedmapcell}{\idxshape{\sequence{\nat_i}}}
  \newcommand{\erasedmapresult}{\erasedliftresult}
  \newcommand{\erasedmapfncell}{\arrlit{\EraseA{\atval_f}}{}}
  \newcommand{\erasedmapargcell}{\arrlit{\sequence{\EraseA{\atval_c}}}{}}
\begin{alltt}
\(\expr = \)(\(\mapfn\)\\
\phantom{\(\expr = \) }\(\maparg\)\\
\phantom{\(\expr = \) }\(\cdots\))\(\sp{\mapresultannot}\)\\
\(\mapsto\sb{\mathit{map}}\)\\
\(\expr' = \)(frame (\(\sequence{\nat\sb{f}}\))\\
\phantom{\(\expr' = \) }(\(\mapfncell\)\\
\phantom{\(\expr' = \) \ }\(\mapargcell\)\\
\phantom{\(\expr' = \) \ }\(\cdots\))\(\sp{\resultcellannot}\)\\
\phantom{\(\expr' = \) }\(\cdots\))\(\sp{\mapresultannot}\)
\end{alltt}
The argument cells' atoms
$\parens{\sequence{\parens{\sequence{\atval_c}}}}$
are given by
$\mathit{Transpose} \llb \sequence{\mathit{Split}_{\nat_c} \llb \sequence{\atval_a} \rrb} \rrb$,
where
each $\nat_c$ is computed as the product of
the corresponding position's expected argument dimensions $\sequence{\nat_i}$,
as in Figure \ref{fig:DynamicSemantics}.
We then consider the erased form of $\expr$:
\begin{alltt}
\(\EraseE{\expr} = \)(\(\erasedmapfn\)
         (\(\erasedmaparg\) \(\erasedmapcell\))
         \(\cdots\) \(\erasedliftresult\))
\end{alltt}
This is a $\mathit{map}$ redex in Erased Remora.
The argument atoms
$\parens{\sequence{\parens{\sequence{\EraseA{\atval_a}}}}}$
are split up into cells in the same way, with
$\parens{\sequence{\parens{\sequence{\eatvalue_c}}}}
= \parens{\sequence{\parens{\sequence{\EraseA{\atval_c}}}}}$
as the result of
$\mathit{Transpose} \llb \sequence{\mathit{Split}_{\nat_c} \llb \sequence{\EraseA{\atval_a}} \rrb} \rrb$.
So $\EraseE{\expr}$ steps to
\begin{samepage}
\begin{alltt}
\(\EraseE{\expr'} = \)(frame (\(\sequence{\nat\sb{f}}\))\\
\phantom{\(\EraseE{\expr'} = \) }(\(\erasedmapfncell\)\\
\phantom{\(\EraseE{\expr'} = \) \ }(\(\erasedmapargcell\) \(\erasedmapcell\))\\
\phantom{\(\EraseE{\expr'} = \) \ }\(\cdots\))\\
\phantom{\(\EraseE{\expr'} = \) }\(\cdots\))
\end{alltt}
\end{samepage}

  \paragraph{Beta redex:}
  Note that $\type_I$ here must be the array type
  $\typearray{\type_i}{\idxshape{\sequence{\nat_a}}}$.
  \newcommand{\betafnannot}
  {\typearray
    {\typefun
      {\sequence{\type_I}}
      {\type_O}}
    {\idxshape{}}}
  \newcommand{\betafn}
  {\notedarrlit[\betafnannot]
    {\lam{\sequence{\notevar{\var}{\type_i}}}{\expr_b}}
    {}}
  \newcommand{\erasedbetafn}
  {\arrlit
    {\lam{\sequence{\var}}{\EraseE{\expr_b}}}
    {}}
  \newcommand{\betaarg}{\notedarrlit[\type_I]{\sequence{\atval_a}}{\sequence{\nat_a}}}
  \newcommand{\erasedbetaarg}{\arrlit{\sequence{\EraseA{\atval_a}}}{\sequence{\nat_a}}}
  \newcommand{\erasedbetacell}{\idxshape{\sequence{\nat_a}}}
\begin{alltt}
\(\expr = \)(\(\betafn\)\\
\phantom{\(\expr = \) }\(\betaarg\)\\
\phantom{\(\expr = \) }\(\sequence{}\))\(\sp{\type\sb{O}}\)\\
\(\mapsto\sb{\beta}\)\\
\(\expr' = \seqsubst{\expr\sb{b}}{\var}{\betaarg}\)
\end{alltt}
Then
\begin{alltt}
\(\EraseE{\expr} = \)(\(\erasedbetafn\)\\
\phantom{\(\EraseE{\expr} = \) }(\(\erasedbetaarg\) \(\erasedbetacell\))\\
\phantom{\(\EraseE{\expr} = \) }\(\sequence{}\))\(\sp{\type\sb{O}}\)\\
\(\mapsto\sb{\beta}\) \(\seqsubst{\EraseE{\expr\sb{b}}}{\var}{\erasedbetaarg}\)
\end{alltt}
By Lemma \ref{EESubstErase} (substitution commutes with erasure),
this result term is equal to $\EraseE{\expr'}$.


  \paragraph{I-Beta redex:}
  \newcommand{\ibetafnannot}
  {\typearray
    {\typedprod
      {\sequence{\notevar{\var}{\sort}}}
      {\type_b}}
    {\idxshape{\sequence{\nat_f}}}}
  \newcommand{\ibetafn}
  {\notedarrlit[\ibetafnannot]
    {\sequence{\ilam{\notevar{\var}{\sort}}{\expr_b}}}
    {\sequence{\nat_f}}}
  \newcommand{\ibetaresult}
  {\notedfrm[\type_R]
    {\sequence{\seqsubst{\expr_b}{\var}{\idx_a}}}
    {\sequence{\nat_f}}}
  \newcommand{\erasedibetafn}
  {\arrlit
    {\sequence{\ilam{\var}{\EraseE{\expr_b}}}}
    {\sequence{\nat_f}}}
  \newcommand{\erasedibetaresult}
  {\notedfrm[\type_R]
    {\sequence{\seqsubst{\EraseE{\expr_b}}{\var}{\idx_a}}}
    {\sequence{\nat_f}}}
\begin{alltt}
\(\expr = \)(i-app \(\ibetafn\)\\
\phantom{\(\expr = \) }\(\idx\sb{a}\) ...)\(\sp{\type\sb{R}}\)\\
\(\mapsto\sb{i\beta}\)\\
\(\expr' = \ibetaresult\)
\end{alltt}
Then
\begin{alltt}
\(\EraseE{\expr} = \)(i-app \(\erasedibetafn\) \(\idx\sb{a}\) ... \(\EraseT{\type\sb{R}}\))\\
\(\mapsto\sb{i\beta} \erasedibetaresult = \EraseE{\expr'}\)
\end{alltt}

  \paragraph{T-Beta redex:}
  \newcommand{\tbetafnannot}
  {\typearray
    {\typeuniv
      {\sequence{\notevar{\var}{\kind}}}
      {\type_b}}
    {\idxshape{\sequence{\nat_f}}}}
  \newcommand{\tbetafn}
  {\notedarrlit[\tbetafnannot]
    {\sequence{\tlam{\notevar{\var}{\kind}}{\expr_b}}}
    {\sequence{\nat_f}}}
  \newcommand{\tbetaresult}
  {\notedfrm[\type_R]
    {\sequence{\seqsubst{\expr_b}{\var}{\type_a}}}
    {\sequence{\nat_f}}}
  \newcommand{\erasedtbetafn}
  {\arrlit
    {\sequence{\ilam{\var}{\EraseE{\expr_b}}}}
    {\sequence{\nat_f}}}
  \newcommand{\erasedtbetaresult}
  {\notedfrm[\type_R]
    {\sequence{\seqsubst{\EraseE{\expr_b}}{\var}{\type_a}}}
    {\sequence{\nat_f}}}
\begin{alltt}
\(\expr = \)(t-app (array (\(\sequence{\nat\sb{f}}\))\\
\phantom{\(\expr = \)(t-app (array }\(\sequence{\tlam{\notevar{\var}{\kind}}{\expr\sb{b}}}\))\(\sp{\tbetafnannot}\)\\
\phantom{\(\expr = \) }\(\type\sb{a}\) ...)\(\sp{\type\sb{R}}\)\\
\(\mapsto\sb{t\beta}\)\\
\(\expr' = \tbetaresult\)
\end{alltt}
Recall that type abstraction and application erase to index abstraction and application.
So in the erased language, we have:
\begin{alltt}
\(\EraseE{\expr} = \)(i-app \(\erasedtbetafn\) \(\type\sb{a}\) ... \(\EraseT{\type\sb{R}}\))\\
\(\mapsto\sb{i\beta} \erasedtbetaresult = \EraseE{\expr'}\)
\end{alltt}

  \paragraph{Unbox redex:}
  \newcommand{\unboxvars}{\sequence{\var_i}\text{ }\var_e}
  \newcommand{\unboxval}{\annotate
    {\arrlit{\sequence{\dsum{\sequence{\idx_s}}{\val_s}{\type_s}}}{\sequence{\nat_s}}}}
  \newcommand{\unboxbody}{\annotate[\type_B]{\expr_b}}
  \newcommand{\unboxresult}
  {\notedfrm[\type_R]
    {\sequence{\multisubst{\expr_b}{\sequence{\singlesubst{\var_i}{\idx_s},},
          \singlesubst{\var_e}{\val_s}}}}
    {\sequence{\nat_s}}}
  \newcommand{\erasedunboxval}
  {\arrlit{\sequence{\dsum{\sequence{\idx_s}}{\EraseE{\val_s}}{}}}{\sequence{\nat_s}}}
  \newcommand{\erasedunboxresult}
  {\frm
    {\sequence{\multisubst{\EraseE{\expr_b}}{\sequence{\singlesubst{\var_i}{\idx_s},},
          \singlesubst{\var_e}{\EraseE{\val_s}}}}}
    {\idxappend{\idxshape{\sequence{\nat_s}} \; \EraseT{\type_B}}}}
  \newcommand{\rewrittenunboxresult}
  {\frm
    {\sequence{\EraseE{\multisubst{\expr_b}{\sequence{\singlesubst{\var_i}{\idx_s},},
            \singlesubst{\var_e}{\val_s}}}}}
    {\idxappend{\idxshape{\sequence{\nat_s}} \; \EraseT{\type_B}}}}
\begin{alltt}
\(\expr = \)(unbox (\(\unboxvars\) \(\unboxval\)) \(\unboxbody\))\(\sp{\type\sb{R}}\)\\
\(\mapsto\sb{\mathit{unbox}}\)\\
\(\expr' = \unboxresult\)
\end{alltt}
Then relying on our earlier result that erasure commutes with substitution,
we can step the erased version of $\expr$ to get the erased version of $\expr'$:
\begin{alltt}
\(\EraseE{\expr} = \) (unbox (\(\unboxvars\) \(\erasedunboxval\)) \(\EraseE{\expr\sb{b}}\) \(\EraseT{\type\sb{B}}\))\\
\(\mapsto\sb{\mathit{unbox}} \erasedunboxresult\)\\
\( = \rewrittenunboxresult = \EraseE{\expr'}\)
\end{alltt}

\paragraph{Mis-applied primitive operator:}
The remaining case is that $\expr \not\mapsto$, but $\expr$ is not a value.
Since $\expr$ is well-typed, Lemma \ref{Progress} (Progress) implies that
$\expr$ must be a mis-application of a primitive operator.
That is, $\expr$ has the form
$\ctxt\left[\misapp\right]$,
where $\Sigref{\primop}$ is
{\tt (-> ($\type_I$ $\sequence{}$) $\type_O$)},
and the types of $\sequence{\val}$ are $\sequence{\type_i}$.
Using Lemma \ref{ErasureInContext} (Erasure in context),
$\EraseE{\expr}$ is
$\EraseC{\ctxt}[${\tt ($\primop$ ($\EraseE{\val}$ $\EraseT{\type_I}$) $\sequence{}$ $\EraseT{\type_r}$)}$]$.
Since each $\val$ has shape $\EraseT{\type_I}$,
we still have function application with a scalar frame,
and the values given as arguments are still out-of-domain for $\primop$.
\end{sproof}

Since we have a deterministic operational semantics for
both explicitly typed Remora and type-erased Remora,
the lockstep lemma also works in reverse.
If an erased term takes an evaluation step,
its preimage cannot be a value form or stuck state.
The preimage must therefore step to some result expression,
which itself erases to the same result.
Similarly, a value form or stuck state in erased Remora
cannot have a preimage which takes an evaluation step.

\begin{corollary}[Reverse lockstep]
If $\EraseE{\expr} \mapsto \EraseE{\expr'}$,
then for any $\expr''$ such that $\expr \mapsto \expr''$, we have $\expr' \EraseEqv \expr''$,
and at least one such $\expr''$ exists.
If $\EraseE{\expr} \not\mapsto$, then $\expr \not\mapsto$.
\end{corollary}

Recall our relation $\EraseEqv$ on the set of machine states $S = \wt{\Expr} \uplus \erased{\wt{\Expr}}$,
where $\wt{\Expr} = \{e \in \Expr | \typeof{\emptyEnv}{\emptyEnv}{\emptyEnv}{\expr}{\type}\}$,
\ie the set of well-typed explicitly-typed terms,
and $\erased{\wt{\Expr}}$ is the image of $\wt{\Expr}$ under type erasure.
$\EraseEqv$ is the equivalence closure of the relation given by the erasure function $\EraseE{\cdot}$.
That is, $\EraseEqv$ is the least relation which
relates two states $s$ and $w$ iff any of the following hold:
\begin{enumerate}
\item{$s \in \wt{Expr}$ and $\EraseE{s} = w$ (erasure proper)}
\item{$s =_\alpha w$ (reflexivity)}
\item{$w \EraseEqv s$ (symmetry)}
\item{$s \EraseEqv s'$ and $s' \EraseEqv w$ (transitivity)}
\end{enumerate}
Expanding the erasure relation based on $\EraseE{\cdot}$
to include both symmetry and transitivity
relates any two explicitly typed expressions
which produce $\alpha$-equivalent erased terms.
A $\EraseEqv$ equivalence class consists of a single erased Remora expression
and all of its preimages.
There can be only one erased Remora expression because
type erasure is a well-defined function
(\ie, no single explicitly typed expression
can erase to multiple different results).
Formally, every $\EraseEqv$ equivalence class must have the form
$$\{\eexpr\} \uplus \{\expr \in \Expr \, | \, \EraseE{\expr} = \eexpr\}$$
\begin{theorem}
  \label{Bisimulation}
  $\EraseEqv$ is a bisimulation.
  That is, for any states $s, w \in S$ if $s \EraseEqv w$,
  either $(s \mapsto u \wedge w \mapsto v \wedge u \EraseEqv v)$
  or $(s \not\mapsto \wedge w \not\mapsto)$.
\end{theorem}
\begin{sproof}[\BODY]
  There are four cases to consider,
  depending on which of $\Expr$ or $\erased{\Expr}$ each related term is drawn from,
  but we can merge the two cases where $s$ and $w$ are drawn from different languages.

  \paragraph{$s \in \erased{\Expr}$ and $w \in \Expr$, or vice versa:}
  Then $s$ is the sole type-erased expression in its equivalence class,
  and $\EraseE{w} = s$
  (or vice versa).
  Our proof obligation is exactly the lockstep lemma (Lemma \ref{Lockstep}).

  \paragraph{$s, w \in \erased{\Expr}$:}
  Since each equivalence class contains only one type-erased expression, $s = w$.
  They must therefore have the same reduction behavior.

  \paragraph{$s, w \in \Expr$:}
  If $s \not\mapsto$, then the lockstep lemma implies $\EraseE{s} = \EraseE{w} \not\mapsto$.
  Then by reverse lockstep, $w \not\mapsto$ as well.
  On the other hand, if $s \mapsto s'$, then $\EraseE{s} = \EraseE{w} \mapsto \EraseE{s'}$.
  Lockstep implies $\EraseE{w} \mapsto \EraseE{w'}$.
  Since Erased Remora has deterministic operational semantics,
  $\EraseE{s'} = \EraseE{w'}$
  (they are both the result of taking an evaluation step from the same expression).
  Therefore, all of their preimages, including $s'$ and $w'$ are erasure-equivalent,
  \ie, $s' \EraseEqv w'$.
\end{sproof}


\section{Related Work}
\label{sec:RelatedWork}

Rank polymorphism originally appeared in APL \cite{APL},
which Iverson designed as a form of mathematical notation,
with the APL interpreter serving to eliminate the semantic ambiguity
found in conventional notation \cite{IversonTuringLecture}.
At first, APL only lifted \emph{scalar} functions
to operate on aggregate data via pointwise application,
either on a scalar argument and an aggregate argument
or on two aggregates of identical shape.
Subsequent development introduced the notion of function rank,
the number of dimensions a function expects its argument to have.
This generalized the scalar function lifting,
\eg, allowing a vector-mean function to produce a vector of results
when given a matrix argument,
and it introduced the ``frame of cells'' view of aggregate arguments
where pointwise lifting generalizes to cellwise lifting.
The next generalization step was to loosen the rule
on frame shape compatibility.
In J \cite{J}, which Iverson created as a successor to APL,
a function can be applied to two arguments of differing frames
as long as one frame---%
viewed as a sequence of dimensions---%
is a prefix of the other.
This was a conscious design decision on Iverson's part:
prefix agreement was chosen over suffix agreement because
it fit better with APL's emphasis on operating along arrays' major axes \cite{RankAndUniformity}.

FISh also made implicit aggregate lifting part of the semantics of function application,
and its static semantics resolves the shapes of all arrays \cite{FISh}.
A conventional type judgment describes
the elements of arrays computed by a FISh program,
and a second judgment ascribes a shape to each array.
In FISh's metatheory,
the shape of a function is a function on shapes.
Thus a function application's shape can be calculated statically
by applying the shape function to the arguments' shapes.

As elegant as this model is in a first-order language,
it is incompatible with first-class functions.
When functions appear in arrays which are themselves applied to arguments,
the shape must describe
the layout of of that collection of functions,
not just the functions' own behavior.
In Remora, a function which checks
whether a point in $\mathbb{R}^3$ is near the origin
might have the type
{\tt (-> ((Arr Float (Shp 3))) (Arr Bool (Shp)))}.
FISh considers this function's shape to be
the function on a singleton domain
(containing only $[3]$)
which returns the empty vector.
However, Remora expressions produce array data,
even in function position.
A function array {\tt near-origin?} containing only that function
has its own ``first-order'' shape {\tt (Shp)},
independent of any function on shapes summarizing its behavior.

In resolving all shapes statically,
FISh is also too restrictive to permit shapes determined from run-time data.
Functions like {\tt iota} and {\tt filter} cannot exist,
nor can the ragged data which would result from lifting them.
By characterizing functions with restricted dependent types,
Remora escapes both of these limitations.



ZPL is a data-parallel language
which was designed to live within a larger language
with a more general parallelism mode \cite{ZPLArraySublang}.
Its programming model is based on
an explicit map operation
over programmer-specified index space within an array.
Several built-in operators modify an index space,
such as shifting a section along some dimension,
adding a new dimension by broadcasting,
or slicing out a particular sub-array.
The set of built in operators is constructed
to make communication cost implications clear to the programmer \cite{ZPLPerformance}.

NESL uses a nested-vector data model,
rather than the rank-polymorphic regular-array model \cite{NESL}.
Programs map operations over vectors using a comprehension notation.
Since a vector's elements may themselves be vectors
with widely varying lengths,
NESL's main performance trick is
turning irregular nested vectors into a flat internal representation.
After flattening, NESL can evenly distribute the computation workload
by splitting at places that user code doe not consider not sub-vector boundaries.



The goal of isolating the cell-level portion of a program
has also led to some domain-specific languages,
targeting specific varieties of regular aggregate data.
Halide is a language designed for writing image processing pipelines \cite{Halide}.
The iteration space is the set of pixels in an image.
A Halide programmer writes code describing
how an individual pixel should be handled
(or a cluster of pixels for stencil computation),
and then in a separate portion of the program,
the programmer describes how the iteration space ought to be traversed.
Since the program does not intermingle
loop-nest control code with loop-body computation code,
it is easier for a maintainer to tune for performance
by adjusting the iteration schedule.

Diderot is a programming language
specifically aimed at processing medical images \cite{Diderot}.
The universe of data consists of tensor fields,
intended as functions on the continuous domain $\mathbb{R}^n$,
rather than any particular discrete index space.
Diderot offers pointwise arithmetic on tensors
and a collection of common operations such as outer product and transposition.
Aggregate lifting appears again,
albeit in a smaller form,
with some operations on tensor fields.

HorseIR uses implicit lifting over arrays
as an intermediate representation for SQL queries \cite{HorseIR}.
HorseIR is lower-level than APL itself,
serving as a three-address IR with vector instructions,
and the IR itself is designed to ease loop fusion.
In contrast with APL, all HorseIR arrays are vectors---%
there are no higher-rank arrays, and scalars are represented as unit-length vectors.
There is also a list datatype for handling heterogeneous aggregate data,
such as one row of a database table.


Our type system's use of restricted dependent types
is inspired by Dependent ML \cite{DependentML}.
While Dependent ML is designed with the expectation
that the index language has a fully decidable theory,
Remora's index language does not \cite{Durnev}.
In subsequent work,
Dependent ML also focused on ensuring safety of array index accesses,
using singleton and range types to ensure
that numbers used for indexing fell within arrays' bounds \cite{DMLBounds}.

Others have applied established type system machinery
to an APL-like computation model.
Thatte's coercion semantics \cite{ThatteCoercion}
uses a form subtyping in which
scalar types are subtypes of aggregates,
and aggregate types in certain situations are subtypes of higher-dimensional aggregates.
The subtyping rules emit coercions which invoke functions such as {\tt map} and {\tt replicate},
automatically adapting scalar functions to aggregate data.
However, the restrictions on treating aggregate types as subtypes of larger aggregates
prohibit lifting for functions which expect non-scalar input
(\ie, those whose expected rank is greater than zero)
and lifting to unequal frame shapes
(\eg, vector-matrix addition).
Gibbons's embedding of in an extended Haskell \cite{APLicative}
extensively uses type-level programming
and defines much of the rank-polymorphic lifting machinery
in terms of transposition.

Single Assignment C (SAC) is a variant of C without mutation \cite{SAC}.
The primary iteration mechanism in SAC is the {\tt with} loop,
in which the programmer describes a traversal of the index space
and builds an output array an element at a time.
The lack of mutation pushes the programmer
to avoid writing loop-carried dependence,
much like in the rank-polymorphic model.
Translation of well-behaved APL programs to SAC
tends to be straightforward \cite{APLSAC},
though the resulting loop structure is
somewhat specific to the function being lifted.
A SAC variant called Qube introduces a Dependent ML-style type system
to ensure the safety of indexed array element accesses \cite{Qube}.
Qube retains SAC's {\tt with} loop-based programming model---%
relying on range types in contrast to
APL's and Remora's implicitly lifted function application semantics---%
and due to its C roots,
it does not support first-class functions
to the extent of permitting arrays to appear in function position.

Since the syntactic structure of a line of APL code,
in particular the meaning of the juxtaposition of two terms,
depends on the meanings of names which appear in the code,
standard APL does not admit a fully static parsing algorithm.
Due to this and other idiosyncratic warts,
past efforts to compile APL have typically targetted
large but well-behaved subsets of the language.

A prominent exception is the APL\textbackslash3000 compiler,
which could produce machine code for individual statements
and used interpretation to manage inter-statement control flow \cite{APL3000Compiler}.
This allowed some intraprocedural optimization,
such as fold-unfold and map-map fusion%
\footnote{In the APL community,
  these transformations are respectively referred to as
``beating'' (imagine a metronome producing a stream of beats)
and ``dragging along'' (amassing a chain of
delayed scalar operations to apply in a single loop body).}
\cite{AbramsAPLMachine},
analogous to deforestation
which arose in mainstream functional languages \cite{BurstallDarlington, WadlerListless}.
It also ensured that names' dynamic meanings would be available
by the time execution reached any statement which used them.

Weiss and Saal instead applied interprocedural data-flow analysis
to resolve the syntactic classes of variable names in APL code \cite{ParsingAPL}.
This analysis is not complete for APL itself
due to the possibility that reassignment of a variable name
will change how some line of code parses.
However, the authors found that
real-world code did not make use of
the full freedom to manipulate
the syntactic structure of an APL program
by dynamically reassigning variables,
suggesting that this is a misfeature which can be discarded for little cost.
However, the common line-at-a-time compilation style
used in other work
also served to ensure that
by the time execution reaches a line of code,
its variable names, and thus their syntactic classes, could be resolved.

Budd describes a compiler for an APL variant
in which the ambiguity in parsing is avoided by
declaring identifiers before use
and name resolution is simplified by adopting lexical scope.
\cite{BuddCompiler}.

APEX parallelizes an APL dialect
with restrictions on many APL features deemed
incompatible with compilation
(such as producing a string representation of an arbitrary function)
or not strictly necessary for practical use (such as {\tt GOTO}) \cite{Apex}.
APEX's shape compatibility rules for implicit aggregate lifting
are less permissive than APL,
\eg, prohibiting most cases of vector-matrix addition.
Instead, both arguments must have the same rank
or at least one argument passed to a scalar-consuming function
be a scalar or singleton vector.

A more recent line of work has focused on
intermediate representations of array programs,
such as the Typed Array Intermediate Language,
which makes aggregate operations explicit
using an {\tt each} primitive operator \cite{APLTAIL}.
The type system tracks arrays' ranks,
which provides enough information to recognize
when an {\tt each} call is needed,
but it is not meant to ensure that
lifting a function application has a well-defined result
(\ie, shape incompatibility is still possible).
$\mathcal{L}_0$ plots out a loop fusion strategy
inspired by control flow graph reduction \cite{T2GraphReduction}.
The $\mathcal{L}_0$ compiler splits array programs
into kernels of fused aggregate operations
based on the applicability of $T_2$ graph reductions
(merging nodes $X$ and $Y$ if $X$ is $Y$'s sole predecessor)
to the program's data flow graph.


\section{Conclusion}

While APL and its descendant languages have attracted a devoted userbase,
there has been little cross talk between
the array-language community and
the broader programming language research community.
Much of the analysis opportunity taken for granted by lambda calculists
has been unavailable in APL---%
despite the many implementations,
such languages lacked formal semantics amenable to proofs.
Meanwhile, implementations of rank-polymorphic languages
have struggled with compilation due to control structure that is ``too dynamic,''
depending on run-time data to determine the structure of a loop nest.

Developing Remora's semantics kills two birds with one stone:
Formally stated reduction rules describe
the results expected from the implicit, data-driven control structure,
and typing rules give enough information about array shapes
to identify that control structure statically.
This necessarily entails recognizing programs
which cannot have such a control structure
due to incompatible array shapes---%
this is not what we set out to do
but a benefit realized by pursuing a larger goal.
By casting the aggregate lifting as an extension to
$\lambda$-calculus's function application semantics,
we escape from APL's limitation on function arity
and can treat functions as first-class values.

There are two things to look for in evaluating a type system.
We want to know the conclusions drawn by type checking reflect reality,
which is handled by a conventional type soundness theorem.
We also want to be sure that the types convey the information we seek.
In Remora's case, that information is
the iteration structure implicit in each function application.
The typing rule for function application
(and similarly, the rules for type and index application)
produces a static characterization of that implicit iteration structure.

Our type erasure can be seen as a compilation pass
which moves the decision about how to break arguments into cells
from the function's type into the application term.
While our primary purpose in presenting type erasure
is to point out type-level information which is not truly needed at run time,
it also serves to make the control structure one step more explicit.
Arguments' full shapes---available both at erasure time
and later by inspecting argument terms---%
suffice to determine the arguments' frames,
the last puzzle piece needed to turn
Remora's implicit iteration structure into
explicit calls to {\tt map} and {\tt replicate} functions.
Our intention for future work is to
demonstrate use of that shape information for
fully static compilation of Remora code.
\bibliography{paper}

\end{document}